\documentclass[12pt]{article}

\usepackage{epsfig,color,amsmath,amssymb,euscript}
\usepackage{mathabx}
\usepackage{subfigure}
\usepackage{fancyhdr}
\usepackage{natbib}
\usepackage{rotating}
\usepackage{amsmath}
\usepackage{amsfonts}
\usepackage{setspace}
\usepackage{fancyhdr}
\usepackage{lscape}
\usepackage{amsthm}
\usepackage{enumerate} 
\usepackage{threeparttable}
\usepackage{diagbox}
\usepackage{multirow}
\usepackage{makecell}
\usepackage{xcolor}
\usepackage{colortbl}
\usepackage[colorlinks,linkcolor = blue,anchorcolor=blue, citecolor= blue]{hyperref}
\usepackage{url} % not crucial - just used below for the URL 

\usepackage{graphicx}
\usepackage{enumerate}
\usepackage{mathrsfs}
\usepackage{verbatim}
\usepackage{blkarray}
\usepackage{graphicx}

\usepackage{authblk}
\usepackage{setspace}
\usepackage{threeparttable}
\usepackage{caption}
\usepackage{algorithm,algcompatible}
\usepackage{pdflscape}

\usepackage[resetlabels]{multibib}
\allowdisplaybreaks
\newcites{S}{References} 

\algnewcommand\INPUT{\item[\textbf{Input:}]}%
\algnewcommand\OUTPUT{\item[\textbf{Output:}]}%

\def\T{{\mathrm{\scriptscriptstyle \top} }}

\theoremstyle{definition}

\newtheorem{remark}{Remark}

\newenvironment{acks}{
	\section*{Acknowledgements}
}{}

\def \P{{\mathbb{P}}}

\newcommand{\bA}{{\mathbf A}}
\newcommand{\bB}{{\mathbf B}}
\newcommand{\bC}{{\mathbf C}}
\newcommand{\bX}{{\mathbf X}}
\newcommand{\bY}{{\mathbf Y}}
\newcommand{\bZ}{{\mathbf Z}}
\newcommand{\bF}{{\mathbf F}}
\newcommand{\bI}{{\mathbf I}}

\newcommand{\bU}{{\mathbf U}}
\newcommand{\bV}{{\mathbf V}}
\newcommand{\bW}{{\mathbf W}}
\newcommand{\bG}{{\mathbf G}}
\newcommand{\bM}{{\mathbf M}}
\newcommand{\bP}{{\mathbf P}}
\newcommand{\bQ}{{\mathbf Q}}
\newcommand{\bH}{{\mathbf H}}
\newcommand{\bS}{{\mathbf S}}
\newcommand{\bL}{{\mathbf L}}
\newcommand{\bD}{{\mathbf D}}
\newcommand{\bE}{{\mathbf E}}
\newcommand{\bR}{{\mathbf R}}
\newcommand{\bu}{{\mathbf u}}
\newcommand{\bv}{{\mathbf v}}
\newcommand{\bw}{{\mathbf w}}

\newcommand{\bh}{{\mathbf h}}
\newcommand{\bg}{{\mathbf g}}
\newcommand{\br}{{\mathbf r}}
\newcommand{\bs}{{\mathbf s}}
\newcommand{\ba}{{\mathbf a}}
\newcommand{\bb}{{\mathbf b}}
\newcommand{\bd}{{\mathbf d}}
\newcommand{\be}{{\mathbf e}}

\newcommand{\bN}{{\mathbf N}}
\newcommand{\bJ}{{\mathbf J}}
\newcommand{\bK}{{\mathbf K}}

\newcommand{\bUpsilon}{{\mathbf \Upsilon}}

\newcommand{\calM}{{\cal M}}

\newcommand{\bx}{\mathbf{x}}

\newcommand{\bbeta}  {\boldsymbol{\beta}}

\newcommand{\bOmega}{\boldsymbol{\Omega}}
\newcommand{\bomega}{\boldsymbol{\omega}}
\newcommand{\bSigma}{\boldsymbol{\Sigma}}
\newcommand{\bDelta}{\boldsymbol{\Delta}}
\newcommand{\bgamma}{\boldsymbol{\gamma}}

\newcommand{\bTheta} {\boldsymbol{\Theta}}
\newcommand{\bPhi} {\boldsymbol{\Phi}}
\newcommand{\bPsi} {\boldsymbol{\Psi}}
\newcommand{\btheta} {\boldsymbol{\theta}}

\newcommand{\bGamma} {\boldsymbol{\Gamma}}
\newcommand{\bLambda} {\boldsymbol{\Lambda}}

\newcommand{\bPi}{\boldsymbol{\Pi}}
\newcommand{\bzero}{{\mathbf 0}}

\newcommand{\bXi}{\boldsymbol{\Xi}}

\newcommand{\beps}{\boldsymbol{\varepsilon}}
\newcommand{\bphi}{\boldsymbol{\phi}}

\newcommand{\vx}{\mathbf{x}}

\newcommand{\E}{\mathbb{E}}
\newcommand{\bT}{\mathbf{T}}

\newtheorem{theorem}{Theorem}
\newtheorem{lemma}{Lemma}
\newtheorem{proposition}{Proposition}
\newtheorem{cd}{Condition}

\makeatletter
\newcommand{\fdsy@scale}{1.0}
\newcommand\fdsy@mweight@normal{Book}
\newcommand\fdsy@mweight@small{Book}
\newcommand\fdsy@bweight@normal{Medium}
\newcommand\fdsy@bweight@small{Medium}

\DeclareFontFamily{U}{FdSymbolC}{}

\DeclareFontShape{U}{FdSymbolC}{m}{n}{
	<-7.1> s * [\fdsy@scale] FdSymbolC-\fdsy@mweight@small
	<7.1-> s * [\fdsy@scale] FdSymbolC-\fdsy@mweight@normal
}{}
\DeclareFontShape{U}{FdSymbolC}{b}{n}{
	<-7.1> s * [\fdsy@scale] FdSymbolC-\fdsy@bweight@small
	<7.1-> s * [\fdsy@scale] FdSymbolC-\fdsy@bweight@normal
}{}

\DeclareSymbolFont{arrows}{U}{FdSymbolC}{m}{n}
\SetSymbolFont{arrows}{bold}{U}{FdSymbolC}{b}{n}

\DeclareMathSymbol{\upvDash}{\mathrel}{arrows}{233}

\DeclareMathSymbol{\upmodels}{\mathrel}{arrows}{237}
\makeatother

\def\spacingset#1{\renewcommand{\baselinestretch}%
	{#1}\small\normalsize} \spacingset{1}

\newcommand{\blind}{1}

\def\singlespace{\def\baselinestretch{1}\@normalsize}

\begin{document}
	
	\if1\blind
	{
		\spacingset{1.25}
		\title{\bf \Large Identification and Estimation for Matrix Time Series CP-factor Models}
		\author[1,2]{Jinyuan Chang}
		\author[1]{Yue Du}
		\author[1]{Guanglin Huang}
		\author[3]{Qiwei Yao}

		\affil[1]{\it \small Joint Laboratory of Data Science and Business
			Intelligence, Southwestern University of Finance and Economics, Chengdu, China}
		\affil[2]{\it \small State Key Laboratory of Mathematical Sciences, Academy of Mathematics and Systems Science, Chinese Academy of Sciences, Beijing, China}
		\affil[3]{\it \small Department of Statistics, The London School of Economics and Political Science, London, U.K.}

		\setcounter{Maxaffil}{0}
		
		\renewcommand\Affilfont{\itshape\small}
		%\date{\today}
		\date{\vspace{-5ex}}
		\maketitle
		%\vspace{-2cm}
	} \fi
	\if0\blind
	{
		\bigskip
		\bigskip
		\bigskip
		\begin{center}
			{
				\Large \bf Identification and Estimation for Matrix Time Series CP-factor Models
			}
		\end{center}
		\medskip
	} \fi
	
	\spacingset{1.5}
	\begin{abstract}
		We propose a new method for identifying and estimating the CP-factor models for matrix time series. Unlike the generalized eigenanalysis-based method of \cite{chang2023modelling} for which the convergence rates of the associated estimators may suffer from small eigengaps as the asymptotic theory is based on some matrix perturbation analysis, the proposed new method enjoys faster convergence rates which are free from any eigengaps. It achieves this by turning the problem into a joint diagonalization of several matrices whose elements are determined by a basis of a linear system, and by choosing the basis carefully to avoid near co-linearity (see Proposition \ref{pro:theta-unique} and Section \ref{sec:theta-hat-est}). Furthermore, unlike the generalized
		eigenanalysis-based method which
		requires the two factor loading matrices to be full-ranked, the proposed new method can handle rank-deficient factor loading matrices. 
		Illustration with both simulated and real matrix time series data shows the advantages of the proposed new method.
		
	\end{abstract}
	
	
	\spacingset{1.69}
	\setlength{\abovedisplayskip}{0.2\baselineskip}
	\setlength{\belowdisplayskip}{0.2\baselineskip}
	\setlength{\abovedisplayshortskip}{0.2\baselineskip}
	\setlength{\belowdisplayshortskip}{0.2\baselineskip}

	\section{Introduction}
	The modern capacity for data collection has resulted in an abundance of time series data, with those in high-dimensional matrix format increasingly prevalent across diverse fields such as economics, finance, engineering, environmental sciences, medical research, network traffic monitoring, image processing and others. The demand of modeling and forecasting high-dimensional matrix time series brings the opportunities with challenges. Let $\bY_t = (y_{i,j,t})$ be a $p \times q$ matrix recorded at time $t$, where  $y_{i,j,t}$ represents the value of, for example, the $j$-th variable on the $i$-th individual at time $t$. A popular approach to model $\bY_t$ in the existing literature is via the so-called Tucker decomposition, namely the matrix Tucker-factor model. See, for example, \cite{wang2019factor}, \cite{chen2019constrained}, \cite{chen_chen_2022}, and \cite{han2024tensor}. It represents a high-dimensional matrix time series as a linear combination of a lower-dimensional matrix process. The Tucker decomposition can be viewed as a natural extension of the factor model for vector time series considered in  \cite{lam2012factor} and \cite{Chang2015}. Similarly we can only identify the factor loading spaces (the linear spaces spanned by the columns of the factor loading matrices) in the matrix Tucker-factor model while the factor loading matrices themselves are not uniquely defined. Parallel to the approaches based on Tucker decomposition, % for matrix time series $\bY_t$,
	\cite{chang2023modelling} and \cite{han2024cp}
	consider to model $\bY_t$ via the so-called canonical polyadic (CP) decomposition, namely the matrix CP-factor model. It provides a more comprehensive dimensionality reduction as the dynamic structure of a matrix time series is driven by a vector process rather than a matrix process. Furthermore the factor loading matrices in the matrix CP-factor model can be identified uniquely up to the column reflection and permutation indeterminacy under some regularity conditions.

	The CP-factor model for matrix time series $\bY_t$ 
	admits the form
	\begin{equation}\label{eq:abm}
		\bY_t =  \bA \bX_{t} \bB^{{\T}} + \beps_t\,, ~~~~ t\geq1\,, 
	\end{equation}
	where $\bX_t={\rm diag}(\bx_t)$ with $\bx_t =(x_{t,1},\ldots, x_{t,d})^\T$ being a $d \times 1$ time series, $\beps_t$ is a $p\times q$ matrix white noise, and $\bA=(\ba_1,\ldots,\ba_d)$ and $\bB=(\bb_1,\ldots,\bb_d)$ are, respectively, $p\times d$ and $q \times d$ constant matrices which are called
	factor loading matrices. See, for example, \cite{chang2023modelling}.
	Without loss of generality, we assume $|\ba_\ell|_2=1= |\bb_\ell|_2$  for each $\ell = 1,\ldots,d$. 
	For matrix  CP-factor model \eqref{eq:abm}, we cannot observe $(\bA,\bB,\bX_t,\beps_t)$ and only assume   $1 \le d < \min(p,q)$ is an unknown fixed integer. 
	Based on the  assumption ${\rm rank}(\bA) =d ={\rm rank}(\bB)  $,   \cite{chang2023modelling} proposes a one-pass estimation procedure for $(d,\bA,\bB)$ which 
	identifies $(\bA, \bB)$ uniquely up to the column reflection and permutation indeterminacy. In contrast to the standard alternating least squares method and its variations \citep{han2022tensor, han2024cp}, 
	the estimation procedure proposed in \cite{chang2023modelling} is based on solving some generalized eigenequations and requires no iterations.
	Note that the incoherence conditions imposed in \cite{han2022tensor} and \cite{han2024cp} also require both $\bA$
	and $\bB$ to be full-ranked. In fact those conditions imply that both $\{ \ba_\ell\}_{\ell=1}^d$ and $\{ \bb_\ell\}_{\ell=1}^d$ are two sets of near-orthogonal vectors. We do not require such an incoherence condition in this paper.
	
	In this paper, we investigate the identification issue of the  CP-factor model \eqref{eq:abm} for matrix time series without imposing the condition ${\rm rank}(\bA)=d={\rm rank}(\bB)$. Let
	\begin{equation*}
		{\rm rank}(\bA)=d_1~~\textrm{and}~~{\rm rank}(\bB)=d_2\,.
	\end{equation*}
	Then $1\le d_1, d_2 \le d$.
	As the CP-decomposition for 3-way tensors often exhibits rank-deficient factor loading matrices
	\citep{kolda2009tensor}, i.e., in model \eqref{eq:abm} it may hold that 
	%$\min(d_1, d_2) < d$ or even
	$\max(d_1, d_2)<d$. 
	We need to identify the condition under which $\bA$ and $\bB$ are uniquely identifiable up to the column reflection and permutation indeterminacy. Our setting allows all scenarios in terms of the relationships among $d_1,\, d_2 $ and $ d$.
	
	%Furthermore, we propose a new one-pass estimation procedure for $(d,\bA,\bB)$without imposing the condition ${\rm rank}(\bA)=d= {\rm rank}(\bB)$. 
	%
	%Figure \ref{fig:identification procedure} states our proposed unified identification procedure for $(d_1,d_2,d)$ and $(\bA,\bB)$ across different scenarios. Roughly speaking, $d_1$, $d_2$ and $d$ can be identified, respectively, as the rank of certain nonnegative-definite matrices $\bM_1$, $\bM_2$ and $\bM$. To identify $(\bA,\bB)$, a $d_1^2d_2^2\times d(d+1)/2$ matrix $\bOmega$ with ${\rm rank}(\bOmega)\le d(d-1)/2$ plays the key role. On one hand, if ${\rm rank}(\bOmega) = d(d-1)/2$,  we can identify $(\bA, \bB)$ uniquely upto the reflection and permutation indeterminacy. See Proposition \ref{pro:theta-unique} in Section \ref{sec:theta-iden} for details. On the other hand, if ${\rm rank}(\bOmega) < d(d-1)/2$, Proposition \ref{pro:Omega-impossible} in Section \ref{sec:unique-ab} shows that we cannot consistently estimate $(\bA, \bB)$ in general in the min-max sense. 
	%\begin{figure}[htbp]
	%\centerline{\includegraphics[width= 14.5cm]%{identification procedure-v2.jpg}}
	%\caption{Our proposed identification procedure for the matrix CP-factor model \eqref{eq:abm}.}
	%\label{fig:identification procedure}
	%\end{figure}
	%In practice,  with the available observations $\{\bY_{t}\}_{t=1}^{n}$, we also introduce  a feasible estimation method for estimating $(d_1,d_2,d)$ and $(\bA,\bB)$ across different scenarios, which only 
	The proposed new estimation procedure consists of several steps (see Section \ref{sec:estimation}). The key idea is to transform the $p\times q$ matrix CP-factor model (\ref{eq:abm}) to a $(d_1d_2)$-vector factor model, and then to identify the columns of $\bA$ and $\bB$ by a joint diagonalization of several symmetric matrices whose elements are determined by a basis, and in fact any basis, of a linear system (see Proposition \ref{pro:theta-unique}). Therefore, we can choose an appropriate basis to avoid near co-linearity such that our estimator enjoys faster convergence rate than those eigenanalysis-based estimators (see Section \ref{sec:theta-hat-est}). 
	Note that the convergence rates of the eigenanalysis-based estimators are derived based on some matrix perturbation analysis, and may suffer from the adverse 
	impact of 
	%the spectral decomposition of several non-negative definite matrices and a matrix  non-orthogonal joint diagonalization. Our theoretical analysis shows that the new estimator for $(\bA, \bB)$ is free from the adverse impact of any
	eigen-gap (i.e., the minimum pairwise gap among a set of eigenvalues). Our newly proposed estimator is free from this adversity.
	For example, the convergence rate of the estimator of
	\cite{chang2023modelling} can be formulated as the product of the rate of our new estimator and the inverse of an eigen-gap (See Remark \ref{rek:chang}). Note that the eigen-gap typically diminishes to 0 when  $p$ or/and $q$ diverge to infinity. %It is also worth pointing out that the convergence rate of the new estimator depends on $\max(p, q)$ instead of $p\times q$. Thus there is a clear technical advantage for treating  $p\times q$ matrix time series $\bY_t$ directly rather than vectorizing it into a $(pq)\times 1$ vector process. 
	% Unlike \cite{han2022tensor} and \cite{han2024cp}, we have achieved the convergence rate without imposing any incoherence conditions  which imply that both $\{ \ba_\ell\}_{\ell=1}^d$ and $\{ \bb_\ell\}_{\ell=1}^d$ are two sets of near-orthogonal vectors.

	The rest of the paper is organized as follows.  Section \ref{sec:pre} gives preliminaries of the matrix CP-factor model \eqref{eq:abm}. A general identification strategy for the matrix CP-factor model is presented in Section \ref{sec:identify-ab}.  Section \ref{sec:estimation} provides a one-pass estimation procedure for $(d_1,d_2,d,\bA,\bB)$. Section \ref{sec:prediction} gives a unified prediction approach for the matrix CP-factor model. We investigate the associated theoretical properties of the proposed method in Section \ref{sec:asymptotics}. Numerical results with simulation studies and real data analysis are given in Section \ref{section:simulation}. The \textsf{R}-function \texttt{CP\_MTS} for implementing our newly proposed method is available publicly in the \texttt{HDTSA} package \citep{chang2024hdtsa}. All technical proofs and some additional simulation studies are relegated in the supplementary material.

	\textit{Notation}. For a positive integer $m$, write $[m] = \{1, \ldots , m\}$, and denote by $\bI_m$ the $m \times m$ identity matrix. Denote by $I(\cdot)$ the indicator function.  For an $m_1 \times m_2$ matrix $\bH = (h_{i,j} )_{m_1 \times m_2}$, let $\mathcal{R}(\bH)=\max\{k:{\text{any}\ k\ \text{columns of the matrix $\bH$ are linearly  independent}}\}$, and denote by $\mathcal{M}(\bH)$ the linear space spanned by the columns of $\bH$.
	Let $\|\bH\|_2$, $\|\bH\|_\text{F}$, ${\rm rank}(\bH)$, $\lambda_{i}(\bH)$, and $\sigma_{i}(\bH)$   be, respectively, the spectral norm, Frobenius norm, rank, $i$-th largest eigenvalue, and $i$-th largest singular value of matrix $\bH$.
	Specifically, if $m_2 = 1$,  we use $|\bH|_1 = \sum_{i=1}^{m_1}|h_{i,1} |$ and $|\bH|_2=(\sum_{i=1}^{m_1}h_{i,1}^2)^{1/2}$
	to denote, respectively, the  $L_1$-norm and $L_2$-norm of the $m_1$-dimensional vector $\bH$. 
	Also, denote by $\bH^{{\T}}$ and $\bH^{+}$, respectively, the transpose and the Moore-Penrose inverse of $\bH$. 
	% When $m_1=m_2$, denote by ${\rm tr}(\bH)$ the trace of $\bH$.
	The operator ${\rm diag}(\cdot)$ stacks a vector into a square diagonal matrix. Let $\otimes$ denote the Kronecker product, and 
	$\odot$ denote the Khatri-Rao product such that $\check{\bH} \odot \tilde{\bH} = (\check{\bh}_1\otimes\tilde{\bh}_1,\ldots,\check{\bh}_m\otimes\tilde{\bh}_m)$ for any matrices $\check{\bH} = (\check{\bh}_1,\ldots,\check{\bh}_m)$ and $\tilde{\bH} = (\tilde{\bh}_1,\ldots,\tilde{\bh}_m)$. %Denote  $\lceil x \rceil$ by the smallest integer larger than $x$.  
	Moreover, for any two sequences of positive numbers $\{\tau_{k}\}$ and $\{\tilde{\tau}_{k}\}$, we write $\tau_{k} \asymp \tilde{\tau}_{k}$ if $\tau_{k}/\tilde{\tau}_{k}=O(1)$ and $\tilde{\tau}_{k}/\tau_{k}=O(1)$ as $k\rightarrow\infty$, and write $\tau_k \ll \tilde{\tau}_k$ or $ \tilde{\tau}_k \gg \tau_k$ if $\lim\sup_{k\to \infty} \tau_k/\tilde{\tau}_k=0$. To simplify our presentation, for a matrix $\bH =(h_{i,j})_{m_1\times m_2}$, we write $\vec\bH$ or ${\rm vec}(\bH)$ as an $(m_1m_2)$-dimensional vector with the $\{(j-1)m_1 +i\}$-th element being $h_{i,j}$, and for a tensor $\mathcal{H}=(h_{i,j,k,l})_{m_1\times m_2\times m_3 \times m_4}$, we write  $\vec{\mathcal{H}}$ as an $(m_1m_2m_3m_4)$-dimensional vector with the $\{(i-1)m_2m_3m_4 + (j-1)m_3m_4 + (k-1)m_4 + l\}$-th element being $h_{i,j,k,l}$.

	\section{Preliminary}\label{sec:pre}
	
	Recall that, in the matrix CP-factor model \eqref{eq:abm}, $\bA$ and $\bB$ are, respectively, $p\times d$ and $q\times d$ matrices with ${\rm rank}(\bA)=d_1$ and ${\rm rank}(\bB)=d_2$, and $d_1,d_2 \in [d]$. %As we will show in the first paragraph of Section \ref{sec:identifyUV}, $(d,d_1,d_2)$ should satisfy the restriction $d_1d_2\geq d$.  We impose the following regularity condition on the matrix CP-factor model \eqref{eq:abm}.
	Model \eqref{eq:abm} can be equivalently represented as 
	\begin{equation*}%\label{eq:frontSlice}
		\vec\bY_t=(\bB \odot \bA)\mathbf{x}_t + \vec\beps_t\,,~~~~t\geq1\,,
	\end{equation*}  
	where $\bB \odot \bA =(\bb_1\otimes \ba_1,\ldots, \bb_{d}\otimes \ba_{d})$ and $\mathbf{x}_t= (x_{t,1},\ldots, x_{t,d})^{\T}$.
	Condition \ref{cd:ra}(i) below holds naturally.  
	If ${\rm rank}(\bB \odot \bA) = \tilde{d} < d$, the matrix $\bB \odot \bA$ has $\tilde{d}$ linearly independent columns that span its column space. Therefore, we can find $\{\bb_{\ell_1}\otimes \ba_{\ell_1},\ldots, \bb_{\ell_{\tilde{d}}}\otimes \ba_{\ell_{\tilde{d}}}\}$ with  some distinct $\ell_1,\ldots,\ell_{\tilde{d}} \in [d]$ such that they provide a basis for $\mathcal{M}(\bB \odot \bA)$.  The remaining columns of $\bB \odot \bA$ can be expressed as linear combinations of this set of basis vectors. Then $\bB \odot \bA = (\bb_{\ell_1}\otimes \ba_{\ell_1},\ldots, \bb_{\ell_{\tilde{d}}}\otimes \ba_{\ell_{\tilde{d}}}) \tilde{\bC}$ for some $\tilde{d} \times d$ matrix $\tilde{\bC}$. Since $(\bA,\bB,\bX_t)$ are unobserved, we can reformulate  $\vec\bY_t$ in a new form  $\vec\bY_t=(\tilde{\bB} \odot \tilde{\bA})\tilde{\mathbf{x}}_t + \vec\beps_t$ with $\tilde{\bA}=(\ba_{\ell_1},\ldots,\ba_{\ell_{\tilde{d}}})$, $\tilde{\bB}= (\bb_{\ell_1},\ldots,\bb_{\ell_{\tilde{d}}})$ and $\tilde{\mathbf{x}}_t=\tilde{\bC}\mathbf{x}_t$. In this new form, the newly defined factor loading matrices $\tilde{\bA}\in \mathbb{R}^{p \times \tilde{d}}$ and $\tilde{\bB} \in \mathbb{R}^{q \times \tilde{d}}$ satisfy  $\textup{rank}(\tilde{\bB}\odot\tilde{\bA}) = \tilde{d}$. 
	On the other hand, since $\beps_t$ is a matrix white noise, Condition \ref{cd:ra}(ii) holds automatically.

	% If ${\rm rank}(\bB\odot \bA)=\tilde{d}<d$, we may then assume, without loss of generality, $\bB \odot \bA= (\bb_1\otimes \ba_1,\ldots, \bb_{\tilde{d}}\otimes \ba_{\tilde{d}})\tilde{\bC}$ for some $\tilde{d}\times d$ matrix $\tilde{\bC}$. Then we have $\vec\bY_t=(\tilde{\bB} \odot \tilde{\bA})\tilde{\mathbf{x}}_t + \vec\beps_t$ with $\tilde{\bA}=(\ba_1,\ldots,\ba_{\tilde{d}})$, $\tilde{\bB}=(\tilde{\bb}_1,\ldots,\tilde{\bb}_{\tilde{d}})$ and $\tilde{\mathbf{x}}_t=\tilde{\bC}\mathbf{x}_t$,  which indicates that the matrix time series $\bY_t$ can be represented as a new matrix CP-factor model with the newly defined factor loading matrices $\tilde{\bA}$ and $\tilde{\bB}$ such that $\tilde{\bB}\odot\tilde{\bA}$ has full column rank. 

	\begin{cd}\label{cd:ra}
		{\rm(i)}  ${\rm rank}(\bB \odot \bA)=d$. {\rm(ii)} $\mathbb{E}(\beps_t)=\bf{0}$  for any $t\geq1$, $\mathbb{E}(\beps_t \otimes \beps_s)=\bf{0}$ for all $t\ne s$, and $\mathbb{E}(x_{t,\ell}\beps_s)=\bf{0}$ for any $\ell \in[d]$ and $t\le s$.
	\end{cd}
	
	%\begin{rek}
	% Condition \ref{cd:ra} is mild. 
	%Write $\mathbf{f}_t=(x_{t,1},\ldots, x_{t,d})^{\T}$. The matrix CP-factor model \eqref{eq:abm} can be also reformulated as  
	%	$
	%		{\rm vec}(\bY_t)=(\bB \odot \bA)\mathbf{f}_t + {\rm vec}(\beps_t)$.
	%	If ${\rm rank}(\bB \odot \bA)=\tilde{d}<d$, we can assume  ${\rm rank}(\bb_1\otimes \ba_1,\ldots, \bb_{\tilde{d}}\otimes \ba_{\tilde{d}})=\tilde{d}$ without loss of generality, then $\bB \odot \bA= (\bb_1\otimes \ba_1,\ldots, \bb_{\tilde{d}}\otimes \ba_{\tilde{d}})\bC$ for some $\tilde{d}\times d$ matrix $\bC$. Hence, we have
	%	$
	%		{\rm vec}(\bY_t)= (\tilde{\bB} \odot \tilde{\bA}) %\tilde{\mathbf{f}}_t + {\rm vec}(\beps_t)
	%	$ 
	%	with $\tilde{\bA}=(\ba_1,\ldots, \ba_{\tilde{d}})$, $\tilde{\bB}=(\bb_1,\ldots, \bb_{\tilde{d}})$ and $\tilde{\mathbf{f}}_t=\bC \mathbf{f}_t$, which indicates that the matrix time series $\{\bY_t\}_{t\geq1}$ can be represented as a new matrix CP-factor model with the newly defined loading matrices $\tilde{\bA}$ and $\tilde{\bB}$ such that $\tilde{\bB}\odot\tilde{\bA}$ has full column rank. Hence, Condition \ref{cd:ra}(i) holds automatically. Since $\{\beps_t\}_{t\geq1}$ is a matrix white noise sequence, Condition \ref{cd:ra}(ii) is satisfied automatically.
	%\end{rek}
	
	%There are three scenarios in terms the relationship among $d$, $d_1$ and $d_2$: (i) $d_1 = d_2 = d$, (ii) $d_2 <d_1 = d$, and (iii) $\max(d_1,d_2) < d$. For scenario (i),
	When $d_1=d_2=d$, \cite{chang2023modelling} provides a one-pass estimator for $(\bA,\bB)$ by solving some generalized eigenequations defined by the matrices   
	\begin{equation}\label{eq:sigma_yxi}
		\bSigma_{\bY,\xi}(k) = \frac{1}{n-k}\sum_{t = k+1}^{n}\E[ \{ \bY_t -\E(\bar{\bY})\}\{ \xi_{t-k} -\E(\bar{\xi})\} ]\,,~~~~k\geq1\,,
	\end{equation}
	where $\bar{\bY} = n^{-1}\sum_{t = 1}^{n}\bY_t$, $\xi_t$ is a scalar defined as a linear combination of the elements of $\bY_t$, and  $\bar{\xi} = n^{-1}\sum_{t = 1}^{n}\xi_t$. For example, we can select $\xi_{t}$ as the first principal component of
	$\vec \bY_t$.

	Recall $\bA^+$ and $ \bB^+$ are, respectively,
	the Moore-Penrose inverse of $\bA$ and $\bB$. The key requirement underlying the results of \cite{chang2023modelling} is  $\bA^+\bA = \bB^+\bB=\bI_d$, which only holds when $d_1=d_2=d$. % If either $d_1$ or $d_2$ is smaller than $d$, then either the Moore-Penrose inverse of $\bA$ or $\bB$ does not exist. 
	Hence, the estimation method of \cite{chang2023modelling} is not applicable when $\min(d_1, d_2) <d$. 
	% Note that the CP-decomposition for 3-way tensors can often exhibit rank-deficient factor loading matrices \citep{kolda2009tensor}, i.e., in model (\ref{eq:abm}) it may hold that  $\min(d_1, d_2) < d$ or even $\max(d_1, d_2) < d$. 
	%See Section 3.1 of \cite{kolda2009tensor} and the references within.
	In this paper, we consider a new approach which
	identifies $(d,\bA,\bB)$ without the condition $d_1=d_2=d$. Furthermore we propose a 
	unified and more efficient one-pass estimation 
	for $(\bA,\bB)$ regardless they are rank-deficient or not.

	\section{Identification of %$(d_1,d_2,d)$ and 
		$(\bA,\bB)$}\label{sec:identify-ab}
	
	We need to identify in model \eqref{eq:abm} the order $d$ and the factor loading pairs $(\ba_1, \bb_1),\ldots,(\ba_d,\bb_d)$. % upto permutation and reflection indeterminacy.
	To carry out this task, we first introduce a reduced model for a $d_1 \times d_2$ matrix time series, and then identify $d$ and the CP-factor loadings for the reduced model via (i) a factor model for a vector time series, and (ii) a non-orthogonal joint diagonalization of $d$ symmetric matrices. %Finally we specify the conditions under which $\bA$ and $\bB$ are uniquely identifiable.
	
	\subsection{A reduced model}
	
	For a prescribed integer $K>1$ and
	$\bSigma_{\bY, \xi}(k)$ specified in \eqref{eq:sigma_yxi}, define
	\begin{equation}\label{eq:M1M2}
		\bM_1=\sum_{k=1}^{K}\bSigma_{\bY, \xi}(k)\bSigma_{\bY, \xi}(k)^{{\T}}~~\textrm{and}~~ \bM_2=\sum_{k=1}^{K}\bSigma_{\bY, \xi}(k)^{{\T}}\bSigma_{\bY, \xi}(k)\,.
	\end{equation}
	Furthermore, due to $\bX_t={\rm diag}(\bx_t)$ with $\bx_t=(x_{t,1},\ldots,x_{t,d})^{{\T}}$, we let 
	$$ \bG_k={\rm diag}(\bg_k)=\frac{1}{n-k}\sum_{t=k+1}^{n}\mathbb{E}[\{\bX_t-\mathbb{E}(\bar{\bX})\}\{\xi_{t-k}-\mathbb{E}(\bar{\xi})\}] \,,~~~~k\in[K]\,,$$
	where $\bar{\bX} = n^{-1}\sum_{t = 1}^{n}\bX_t$. It follows from \eqref{eq:abm} and Condition \ref{cd:ra} that
	\begin{equation*}%\label{eq:M1M2-G}
		\bM_1 = \bA \bigg(\sum_{k = 1}^K\bG_k \bB^{\T} \bB \bG_k\bigg) \bA^{{\T}} ~~ \text{and} ~~  \bM_2 = \bB \bigg(\sum_{k = 1}^K\bG_k \bA^{\T} \bA \bG_k\bigg) \bB^{{\T}}\,.
	\end{equation*}
	Let $\bG=\sum_{k=1}^K \bg_{k}\bg_{k}^{\T}$. Proposition \ref{pro:m1-rank-con} shows that $d_1$ and $d_2$ can be identified, repsectively, by ${\rm rank}(\bM_1)$ and ${\rm rank}(\bM_2)$. 
	
	\begin{proposition}\label{pro:m1-rank-con}
		Let Condition \ref{cd:ra} hold and all the main diagonal elements of ${\bG}$ 
		are nonzero.  The following two assertions hold.
		\begin{enumerate}[(i)]
			\item If $\max\{ \mathcal{R}({\bG}) + d_2,\mathcal{R}(\bB^\T \bB) + {\rm rank}({\bG}) \}>d$, then ${\rm rank}(\bM_1)=d_1$.
			\item If $\max\{ \mathcal{R}({\bG}) + d_1,\mathcal{R}(\bA^\T\bA) + {\rm rank}({\bG}) \}>d$, then ${\rm rank}(\bM_2)=d_2$. 
		\end{enumerate}
		%     \begin{enumerate}[(i)]
			%     \item If $\max\{ \mathcal{R}({\bG}) + d_2,\mathcal{R}(\bB^\T \bB) + {\rm rank}({\bG}) \}>d$, then ${\rm rank}(\bM_1)=d_1$.
			%     \item If $\max\{ \mathcal{R}({\bG}) + d_1,\mathcal{R}(\bA^\T \bA) + {\rm rank}({\bG}) \}>d$, then ${\rm rank}(\bM_2)=d_2$. 
			% \end{enumerate}
	\end{proposition}

	The conditions required in Proposition \ref{pro:m1-rank-con} are mild.  
	% Notice that all the main diagonal elements of $\bG$ are positive if all the components of some $\bg_k$ are nonzero with $k\in [K]$, which implies $\mathcal{R}(\bG)\geq1$. 
	Note that if all the components of some
	$\bg_k$ are nonzero, then all the main diagonal elements of $\bG$ are positive and $\mathcal{R}(\bG)\geq1$.
	For the scenario $d_1=d_2=d$,  Proposition \ref{pro:m1-rank-con} holds automatically.  When $\min(d_1, d_2) < d$,  we suppose that $d_2 \le d_1$ without loss of generality. For the scenario $d_2 <d_1 = d$, we only need to identify $d_2$. Proposition \ref{pro:m1-rank-con}(ii) holds automatically in this scenario, which implies $d_2$ could be identified trivially.   For the scenario $\max(d_1, d_2) < d$, by Condition \ref{cd:ra}(i), we know $\ba_\ell\neq \bzero$ and $\bb_\ell\neq \bzero$ for each $\ell\in[d]$, which implies $\mathcal{R}(\bA^{\T}\bA)\geq1$ and $\mathcal{R}(\bB^{\T}\bB)\geq1$.  Proposition \ref{pro:rankwith-xt} proposes some sufficient conditions such that $ {\rm rank}({\bG})=d$,  which make Proposition \ref{pro:m1-rank-con} hold automatically.  Define  $$\bSigma_{\bx}(k) =   \frac{1}{n-k}\sum_{t=k+1}^{n}  \mathbb{E}[ \{\bx_{t}-\mathbb{E}(\bar{\bx})\} \{\bx_{t-k}-\mathbb{E}(\bar{\bx})\}^{\T} ]\,,~~~~k\in[K]\,,$$ where $\bar{\bx} =n^{-1}\sum_{t=1}^{n}\bx_t$. Write $\xi_{t} = \bomega^{\T}\vec{\bY}_{t} $ and  $\bSigma_{\bx,K}=\{\vec{\bSigma}_{\bx} (1) , \ldots, \vec{\bSigma}_{\bx} (K) \} \in \mathbb{R}^{d^2 \times K}$. 
	\begin{proposition}\label{pro:rankwith-xt}
		Assume that $\mathbb{E}(\bx_t \otimes  \vec{\beps}_{t-k} ) =\bzero$ for any $k\in[K]$ with $K\ge d^2$. If $\bomega^{\T} (\bB \odot \bA) \ne \bzero$ and  ${\rm rank} (\bSigma_{\bx,K}) =d^2 $, then ${\rm rank}(\bG) =d$.  
	\end{proposition}
	
	Due to $\ba_\ell\neq \bzero$ and $\bb_\ell\neq \bzero$ for each $\ell\in[d]$, 
	the requirement $\bomega^{\T} (\bB \odot \bA) \ne \bzero$ is generally mild and can be satisfied by appropriately choosing a nonzero vector $\bomega$. If $\bx_t$ satisfies ${\rm rank} (\bSigma_{\bx,K}) =d^2 $, and  $\bx_t$ and $ \vec{\beps}_{t-k}$ are uncorrelated for $k\in[K]$, 
	% the selection of $\xi_{t} = \bomega^{\T}\vec{\bY}_{t} $ satisfying  $\bomega^{\T} (\bB \odot \bA) \ne \bzero$,
	Proposition \ref{pro:rankwith-xt} shows that ${\rm rank}(\bG) =d$. Combining with Proposition \ref{pro:m1-rank-con}, % Condition \ref{cd:M1M2eigenvalues} holds automatically. Notice that the assumption $\mathbb{E}(\bx_t \otimes  \vec{\beps}_{t-k} ) =\bzero$ is mild which is also  assumed in \cite{wang2019factor}. Hence, 
	it is reasonable to assume Condition \ref{cd:M1M2eigenvalues}, which ensures $\calM(\bM_1) = \calM(\bA)$ and 
	$\calM(\bM_2) = \calM(\bB),$ i.e., the information on the loadings $\{\ba_\ell\}_{\ell=1}^d$ and $\{\bb_\ell\}_{\ell=1}^d$ is, respectively, kept in $\bM_1$ and $\bM_2$. 
	\begin{cd}\label{cd:M1M2eigenvalues}
		${\rm rank}(\bM_1)=d_1$ and ${\rm rank}(\bM_2)=d_2$. 
	\end{cd} 
	Now perform the spectral decomposition for $\bM_1$ and $\bM_2$: 
	\begin{align}\label{eq:definition of PQ}
		\bM_1 = \bP\bD_1\bP^{{\T}}~~\textrm{and}~~
		\bM_2 = \bQ\bD_2\bQ^{{\T}}\,, 
	\end{align}
	where $\bD_1$ and $\bD_2$ are, respectively, $d_1\times d_1$ and $d_2 \times d_2$ full-ranked diagonal matrices, $\bP^\T \bP=\bI_{d_1}$ and $\bQ^\T \bQ = \bI_{d_2}$.
	As $\calM(\bP)=\calM(\bM_1)= \calM(\bA)$ and
	$\calM(\bQ)=\calM(\bM_2)=\calM(\bB)$, then
	%
	%where the columns of $\bP \in \mathbb{R}^{p\times d_1}$  are  the $d_1$  orthonormal eigenvectors corresponding
	%to the $d_1$ nonzero eigenvalues of $\bM_1$, and $\bD_1 \in \mathbb{R}^{d_1\times d_1}$ is a diagonal matrix with the corresponding eigenvalues as the diagonal elements. Due to $\bA = (\ba_1,\ldots,\ba_d)$ with $|\ba_\ell|_2 = 1$ for each $\ell\in [d]$, by \eqref{eq:M1-rank-dec} and \eqref{eq:definition of PQ}, we have
	\begin{align}\label{eq:FA}
		\bA =%\bP\bD_1\bP^{{\T}} (\bC_{\rm A}^{+})^{{\T}} %\tilde{\bM}_{1}^{-1}\bF_{\rm A} : = 
		\bP \bU ~~{\rm and}~~
		\bB= \bQ \bV\,,
	\end{align}
	where  $\bU$ and $\bV$ are, respectively,
	${d_1\times d}$ and $d_2 \times d$ matrices with unit column vectors. Since $\bP$ and $\bQ$ are determined by
	the spectral decomposition (\ref{eq:definition of PQ}), we only need to identify $(\bU, \bV)$ in order to identify
	$(\bA, \bB)$. When $d=1$, we may take $\ba_1 = \bA = \bP$ and $\bb_1 = \bB = \bQ$. Therefore only the non-trivial case with $d\ge 2$ will be considered in the sequel. 

	Define a $d_1 \times d_2$ process $\bZ_t = \bP^\T \bY_t \bQ$. It follows from \eqref{eq:abm} and \eqref{eq:FA} that
	\begin{align}\label{eq:zt}
		\bZ_t=\bU \bX_t \bV^{{\T}} +\bDelta_t\,,~~~~t\geq1\,,
	\end{align}
	where $\bDelta_t=\bP^{{\T}} \beps_t\bQ$ is a matrix white noise. This is a reduced form of the CP-factor model \eqref{eq:abm} for the matrix time series $\bY_t$.
	We will identify $(\bU,\bV)$ based on this reduced model.

	\subsection{%Identification of $d$ and $(\bU,\bV)$ -- 
		A vector factor model}
	\label{sec:identifyUV}
	
	Recall $\bX_t={\rm diag}(\bx_t)$ with $\bx_t =(x_{t,1},\ldots, x_{t,d})^\T$. It follows from \eqref{eq:zt}  that  
	\begin{equation}    
		\label{eq:facm}
		\vec\bZ_t = (\bV\odot \bU)\mathbf{x}_t + \vec\bDelta_t\,,~~~~t\geq1\,.
	\end{equation}
	%where $\mathbf{f}_t=(x_{t,1},\ldots, x_{t,d})^{\T}$. Due to $\bA=\bP\bU$ and  $\bB=\bQ\bV$,  then 
	This is the standard factor model for vector time series
	considered by \cite{lam2012factor} and \cite{Chang2015}. Note that $\bB \odot \bA = (\bQ \otimes \bP)(\bV \odot \bU)$, where $\bP\in\mathbb{R}^{p\times d_1}$ and $\bQ\in\mathbb{R}^{q\times d_2}$ with ${\rm rank}(\bP)=d_1$ and ${\rm rank}(\bQ)=d_2$. By Condition \ref{cd:ra}(i), we know the  dimension of the factor loading space $\calM(\bV\odot\bU)$ in \eqref{eq:facm} is $d$, as ${\rm rank}(\bV\odot \bU) = {\rm rank}(\bB\odot\bA) =d$.
	Using the techniques developed in \cite{Chang2015}, we can identify $d$ and
	$\calM(\bV\odot\bU)$ uniquely based on an eigenanalysis. More precisely, we can find a $(d_1 d_2)\times d$ matrix $\bW$, with
	$\bW^\T \bW =\bI_d$, such that
	\begin{equation}\label{eq:iduv}
		\bV\odot \bU \equiv (\bv_1 \otimes \bu_1, \ldots, \bv_d\otimes \bu_d) = \bW\bTheta\,,
	\end{equation}
	where $\bTheta$ is an unknown $d\times d$ invertible matrix with unit column vectors. Since the $d$ columns of $\bW$ are the orthogonal basis of  $\calM(\bV\odot\bU)$, we can select $\bW$ in \eqref{eq:iduv} as an arbitrary $(d_1d_2)\times d$ matrix such that $\bW^{\T}\bW=\bI_d$ and $\mathcal{M}(\bW)=\mathcal{M}(\bU\odot\bV)$. In \eqref{eq:iduv}, different selections of $\bW$ will lead to different $\bTheta$. As we will show in Section \ref{sec:theta-iden}, for any given $\bW$, the associated rotation matrix $\bTheta$ can be uniquely identified up to the column reflection and permutation indeterminacy. Write 
	\begin{equation}
		\label{eq:c=wtheta}   
		\bC\equiv (\vec\bC_1, \ldots, \vec\bC_d) = \bW\bTheta \equiv (\vec \bW_1, \ldots, \vec\bW_d)\bTheta\,,
	\end{equation}
	where $\bC_\ell$ and $\bW_\ell$ are $d_1\times d_2$ matrices. 
	We put the columns of both $\bC$ and $\bW$ in the form of vectorized $d_1\times d_2$ matrices for some technical convenience which will be obvious soon. It follows from 
	(\ref{eq:iduv}) and \eqref{eq:c=wtheta} that $\vec\bC_\ell = \bv_\ell\otimes \bu_\ell $, which implies $\bu_\ell \bv_\ell^\T = \bC_\ell$.  % as $\overrightarrow{(\bu_\ell \bv_\ell^\T)}= \bv_\ell\otimes \bu_\ell$. 
	Given $\bW$ and its associated rotation matrix $\bTheta$, the $(d_1d_2)\times d$ matrix $\bC$ specified in \eqref{eq:c=wtheta} is uniquely identified, which can be used to identify $(\bU,\bV)$. See Proposition \ref{pro:rank-1} for details.
	
	\begin{proposition}
		\label{pro:rank-1}
		Let Conditions \ref{cd:ra} and
		\ref{cd:M1M2eigenvalues} hold. Then
		matrices
		$\bC_1, \ldots, \bC_d$ specified in \eqref{eq:c=wtheta} are all of rank 1 with the nonzero singular value equal to 1, and $(\bu_\ell, \bv_\ell)$ are the unit singular vectors of $\bC_\ell$ for each $\ell \in [d]$. 
	\end{proposition}
	
	\subsection{A non-orthogonal joint diagonalization}\label{sec:theta-iden}
	
	For given $\bW$ in \eqref{eq:iduv}, Proposition \ref{pro:rank-1} implies that the task of identifying $(\bU,\bV)$ boils down to identifying $\bTheta$ specified in \eqref{eq:iduv} such that $\bC_1,\ldots,\bC_d$ defined in \eqref{eq:c=wtheta} satisfying ${\rm rank}(\bC_\ell)=1$ for each $\ell \in [d]$. 
	By \eqref{eq:c=wtheta}, it holds that 
	\begin{equation}
		\label{eq:bi-representation}    
		\bC_\ell = \sum_{i=1}^d \theta_{i,\ell} \bW_i ~~
		{\rm and} ~~
		\bW_\ell = \sum_{i=1}^d \theta^{i,\ell} \bC_i\,, ~~~~ \ell\in[d]\,,
	\end{equation}
	where $\theta_{i,j}$ and $\theta^{i,j}$ denote, respectively, the $(i,j)$-th elements of $\bTheta$ and
	$\bTheta^{-1}$.
	
	For any two matrices  $\bD=(d_{i,j})$ and $\bF=(f_{i,j})$  of the same
	size,  define  $\bPsi(\bD, \bF)$ to be a 4-way tensor  with the
	$(i,j,k,\ell)$-th element
	$
	d_{i,k}f_{j,\ell} + d_{j,\ell}f_{i,k} - d_{i,\ell} f_{j,k} - d_{j,k}f_{i,\ell}.
	$ 
	By Theorem 2.1 of \cite{de2006link},
	for any  matrix $\bD\ne \bf0$, ${\rm rank}(\bD)=1$ if and only if $\bPsi(\bD, \bD)= \bf0$.  
	Hence, by \eqref{eq:bi-representation}, for given $\bW$ in \eqref{eq:iduv}, we know $\bTheta = (\theta_{i,j})$ is the solution of
	\begin{equation}\label{eq: quadratic solution}
		{\bf0} = \bPsi(\bC_\ell, \bC_\ell) = \sum_{i,j=1}^d \theta_{i,\ell} \theta_{j, \ell}
		\bPsi(\bW_i, \bW_j)\,, ~~~~ \ell\in[d]\,. 
	\end{equation}
	This is a set of quadratic equations. 
	% {\color{red}Solving such a system of quadratic equations as in \eqref{eq: quadratic solution} often leads to a non-convex optimization problem, which is generally NP-hard. To address this, we propose the following approach for identifying $\bTheta$.}
	Consider a $(d_1^2d_2^2)\times d(d+1)/2$
	matrix 
	\begin{align}\label{eq:omega-D}
		\bOmega=\big(\vec \bPsi(\bW_1, \bW_1), \ldots, \vec \bPsi(\bW_1, \bW_d),
		\vec\bPsi(\bW_2, \bW_2), \ldots, \vec\bPsi(\bW_d, \bW_d)\big)\,.
	\end{align} 
	Proposition \ref{pro:rankomega} is instrumental in solving those quadratic equations.

	\begin{proposition}\label{pro:rankomega}
		Let $d\ge 2$ and Condition \ref{cd:M1M2eigenvalues} hold. The following three assertions hold. 
		\begin{enumerate}[(i)] 
			\item  ${\rm rank}(\bOmega) \le d(d-1)/2$.
			\item  ${\rm rank}(\bOmega)=d(d-1)/2$ if and only if
			the ${d(d-1)/2}$ vectors $\vec \bPsi(\bC_1, \bC_2), \ldots, \vec \bPsi(\bC_1, \bC_d),$
			$\vec\bPsi(\bC_2, \bC_3), \ldots, \vec\bPsi(\bC_{d-1}, \bC_d)$ are linearly independent.
			\item  Let ${\rm ker}(\bOmega) = \{\bh \in \mathbb{R}^{d(d+1)/2} : \bOmega \bh = \mathbf{0}\}$. Then ${\rm dim}\{{\rm ker}(\bOmega)\} = d$ if and only if ${\rm rank}(\bOmega)=d(d-1)/2$.  
		\end{enumerate}
	\end{proposition}
	
	Now assume ${\rm rank}(\bOmega) ={d(d-1)/2}$. Let
	$\bh_m =
	(h_{1,1}^m, \ldots, h_{1,d}^m, h_{2,2}^m, \ldots, h_{d,d}^m)^\T$, $m\in [d]$, 
	be a set of basis vectors of ${\rm ker}(\bOmega)$. 
	Recall $\bPsi(\bC_{\ell}, \bC_{\ell})=\bf0$ for any $\ell\in[d]$. By  (\ref{eq:bi-representation}), it holds that
	\begin{align*}
		\bf0 &= \sum_{1\le i\le j\le d} h^m_{i,j} \bPsi(\bW_i, \bW_j)
		= \sum_{1\le i\le j\le d} h^m_{i,j} \sum_{k,\ell=1}^d \theta^{k,i} \theta^{\ell,j}\bPsi(\bC_k, \bC_\ell)\\
		&= \sum_{1\le k<\ell \le d} \bPsi(\bC_k, \bC_\ell) \sum_{1\le i\le j\le d}
		(\theta^{k,i} \theta^{\ell,j} + \theta^{k,j} \theta^{\ell,i})h^m_{i,j}\,.
	\end{align*}
	By Proposition \ref{pro:rankomega}(ii), we have
	\begin{equation}
		\label{eq:off-diagonal}
		\sum_{1\le i\le j\le d}
		(\theta^{k,i} \theta^{\ell,j} + \theta^{k,j} \theta^{\ell,i}) h^m_{i,j} =0 ~~\mbox{for all}~~ 1\le k < \ell \le d\,.
	\end{equation}
	Let $\bH_{m}$ be a $d\times d$ matrix with the $(i,i)$-th element being $h_{i,i}^m$ for any $i$, and the $(i,j)$-th and $(j,i)$-th elements being $h_{i,j}^m/2$ for any $i<j$. Based on \eqref{eq:off-diagonal}, we know $\bGamma_m \equiv \bTheta^{-1} \bH_m (\bTheta^{-1})^\T$  is a diagonal matrix, i.e., 
	we can find $\bTheta^{-1}$ which diagonalizes jointly $\bH_m = \bTheta \bGamma_m \bTheta^\T$ for each $m\in[d]$.
	
	It is also clear from \eqref{eq:off-diagonal} that
	the diagonal property is independent of the norms
	of row vectors of $\bTheta^{-1}$. Hence all the columns of $\bTheta$ can be set as unit vectors. Proposition \ref{pro:theta-unique} shows that $\bTheta$ is invariant with respect to the choice of the basis vectors
	for ${\rm ker}(\bOmega)$.
	The available algorithms for this joint diagonalization
	include the joint approximate diagonalization of \cite{pham2001blind}, the fast Frobenius diagonalization of \cite{ziehe2004fast}, and the quadratic diagonalization of \cite{vollgraf2006quadratic}.

	\begin{proposition}\label{pro:theta-unique}
		Let Condition \ref{cd:M1M2eigenvalues} hold. For a given $(d_1d_2) \times d$ matrix $\bW$ in \eqref{eq:iduv} such that $\bW^{\T}\bW=\bI_d$ and $\mathcal{M}(\bW)=\mathcal{M}(\bU\odot\bV)$, if $\bOmega$ defined in \eqref{eq:omega-D} satisfies ${\rm rank}(\bOmega)=d(d-1)/2$, then $\bTheta$ in \eqref{eq:iduv} can be uniquely identified by the  non-orthogonal joint diagonalization $\bH_m =\bTheta \bGamma_m \bTheta^\T$, $m\in [d]$, 
		up to the column reflection and permutation indeterminacy, and $\bTheta$ is invariant with respect to the choice of the basis vectors of 
		${\rm ker}(\mathbf{\Omega})$. 
	\end{proposition}

	Proposition
	\ref{pro:Psitouniqueness} provides a sufficient condition under which ${\rm rank}(\bOmega)=d(d-1)/2$. Such sufficient condition holds automatically when $d_1=d_2=d$, as then $\mathcal{R}(\bA) = \mathcal{R}(\bB)=d$. When $d_1\neq d_2$, we assume $d_2<d_1$ without loss of generality. For the scenario $d_2<d_1=d$, since $\mathcal{R}(\bA)=d$,   Proposition \ref{pro:Psitouniqueness} indicates that ${\rm rank}(\bOmega)= d(d-1)/2$ if $\mathcal{R}(\bB) \ge 2$. Actually, the requirement  $\mathcal{R}(\bB) \ge 2$ is necessary for the identification of $(\bA, \bB)$ when $d_2<d_1=d$.    Recall $\vec \bY_{t}=(\bB\odot \bA)\bx_{t} +  \vec \beps_{t}$. If $\mathcal{R}(\bB)=1$, since $|\bb_{\ell}|_2=1$ for each  $\ell\in[d]$, we can assume $\bb_{2}=\bb_{1}$ without loss of generality.  
	Let $\tilde{\bB}=\bB$ and $\tilde{\bA}=(\tilde{\ba}_1,\ldots, \tilde{\ba}_d)$, where $\tilde{\ba}_{\ell}=\ba_{\ell}$ for  any $\ell \ge 2$, and  $\tilde{\ba}_1=c_1\ba_1+c_2\ba_2$ for some nonzero constants $c_1, c_2$ such that $|\tilde{\ba}_1|_2=1$.
	Select $\bXi=(\xi_{i,j})$ with $\xi_{1,1}=c_1$, $\xi_{2,1}=c_2$, $\xi_{i,i}=1$ for any $ 2 \le i \le d$,  and $\xi_{i,j}=0$ otherwise. Then $\bY_t$ can be also formulated by another matrix CP-factor model $ \vec\bY_{t} = (\tilde{\bB} \odot \tilde{\bA}) \bXi^{-1}\bx_{t} + \vec \beps_{t}$.
	
	\begin{proposition}\label{pro:Psitouniqueness}
		Let $d\ge 2$, and Conditions \ref{cd:ra} and
		\ref{cd:M1M2eigenvalues} hold.
		Then ${\rm rank}(\bOmega)= d(d-1)/2$ provided that
		$\mathcal{R}(\bA) + d_2 \ge d + 2$ and $\mathcal{R}(\bB) + d_1 \ge d + 2$.	\end{proposition}

	By Propositions \ref{pro:rank-1} and \ref{pro:theta-unique}, if ${\rm rank}(\bOmega)=d(d-1)/2$, then $\bU$ and $\bV$ specified in \eqref{eq:FA} can be uniquely defined up to the column reflection and permutation indeterminacy, which implies $\bA$ and $\bB$ can be uniquely defined up to the column reflection and permutation indeterminacy. For $d\geq2$, Proposition 
	\ref{pro:Omega-impossible} shows that the requirement ${\rm rank}(\bOmega)=d(d-1)/2$ is necessary for identifying $(\bA,\bB)$, and it is impossible to obtain the consistent estimators for $(\bA, \bB)$ without such requirement.

	\begin{proposition}
		\label{pro:Omega-impossible}
		Let $d\ge 2$. Consider the following parameter space for the matrix CP-factor model \eqref{eq:abm}:
		\begin{align*}
			&\mathcal{U} = \big\{(\bA,\bB): \bA=(\ba_1,\ldots, \ba_d)~\textrm{and}~\bB=(\bb_1,\ldots, \bb_d)~\textrm{with}~|\ba_\ell|_2 =1= |\bb_\ell|_2 \\
			&~~~~~~~~~~~~~~~~~~~~~~~~\, \textrm{for each}~\ell \in [d],  ~\textrm{and}~ {\rm rank}(\bOmega) < d(d-1)/2 ~\textrm{with}~\bOmega~\textrm{defined as}~\eqref{eq:omega-D} \big\}\,.
		\end{align*}
		Write $\mathcal{G}=\{(\breve{\bA}, \breve{\bB}):\breve{\bA}=(\breve{\ba}_1, \ldots, \breve{\ba}_d) \in \mathbb{R}^{p\times d}, \breve{\bB}=(\breve{\bb}_1, \ldots,\breve{\bb}_d) \in \mathbb{R}^{q\times d} \}$ for the class of all measurable estimators of $(\bA,\bB)$ based on the data $\{\bY_t\}_{t = 1}^n$. Under Conditions \ref{cd:ra} and
		\ref{cd:M1M2eigenvalues}, it holds that
		\begin{equation*}
			\inf_{(\breve{\bA},\breve{\bB}) \in \mathcal{G}} \sup_{(\bA,\bB) \in 
				\mathcal{U}} \mathbb{P} \bigg[ \max \{\mathscr{D}(\breve{\bA},\bA), \mathscr{D}(\breve{\bB},\bB)\} \ge \frac{1}{8} \bigg] \ge \frac{1}{2} \,,
		\end{equation*}
		where $\mathscr{D}(\breve{\bA},\bA) = \max_{\ell \in[d]} | \breve{\ba}_{\ell} - \ba_{\ell}|_2$ and $\mathscr{D}(\breve{\bB},\bB) = \max_{\ell \in[d]} | \breve{\bb}_{\ell} - \bb_{\ell}|_2$.
	\end{proposition}

	\section{Estimation}\label{sec:estimation}
	
	Based on Section \ref{sec:identify-ab}, we can estimate $d$ and
	$(\ba_{\ell}, \bb_{\ell})$ for $\ell \in[d]$ via the following five steps: 
	\begin{enumerate}[{\it Step 1.}]
		\item  Based on \eqref{eq:definition of PQ}, we can obtain the estimates for $d_1, \, d_2$, $\bP$ and   $\bQ$, denoted by $\hat{d}_1$, $\hat{d_2}$, $\hat{\bP}$ and ${\hat{\bQ}}$, respectively.
		\item  Based on \eqref{eq:facm}, we can obtain the estimates for $d$ and $\bW$ (the orthogonal basis of $\mathcal{M}(\bV\odot\bU)$) with replacing $\bZ_t$ by $\hat{\bZ}_t =  \hat{\bP}^{\T}\bY_{t}\hat{\bQ} $.  Denote by $\hat{d}$ and $\hat{\bW} $ the associated estimators. 
		\item  With replacing $\bW$ involved in \eqref{eq:iduv} by $\hat{\bW}$, we can use the joint diagonalization algorithm mentioned in Section \ref{sec:theta-iden} to obtain $\hat{\bTheta}$, the estimate of $\bTheta$ involved in \eqref{eq:iduv}. % such that $\hat{\bC}_1,\ldots,\hat{\bC}_{\hat{d}}$ defined as $\hat{\bC} =({\rm vec}(\hat{\bC}_1), \ldots, {\rm vec}(\hat{\bC}_{\hat{d}})) =\hat{\bW} \hat{\bTheta}$ satisfy ${\rm rank}(\hat{\bC}_\ell)=1$ for each $\ell\in[\hat{d}]$.
		\item  Let $\hat{\bC} =({\rm vec}(\hat{\bC}_1), \ldots, {\rm vec}(\hat{\bC}_{\hat{d}})) =\hat{\bW} \hat{\bTheta}$. For each $\ell\in[\hat{d}]$, we select $\hat{\bu}_{\ell}$ and $\hat{\bv}_{\ell}$, respectively, as the unit eigenvectors corresponding to the largest eigenvalues of $\hat{\bC}_\ell\hat{\bC}_\ell^{\T}$ and $\hat{\bC}_\ell^{\T}\hat{\bC}_\ell$. Based on Proposition \ref{pro:rank-1}, we can estimate $(\bu_\ell,\bv_\ell)$ by $(\hat{\bu}_\ell,\hat{\bv}_\ell)$ for each $\ell\in[\hat{d}]$.    
		\item   Based on 
		\eqref{eq:FA}, we can estimate $\bA $ and $\bB$, respectively, by $\hat{\bA}=\hat{\bP}(\hat{\bu}_1, \ldots, \hat{\bu}_{\hat{d}})$ and $\hat{\bB}=\hat{\bQ}(\hat{\bv}_1, \ldots, \hat{\bv}_{\hat{d}})$.
	\end{enumerate}

	Steps 4 and 5 are straightforward. More details of Steps 1--3 are given, respectively, in Sections \ref{sec:phqh-est}--\ref{sec:theta-hat-est}. Especially Step 3 involves a further rotation to improve the convergence rate of the estimation. All the estimation is based on observations $\{\bY_t\}_{t=1}^{n}$.

	\subsection{Estimating $d_1, \, d_2, \, \bP$ and $\bQ$}\label{sec:phqh-est}
	Let $\xi_t$ be a prescribed linear combination of $\bY_t$ (e.g. the first principal component of
	$\vec \bY_t$), and $K> 1$ be a prescribed integer. Based on \eqref{eq:M1M2}, we put
	\begin{align}
		\label{eq:m1h-m2h}
		\hat{\bM}_1=&~\sum_{k=1}^{K}T_{\delta_1}\{\hat{\bSigma}_{\bY,\xi}(k)\}T_{\delta_1}
		\{\hat{\bSigma}_{\bY,\xi}(k)^{{\T}}\} \,, \notag\\ \hat{\bM}_2=&~\sum_{k=1}^{K}T_{\delta_1}\{\hat{\bSigma}_{\bY,\xi}(k)^{{\T}}\}T_{\delta_1}\{\hat{\bSigma}_{\bY,\xi}(k)\}\,, 
	\end{align}
	where $T_{\delta_1}(\cdot)$ is a truncation operator with the threshold level $\delta_1\geq0$, i.e., $T_{\delta_1}(\bS)=(s_{i,j}I(|s_{i,j}|\ge \delta_1))$ for any matrix $\bS =(s_{i,j})$,  and 
	\begin{equation}
		\label{eq:esthatsig}
		\hat{\bSigma}_{\bY,\xi}(k) =\frac{1}{n-k}\sum_{t=k+1}^{n}(\bY_t
		-\bar{\bY})(\xi_{t-k} -\bar{\xi})\,,~~~~ k\in[K]\,.
	\end{equation}
	We set $\delta_1>0$ in \eqref{eq:m1h-m2h} when $pq\ge n$. Note that $\hat \bM_1$ is a $p\times p$ matrix, and $\hat \bM_2$ is a $q\times q$ matrix.
	By Condition \ref{cd:M1M2eigenvalues}, we can estimate $d_1$ and $d_2$ by the eigenvalue-ratio based method \citep{Chang2015} as follows:
	\begin{equation}
		\label{eq:d1h}
		\hat{d}_1=\arg\min_{j\in[
			p]}\frac{\lambda_{j+1}(\hat{\bM}_1)+c_{1,n}}
		{\lambda_{j}(\hat{\bM}_1)+c_{1,n}}~~ {\rm and}
		~~
		\hat{d}_2=\arg\min_{j\in[q]}\frac{\lambda_{j+1}(\hat{\bM}_2)+c_{2,n}}{\lambda_{j}(\hat{\bM}_2)+c_{2,n}}
	\end{equation}
	for some $c_{1,n}, c_{2,n} \to 0^+$ as $n\rightarrow\infty$. The proposed eigenvalue-ratio based method here is an extension of that in \cite{lam2012factor}. Adding  $c_{1,n}$ and $c_{2,n}$ is to avoid the technical difficulties associated with handling potential ``0/0'' cases and can lead to consistent estimates for $d_1$ and $d_2$. See Theorem \ref{thm:rank} in Section \ref{sec:asymptotics} for details. In contrast, the eigenvalue-ratio based method proposed in \cite{lam2012factor} without adding $c_{1,n}$ and $c_{2,n}$ only ensures that the numbers of factors are not underestimated, without providing consistency.

	%, which are introduced to control ${\color{red}``}0/0"$.
	
	Perform the spectral decomposition for the non-negative definite matrices $\hat \bM_1$ and $\hat \bM_2$. Let $\hat \bP$  be the $p\times \hat d_1$ matrix of which the columns are the $\hat d_1$ orthonormal eigenvectors of $\hat \bM_1$ corresponding to its $\hat d_1$ largest eigenvalues,
	and $\hat \bQ$  be the $q\times \hat d_2$ matrix of which the columns are the $\hat d_2$ orthonormal eigenvectors of $\hat \bM_2$ corresponding to its $\hat d_2$ largest eigenvalues. Now we are ready to reduce the original $p\times q$ process $\bY_t$ to the $\hat d_1 \times \hat d_2$ process
	\begin{equation*}
		%\label{eq:zthat}
		\hat \bZ_t = \hat \bP^\T \bY_t \hat \bQ \,,~~~~t\geq1\,.
	\end{equation*}
	
	\subsection{Estimating $d$ and $\bW_1, \ldots, \bW_d$}\label{sec:wh-est}
	
	Based on \eqref{eq:facm} and \eqref{eq:iduv},  we can reformulate  \eqref{eq:facm} as  $\vec \bZ_{t} = \bW \bx_{t}^{*} +\vec \bDelta_{t}$ with $\bx_{t}^{*} = \bTheta\bx_{t}$. 
	Hence, we can estimate a factor loading matrix $\bW = (\vec \bW_1, \ldots, \vec\bW_d)$ based on the method proposed in  \cite{lam2011estimation}, \cite{lam2012factor} and \cite{Chang2015}.  To do this, we put
	\begin{align}\label{eq:mh}
		\hat{\bM} =  \sum_{k=1}^{\tilde{K}} \hat{\bSigma}_{\vec{\bZ}}(k)\hat{\bSigma}_{\vec{\bZ}}(k)^{{\T}}  
	\end{align}  
	with a prescribed integer $\tilde{K}\ge 1$ and
	\begin{align}\label{eq:zh}
		\hat{\bSigma}_{\vec{\bZ}}(k)= (\hat{\bQ}^{{\T}} \otimes \hat{\bP}^{{\T}}) T_{\delta_2}\{\hat{\bSigma}_{\vec{\bY}}(k)\}  (\hat{\bQ} \otimes \hat{\bP})\,,~~~~k\in[\tilde{K}]\,,
	\end{align}
	where  $T_{\delta_2}(\cdot) $ is a truncation operator with the threshold level $\delta_{2} \ge 0$, and 
	\begin{align*}
		\hat{\bSigma}_{\vec{\bY}}(k)=\frac{1}{n-k}\sum_{t=k+1}^{n}(\vec{\bY}_{t}-\bar{\vec{\bY}})(\vec{\bY}_{t-k}-\bar{\vec{\bY}})^{\T} \,, ~~~~k\in[\tilde{K}]\,,
	\end{align*} 
	with  $\bar{\vec{\bY}}= n^{-1}\sum_{t=1}^{n}\vec{\bY}_{t}$.
	Analogous to \eqref{eq:d1h}, we can estimate $d$ as  
	\begin{equation}
		\label{eq:dh}
		\hat{d}=\bigg\{\arg\min_{j\in [\hat d_1 \hat d_2]}\frac{\lambda_{j+1}(\hat{\bM})+c_{3,n}}{\lambda_{j}(\hat{\bM})+c_{3,n}}\bigg\}I(\hat{d}_1\hat{d}_2 \ge 2) + I(\hat{d}_1\hat{d}_2 =1) 
	\end{equation}
	for some $c_{3,n} \to 0^+$ as $n\to \infty$.
	Furthermore we let
	$\hat \bW \equiv ( {\rm vec}(\hat \bW_1), \ldots, {\rm vec}(\hat
	\bW_{\hat d}))$ be the $(\hat d_1 \hat d_2)\times \hat d$ matrix of which the columns are 
	the $\hat d$ orthonormal eigenvectors of $\hat \bM$ corresponding to its largest $\hat d$ eigenvalues.

	\begin{remark}\label{rek:two-stage}
		We can also consider an alternative two-stage procedure to estimate  $d$ and $\bW$ in Step 2. Notice that $\vec\bY_t=(\bB \odot \bA)\mathbf{x}_t + \vec\beps_t$ for $t \ge 1$. We can firstly obtain the estimates of $d$ and $\bT$ (the orthogonal basis of $\mathcal{M}(\bB \odot \bA)$), denoted by $\hat{d}$ and $\hat{\bT}$, based on the method proposed in \cite{lam2011estimation}, \cite{lam2012factor} and \cite{Chang2015}. Recall $\bV \odot \bU = (\bQ \otimes \bP)^{\T}(\bB \odot \bA)$ and $\bW$ is an orthogonal basis of $\mathcal{M}(\bV \odot \bU)$. Based on $(\hat{\bP},\hat{\bQ})$, the estimates of $\bP$ and $\bQ$ obtained in Step 1, we can then estimate $\bW$ by $ (\hat{\bQ} \otimes \hat{\bP})^{\T}\hat{\bT}$. Figure \ref{fig: compare-W} in the supplementary material shows that, although this alternative two-stage approach yields estimation errors nearly identical to those of our proposed method, it is considerably more computationally expensive when $p$ and $q$ are large. This is because the first stage of this alternative approach requires an eigen-decomposition of a $(pq) \times (pq)$ matrix defined based on the sample auto-covariance matrices of $\{\vec{\mathbf{Y}}_t\}_{t=1}^{n}$, whereas Step 2 of our proposed method only involves an eigen-decomposition of a $(\hat{d}_1\hat{d}_2) \times (\hat{d}_1\hat{d}_2)$ matrix. This indicates that our proposed method can significantly reduce the computational complexity, especially in high-dimensional settings.
	\end{remark}

	\subsection{Estimating $\mathbf{\Theta}$ via joint diagonalization}\label{sec:theta-hat-est}
	
	For 4-way tensor $\bPsi(\cdot,\cdot)$ defined in Section \ref{sec:theta-iden}, we define a $(\hat d_1^2\hat d_2^2)\times \hat{d}(\hat d+1)/2$ matrix  $\hat \bOmega$ as follows:
	\begin{align}\label{eq:omega-h}
		\hat{\bOmega} = \big(\vec \bPsi(\hat\bW_1, \hat\bW_1), \ldots, \vec \bPsi(\hat\bW_1, \hat\bW_{ \hat{d} }),
		\vec\bPsi(\hat\bW_2, \hat\bW_2), \ldots, \vec\bPsi(\hat\bW_{ \hat{d} }, \hat\bW_{ \hat{d} }) \big)\,,
	\end{align}
	which is an estimate of $\bOmega$ defined as in  \eqref{eq:omega-D}. Let
	\begin{equation}
		\label{eq:init-basis}
		\tilde \bh_m=
		(\tilde h_{1,1}^m, \ldots, \tilde h_{1, \hat d}^m, \tilde h_{2,2}^m, \ldots, \tilde h_{\hat d,\hat d}^m)^\T\,, ~~~~ m\in [\hat{d}]\,,
	\end{equation}
	be  the right-singular vectors of $\hat \bOmega$ corresponding to the $\hat d$ smallest singular values. Such selected $\{\tilde{\bh}_m\}_{m=1}^{\hat{d}}$  provides the estimate for a basis of ${\rm ker}(\bOmega)$. By Proposition \ref{pro:theta-unique}, an estimator for $\bTheta$ can be obtained by the joint diagonalization of $\tilde \bH_1, \ldots, \tilde{\bH}_{\hat{d}}$,  which are constructed in the same manner as $\bH_m$ with $\bh_m$ replaced by $\tilde \bh_m$. See the statement below \eqref{eq:off-diagonal}. 
	
	Though $\bTheta$ can be uniquely identified by any set of basis $\{ \bh_m\}_{m=1}^{d}$ of ${\rm ker}(\bOmega)$ (see Proposition \ref{pro:theta-unique}), the accuracy of its estimator depends on the choice of $\{ \bh_m\}_{m=1}^{d}$ sensitively. Motivated by Proposition \ref{pro:rotation} at the end of this section, a good choice  is to rotate
	the basis vectors $\{\tilde{\bh}_m\}_{m=1}^{\hat{d}}$ in \eqref{eq:init-basis} first. More specifically,  let $(\hat{\bh}_1, \ldots, \hat{\bh}_{\hat{d}}) = (\tilde{\bh}_1,\ldots, \tilde{\bh}_{\hat{d}}) \hat \bPi$ with 
	\begin{align*}
		\hat \bPi = \{2(\hat \bUpsilon_0^\T \hat\bUpsilon_2)( \hat\bUpsilon_1^\T \hat\bUpsilon_2 +   \hat\bUpsilon_2^\T \hat\bUpsilon_1)^{-1}( \hat\bUpsilon_2^\T \hat\bUpsilon_0) \}^{-1/2}\,,
	\end{align*}
	where
	\begin{gather*}
		\hat\bUpsilon_0   =
		\big(\textup{vec}(\tilde\bH_1),\ldots,\textup{vec}(\tilde\bH_{\hat{d}})\big) \, , ~~
		\hat\bUpsilon_1   =
		\big(\textup{vec}(\tilde\bH^{-1}\tilde\bH_1),\ldots,\textup{vec}(\tilde\bH^{-1}\tilde\bH_{\hat d})\big)\,,\\
		\hat\bUpsilon_2   = \big(\textup{vec}(\tilde\bH_1\tilde\bH^{-1}),\ldots,\textup{vec}(\tilde\bH_{\hat{d}}\tilde\bH^{-1})\big)\,, ~~
		\tilde \bH = \sum_{m=1}^{\hat d} \phi_m \tilde \bH_m
	\end{gather*}
	for some $\hat d$-dimensional vector $(\phi_1, \ldots, \phi_{\hat d})^\T \ne \bzero$ such that $\tilde{\bH}$ is invertible. Section \ref{sec:simulation setting up} specifies how to select $(\phi_1, \ldots, \phi_{\hat d})^\T$ in practice.   Define $\hat{\bH}_{1},\ldots, \hat{\bH}_{\hat{d}}$ in the same manner as $\bH_{m}$ but with replacing  $\bh_m$ by $\hat{\bh}_m$. Utilizing the fast Frobenius diagonalization  algorithm introduced by \cite{ziehe2004fast}, we can obtain the non-orthogonal joint diagonalizer $\bPhi$ for $\hat{\bH}_1, \ldots, \hat{\bH}_{\hat{d}}$ such that all the columns of  $\bPhi^{-1}$ are unit vectors. Then, $\bTheta$ involved in \eqref{eq:iduv} can be estimated by  $\hat{\bTheta} =\bPhi^{-1}$.

	Now we give some illustrations on how the set of basis $\{\bh_m\}_{m=1}^d$ of ${\rm ker}(\bOmega)$ used to identify $\bTheta$ affects the convergence rate of the associated estimator of $\bTheta$ based on the non-orthogonal joint diagonalization. Recall
	\[
	\bH_m= \bTheta \,{\rm diag}(\gamma_{1,m} ,
	\ldots, \gamma_{d,m}) \,\bTheta^\T, ~~~~ m\in[d]\,.
	\]
	By Theorem 3 of \cite{afsari2008sensitivity},  the convergence
	rate of the estimator for $\bTheta$ based on the fast Frobenius diagonalization algorithm  is bounded by 
	\[
	\frac{\eta({\bh_1}, \ldots, \bh_d)}{1-\rho^2({\bh}_1, \ldots, \bh_d) } \times K_n(d,\bTheta) \,,
	\]
	where $K_{n}(d, \bTheta)$ is a universal  quantity only depending on $(n,d, \bTheta)$, and
	\begin{align}\label{eq:alpha and rho}
		\rho({\bh}_1, \ldots, \bh_d) &= \max_{k,\ell\in[d]:\,k\ne \ell}\frac{|\sum_{m =
				1}^d\gamma_{k,m}\gamma_{\ell,m}|}{(\sum_{m =
				1}^d\gamma_{k,m}^2)^{1/2} (\sum_{m = 1}^d\gamma_{\ell,m}^2)^{1/2}}
		\, , \notag\\
		\eta({\bh}_1, \ldots, \bh_d) &= \max_{k,\ell\in[d]:\,k\ne\ell}\bigg(\frac{1}{\sum_{m = 1}^d\gamma_{\ell,m}^2} + \frac{1}{\sum_{m = 1}^d\gamma_{k,m}^2}\bigg)\,.  
	\end{align}
	Ideally we should choose $\{ \bh_m\}_{m=1}^{d}$ such that
	$\rho(\bh_1, \ldots, \bh_d)=0$ and $\eta(\bh_1, \ldots, \bh_d)$ as
	small as possible. For any given set of basis $\{\bh_m\}_{m=1}^d$ of ${\rm ker}(\bOmega)$, Proposition \ref{pro:rotation} indicates that we should replace $\{\bh_m\}_{m=1}^d$ by its rotation $\{\bh_m^*\}_{m=1}^d$ such that 
	$(\bh_1^*, \ldots, \bh_d^{*}) = (\bh_1, \ldots, \bh_d) \bPi $ with 
	\begin{align*}
		\bPi = \{2(\bUpsilon_0^\T\bUpsilon_2)(\bUpsilon_1^\T\bUpsilon_2 +  \bUpsilon_2^\T\bUpsilon_1)^{-1}(\bUpsilon_2^\T\bUpsilon_0) \}^{-1/2} \,,
	\end{align*}
	where
	\begin{gather}\label{eq:bUpsilon_k}
		\bUpsilon_0   =
		\big(\textup{vec}(\bH_1),\ldots,\textup{vec}(\bH_d)\big) \, ,  ~~
		\bUpsilon_1   =
		\big(\textup{vec}(\bH^{-1}\bH_1),\ldots,\textup{vec}(\bH^{-1}\bH_d)\big)\,, \notag\\
		\bUpsilon_2   = \big(\textup{vec}(\bH_1\bH^{-1}),\ldots,\textup{vec}(\bH_d\bH^{-1})\big)\,, ~~ \bH = \sum_{m=1}^d \phi_m \bH_m 
	\end{gather}
	for some $d$-dimensional vector $(\phi_1, \ldots, \phi_d)^\T \ne \bzero$ such that $\bH$ is invertible. 
	%where $\phi_1, \ldots, \phi_d$ are any constants and at least one of them is nonzero. 
	
	\begin{proposition}
		\label{pro:rotation}
		$\rho(\bh^*_1, \ldots, \bh_d^*) =0$ and $\eta(\bh^*_1, \ldots, \bh_d^*) =2$.
	\end{proposition}

	\section{Prediction}\label{sec:prediction} 
	Given observations $\{\bY_t\}_{t=1}^n$, we can also use the matrix CP-factor model \eqref{eq:abm} to forecast the future values $\bY_{n+h}$ for $h\ge 1$. More specifically, we can predict $\bY_{n+h}$ by recovering the latent process $\{\bX_t\}_{t=1}^n$. Let  
	$\hat{\bL} = \hat{\bB} \odot \hat{\bA} $ with $\hat{\bA}\in\mathbb{R}^{p\times\hat{d}}$ and $\hat{\bB}\in\mathbb{R}^{q\times\hat{d}}$ being, respectively, the estimates of the factor loading matrices $\bA$ and $\bB$ in the matrix CP-factor model \eqref{eq:abm}. If ${\rm rank}(\hat{\bL})=\hat{d}$, we can recover $\bX_t$ by $\hat{\bX}_t = \text{diag}(\hat{\bx}_t)$ with $\hat{\bx}_t=\hat{\bL}^{+}\bY_t = (\hat{x}_{t,1},\ldots,\hat{x}_{t,\hat{d}})^\T$. In order to predict $\bY_{n+h}$, we only need to fit a $\hat{d}$-dimensional multivariate time series model for $\{\hat{\bx}_t\}^n_{t=1}$. Then we can predict $\bY_{n+h}$ by $\hat{\bY}_{n+h} = \hat{\bA}\tilde{\hat{\bX}}_{n+h}\hat{\bB}^\T$ with $\tilde{\hat{\bX}}_{n+h} = \textup{diag} (\tilde{\hat{\bx}}_{n+h})$, where $\tilde{\hat{\bx}}_{n+h}$ is the $h$-step ahead forecast of  $\hat{\bx}_{n+h}$ based on the fitted model for  $\{\hat{\bx}_t\}^n_{t=1}$.  \cite{chang2023modelling} uses this idea to predict $\bY_{n+h}$ under the assumption $d_1 = d_2 = d$ based on the CP-refined estimate considered there for $(\bA,\bB)$. Since the CP-refined estimate \citep{chang2023modelling} does not work if the assumption $d_1=d_2=d$ is not satisfied, we cannot select $(\hat{\bA},\hat{\bB})$ as the CP-refined estimate to recover $\bX_t$ in these cases. When $d_1=d_2=d$ is not satisfied, if the factor loading matrices $\bA$ and $\bB$ can be uniquely identified, we can select $(\hat{\bA},\hat{\bB})$ as our newly proposed estimate specified in Section \ref{sec:estimation}, and use the same idea to predict $\bY_{n+h}$.
	
	As shown in Proposition \ref{pro:Omega-impossible}, if ${\rm rank}(\bOmega) < d(d-1)/2$, the factor loading matrices $\bA$ and $\bB$ cannot be uniquely identified, which implies that we cannot recover $\bX_t$ successfully. Hence, above mentioned strategy for predicting $\bY_{n+h}$ does not always work. A natural question is that whether we can propose a unified prediction procedure for $\bY_{n+h}$ based on the matrix CP-factor model \eqref{eq:abm} without any assumption on the relationship among $d_1$, $d_2$ and $d$. By \eqref{eq:FA} and \eqref{eq:zt}, we have 
	\begin{align}\label{eq:newmodel}
		\bY_t=\bP\bZ_t\bQ^{\T}+\underbrace{\beps_t - \bP\bP^{\T}\beps_t\bQ\bQ^{\T}}_{{\rm white~noise}}
	\end{align}
	with $\bZ_t = \bP^{\T}\bY_t\bQ$. In order to predict $\bY_{n+h}$, we only need to predict $\bZ_{n+h}$. For $(\hat{\bP},\hat{\bQ},\hat{\bW})$ specified in Sections \ref{sec:phqh-est} and \ref{sec:wh-est}, we define 
	\begin{align*}
		\hat{\bx}^*_t   =  \hat{\bW} ^{\T}{\rm vec}(\hat{\bP}^{\T}\bY_{t}\hat{\bQ})\,,~~~~ t\in[n]\,.
	\end{align*}
	Proposition \ref{pro:PQW} in Section \ref{sec:asymptotics} indicates that such defined $\hat{d}$-dimensional vector $\hat{\bx}_t^*$ provides a  recovery of $\mathbf{E}_3\bx_t^*$, where  $ \bx_t^* = \bTheta\bx_{t} $,  and $\mathbf{E}_3$ is an orthogonal matrix specified in Proposition \ref{pro:PQW}. Hence, we can fit a $\hat{d}$-dimensional vector time series model for $\{\hat{\bx}_t^*\}^n_{t = 1}$.  
	Let  $\tilde{\hat{\bx}}_{n+h}^*$ be the $h$-step ahead forecast of $\hat{\bx}_{n+h}^*$. By \eqref{eq:facm} and Proposition \ref{pro:PQW}, we know $\hat{\bW}\tilde{\hat{\bx}}_{n+h}^*$ provides a prediction of $(\bE_2 \otimes \bE_1)\vec{\bZ}_{n+h}$, where $\bE_1$ and $\bE_2$ are two orthogonal matrices specified in Proposition \ref{pro:PQW}. Let $\hat{\bZ}_{n+h}$ satisfy ${\rm vec}(\hat{\bZ}_{n+h})= \hat{\bW} \tilde{\hat{\bx}}_{n+h}^*$. Applying Proposition \ref{pro:PQW} again, by \eqref{eq:newmodel}, we know 
	$\hat{\bY}_{n+h} = \hat{\bP}\hat{\bZ}_{n+h}\hat{\bQ}^{\T}$ provides a prediction of $\bY_{n+h}$. This new prediction idea only depends on the calculation of three matrices $\hat{\bP}$, $\hat{\bQ}$ and $\hat{\bW}$. As we have discussed in Sections \ref{sec:phqh-est} and \ref{sec:wh-est}, determining $\hat{\bP}$, $\hat{\bQ}$ and $\hat{\bW}$ only involves the spectral decomposition of $\hat{\bM}_1$, $\hat{\bM}_2$ and $\hat{\bM}$, respectively, which does not require any additional assumption on the relationship among $d_1$, $d_2$ and $d$. Hence, our newly proposed prediction strategy provides a unified prediction procedure for $\bY_{n+h}$ based on the matrix CP-factor model \eqref{eq:abm} regardless of the relationship among $d_1$, $d_2$ and $d$. When the linear dynamic structure is concerned for the latent process $\bX_t$, our numerical studies in Section \ref{sec:sim-est} indicate that if the factor loading matrices $\bA$ and $\bB$ can be uniquely identified, the finite-sample performance of our newly proposed prediction method is almost identical to the prediction idea considered in \cite{chang2023modelling} with selecting $(\hat{\bA},\hat{\bB})$ as our proposed estimate of $(\bA,\bB)$ specified in  Section \ref{sec:estimation}. However, if the factor loading matrices $\bA$ and $\bB$ cannot be uniquely identified, our newly proposed prediction method outperforms that of \cite{chang2023modelling}.

	\section{Asymptotic properties}\label{sec:asymptotics}
	
	As we do not impose the stationarity on $\{\bY_t\}$, we use the concept of ``$\alpha$-mixing'' to characterize
	the serial dependence of $\{\bY_t\}$ with the $\alpha$-mixing coefficients defined as
	\begin{align}\label{eq:mixinga}
		\alpha(k)=\sup_{r}\sup_{A\in \mathcal{F}_{-\infty}^{r},B\in \mathcal{F}_{r+k}^{\infty} } |\mathbb{P}(AB)-\mathbb{P}(A)\mathbb{P}(B)|\,, ~~~~ k\ge 1\,,
	\end{align}
	where $\mathcal{F}_{r}^{s}$ is the $\sigma$-filed generated by $\{\bY_t: r\le t\le s\}$. Write 
	\begin{align*}
		\bSigma_{\vec{\bY}}(k) = \frac{1}{n-k}\sum_{t=k+1}^{n} \mathbb{E}[\{\vec{\bY}_{t}-\mathbb{E}(\bar{\vec{\bY}})\}\{\vec{\bY}_{t-k}-\mathbb{E}(\bar{\vec{\bY}})\} ^{\T}]\,,~~~~ k\ge 1\,,
	\end{align*}
	where $\bar{\vec{\bY}}= n^{-1}\sum_{t=1}^{n}\vec{\bY}_{t}$. Define $\bM = \sum_{k=1}^{\tilde{K}} \bSigma_{\vec{\bZ}}(k) \bSigma_{\vec{\bZ}}(k)^{\T} $ with $\tilde{K}$ given in \eqref{eq:mh} and
	\begin{align*}
		\bSigma_{\vec{\bZ}}(k) =\frac{1}{n-k}\sum_{t=k+1}^{n} \mathbb{E} [\{ \vec{\bZ}_t - \mathbb{E}(\bar{\vec{ \bZ}} )  \} \{ \vec{\bZ}_{t-k}- \mathbb{E}(\bar{\vec{\bZ}} )\}^{\T} ]\,,~~~~k\ge1\,,
	\end{align*}
	where $\bar{\vec{\bZ}}=n^{-1}\sum_{t=1}^{n}\vec \bZ_{t}$.   Following the arguments in \cite{Chang2015},   we can identify $d$ as $d={\rm rank}(\bM)$, and  select $\vec\bW_1,\ldots, \vec\bW_d$ involved in \eqref{eq:c=wtheta} as the $d$ orthonormal eigenvectors of $\bM$ corresponding to the $d$ nonzero eigenvalues  $\lambda_1(\bM) \ge \cdots \ge \lambda_d(\bM) > 0 $, i.e.,  $\vec\bW_{\ell}$ is the eigenvector associated with the eigenvalue $\lambda_{\ell}(\bM)$ for $\ell\in[d]$.  We need the following regularity conditions in our theoretical analysis.
	
	\begin{cd}\label{cd:bounded-value}
		{\rm(i)} The nonzero singular values of $\bB \odot \bA$ are uniformly bounded away from zero. {\rm(ii)} The nonzero eigenvalues of $\bM_1$,  $\bM_2$ and $\bM$ are uniformly bounded away from zero. 
		% (iii) The nonzero eigenvalues of $\bM$ are distinct and uniformly bounded away from zero.
	\end{cd}
	
	\begin{cd}\label{cd:tail}
		{\rm (i)} There exist some universal constants $K_1>0$, $K_2>0$ and $r_1\in(0,2]$ such that 
		$
		\mathbb{P}(|y_{i,j,t}|>x)\le K_1 \exp(-K_2x^{r_1})$ and $\mathbb{P}(|\xi_t|>x)\le K_1 \exp(-K_2x^{r_1})$
		for any $x>0$,  $i\in[p] $, $j\in[q]$ and $t\in[n]$. {\rm (ii)} There exist some universal constants $K_3>0$, $K_4>0$ and $r_2\in (0,1]$ such that the $\alpha$-mixing coefficients $\alpha(k)$ defined as in \eqref{eq:mixinga} satisfy 
		$
		\alpha(k)\le K_3\exp(-K_4k^{r_2})$
		for $k\ge 1$.
	\end{cd}
	\begin{cd}\label{cd:sgm_yxi}
		{\rm (i)}	There exists a universal constant  $K_5>0$ such that
		$ \|\bSigma_{\bY,\xi}(k)\|_2\le K_5 $ for any $k\in[K]$, and $ \|\bSigma_{\vec{\bY}}(k)\|_2 \le K_5$ for any $k \in [\tilde{K}]$. 
		{\rm (ii)} Write $\bSigma_{\bY,\xi}(k)= (\sigma_{y,\xi,i,j}^{(k)})_{p\times q}$ and $\bSigma_{\vec{\bY}} (k) = (\sigma_{i,j}^{(k)})_{pq\times pq}$. There exists a universal constant $\iota  \in[0,1)$ such that
		\begin{align*}
			&\max_{i_1 \in [p]}\sum_{j_1=1}^{q}|\sigma_{y,\xi,i_1,j_1}^{(k)}|^{\iota} \le s_1\,,~~\max_{j_1 \in [q]}\sum_{i_1=1}^{p}|\sigma_{y,\xi,i_1,j_1}^{(k)}|^{\iota}  \le s_2\,,\\
			&\max_{i_2 \in [pq]}\sum_{j_2=1}^{pq}|\sigma_{i_2,j_2}^{(k)}|^{\iota} \le s_3~~\textup{and}~~\max_{j_2 \in [pq]}\sum_{i_2=1}^{pq}|\sigma_{i_2,j_2}^{(k)}|^{\iota} \le s_4\,,
		\end{align*}
		%     $
		% \max_{i_1 \in [p]}\sum_{j_1=1}^{q}|\sigma_{y,\xi,i_1,j_1}^{(k)}|^{\iota} \le s_1$, $\max_{j_1 \in [q]}\sum_{i_1=1}^{p}|\sigma_{y,\xi,i_1,j_1}^{(k)}|^{\iota}  \le s_2$, $\max_{i_2 \in [pq]}\sum_{j_2=1}^{pq}|\sigma_{i_2,j_2}^{(k)}|^{\iota} \le s_3$ and $\max_{j_2 \in [pq]}\sum_{i_2=1}^{pq}|\sigma_{i_2,j_2}^{(k)}|^{\iota} \le s_4$
		%     for any   $i_1 \in [p]$, $j_1 \in [q]$ and   $i_2,j_2 \in [pq]$,  
		where $s_1$, $s_2$, $s_3$ and $s_4$  may diverge together with $p$ and $q$,  respectively. 
	\end{cd}
	
	% \begin{cd}\label{cd:sy}
		% 	{\rm (i)}	There exists a universal constant $K_6>0$ such that $\|\bSigma_{\check{\bY}}(k)\|_2 \le K_6$ for any $k \in [\tilde{K}]$. {\rm (ii)} Write $\bSigma_{\check{\bY}}(k)= (\sigma_{i,j}^{(k)})_{pq\times pq}$. There exists a universal constant $\iota \in [0,1)$  such  that $\sum_{j=1}^{pq}|\sigma_{i,j}^{(k)}|^{\iota} \le s_3$ and $\sum_{i=1}^{pq}|\sigma_{i,j}^{(k)}|^{\iota} \le s_4$ for any $i,j \in [pq]$, where $s_3$ and $s_4$ may, respectively, diverge together with $p$ and $q$. 
		% \end{cd}
	Condition \ref{cd:bounded-value} is used to
	simplify the presentation for the results.  Our technical proofs indeed allow the nonzero singular   values of $\bB \odot \bA$, and the nonzero eigenvalues of $\bM_1$, $\bM_2$ and $\bM$   decay to zero as $p$ and/or $q$ grow to infinity.
	%{\color{blue} Furthermore, Condition \ref{cd:bounded-value}(iii) guarantees the uniqueness of $\bW$ and $\bW^{*}$ given in \eqref{eq:facm} and \eqref{eq:facorth}, which implies $\bTheta$ is uniquely defined. As shown in Lemma \ref{lemma:theta} in the supplementary material, we can derive the convergence rate  of the estimation error of $\bTheta$ directly. }
	Condition \ref{cd:tail} is also used in \cite{chang2023modelling}, which is a common assumption in the literature on ultrahigh-dimensional data analysis. See \cite{chang2023modelling} for the discussion of their validity. We impose Condition \ref{cd:sgm_yxi}(i) just for simplifying the presentation. Our technical proofs indeed allow $\max_{k \in [K]} \|\bSigma_{\bY,\xi}(k)\|_2$  and $ \max_{k \in [K]}\|\bSigma_{\vec{\bY}}(k)\|_2$ to diverge as $p$ and/or $q$ grow to infinity. Condition \ref{cd:sgm_yxi}(ii) imposes some sparsity requirement on $\bSigma_{\bY,\xi}(k)$ and $\bSigma_{\vec{\bY}}(k)$. 
	Under some sparsity condition on $\bA$ and $\bB$, applying the technique used to derive Lemma 5 of 
	\cite{Chang2018}, we can show that Condition \ref{cd:sgm_yxi}(ii) holds for certain $(s_1,s_2,s_3,s_4)$. Let 
	\begin{equation}\label{eq:Pi1 and Pi2}
		\Pi_{1,n}=(s_1s_2)^{1/2}\bigg\{\frac{\log(pq)}{n}\bigg\}^{(1-\iota)/2} ~~ \text{and} ~~ \Pi_{2,n}= (s_3s_4)^{1/2} \bigg\{\frac{\log(pq)}{n}\bigg\}^{(1-\iota)/2} \,.
	\end{equation}
	
	Theorem \ref{thm:rank} shows that the eigenvalue-ratio based estimators $\hat{d}_1$, $\hat{d}_2$ and $\hat{d}$ provide consistent estimates for $d_1$, $d_2$ and $d$, respectively.
	
	\begin{theorem}\label{thm:rank}
		
		Let Conditions \ref{cd:ra}--\ref{cd:sgm_yxi} hold. Select the threshold levels in \eqref{eq:m1h-m2h} and  \eqref{eq:zh} as
		\[
		\delta_1=\breve{C}\sqrt{\frac{\log(pq)}{n}}~~\textrm{and}~~   \delta_2=\tilde{C}\sqrt{\frac{\log(pq)}{n}}
		\]
		for some sufficiently large constants $\breve{C}, \tilde{C}>0$. For any $(c_{1,n},c_{2,n}, c_{3,n})$ given in \eqref{eq:d1h} and  \eqref{eq:dh} satisfying $\Pi_{1,n}  \ll c_{1,n},c_{2,n} \ll 1$ and $\max(\Pi_{1,n}, \Pi_{2,n}) \ll c_{3,n} \ll 1$, it holds that  
		\[
		\mathbb{P}\{(\hat{d}_1,\hat{d}_2,\hat{d})=(d_1,d_2,d)\} \to 1
		\]
		as $n \to \infty$,
		provided that $\Pi_{1,n}+\Pi_{2,n}\ll1$ and  $\log(pq)\ll n^c$ for some constant $c \in (0,1)$ depending only on $r_1$ and $r_2$.
		% (i) Let Conditions \ref{cd:ra}--\ref{cd:sgm_yxi} hold and the threshold level $\delta_1=\breve{C}\{n^{-1}\log(pq)\}^{1/2}$ for some sufficiently large constant $\breve{C}>0$. For any $c_{1,n}$ and $c_{2,n}$ in \eqref{eq:d1h} satisfying $\Pi_{1,n}  \ll c_{1,n} \ll 1$ and $\Pi_{1,n}  \ll c_{2,n} \ll 1$, it holds that $\mathbb{P}(\hat{d_1}=d_1) \to 1$ and $\mathbb{P}(\hat{d_2}=d_2) \to 1$ as $n \to \infty$ provided that $\Pi_{1,n}=o(1)$ and  $\log(pq)=o(n^c)$ for some constant $c \in (0,1)$ depending only on $r_1$ and $r_2$. (ii) Let Conditions \ref{cd:ra}--\ref{cd:sgm_yxi}  hold and the threshold level $\delta_2=\bar{C}\{n^{-1}\log(pq)\}^{1/2}$ for some sufficiently large constant $\bar{C} > 0$. For any $c_{3,n}$ in \eqref{eq:dh} satisfying $\max\{\Pi_{1,n}, \Pi_{2,n}\} \ll c_{3,n} \ll 1$, it holds that $\mathbb{P}(\hat{d}=d) \to 1$ as $n \to \infty$ provided that $\Pi_{1,n}=o(1)$, $\Pi_{2,n}=o(1)$ and  $\log(pq)=o(n^c)$ for some constant $c \in (0,1)$ depending only on $r_1$ and $r_2$.
	\end{theorem}

	% Recall the columns of $\bP$, $\bQ$ and $\bW^*$ are, respectively, the $d_1$, $d_2$ and $d$ orthonormal eigenvectors of $\bM_1$, $\bM_2$ and $\bM$ corresponding to the $d_1$, $d_2$ and $d$ nonzero eigenvalues of $\bM_1$, $\bM_2$ and $\bM$, where $\bM_1$, $ \bM_2$ and $\bM$ defined in \eqref{eq:M1M2} and \eqref{eq:matrixM}. 
	Proposition \ref{pro:PQW} states the asymptotic performance of $\hat{\bP}$, $\hat{\bQ}$ and $\hat{\bW}$.    
	
	\begin{proposition}\label{pro:PQW}
		Let Conditions \ref{cd:ra}--\ref{cd:sgm_yxi} hold. Select the threshold levels in \eqref{eq:m1h-m2h} and  \eqref{eq:zh} as 
		\[
		\delta_1=\breve{C}\sqrt{\frac{\log(pq)}{n}}~~\textrm{and}~~   \delta_2=\tilde{C}\sqrt{\frac{\log(pq)}{n}}
		\]
		for some sufficiently large constants $\breve{C}, \tilde{C}>0$. Assume that $\Pi_{1,n} + \Pi_{2,n}\ll1$ and  $\log(pq)\ll n^c$ for some constant $c \in (0,1)$ depending only on $r_1$ and $r_2$. If $(\hat{d}_1,\hat{d}_2)=(d_1,d_2)$, there exist some orthogonal matrices $\bE_1 \in \mathbb{R}^{d_1 \times d_1}$ and $\bE_2 \in \mathbb{R}^{d_2 \times d_2}$ such that 
		\[
		\|\hat{\bP}\bE_1  - \bP  \|_2=O_{\rm p}(\Pi_{1,n})=\|\hat{\bQ}\bE_2 - \bQ \|_2\,.
		\]
		Furthermore, if $(\hat{d}_1,\hat{d}_2,\hat{d})=(d_1,d_2,d)$, there exists an orthogonal matrix $\bE_3 \in \mathbb{R}^{d \times d}$ such that 
		\[
		\|(\bE_2 \otimes \bE_1)^{{\T}} \hat{\bW}\bE_3 -\bW \|_2=O_{\rm p}(\Pi_{1,n} + \Pi_{2,n})\,.
		\]
	\end{proposition}
	% Notice that the orthogonal matrices $\bE_1$, $\bE_2$ and $\bE_3$ represents the indeterminacy of these eigenvectors, which arises from reflections and/or possible tied (nonzero) eigenvalues. 

	If the nonzero eigenvalues of $\bM$ are distinct, $\bE_3$ will be a diagonal matrix with its diagonal elements being $1$ or $-1$. For the trivial case $d=1$, if $(\hat{d}_1,\hat{d}_2,\hat{d})=(d_1,d_2,d)$, we have $\bE_1 =\pm 1$ and $ \bE_2 = \pm 1$. Following the discussion below \eqref{eq:FA}, it holds in this trivial case that $|\hat{\bA}\bE_1  - \bA  |_2=O_{\rm p}(\Pi_{1,n})=|\hat{\bB}\bE_2 - \bB |_2 $   provided that $\Pi_{1,n} \ll 1$ and  $\log(pq)\ll n^c$ for some constant $c \in (0,1)$ depending only on $r_1$ and $r_2$. Note that $\mathbb{P}\{(\hat{d}_1,\hat{d}_2,\hat{d})=(d_1,d_2,d)\} \to 1$ as $n\to \infty$. Hence, in the trivial case $d=1$, $(\bA,\bB)$ can be consistently estimated up to the reflection indeterminacy.
	For the non-trivial case $d\ge 2$, the convergence rates of the estimation errors for $ \bA $ and $ \bB $ will be shown in Theorem   \ref{thm:hat_a_hat_b}. 
	
	To present Theorem \ref{thm:hat_a_hat_b}, we need to introduce some notation first. For $(\bE_1,\bE_2,\bE_3)$ specified in Proposition \ref{pro:PQW}, let $\breve{\bW} = (\bE_2 \otimes \bE_1) \bW  \bE_3^{\T}$. Define $\breve{\bOmega}$ in the same manner as $\bOmega$ given in \eqref{eq:omega-D} but with replacing $\bW$ by $\breve{\bW}$. 
	% For $\{\hat{\bh}_1^{\rm (init)}, \ldots, \hat{\bh}_{\hat{d}}^{\rm (init)}\}$ specified in \eqref{eq:h^(init)}, when $(\hat{d}_1,\hat{d}_2,\hat{d}) = (d_1,d_2,d)$, let $\{\breve{\bh}_{1} ,\ldots, \breve{\bh}_{d} \}$ be the basis of ${\rm ker}(\breve{\bOmega})$ which can be consistently estimated by $\{\hat{\bh}_1^{\rm (init)}, \ldots, \hat{\bh}_{\hat{d}}^{\rm (init)}\}$.  Define $\bar{\breve{\bGamma}}$  in the same manner as $\bar{\bGamma}$ specified in Section \ref{sec:sel-basis} but with replacing $\{\tilde{\bh}_{i}\}_{i=1}^{d}$ by $\{\breve{\bh}_{i}\}_{i=1}^{d}$, where $\{\tilde{\bh}_{i}\}_{i=1}^{d}$ is given in Section \ref{sec:theta-iden}.
	Following the discussions of  Propositions \ref{pro:theta-unique}--\ref{pro:Omega-impossible}, the  requirement ${\rm rank}(\bOmega) =d(d-1)/2$ is crucial for the identification of $(\bA, \bB)$ when $d\ge 2$. 
	As shown in Section \ref{sec:sub-theta} in the supplementary material for the proof of Lemma \ref{lemma:theta}, we know ${\rm rank}(\breve{\bOmega}) ={\rm rank}(\bOmega)$.  By Proposition \ref{pro:rankomega}(i), it holds that ${\rm rank}(\bOmega) = d(d-1)/2$ if and only if $\lambda_{d(d-1)/2}(\breve{\bOmega}^{\T}\breve{\bOmega}) >0$. 
	Note that $\breve{\bOmega}^{\T}\breve{\bOmega}$ is a $\{d(d+1)/2\} \times \{d(d+1)/2\}$ matrix. We require the following mild condition in our theoretical analysis.
	\begin{cd}\label{cd:eigen-bomega}
		$\lambda_{d(d-1)/2}(\breve{\bOmega}^{\T}\breve{\bOmega})$  is uniformly bounded away from zero.
	\end{cd}

	Write $\hat{\bA} = (\hat{\ba}_1,\ldots,\hat{\ba}_{\hat{d}})$ and $\hat{\bB} = (\hat{\bb}_1,\ldots,\hat{\bb}_{\hat{d}})$, where  $(\hat{\bA},\hat{\bB})$ are specified in Section \ref{sec:estimation}. Recall  $\bA = (\ba_1,\ldots,\ba_{d})$ and $\bB = (\bb_1,\ldots,\bb_{d})$. Theorem \ref{thm:hat_a_hat_b}  indicates that the columns of  $\hat{\bA} $ and $\hat{\bB}$ are, respectively, consistent to those of $\bA$ and $\bB$ up to the reflection and permutation indeterminacy.

	\begin{theorem}\label{thm:hat_a_hat_b}
		Let $d\ge2$ and Conditions \ref{cd:ra}--\ref{cd:eigen-bomega} hold. Select the threshold levels in \eqref{eq:m1h-m2h} and  \eqref{eq:zh} as
		\[
		\delta_1=\breve{C}\sqrt{\frac{\log(pq)}{n}}~~\textrm{and}~~   \delta_2=\tilde{C}\sqrt{\frac{\log(pq)}{n}}
		\]
		for some sufficiently large constants $\breve{C}, \tilde{C}>0$. If   $(\hat{d}_1, \hat{d}_2, \hat{d}) = (d_1,d_2,d)$,  there exists a permutation of $(1,\ldots,d)$, denoted by $(j_1,\ldots, j_d)$, such that 
		\begin{align*}
			\max_{\ell \in [d]}| {\kappa}_{1,\ell}\hat{\ba}_{j_\ell }-\ba_\ell |_2=   O_{\rm p} (\Pi_{1,n} + \Pi_{2,n})  =  \max_{\ell \in[d]}| {\kappa}_{2,\ell}\hat{\bb}_{j_\ell }-\bb_\ell|_2
		\end{align*}
		with some $ {\kappa}_{1,\ell}, {\kappa}_{2,\ell} \in \{1,-1\}$,  provided that $ \Pi_{1,n}+\Pi_{2,n}  \ll1$ and  $\log(pq)\ll n^c$ for some constant $c \in (0,1)$ depending only on $r_1$ and $r_2$. 
	\end{theorem}
	
	\begin{remark}\label{rek:chang}
		The convergence rates of the estimates for $\ba_{\ell}$ and $\bb_{\ell}$ suggested in \cite{chang2023modelling} are, respectively, $(1+\vartheta_{\ell
		}^{-1})\cdot O_{\rm p}(\tilde{\Pi}_{1,n} + \tilde{\Pi}_{2,n})$ and $\{1+(\vartheta_{\ell
		}^{*})^{-1}\}\cdot O_{\rm p}(\tilde{\Pi}_{1,n} + \tilde{\Pi}_{2,n})$, where  $\vartheta_\ell$ and $\vartheta^*_\ell$ are the eigen-gaps defined as in Equation (38) of  \cite{chang2023modelling},  $\tilde{\Pi}_{1,n} = \Pi_{1,n}$, and $\tilde{\Pi}_{2,n}=(\tilde{s}_3\tilde{s}_4)^{1/2} \{n^{-1}\log(pq)\}^{(1-\iota)/2} $. Here, $\tilde{s}_3$ and $\tilde{s}_4$ control the sparsity of the matrix
		\begin{align*}
			\bSigma_{\mathring{\bY}}(k) =  \frac{1}{n-k} \sum_{t=k+1}^{n} \mathbb{E}[\{\bY_{t}-\mathbb{E}(\bar{\bY})\} \otimes {\rm vec}\{\bY_{t-k}-\mathbb{E}(\bar{\bY})\} ]  =:\big(\sigma_{\mathring{y}, r,s}^{(k)}\big)_{(p^2q)\times q} 
		\end{align*}
		in the sense that  $\sum_{s=1}^{q}|\sigma_{\mathring{y},r,s}^{(k)}|^{\iota} \le \tilde{s}_3$ and   $\sum_{r=1}^{p^2q}|\sigma_{\mathring{y},r,s}^{(k)}|^{\iota} \le \tilde{s}_4$ for any $r\in[p^2q]$ and $s\in[q]$. Recall $\Pi_{2,n}=(s_3s_4)^{1/2} \{n^{-1}\log(pq)\}^{(1-\iota)/2} $ with $(s_3, s_4)$ specified  in Condition \ref{cd:sgm_yxi}(ii). By direct calculation, we have $s_3 \le p \tilde{s}_3$ and $\tilde{s}_4 \le p s_4$. Under some mild conditions,  it holds that $s_3s_4 \asymp \tilde{s}_3\tilde{s}_4$, which implies $\tilde{\Pi}_{2,n} \asymp \Pi_{2,n}$.  Hence, if  $\vartheta_\ell$ and $\vartheta^*_\ell$ are uniformly bounded away from zero, Theorem \ref{thm:hat_a_hat_b}  indicates that our new estimators share the same convergence rates of those proposed in \cite{chang2023modelling}. If  $\vartheta_{\ell} \to 0$ or $\vartheta_{\ell}^{*} \to 0$, our new estimators will have faster convergence rates than the estimators considered in \cite{chang2023modelling}. 
	\end{remark}
	
	\begin{remark}
		The model considered in \cite{han2024cp} for order 2 tensor is in the same form as our CP-factor model \eqref{eq:abm}. Therefore, the two estimation procedures proposed in \cite{han2024cp}, the composite PCA (denoted by cPCA) and the High-Order Projection Estimators (denoted by HOPE), can also be used to estimate the loading matrices $\bA$ and $\bB$ in our CP-factor model \eqref{eq:abm}, where cPCA is a one-pass estimation and HOPE is an iterative refinement initialized at the cPCA solution. \cite{han2024cp} assumes each latent factor $x_{t,\ell}=w_\ell f_{t,\ell}$ where $\{f_{t,\ell}\}_{t\ge 1}$ is stationary with $\mathbb{E}(f_{t,\ell})=0$ and $\mathbb{E}(f_{t,\ell}^2)=1$, and $w_\ell$ represents the signal strength. Under the model setting of \cite{han2024cp}, the latent factor process $\{x_{t,\ell}\}_{t\ge 1}$ is stationary for each $\ell\in[d]$. Moreover, \cite{han2024cp} also assumes $\mathbb{E}(f_{t-h,\ell_1}f_{t,\ell_2}) = 0$ for all $\ell_1 \neq \ell_2$ and $h \ge 1$, which implies $\mathbb{E}(x_{t-h,\ell_1}x_{t,\ell_2}) = 0$ for all $\ell_1 \neq \ell_2$ and $h \ge 1$. However,  these assumptions imposed on the latent factors are not necessary in our proposed method.  Write $\delta = \| (\bB \odot \bA)^{\T}(\bB \odot \bA) - \mathbf{I}_d \|_2$,  $\psi_{\ell}=w_{\ell}^2\mathbb{E}(f_{t-h,\ell} f_{t,\ell})$ with some fixed lag $h\geq1$, and $\psi_{*} =\min_{\ell \in[d+1]}(\psi_{\ell-1} -\psi_{\ell})$ with $\psi_{0}=\infty$ and $\psi_{d+1}=0$. To simplify the comparison between the theoretical results of \cite{han2024cp} and our proposed method, we ignore the permutation indeterminacy among the estimators. Theorem 1 of \cite{han2024cp} shows that the cPCA estimators $\hat{\ba}^{\textup{cpca}}_1,\ldots,\hat{\ba}^{\textup{cpca}}_d$, $\hat{\bb}^{\textup{cpca}}_1,\ldots,\hat{\bb}^{\textup{cpca}}_d$ satisfy
		\begin{align*} \label{eq: cpca error bound}
			& \max_{\ell\in[d]}\{1 - (\ba_\ell^{\T}\hat{\ba}^{\textup{cpca}}_\ell)^2\}^{1/2} + \max_{\ell\in[d]}\{1 - (\bb_\ell^{\T}\hat{\bb}^{\textup{cpca}}_\ell)^2\}^{1/2}\\
			&~~~~~~~~\lesssim \bigg(1+\frac{2\psi_1}{\psi_{*}}\bigg) \delta + \psi_{*}^{-1}\bigg\{\max_{\ell\in[d]}w_{\ell}^2\sqrt{\frac{\log n}{n}} + \bigg(1+\max_{\ell\in[d]}w_{\ell} \bigg)\sqrt{\frac{pq}{n}}\bigg\} 
			% \max_{\ell\in[d]}\{1 - (\bb_\ell^{\top}\hat{\bb}^{\textup{cpca}}_\ell)^2\}^{1/2} &\lesssim \bigg(1+\frac{2\psi_1}{\psi_{*}}\bigg) \delta + \psi_{*}^{-1}\bigg\{\max_{\ell\in[d]}w_{\ell}^2\sqrt{\frac{\log n}{n}} + \bigg(1+\max_{\ell\in[d]}w_{\ell} \bigg)\sqrt{\frac{pq}{n}}\bigg\}\,, 
		\end{align*}
		with probability at least $1 - (nd)^{-C_1} - e^{-pq}$, where  $C_1$ is a positive constant.  Theorem 2 of \cite{han2024cp} shows that, after a sufficient number of iterations, the HOPE estimators $\hat{\ba}^{\textup{iso}}_1,\ldots,\hat{\ba}^{\textup{iso}}_d$, $\hat{\bb}^{\textup{iso}}_1,\ldots,\hat{\bb}^{\textup{iso}}_d$ satisfy
		\begin{align*} 
			\max_{\ell\in[d]}\{1 - (\ba_\ell^{\T}\hat{\ba}^{\textup{iso}}_\ell)^2\}^{1/2}+  \max_{\ell\in[d]}\{1 - (\bb_\ell^{\T}\hat{\bb}^{\textup{iso}}_\ell)^2\}^{1/2} &\lesssim  (\psi_{d}^{-1} +\psi_{d}^{-1/2} )  \sqrt{\frac{\max(p,q)}{n}}   
		\end{align*}
		with probability at least $1 - (nd)^{-C_2} - e^{-p} - e^{-q}$, provided that the cPCA estimators satisfy certain convergence rates, where  $C_2$ is a positive constant. For our proposed estimators $\hat{\ba}_1,\ldots,\hat{\ba}_d,\hat{\bb}_1,\ldots,\hat{\bb}_d$, due to $1 - (\hat{\ba}_\ell^{\T}\ba_\ell)^2  \le  | \kappa_{1,\ell} \hat{\ba}_{\ell} - \ba_\ell |_2^2 $ and $ 1 - (\hat{\bb}_\ell^{\T}\bb_\ell)^2  \le  | \kappa_{2,\ell} \hat{\bb}_{\ell} - \bb_\ell |_2^2 $ for $\kappa_{1,\ell},\kappa_{2,\ell} \in \{1,-1\}$, then   
		\begin{align*}
			\max_{\ell\in[d]}\{1 - (\ba_\ell^{\T}\hat{\ba}_\ell)^2\}^{1/2} + \max_{\ell\in[d]}\{1 - (\bb_\ell^{\T}\hat{\bb}_\ell)^2\}^{1/2} &\lesssim \Pi_{1,n} + \Pi_{2,n} 
			% \{(s_1s_2)^{1/2}+(s_3s_4)^{1/2}\}\bigg\{\frac{\log (pq)}{n}\bigg\}^{(1-\iota)/2}\,,\\
			% \max_{\ell\in[d]}\{1 - (\bb_\ell^{\top}\hat{\bb}_\ell)^2\}^{1/2} &\lesssim   \{(s_1s_2)^{1/2}+(s_3s_4)^{1/2}\}\bigg\{\frac{\log (pq)}{n}\bigg\}^{(1-\iota)/2} 
		\end{align*}
		with probability approaching one, where $\Pi_{1,n}$ and $\Pi_{2,n}$ are specified in \eqref{eq:Pi1 and Pi2}. Hence, the two estimation procedures proposed in \cite{han2024cp} can only work for $pq \ll n$,  while our proposed method allows $p,q\gg n$. More importantly, in order to obtain the consistency of the cPCA estimators, we need to require $\bB\odot \bA$ to be very close to an orthonormal matrix ($\delta\rightarrow0$ as $n\rightarrow\infty$). However, such requirement may be too restrictive in practice. The larger $\delta$ is, or the smaller $\psi_*$ is, the worse convergence rate of the cPCA estimators will be. Since the HOPE estimators are obtained through an iterative method initialized with the cPCA estimators, the HOPE estimators will perform poorly if the cPCA estimators have large estimation errors. However, the convergence rate of our proposed method does not depend on these quantities.
	\end{remark}

	%\end{rek}
	Theorem \ref{thm:hat_a_hat_b} requires $(\hat{d}_1, \hat{d}_2, \hat{d}) = (d_1,d_2,d)$. By Theorem \ref{thm:rank}, we have $\mathbb{P}\{(\hat{d}_1,\hat{d}_2,\hat{d})=(d_1,d_2,d)\}\rightarrow1$ as $n\rightarrow\infty$. Hence, such requirement is reasonable in our theoretical analysis. More generally, without assuming $(\hat{d}_1, \hat{d}_2, \hat{d}) =(d_1, d_2,d)$, we can consider to measure the difference between $\bA = (\ba_1,\ldots,\ba_d)$ and $\hat{\bA} = (\hat{\ba}_1,\ldots,\hat{\ba}_{\hat{d}})$ by
	\begin{equation}\label{eq:rho_A}
		\varpi^2(\bA,\hat \bA) = \max_{\ell \in [d]}\min_{j \in [\hat{d}]}(1 - |\hat{\ba}_j^{\T} \ba^{ }_\ell|^2)\,.
	\end{equation}
	Also, we can measure the difference between $\bB = (\bb_1,\ldots,\bb_d)$ and $\hat{\bB} = (\hat{\bb}_1,\ldots,\hat{\bb}_{\hat{d}})$ by
	\begin{equation}\label{eq:rho_B} 
		\varpi^2(\bB,\hat \bB) = \max_{\ell \in [d]}\min_{j \in [\hat{d}]}(1 - |\hat{\bb}_j^{\T} \bb^{ }_\ell|^2)\,.
	\end{equation}
	Consider the event $\mathcal{G} =\{(\hat{d}_1, \hat{d}_2, \hat{d}) =(d_1, d_2,d)\}$. Due to $|\hat{\ba}_{j}|_{2} = 1=|\ba_{\ell}|_{2}$ and $| \kappa_{1,\ell}\hat{\ba}_{j_\ell}  -\ba_\ell |^2_2 \ge 2 - 2 | \hat{\ba}_{j_\ell}^{\T}  \ba_\ell  |$ for any $ \kappa_{1,\ell} \in \{1, -1\}$, restricted on $\mathcal{G} $, Theorem \ref{thm:hat_a_hat_b} indicates that 
	$1 - | \hat{\ba}_{j_\ell}^{\T}\ba_\ell  |^2 \le 2( 1 - |\hat{\ba}_{j_\ell}^{\T}\ba_\ell  |) = O_{\rm p}(\Pi_{1,n}^2 + \Pi_{2,n}^2)$
	provided that $ \Pi_{1,n} + \Pi_{2,n} \ll 1$ and  $\log(pq)\ll n^c$ for some constant $c \in (0,1)$ depending only on $r_1$ and $r_2$. Hence, restricted on $\mathcal{G} $, for any $\epsilon >0$, there exists some constant $C_{\epsilon}>0$ such that $\mathbb{P}\{\varpi^2(\bA,\hat \bA) > C_{\epsilon} (\Pi_{1,n}^2 + \Pi_{2,n}^2)\,|\, \mathcal{G} \} \le \epsilon$. Together with Theorem \ref{thm:rank}, we have 
	\begin{align*}
		&\mathbb{P}\{\varpi^2(\bA,\hat \bA) > C_{\epsilon} (\Pi_{1,n}^2 + \Pi_{2,n}^2) \} \\
		&~~~~~~~\le \mathbb{P}\{\varpi^2(\bA,\hat \bA) > C_{\epsilon} (\Pi_{1,n}^2 + \Pi_{2,n}^2)  \,|\, \mathcal{G} \} \, \mathbb{P}(\mathcal{G} )+  \mathbb{P}(\mathcal{G}^{\rm c})\\
		&~~~~~~~\le \mathbb{P}\{\varpi^2(\bA,\hat \bA) > C_{\epsilon} (\Pi_{1,n}^2 + \Pi_{2,n}^2) \,|\, \mathcal{G} \} + \mathbb{P}(\hat{d_1}\ne d_1) + \mathbb{P}(\hat{d_2}\ne d_2)+ \mathbb{P}(\hat{d}\ne d)\\
		&~~~~~~~\le \epsilon + o(1)\to \epsilon
	\end{align*}
	as $n\rightarrow\infty$, which implies  $\varpi^2(\bA,\hat \bA) =O_{\rm p}(\Pi_{1,n}^2 + \Pi_{2,n}^2)$.  Also, we can show $\varpi^2(\bB,\hat \bB)=O_{\rm p}(\Pi_{1,n}^2 + \Pi_{2,n}^2)$.

	\section{Numerical studies}\label{section:simulation}
	In this section, we will evaluate the finite-sample performance of our proposed method  by simulation and real data analysis. The simulation setup is given in Section \ref{sec:simulation setting up}, and the analysis of the simulation results is presented in Section \ref{sec:sim-est}. The real data analysis is given in Section \ref{sec:application}.

	\subsection{Setting up}\label{sec:simulation setting up}
	Let $\bA^\dag \equiv (a^\dag_{i,j})_{p \times d}$ and $\bB^\dag \equiv (b^\dag_{i,j})_{q \times d}$ with the elements drawn from the uniform distribution on $[-3,3]$ independently satisfying ${\rm rank}(\bA^\dag) =d = {\rm rank}(\bB^\dag)$. Define $\bP \in \mathbb{R}^{p\times d_1}$ and $\bQ\in\mathbb{R}^{q\times d_2}$ such that the columns of $\bP$ and $\bQ$ are, respectively, the $d_1$ and $d_2$ left-singular vectors corresponding to the $d_1$ and $d_2$ largest singular values of $\bA^{\dag}$ and $\bB^{\dagger}$. 
	% Performing the singular value decomposition such that $\bA^\dag = \bP^\dag \Sigma_1^\dag (\bU^{\dag})^{{\T}} $ and $\bB^\dag = \bQ^\dag \Sigma_2^\dag (\bV^{\dag })^{\T}$, we then take the first $d_1$ and $d_2$ columns of $\bP^\dag$ and $\bQ^\dag$, respectively, denoted as $\bP$ and $\bQ$.
	Let $\bU^* = \bP^\T \bA^\dagger =(\bu_1^*,\ldots,\bu_d^* )$ and $\bV^* = \bQ^\T \bB^\dagger =(\bv_1^*,\ldots,\bv_d^* )$. Derive $\bU = (\bu_1,\ldots,\bu_d )$ and $\bV = (\bv_1,\ldots,\bv_d )$ with $\bu_j = \bu_j^*/ |\bu_j^*|_2$ and $\bv_j = \bv_j^*/ | \bv_j^* |_2$ for any $j \in [d]$. Write $\vx^*_j = (x^*_{1,j},\ldots,x^*_{n,j})^{\T}$ and let  $\vx^*_1,\ldots,\vx^*_d$ be $d$ independent AR(1) processes with independent $\mathcal{N}(0,1)$ innovations, and the autoregressive coefficients drawn from the uniform distribution on $[-0.95,-0.6]\cup [0.6,0.95]$. Let $\bX_t = \text{diag}( x_{t,1},\ldots,  x_{t,d} )$ with $ x_{t,j} = x^*_{t,j}| \bv_j^* |_2 | \bu_j^* |_2 $ for each $t\in [n]$. The elements of the error term $\beps_t$ are drawn from $\mathcal{N}(0,1)$ independently. 
	Finally, we generate $\bY_t = \bA \bX_t \bB^{\T} + \beps_t$ for any $t \in [n]$ with $\bA=\bP\bU$ and $\bB=\bQ\bV$.  We set $n \in \{300, 600, 900\}$, $d \in \{3,5,7\}$ and $p,q$ taking values between 10 and 160. We consider three different scenarios for $(d,d_1,d_2)$:
	\begin{enumerate} 
		\item[(R1)]  Let $d_1=d_2=d$. In this scenario, $\bA$ and $\bB$ are full rank. 
		\item[(R2)]  Let $d_1 = d - 1$ and $d_2 = d$. In this scenario, only $\bB$ is full rank.
		\item[(R3)]  Let $d_1 = d_2 = d - 1$. In this scenario, both  $\bA$ and $\bB$ are not full rank.
	\end{enumerate}
	
	We follow  \cite{chang2023modelling} to specify $\xi_t$ involved in \eqref{eq:esthatsig}.  Let $\bY=(\vec\bY_1,\ldots, \vec \bY_{n})^{\T}$.  Perform the principal component analysis for $\bY$ and select $\xi_t$ as the average of the first $m$ principal components corresponding to the eigenvalues which count for at least 99\% of the total variations.  Let $\hat\sigma_0^2=(npq)^{-1}\| \bY\|_{\rm F}^2$. We set $\delta_1 = \delta_2 = \hat\sigma_0 \{n^{-1}\log(pq)\}^{1/2}$ in \eqref{eq:m1h-m2h} and  \eqref{eq:zh}, and set $c_{1,n} = c_{2,n} = c_{3,n} = \hat\sigma_0 n^{-1}$ in \eqref{eq:d1h} and \eqref{eq:dh}. We also   choose $K = 20$ and $\tilde{K} = 10$ with $K$ and $\tilde{K}$ given in \eqref{eq:m1h-m2h} and \eqref{eq:mh}, respectively. Here,  using a relatively large value for $K$ is to ensure that $\bM_1$ and $\bM_2$ defined in \eqref{eq:M1M2} satisfy $\textup{rank}(\bM_1) = d_1$ and $\textup{rank}(\bM_2) = d_2$. These two requirements are essential for our proposed method. See Propositions  \ref{pro:m1-rank-con} and \ref{pro:rankwith-xt}. As shown in \eqref{eq:mh}, $\tilde{K}$ is the number of lags used in the methods of  \cite{lam2011estimation}, \cite{lam2012factor} and \cite{Chang2015} to estimate the linear space spanned by the columns of the factor loading matrix in the standard factor model. In practice, a small $\tilde{K}$ (i.e., $1\leq \tilde{K}\leq 10$) is enough and the estimation results are generally robust to the specific choice of $\tilde{K}$.  See our sensitivity analysis with respect to the tuning parameters $K$ and $\tilde{K}$ in Figures \ref{fig:K-estimation}--\ref{fig:Ktilde-rank} of the supplementary material for more details.
	% With some prior knowledge based on the sensitivity analysis of the tuning parameters $K$ and $\tilde{K}$  in Section \ref{sec:sim-est}, we set $K = 20$ and $\tilde{K} = 10$.
	As mentioned in Section \ref{sec:theta-hat-est},  we need to select an appropriate constant vector 
	$\bphi = (\phi_1, \ldots, \phi_{\hat{d}})^{\T}$  to ensure that $\tilde{\bH} = \sum_{i=1}^{\hat{d}} \phi_i \tilde{\bH}_i$ is an invertible matrix with $\tilde{\bH}_i$ defined below \eqref{eq:init-basis}. Let $\phi_{i} = I\{\sigma_{\hat{d}}(\tilde{\bH}_{i}) = \max_{j\in[\hat{d}]}\sigma_{\hat{d}}(\tilde{\bH}_{j})  > 0\}$ for any $i\in[\hat{d}]$. If $|\bphi|_1  = 0$, we randomly generate a unit vector $\bphi$ such that $\sigma_{\hat{d}}(\tilde{\bH}) > 0$. If $|\bphi|_1 \ge 1$,  we arbitrarily keep one nonzero element in $\bphi$ and set all other elements to zero. The simulation results show that our  proposed procedure based on such selected $\bphi$ exhibits good finite-sample performance. We also compare our proposed method with the refined method (denoted by CP-refined) introduced by \cite{chang2023modelling}, and the cPCA and the HOPE methods proposed by \cite{han2024cp} with the recommended tuning parameter $h = 1$ therein. 
	% Notice that our proposed method, the CP-refined and the cPCA are all one-pass estimation procedures, while the HOPE  is an iterative refinement initialized at the cPCA solution. 
	All simulations are implemented in \textsf{R}. Our proposed method is available in \textsf{R}-package \texttt{HDTSA}, which is implemented by calling the \textsf{R}-function \texttt{CP\_MTS} with setting \texttt{method = `CP.Unified'}. The CP-refined method of \cite{chang2023modelling} can also be implemented by calling the \textsf{R}-function \texttt{CP\_MTS}  with setting \texttt{method = `CP.Refined'}.  All simulation results are based on 2000 replications.

	\subsection{Simulation results }\label{sec:sim-est} 
	We first consider the finite-sample performance of the estimation $(\hat{d}_1, \hat{d}_2,\hat{d})$ given in \eqref{eq:d1h} and \eqref{eq:dh}.  
	Note that the CP-refined method of \cite{chang2023modelling} is developed under the assumption $d_1=d_2=d$.  To fairly compare our proposed method and the CP-refined method,  we compare the relative frequency estimate of $\P_{d}: =\P(\hat{d} = d)$ with $\hat{d}$ specified in \eqref{eq:dh}, and the relative frequency estimate of $\P_c :=\P( \hat{d} = d )$ with $\hat{d} $ estimated by the CP-refined method. 
	Note that $\P_{1,2,d}: =\P\{(\hat{d}_1 , \hat{d}_2
	, \hat{d} ) = (d_1,d_2,d)\}\leq \mathbb{P}_d$ with $(\hat{d}_1, \hat{d}_2,\hat{d})$ estimated by our proposed method. Table \ref{table:rf-all} indicates that (i) our proposed method outperforms the CP-refined method across Scenarios R1--R3, and (ii) $(d_1,d_2,d)$ can be consistently estimated by our proposed method. To conserve space, we omit the results for $p<q$ in Scenarios R1 and R3, as the symmetry in the data-generating process leads to results that are nearly identical to those obtained when $p> q$.

	Figure \ref{fig:est-case1} reports the averages of the estimation errors $\varpi^2(\bA,\hat{\bA})$ and $\varpi^2(\bB,\hat{\bB})$  defined in \eqref{eq:rho_A} and \eqref{eq:rho_B} based on 2000 repetitions across different scenarios.
	Our proposed method consistently outperforms all competing methods, except in Scenario R1 with $p > q$, where it performs comparably to the HOPE method in estimating $\bB$. In contrast, the estimation errors of the CP-refined method are very large in Scenarios R2 and R3,  which indicates that the CP-refined method does not work for the matrix CP-factor model \eqref{eq:abm} with rank-deficient factor loading matrices $\bA$ and $\bB$. Also, in Scenarios R2 and R3, the HOPE method offers no notable improvement over the cPCA method and even underperforms the cPCA method in some settings, suggesting that the iterative  method HOPE is ineffective when the factor loading matrices $\bA$ and $\bB$ are rank-deficient. Additionally, all methods lose efficiency when $(d,d_1,d_2) = (3,2,2)$ since  $(\bA,\bB)$ cannot be identified uniquely, but our proposed method still yields the smallest estimation errors.  
	The averages and standard deviations of the estimation errors $\varpi^2(\bA,\hat{\bA})$ and $\varpi^2(\bB,\hat{\bB})$ based on 2000 repetitions are summarized  in Tables \ref{table:varpi-case1}--\ref{table:varpi-case3} in the supplementary material.

	Next, we evaluate the finite-sample performance of our proposed prediction method introduced in Section \ref{sec:prediction}. We generate a sequence $\{\bY_t\}_{t=1}^{n+m+1}$ defined in  Section \ref{sec:simulation setting up} with $m=20$. For any $s \in[m]$, we apply our proposed prediction method to the data $\{\bY_t\}^{n+s-1}_{t=s}$ 
	and then, respectively, obtain the one-step forecast of $\bY_{n+s}$ (denoted by $\hat{\bY}^{(1)}_{n+s}$) and the two-step forecast of $\bY_{n+s+1}$ (denoted by $\hat{\bY}^{(2)}_{n+s+1}$). 
	We also consider the prediction method introduced in \cite{chang2023modelling} to obtain the one-step ahead forecast of $\bY_{n+s}$ and the two-step ahead forecast of $\bY_{n+s+1}$ using $(\hat{\bA}, \hat{\bB})$ estimated from the data  $\{\bY_t\}^{n+s-1}_{t=s}$ for each $s \in [m]$, where $(\hat{\bA}, \hat{\bB})$ can be selected as either (i) our proposed estimate of $(\bA, \bB)$ specified in Section \ref{sec:estimation}, or (ii) the CP-refined estimate of $(\bA, \bB)$ given in \cite{chang2023modelling}.
	% We also consider two alternative methods to obtain the one-step forecast of  $\bY_{n+s}$ and the two-step forecast of $\bY_{n+s+1}$ based on $\{\bY_t\}^{n+s-1}_{t=s}$  for $s \in[m]$. Specifically, for each $s\in [m]$, we adopt the prediction method introduced in \cite{chang2023modelling} based on the data $\{\bY_t\}^{n+s-1}_{t=s}$ by selecting  $(\hat{\bA}, \hat{\bB})$ as our proposed estimates and the CP-refined estimates of \cite{chang2023modelling}, respectively, and then obtain one-step and two-step ahead forecasts for $\bY_{n+s}$ and $\bY_{n+s+1}$, respectively. 
	Here, for the obtained univariate time series, we fit it by an autoregressive (AR) model with the order determined by  the Akaike information criterion (AIC). For the obtained multivariate time series, we fit it by a vector autoregressive (VAR) model with the order determined by the AIC. Based on 2000 repetitions, 
	Figure \ref{fig:fore-case1} plots the averages of the one-step ahead $${\textup{RMSE} } := \frac{1}{m\sqrt{pq}}  \sum_{s=1}^{m} \| \hat{\bY}^{(1)}_{n+s} - \bY_{n+s} \|_\text{F}\,.$$  
	
	It can be observed that (i) in all cases, the finite-sample performance of our newly proposed prediction method is better than the prediction method introduced in \cite{chang2023modelling} with selecting $(\hat{\bA},\hat{\bB})$ as the CP-refined estimate, (ii) in the cases expect $(d,d_1,d_2) = (3,2,2)$, the averages of the one-step ahead $\textup{RMSE}$ of our newly proposed prediction method are almost identical to those of the prediction method introduced in \cite{chang2023modelling} with selecting $(\hat{\bA},\hat{\bB})$ as our proposed estimate, and (iii) in the case $(d,d_1,d_2) = (3,2,2)$, our newly proposed prediction method outperforms the prediction method introduced in \cite{chang2023modelling} with selecting $(\hat{\bA},\hat{\bB})$ as our proposed estimate. Note that the factor loading matrices $\bA$ and $\bB$ cannot be uniquely identified in the case $(d,d_1,d_2) = (3,2,2)$. Hence, we can conclude that (i) when $(\bA,\bB)$ can be uniquely identified, the prediction method introduced in \cite{chang2023modelling} with selecting $(\hat{\bA},\hat{\bB})$ as our proposed estimate works quite well, which has almost identical performance as our newly proposed prediction method; and (ii) our newly proposed prediction method works very well regardless of whether $(\bA,\bB)$ can be uniquely identified or not. The results of two-step ahead forecasting are similar to that of one-step ahead forecasting. See Figure \ref{fig:fore2-case1} in the supplementary material for details.

	\subsection{Real data analysis}\label{sec:application}
\begin{figure}[htbp]
		\centerline{\includegraphics[width= 14cm]{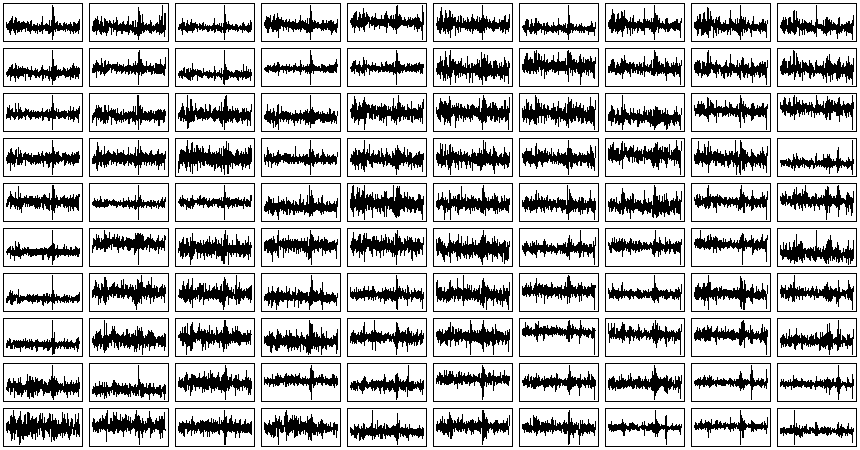}}
		\caption{The time series plots of 100 market-adjusted returns formed on different levels of size (by rows) and book equity to market equity ratio (by columns). The horizontal axis represents time and the vertical axis represents the monthly returns.}
		\label{fig:app-timeseries}
	\end{figure}

	In this section, we illustrate the proposed method for the matrix CP-factor model \eqref{eq:abm} by using the Fama-French $10 \times 10$ return series. We collect the monthly returns from January 1964 to December 2021, which contains 69600 observations for total 696 months. The data are downloaded from \url{http://mba.tuck.dartmouth.edu/pages/faculty/ken.french/data_library.html}. The portfolios are formed by the intersections of 10 levels of size, denoted by (${\rm S}_{1},\ldots,{\rm S}_{10}$), and 10 levels of the book equity to market equity ratio (BE), denoted by  $({\rm BE}_{1}, \ldots, {\rm BE}_{10}) $.  The data contain a small number of missing values in the early years and we transform them to zeros.  Since all the 100 series are clearly related to the overall market condition,  following \cite{wang2019factor}, we decide to remove the influence of market effects before empirical analysis. Two filtering approaches are considered: (i) (CAPM filtering) fitting a standard CAPM model \citep{fama1973risk} to each of the series to remove the market effect, (ii) (Demean filtering) subtracting the corresponding monthly excess market return from each of the series. The market return data are obtained from the same website above. Based on each filtering approach, we finally obtain 100 market-adjusted return series.  
	The 100 market-adjusted return series can be represented as a $10\times10$ matrix time series $\bY_t = (y_{i,j,t})$ for $t\in[696]$ (i.e., $p=q=10$, $n=696$), where $y_{i,j,t}$ is the market-adjusted return at the $i$-th level of size ${\rm S}_{i}$ and the $j$-th level of the BE-ratio ${\rm BE}_{j}$ at time $t$. Figure \ref{fig:app-timeseries} shows the time series plots of the market-adjusted return series $\{y_{i,j,t}\}_{t=1}^n$ based on the CAPM filtering  for $i,j \in[10]$. The rows in Figure \ref{fig:app-timeseries} correspond to the ten levels of size and the columns correspond to the ten levels of the BE-ratio. All series are stationary because they reject the null hypothesis of Augmented Dickey-Fuller test at 5\% significance level.

\begin{figure}[htbp]
		\centering
		\subfigure[Proposed method ($\hat{\bM}_1$)]{\includegraphics[width=0.32\textwidth]{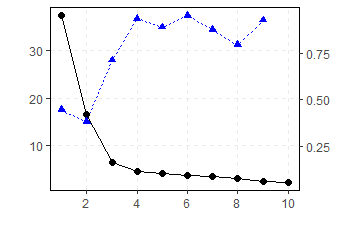}}
		\subfigure[Proposed method ($\hat{\bM}_2$)]{\includegraphics[width=0.32\textwidth]{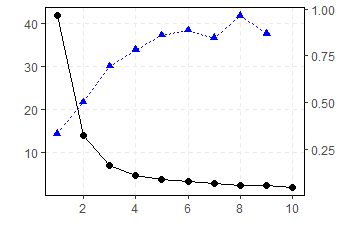}}
		\subfigure[CP-refined method]{\includegraphics[width=0.32\textwidth]{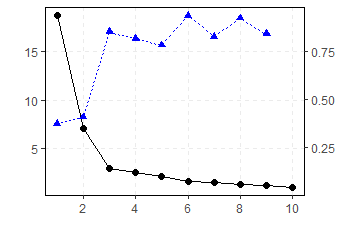}}
		\vspace{1em}
		\subfigure[FAC (column matrix)]{\includegraphics[width=0.33\textwidth]{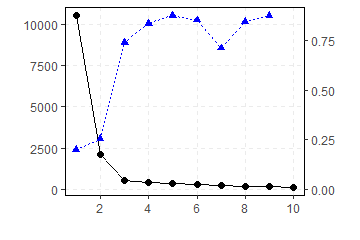}} \hspace{2em}
		\subfigure[FAC (row matrix)]{\includegraphics[width=0.33\textwidth]{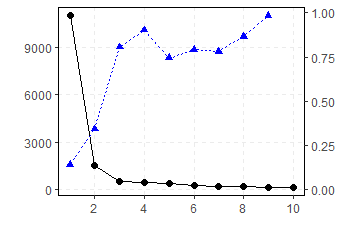}}
		\caption{Scree plots for our proposed method, the CP-refined method and the FAC based on $\{\bY_t\}_{t=1}^{456}$. The black solid line represents the eigenvalues, while the blue dashed line indicates the ratios of adjacent eigenvalues. }
		\label{fig: app-eigenall}
	\end{figure}

	We evaluate the post-sample forecasting performance of our proposed method introduced in Section \ref{sec:prediction} by performing the one-step and two-step ahead rolling forecasts for the 240 monthly readings in the last twenty years (2002--2021). To do this, we first use the data $\{\bY_t\}_{t=1}^{456}$ to determine the rank parameters $(d,d_1,d_2)$. With the tuning parameters selected as those in Section \ref{sec:simulation setting up}, our proposed method  obtains $(\hat{d}, \hat{d}_1, \hat{d}_2) = (2, 2, 1)$, which aligns with the conventional scree plots of $\hat{\bM}_1$ and $\hat{\bM}_2$ given in Figures \ref{fig: app-eigenall}(a) and \ref{fig: app-eigenall}(b), respectively. We adopt  $(\hat{d},\hat{d}_1,\hat{d}_2) = (2,2,1)$ in the rolling forecasts. For each $s\in [240]$, we apply our proposed prediction method to the data $\{\bY_t\}_{t=s}^{455+s}$ and then obtain the one-step forecast of $\bY_{456+s}$, denoted by $\hat{\bY}^{(1)}_{456+s} = (\hat{y}^{(1)}_{i,j,456+s}) $. For the two-step ahead forecast, we apply our proposed prediction method to the data $\{\bY_{t}\}_{t=s}^{454+s}$, and the two-step ahead forecast $\hat{\bY}^{(2)}_{456+s} = (\hat{y}^{(2)}_{i,j,456+s}) $ can be obtained by plug-in the one-step forecast into the fitted model. More specifically, for each $s\in[240]$, we fit the obtained 2-dimensional time series by a VAR model with the order determined by the AIC.  For comparison, we can also fit $\{\bY_{t}\}_{t = s}^{455+s}$ and $\{\bY_{t}\}_{t = s}^{454+s}$ by the following methods and obtain the associated one-step  and two-step ahead forecasts:

	\begin{itemize}
		\item (CP-refined) The CP-refined method of \cite{chang2023modelling} with the pre-determined parameter $K = 10$ therein. The associated rank in this method is estimated as $\hat{d} = 1$ based on   $\{\bY_t\}_{t=1}^{456}$ and then fixed in the rolling forecasts. Motivated by the scree plot in Figure \ref{fig: app-eigenall}(c), we also consider $\hat{d} = 2$  as an alternative. For $\hat{d} = 1$, we fit the obtained univariate time series by an AR model with the order determined by the AIC. For $\hat{d} = 2$, we fit the obtained 2-dimensional time series by a VAR model with the order determined by the AIC. The methods with $\hat{d} = 1$ and $\hat{d} = 2$ are referred to as CP-refined(1) and CP-refined(2), respectively.
		
		\item (cPCA, HOPE) The composite PCA and High-Order Projection Estimators in \cite{han2024cp} with the recommended tuning parameter $h = 1$ therein. Following the same rank specification strategy as in the CP-refined method, we consider both $\hat{r} = 1$ and $\hat{r} = 2$ for the associated rank in these two methods. For $\hat{r} = 1$, we fit the obtained univariate time series by an AR model with the order determined by the AIC. For $\hat{r} = 2$, we fit the obtained 2-dimensional time series by a VAR model with the order determined by the AIC. The methods with $\hat{r} = 1$ are referred to as cPCA(1) and HOPE(1), while those with $\hat{r} = 2$ are denoted as cPCA(2) and HOPE(2).

		\item (FAC) The matrix Tucker-factor model with the FAC method proposed by \cite{wang2019factor}  with the pre-determined parameter $h_0 = 1$ as suggested therein. The associated ranks in this model are estimated as $(\hat{k}_1,\hat{k}_2) = (1,1)$ by the ratio estimators suggested therein based on $\{\bY_{t}\}_{t = 1}^{456}$, and are fixed in the rolling forecasts. Motivated by the scree plots in Figures \ref{fig: app-eigenall}(d) and \ref{fig: app-eigenall}(e), we also consider an alternative setting with  $(\hat{k}_1,\hat{k}_2) = (2,1)$. For $(\hat{k}_1,\hat{k}_2) = (1,1)$, we fit the obtained univariate time series by an AR model with the order determined by the AIC. For $(\hat{k}_1,\hat{k}_2) = (2,1)$, we fit the obtained 2-dimensional time series by a VAR model with the order determined by the AIC. The methods with $(\hat{k}_1,\hat{k}_2) = (1,1)$ and $(\hat{k}_1,\hat{k}_2) = (2,1)$ are referred to as FAC(1,1) and FAC(2,1), respectively.

		\item (TOPUP, TIPUP) The Time series Outer-Product Unfolding Procedure and the Time series Inner-Product Unfolding Procedure proposed by \cite{han2024tensor} for the matrix Tucker-factor model.  The associated ranks in this model are estimated as $(\hat{k}_1,\hat{k}_2) = (2,2)$  by the information criterion considered in \cite{han2022rank} based on $\{\bY_{t}\}_{t = 1}^{456}$, and are fixed in the rolling forecasts. We fit the obtained 4-dimensional time series by a VAR model with the order determined by the AIC. The methods are implemented using  the R package \texttt{tensorTS}.
		
		\item (MAR) The matrix-AR(1) model of \cite{chen2021autoregressive}.

		\item (TS-PCA) Apply the principal component analysis for time series proposed by \cite{Chang2018} to the 100-dimensional time series $\{\vec{\bY}_{t}\}_{t = s}^{455+s}$ and $\{\vec{\bY}_{t}\}_{t = s}^{454+s}$, respectively, to obtain the associated one-step and two-step ahead forecasts. The method is implemented using the R package \texttt{HDTSA}. For the obtained univariate time series, we fit it by an AR model with the order determined by the AIC. For the obtained multivariate time series, we fit it by a VAR model with the order determined by the AIC.

		\item (UniAR) Fit each of 100 component time series by an AR model with the order determined by the AIC.
	\end{itemize}

	For each $s \in [240]$, the one-step ahead forecasting performance  is evaluated by the $\textup{rRMSE}(s)$ and $\textup{rMAE}(s)$ defined as
	\begin{gather*}
		\text{rRMSE}(s)  = \bigg\{ \frac{1}{100} \sum_{i = 1}^{10}\sum_{j = 1}^{10}|\hat{y}^{(1)}_{i,j,456+s} - y_{i,j,456+s}|^2  \bigg\}^{1/2}\,,\\
		\text{rMAE}(s)   =  \frac{1}{100} \sum_{i = 1}^{10}\sum_{j = 1}^{10} |\hat{y}^{(1)}_{i,j,456+s} - y_{i,j,456+s}| \,.
	\end{gather*}
	For the two-step ahead forecast, we can evaluate it by the associated $\textup{rRMSE}(s)$ and $\textup{rMAE}(s)$ analogously. Table \ref{table:app-forecast} reports the averages  of $\{\textup{rRMSE}(s)\}_{s=1}^{240}$ and $\{\textup{rMAE}(s)\}_{s=1}^{240}$, denoted by $\textup{rRMSE}$ and $\textup{rMAE}$, respectively. The standard deviations of $\{\textup{rRMSE}(s)\}_{s=1}^{240}$ and $\{\textup{rMAE}(s)\}_{s=1}^{240}$ are reported in parentheses.
	% Table \ref{table:app-forecast} reports the averages of $\textup{rRMSE}(s)$ and $\textup{rMAE}(s)$ over $s \in [240]$, denoted as $\textup{rRMSE}$ and $\textup{rMAE}$, respectively. We also report the standard deviations of $\{\textup{rRMSE}(s)\}_{s=1}^{240}$ and $\{\textup{rMAE}(s)\}_{s=1}^{240}$ in Table \ref{table:app-forecast}, with the values shown in parentheses. 
	As shown in Table \ref{table:app-forecast}, under CAPM filtering (Panel A), our proposed method achieves the lowest rRMSE and rMAE for both one- and two-step ahead forecasts, outperforming all competing methods.  Under Demean filtering (Panel B), although the HOPE(2) achieves the lowest rRMSE and rMAE, our proposed method performs comparably and yields smaller standard deviations than the HOPE(2).
	Overall, the results show that our proposed method delivers robust and accurate forecasts across different market-adjustment schemes, often outperforming alternatives in both accuracy and stability.

	%%%%%%%%%%%%%%%%%%%%%%%%%%%%%%%%%%%%%%%%%%%%%%
	%% Support information, if any,             %%
	%% should be provided in the                %%
	%% Acknowledgements section.                %%
	%%%%%%%%%%%%%%%%%%%%%%%%%%%%%%%%%%%%%%%%%%%%%%
	\begin{acks} 
		% The authors are grateful to the Editor, an Associate Editor and three referees for their helpful suggestions. 
		
		The authors thank Yuefeng Han for sharing code for implementing the methods proposed in \cite{han2024cp}.  
	\end{acks}
	
	%%%%%%%%%%%%%%%%%%%%%%%%%%%%%%%%%%%%%%%%%%%%%%
	%% Funding information, if any,             %%
	%% should be provided in the                %%
	%% funding section.                         %%
	%%%%%%%%%%%%%%%%%%%%%%%%%%%%%%%%%%%%%%%%%%%%%%
	\section*{Funding}
	J. Chang, Y. Du and G. Huang were supported in part by the National Natural Science
	Foundation of China (Grant nos. 72125008 and  72495122).	Q. Yao was supported in part by the U.K. Engineering and Physical Sciences Research
	Council (Grant nos. EP/V007556/1 and EP/X002195/1).

	% \begin{supplement}
		% 		\stitle{Supplement to ``Identification and Estimation for Matrix Time Series CP-factor Models''.} \sdescription{This supplement contains additional simulation studies and all technical proofs.}
		% \end{supplement}
	\section*{Supplement Material}
	
	{\bf Supplement to ``Identification and Estimation for Matrix Time Series CP-factor Models''.}
	
	This supplement contains additional simulation studies and all technical proofs.

	\bibliographystyle{jasa}

	\spacingset{0.95}\selectfont
	\bibliography{mybibfile}

	\begin{landscape}
		$ $\\
		$ $\\
		
		\begin{table}[htbp]
			\scriptsize
			\caption{
				Relative frequency estimates of $\P_{1,2,d} =\P\{(\hat{d}_1 , \hat{d}_2
				, \hat{d} ) = (d_1,d_2,d)\}$ and $\P_{d} = \P(\hat{d} = d)$ with $(\hat{d}_1, \hat{d}_2,\hat{d})$ estimated by our proposed method, and the relative frequency estimate of $\P_c=\P( \hat{d} = d )$ with $\hat{d} $ estimated by the CP-refined method of \cite{chang2023modelling} in Scenarios R1--R3. All numbers reported below are multiplied by 100.
			}
			\renewcommand\tabcolsep{4.5pt}
			\label{table:rf-all}
			
			\resizebox{22.9cm}{!}{
				\begin{tabular}{c|c|cccccccc|ccccccccc|cccccc}
					\hline\hline
					\multirow{3}{*}{$d$} & \multirow{3}{*}{$n$} & \multicolumn{8}{c|}{R1}                                                                                                                                                                                 & \multicolumn{9}{c|}{R2}                                                                                                                                                                                                                                        & \multicolumn{6}{c}{R3}                                                                                                                                            \\ \cline{3-25} 
					&                      & \multicolumn{4}{c|}{$p = q$}                                                                                 & \multicolumn{4}{c|}{$p > q$}                                                             & \multicolumn{3}{c|}{$p = q$}                                                              & \multicolumn{3}{c|}{$p > q$}                                                               & \multicolumn{3}{c|}{$p < q$}                                          & \multicolumn{3}{c|}{$p = q$}                                                              & \multicolumn{3}{c}{$p > q$}                                           \\
					&                      & $(p,q)$                    & $\mathbb{P}_{1,2,d}$ & $\mathbb{P}_{d}$ & \multicolumn{1}{c|}{$\mathbb{P}_{c}$} & $(p,q)$                     & $\mathbb{P}_{1,2,d}$ & $\mathbb{P}_{d}$ & $\mathbb{P}_{c}$ & $(p,q)$                    & $\mathbb{P}_{1,2,d}$ & \multicolumn{1}{c|}{$\mathbb{P}_{c}$} & $(p,q)$                     & $\mathbb{P}_{1,2,d}$ & \multicolumn{1}{c|}{$\mathbb{P}_{c}$} & $(p,q)$                     & $\mathbb{P}_{1,2,d}$ & $\mathbb{P}_{c}$ & $(p,q)$                    & $\mathbb{P}_{1,2,d}$ & \multicolumn{1}{c|}{$\mathbb{P}_{c}$} & $(p,q)$                     & $\mathbb{P}_{1,2,d}$ & $\mathbb{P}_{c}$ \\ \hline
					\multirow{9}{*}{3}   & 300                  & \multirow{3}{*}{$(20,20)$} & 96.59                & 97.04            & \multicolumn{1}{c|}{94.58}            & \multirow{3}{*}{$(40,10)$}  & 95.11                & 96.74            & 95.52            & \multirow{3}{*}{$(20,20)$} & 85.38                & \multicolumn{1}{c|}{77.02}            & \multirow{3}{*}{$(40,10)$}  & 77.51                & \multicolumn{1}{c|}{0.00}             & \multirow{3}{*}{$(10,40)$}  & 86.57                & 79.05            & \multirow{3}{*}{$(20,20)$} & 88.86                & \multicolumn{1}{c|}{0.00}             & \multirow{3}{*}{$(40,10)$}  & 87.44                & 0.00             \\
					& 600                  &                            & 97.35                & 97.70            & \multicolumn{1}{c|}{96.15}            &                             & 95.16                & 97.02            & 96.01            &                            & 86.36                & \multicolumn{1}{c|}{79.23}            &                             & 76.70                & \multicolumn{1}{c|}{0.00}             &                             & 87.91                & 81.21            &                            & 89.86                & \multicolumn{1}{c|}{0.00}             &                             & 89.09                & 0.00             \\
					& 900                  &                            & 97.65                & 98.00            & \multicolumn{1}{c|}{97.25}            &                             & 96.41                & 97.88            & 97.07            &                            & 84.23                & \multicolumn{1}{c|}{77.54}            &                             & 78.55                & \multicolumn{1}{c|}{0.00}             &                             & 89.29                & 82.90            &                            & 90.97                & \multicolumn{1}{c|}{0.00}             &                             & 89.32                & 0.00             \\ \cline{2-25} 
					& 300                  & \multirow{3}{*}{$(40,40)$} & 98.75                & 98.80            & \multicolumn{1}{c|}{97.90}            & \multirow{3}{*}{$(80,10)$}  & 94.99                & 97.04            & 96.07            & \multirow{3}{*}{$(40,40)$} & 90.88                & \multicolumn{1}{c|}{83.56}            & \multirow{3}{*}{$(80,10)$}  & 78.11                & \multicolumn{1}{c|}{0.00}             & \multirow{3}{*}{$(10,80)$}  & 91.00                & 83.13            & \multirow{3}{*}{$(40,40)$} & 92.83                & \multicolumn{1}{c|}{0.00}             & \multirow{3}{*}{$(80,10)$}  & 90.02                & 0.00             \\
					& 600                  &                            & 99.50                & 99.55            & \multicolumn{1}{c|}{99.05}            &                             & 96.56                & 97.67            & 97.06            &                            & 90.74                & \multicolumn{1}{c|}{85.34}            &                             & 78.31                & \multicolumn{1}{c|}{0.00}             &                             & 90.85                & 84.11            &                            & 93.68                & \multicolumn{1}{c|}{0.00}             &                             & 90.51                & 0.00             \\
					& 900                  &                            & 99.80                & 99.80            & \multicolumn{1}{c|}{99.40}            &                             & 96.97                & 98.69            & 97.88            &                            & 91.17                & \multicolumn{1}{c|}{85.51}            &                             & 76.49                & \multicolumn{1}{c|}{0.00}             &                             & 90.19                & 84.41            &                            & 93.52                & \multicolumn{1}{c|}{0.00}             &                             & 91.54                & 0.00             \\ \cline{2-25} 
					& 300                  & \multirow{3}{*}{$(80,80)$} & 99.75                & 99.80            & \multicolumn{1}{c|}{99.35}            & \multirow{3}{*}{$(160,10)$} & 96.55                & 98.12            & 97.11            & \multirow{3}{*}{$(80,80)$} & 94.31                & \multicolumn{1}{c|}{88.58}            & \multirow{3}{*}{$(160,10)$} & 80.50                & \multicolumn{1}{c|}{0.00}             & \multirow{3}{*}{$(10,160)$} & 92.07                & 84.92            & \multirow{3}{*}{$(80,80)$} & 95.74                & \multicolumn{1}{c|}{0.00}             & \multirow{3}{*}{$(160,10)$} & 90.92                & 0.00             \\
					& 600                  &                            & 100.00               & 100.00           & \multicolumn{1}{c|}{99.50}            &                             & 97.78                & 99.04            & 98.38            &                            & 94.34                & \multicolumn{1}{c|}{89.89}            &                             & 79.97                & \multicolumn{1}{c|}{0.00}             &                             & 91.85                & 86.37            &                            & 95.08                & \multicolumn{1}{c|}{0.00}             &                             & 91.35                & 0.00             \\
					& 900                  &                            & 100.00               & 100.00           & \multicolumn{1}{c|}{99.55}            &                             & 97.84                & 99.14            & 98.49            &                            & 94.87                & \multicolumn{1}{c|}{90.66}            &                             & 77.69                & \multicolumn{1}{c|}{0.00}             &                             & 91.17                & 84.95            &                            & 95.77                & \multicolumn{1}{c|}{0.00}             &                             & 92.01                & 0.00             \\ \hline
					\multirow{9}{*}{5}   & 300                  & \multirow{3}{*}{$(20,20)$} & 97.49                & 98.04            & \multicolumn{1}{c|}{94.88}            & \multirow{3}{*}{$(40,10)$}  & 94.14                & 98.41            & 95.42            & \multirow{3}{*}{$(20,20)$} & 97.25                & \multicolumn{1}{c|}{91.30}            & \multirow{3}{*}{$(40,10)$}  & 89.70                & \multicolumn{1}{c|}{0.00}             & \multirow{3}{*}{$(10,40)$}  & 97.12                & 93.58            & \multirow{3}{*}{$(20,20)$} & 97.99                & \multicolumn{1}{c|}{0.00}             & \multirow{3}{*}{$(40,10)$}  & 96.91                & 0.00             \\
					& 600                  &                            & 98.20                & 98.55            & \multicolumn{1}{c|}{96.29}            &                             & 94.69                & 98.62            & 97.24            &                            & 97.19                & \multicolumn{1}{c|}{93.32}            &                             & 89.86                & \multicolumn{1}{c|}{0.00}             &                             & 97.36                & 94.37            &                            & 98.64                & \multicolumn{1}{c|}{0.00}             &                             & 97.68                & 0.00             \\
					& 900                  &                            & 98.10                & 98.55            & \multicolumn{1}{c|}{96.14}            &                             & 95.63                & 98.78            & 97.51            &                            & 96.40                & \multicolumn{1}{c|}{91.79}            &                             & 90.33                & \multicolumn{1}{c|}{0.00}             &                             & 97.26                & 95.39            &                            & 98.49                & \multicolumn{1}{c|}{0.00}             &                             & 97.53                & 0.00             \\ \cline{2-25} 
					& 300                  & \multirow{3}{*}{$(40,40)$} & 99.60                & 99.60            & \multicolumn{1}{c|}{99.05}            & \multirow{3}{*}{$(80,10)$}  & 95.24                & 99.02            & 97.15            & \multirow{3}{*}{$(40,40)$} & 98.99                & \multicolumn{1}{c|}{96.88}            & \multirow{3}{*}{$(80,10)$}  & 91.13                & \multicolumn{1}{c|}{0.00}             & \multirow{3}{*}{$(10,80)$}  & 97.93                & 94.61            & \multirow{3}{*}{$(40,40)$} & 99.40                & \multicolumn{1}{c|}{0.00}             & \multirow{3}{*}{$(80,10)$}  & 97.66                & 0.00             \\
					& 600                  &                            & 99.45                & 99.55            & \multicolumn{1}{c|}{99.10}            &                             & 96.12                & 99.23            & 98.16            &                            & 99.35                & \multicolumn{1}{c|}{97.39}            &                             & 91.48                & \multicolumn{1}{c|}{0.00}             &                             & 98.14                & 95.98            &                            & 99.55                & \multicolumn{1}{c|}{0.00}             &                             & 97.93                & 0.00             \\
					& 900                  &                            & 99.70                & 99.70            & \multicolumn{1}{c|}{99.35}            &                             & 96.89                & 99.49            & 98.73            &                            & 99.05                & \multicolumn{1}{c|}{97.39}            &                             & 91.32                & \multicolumn{1}{c|}{0.00}             &                             & 98.54                & 96.48            &                            & 99.55                & \multicolumn{1}{c|}{0.00}             &                             & 98.08                & 0.00             \\ \cline{2-25} 
					& 300                  & \multirow{3}{*}{$(80,80)$} & 99.90                & 99.90            & \multicolumn{1}{c|}{99.70}            & \multirow{3}{*}{$(160,10)$} & 96.26                & 99.18            & 98.05            & \multirow{3}{*}{$(80,80)$} & 99.75                & \multicolumn{1}{c|}{98.50}            & \multirow{3}{*}{$(160,10)$} & 92.28                & \multicolumn{1}{c|}{0.00}             & \multirow{3}{*}{$(10,160)$} & 98.29                & 96.28            & \multirow{3}{*}{$(80,80)$} & 99.80                & \multicolumn{1}{c|}{0.00}             & \multirow{3}{*}{$(160,10)$} & 98.49                & 0.00             \\
					& 600                  &                            & 99.95                & 99.95            & \multicolumn{1}{c|}{99.65}            &                             & 96.84                & 99.23            & 98.52            &                            & 99.70                & \multicolumn{1}{c|}{98.85}            &                             & 91.21                & \multicolumn{1}{c|}{0.00}             &                             & 98.59                & 96.83            &                            & 100.00               & \multicolumn{1}{c|}{0.00}             &                             & 98.04                & 0.00             \\
					& 900                  &                            & 99.95                & 99.95            & \multicolumn{1}{c|}{99.70}            &                             & 97.01                & 99.70            & 99.24            &                            & 99.85                & \multicolumn{1}{c|}{99.00}            &                             & 90.56                & \multicolumn{1}{c|}{0.00}             &                             & 99.30                & 97.79            &                            & 99.95                & \multicolumn{1}{c|}{0.00}             &                             & 97.63                & 0.00             \\ \hline
					\multirow{9}{*}{7}   & 300                  & \multirow{3}{*}{$(20,20)$} & 97.94                & 98.60            & \multicolumn{1}{c|}{95.84}            & \multirow{3}{*}{$(40,10)$}  & 84.63                & 98.69            & 96.64            & \multirow{3}{*}{$(20,20)$} & 98.53                & \multicolumn{1}{c|}{94.99}            & \multirow{3}{*}{$(40,10)$}  & 83.97                & \multicolumn{1}{c|}{0.00}             & \multirow{3}{*}{$(10,40)$}  & 97.68                & 96.36            & \multirow{3}{*}{$(20,20)$} & 99.00                & \multicolumn{1}{c|}{0.00}             & \multirow{3}{*}{$(40,10)$}  & 98.23                & 0.00             \\
					& 600                  &                            & 98.59                & 99.39            & \multicolumn{1}{c|}{97.33}            &                             & 87.07                & 99.00            & 97.84            &                            & 99.19                & \multicolumn{1}{c|}{95.76}            &                             & 84.77                & \multicolumn{1}{c|}{0.00}             &                             & 98.24                & 96.78            &                            & 99.70                & \multicolumn{1}{c|}{0.00}             &                             & 98.47                & 0.00             \\
					& 900                  &                            & 98.85                & 99.35            & \multicolumn{1}{c|}{97.25}            &                             & 89.27                & 98.95            & 98.06            &                            & 98.74                & \multicolumn{1}{c|}{96.38}            &                             & 85.96                & \multicolumn{1}{c|}{0.00}             &                             & 98.23                & 96.87            &                            & 99.40                & \multicolumn{1}{c|}{0.00}             &                             & 97.93                & 0.00             \\ \cline{2-25} 
					& 300                  & \multirow{3}{*}{$(40,40)$} & 99.85                & 99.85            & \multicolumn{1}{c|}{99.30}            & \multirow{3}{*}{$(80,10)$}  & 85.49                & 98.85            & 97.39            & \multirow{3}{*}{$(40,40)$} & 99.90                & \multicolumn{1}{c|}{98.60}            & \multirow{3}{*}{$(80,10)$}  & 83.71                & \multicolumn{1}{c|}{0.00}             & \multirow{3}{*}{$(10,80)$}  & 98.13                & 97.73            & \multirow{3}{*}{$(40,40)$} & 100.00               & \multicolumn{1}{c|}{0.00}             & \multirow{3}{*}{$(80,10)$}  & 98.53                & 0.00             \\
					& 600                  &                            & 99.90                & 99.90            & \multicolumn{1}{c|}{99.50}            &                             & 89.37                & 99.32            & 98.27            &                            & 99.80                & \multicolumn{1}{c|}{99.05}            &                             & 84.98                & \multicolumn{1}{c|}{0.00}             &                             & 98.39                & 97.99            &                            & 99.90                & \multicolumn{1}{c|}{0.00}             &                             & 98.13                & 0.00             \\
					& 900                  &                            & 99.85                & 99.90            & \multicolumn{1}{c|}{99.30}            &                             & 89.15                & 99.53            & 98.81            &                            & 99.75                & \multicolumn{1}{c|}{98.65}            &                             & 85.63                & \multicolumn{1}{c|}{0.00}             &                             & 99.15                & 98.49            &                            & 99.95                & \multicolumn{1}{c|}{0.00}             &                             & 98.79                & 0.00             \\ \cline{2-25} 
					& 300                  & \multirow{3}{*}{$(80,80)$} & 100.00               & 100.00           & \multicolumn{1}{c|}{99.70}            & \multirow{3}{*}{$(160,10)$} & 87.47                & 99.53            & 99.06            & \multirow{3}{*}{$(80,80)$} & 99.95                & \multicolumn{1}{c|}{99.70}            & \multirow{3}{*}{$(160,10)$} & 84.41                & \multicolumn{1}{c|}{0.00}             & \multirow{3}{*}{$(10,160)$} & 99.15                & 98.24            & \multirow{3}{*}{$(80,80)$} & 100.00               & \multicolumn{1}{c|}{0.00}             & \multirow{3}{*}{$(160,10)$} & 98.83                & 0.00             \\
					& 600                  &                            & 99.95                & 99.95            & \multicolumn{1}{c|}{99.85}            &                             & 89.06                & 99.74            & 99.11            &                            & 99.90                & \multicolumn{1}{c|}{99.65}            &                             & 86.38                & \multicolumn{1}{c|}{0.00}             &                             & 99.25                & 98.79            &                            & 100.00               & \multicolumn{1}{c|}{0.00}             &                             & 98.48                & 0.00             \\
					& 900                  &                            & 100.00               & 100.00           & \multicolumn{1}{c|}{99.80}            &                             & 92.05                & 100.00           & 99.38            &                            & 100.00               & \multicolumn{1}{c|}{99.70}            &                             & 87.72                & \multicolumn{1}{c|}{0.00}             &                             & 99.35                & 99.04            &                            & 100.00               & \multicolumn{1}{c|}{0.00}             &                             & 98.79                & 0.00             \\ \hline\hline
				\end{tabular}
			} 
			
		\end{table}    
		
	\end{landscape}

	\begin{landscape}
		$ $\\
		\begin{figure}[htbp]
			\centering
			\centerline{\includegraphics[width= 23cm]{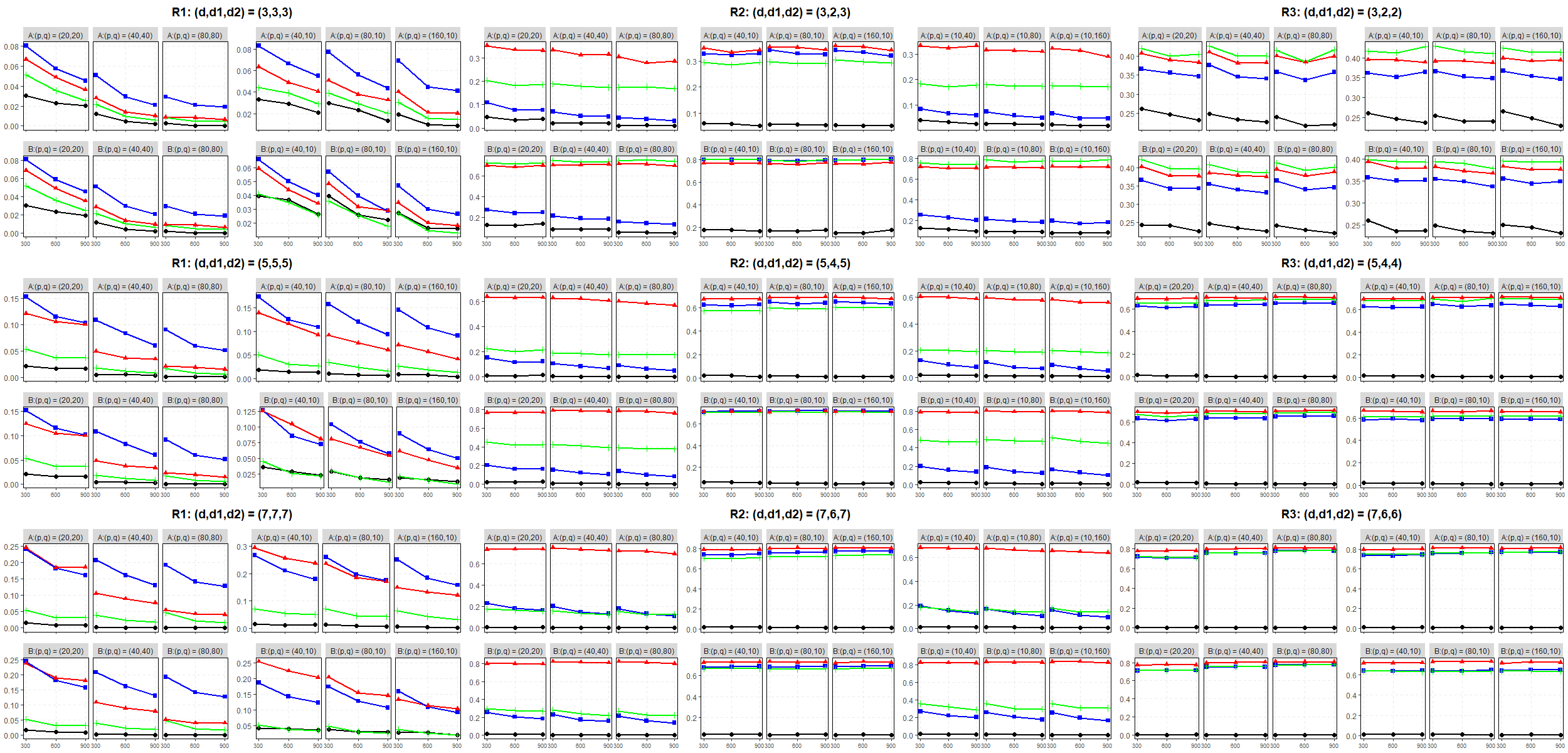}}
			\caption{The lineplots for the averages of estimation errors $\varpi^2(\bA,\hat{\bA})$ and $\varpi^2(\bB,\hat{\bB})$ based on 2000 repetitions in   Scenarios R1--R3. The legend is defined as follows: \textup{(i)} our proposed method ($\color{black}{-\bullet-}$), \textup{(ii)} the CP-refined method of \cite{chang2023modelling} ($\color{red}{-\blacktriangle-}$), \textup{(iii)} the cPCA of \cite{han2024cp} ($\color{blue}{-\blacksquare-}$), and \textup{(iv)} the HOPE  of \cite{han2024cp} ($\color{green}{-+-}$).}
			\label{fig:est-case1}
		\end{figure}
	\end{landscape}

	\begin{landscape}
		$ $\\
		$ $\\
		\begin{figure}[htbp]
			\centering
			\centerline{\includegraphics[width= 23cm]{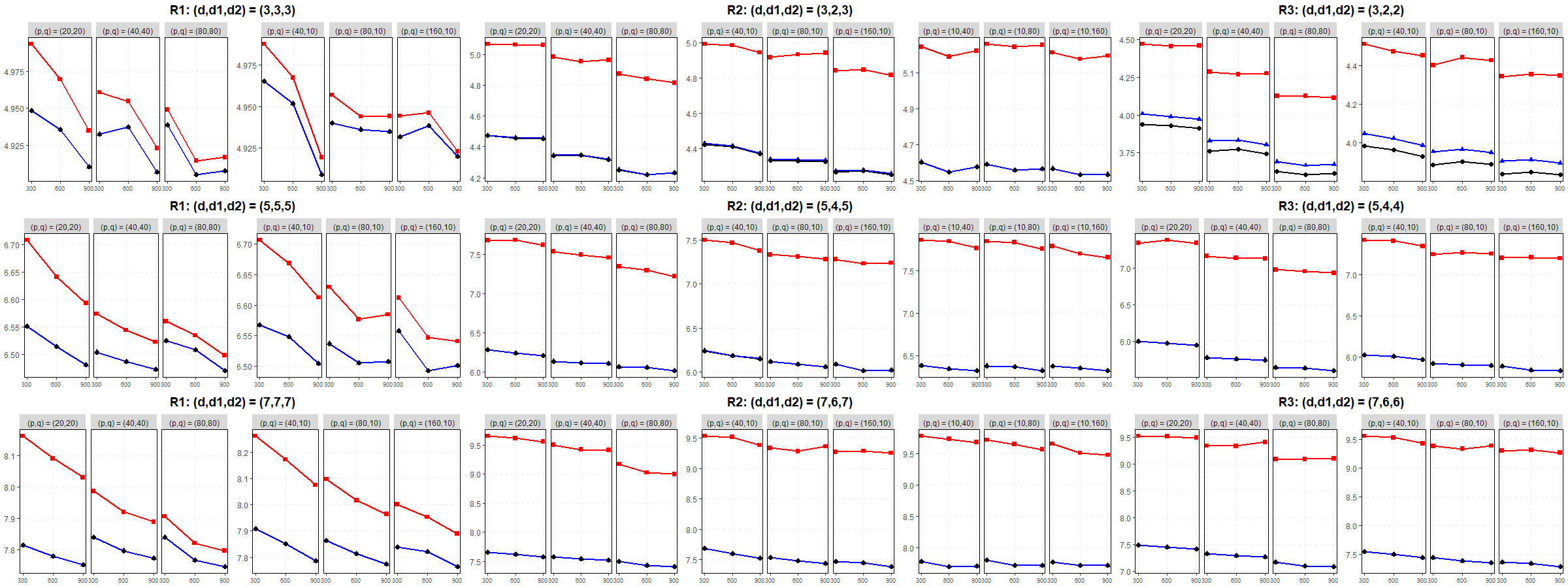}}
			\caption{The lineplots for the averages of one-step ahead forecast RMSE based on 2000 repetitions in Scenarios R1--R3. The legend is defined as follows: \textup{(i)} our proposed prediction method ($\color{black}{-\bullet-}$), \textup{(ii)} the prediction method introduced in \cite{chang2023modelling} with $(\hat{\bA},\hat{\bB})$ selected as our proposed  estimate ($\color{blue}{-\blacktriangle-}$), \textup{(iii)} the prediction method introduced in \cite{chang2023modelling} with $(\hat{\bA},\hat{\bB})$ selected as the CP-refined estimate ($\color{red}{-\blacksquare-}$).}
			\label{fig:fore-case1}
		\end{figure}
	\end{landscape}

	\begin{landscape}
		$ $\\
		$ $\\
		\begin{table}[htbp]
			\footnotesize
			\centering
			\renewcommand\tabcolsep{1.3pt}
			\caption{The averages and standard deviations (in parentheses) of one-step and two-step ahead forecasting errors of the market-adjusted returns from January, 2002 to December, 2021. Panel A and Panel B represent the market-adjusted returns obtained by CAPM filtering and demean filtering, respectively.
			} 
			\label{table:app-forecast}
			\begin{tabular}{ccccccccccccccc}
				\hline\hline
				& Proposed        & CP-refined(1)   & CP-refined(2)   & cPCA(1)  & cPCA(2)  & HOPE(1)  & HOPE(2)         & FAC(1,1) & FAC(2,1) & TOPUP    & TIPUP    & MAR      & TS-PCA   & UniAR    \\ \hline
				\multicolumn{15}{l}{Panel A: CAPM filtering}                                                                                                                                  \\ \hline
				\multicolumn{15}{c}{one-step ahead forecast}                                                                                                                                  \\
				{rRMSE} & \textbf{3.4302} & 3.4485   & 3.4408   & 3.4402   & 3.4361   & 3.4423   & 3.4373          & 3.4482   & 3.4610   & 3.4597   & 3.4669   & 3.4669   & 3.4710   & 3.4895   \\
				& (\textbf{1.5062}) & (1.5223) & (1.4992) & (1.5254) & (1.5154) & (1.5299) & (1.5163)        & (1.5226) & (1.5271) & (1.5277) & (1.5303) & (1.5364) & (1.5028) & (1.5099) \\
				{rMAE}  & \textbf{2.6218} & 2.6424   & 2.6325   & 2.6340   & 2.6294   & 2.6340   & 2.6307          & 2.6433   & 2.6541   & 2.6534   & 2.6566   & 2.6600   & 2.6579   & 2.6711   \\
				& (\textbf{1.0522}) & (1.0746) & (1.0468) & (1.0712) & (1.0561) & (1.0755) & (1.0569)        & (1.0751) & (1.0791) & (1.0763) & (1.0764) & (1.0849) & (1.0578) & (1.0601) \\ \hline
				\multicolumn{15}{c}{two-step ahead forecast}                                                                                                                                  \\
				rRMSE & \textbf{3.4297} & 3.4488   & 3.4378   & 3.4436   & 3.4362   & 3.4447   & 3.4370          & 3.4505   & 3.4627   & 3.4602   & 3.4641   & 3.4401   & 3.4626   & 3.4873   \\
				& (\textbf{1.5003}) & (1.5238) & (1.4972) & (1.5432) & (1.5178) & (1.5477) & (1.5161)        & (1.5254) & (1.5291) & (1.5303) & (1.5331) & (1.5162) & (1.4947) & (1.5111) \\
				rMAE  & \textbf{2.6241} & 2.6444   & 2.6317   & 2.6378   & 2.6327   & 2.6374   & 2.6328          & 2.6456   & 2.6558   & 2.6538   & 2.6552   & 2.6353   & 2.6545   & 2.6705   \\
				& (\textbf{1.0526}) & (1.0767) & (1.0494) & (1.0896) & (1.0684) & (1.0942) & (1.0668)        & (1.0774) & (1.0805) & (1.0789) & (1.0823) & (1.0694) & (1.0575) & (1.0640) \\ \hline
				\multicolumn{15}{l}{Panel B: Demean filtering}                                                                                                                                \\ \hline
				\multicolumn{15}{c}{one-step ahead forecast}                                                                                                                                  \\
				rRMSE & 3.4905          & 3.5146   & 3.4947   & 3.5293   & 3.4987   & 3.5255   & \textbf{3.4862} & 3.5057   & 3.5165   & 3.5268   & 3.5303   & 3.5154   & 3.5244   & 3.5470   \\
				& (1.5698)        & (1.5829) & (1.5725) & (1.5883) & (1.5878) & (1.5876) & (\textbf{1.5824}) & (1.5952) & (1.5884) & (1.5826) & (1.5891) & (1.6093) & (1.6005) & (1.5789) \\
				rMAE  & 2.6676          & 2.6910   & 2.6718   & 2.7047   & 2.6741   & 2.7008   & \textbf{2.6622} & 2.6834   & 2.6923   & 2.7022   & 2.7036   & 2.6923   & 2.6999   & 2.7143   \\
				& (1.1133)        & (1.1288) & (1.1250) & (1.1333) & (1.1273) & (1.1327) & (\textbf{1.1182}) & (1.1402) & (1.1365) & (1.1283) & (1.1319) & (1.1597) & (1.1586) & (1.1250) \\ \hline
				\multicolumn{15}{c}{two-step ahead forecast}                                                                                                                                  \\
				rRMSE & 3.4869          & 3.5142   & 3.4912   & 3.5239   & 3.4920   & 3.5208   & \textbf{3.4700} & 3.5018   & 3.5105   & 3.5269   & 3.5294   & 3.4959   & 3.5124   & 3.5413   \\
				& (1.5685)        & (1.5863) & (1.5721) & (1.5926) & (1.5939) & (1.5943) & (\textbf{1.5954}) & (1.5954) & (1.5894) & (1.5899) & (1.5961) & (1.6057) & (1.5881) & (1.5817) \\
				rMAE  & 2.6674          & 2.6922   & 2.6703   & 2.7013   & 2.6721   & 2.6982   & \textbf{2.6511} & 2.6805   & 2.6885   & 2.7036   & 2.7038   & 2.6765   & 2.6919   & 2.7130   \\
				& (1.1170)        & (1.1358) & (1.1237) & (1.1408) & (1.1391) & (1.1422) & (\textbf{1.1376}) & (1.1403) & (1.1368) & (1.1367) & (1.1406) & (1.1539) & (1.1483) & (1.1323) \\ \hline\hline
			\end{tabular}
		\end{table}

	\end{landscape}

	%%%%%%%%%%%%%%%%%%%%%%%%%%%%%%%%%%%%%%%%%%%%%%%%%%%%%%%%%%%%%%%%%%%%%

	\newpage
	\appendix
	\spacingset{1.7}\selectfont
	\setlength{\abovedisplayskip}{0.2\baselineskip}
	\setlength{\belowdisplayskip}{0.2\baselineskip}
	\setlength{\abovedisplayshortskip}{0.2\baselineskip}
	\setlength{\belowdisplayshortskip}{0.2\baselineskip}
	\begin{center}
		{\noindent \bf \Large Supplementary Material for ``Identification and Estimation for Matrix Time Series CP-factor Models"}\\
	\end{center}
	% \begin{center}
		% 	{\noindent \large Jinyuan Chang, Yue Du, Guanglin Huang and Qiwei Yao}
		% \end{center}
	\bigskip

	\setcounter{page}{1}
	\setcounter{section}{0}
	\renewcommand\thesection{\Alph{section}}
	\setcounter{lemma}{0}
	\renewcommand{\thelemma}{\Alph{section}\arabic{lemma}}
	\setcounter{equation}{0}
	\renewcommand{\theequation}{S.\arabic{equation}}
	
	\setcounter{table}{0}
	\setcounter{figure}{0}
	\renewcommand{\thefigure}{S\arabic{figure}}
	\renewcommand{\thetable}{S\arabic{table}}

	% We first introduce some notation which will be used throughout the supplementary material. 
	We use $C,C_1,\ldots$ to denote some generic positive constants do not depend on $(n,p,q,d)$ which may be different in different uses. For any two sequences of positive numbers $\{\tau_{k}\}$ and $\{\tilde{\tau}_{k}\}$, we write $\tau_{k} \lesssim \tilde{\tau}_{k}$ if $\lim \sup_{k\to \infty} \tau_{k} /\tilde{\tau}_{k} < \infty$. Let $\circ$ denote the Hadamard product. For an $m_1\times \cdots \times m_k$ tensor $\mathcal{H} = (h_{i_1,\ldots,i_k})_{m_1\times\cdots \times m_k}$, let $[\mathcal{H}]_{i_1,\ldots,i_k} = h_{i_1,\ldots,i_k}$.

	\section{Proofs of Propositions \ref{pro:m1-rank-con}, and \ref{pro:rankomega}--\ref{pro:rotation}}

	\subsection{Proof of Proposition \ref{pro:m1-rank-con}}
	Recall $\bG=\sum_{k=1}^{K} \bg_{k}\bg_{k}^{\T}$ with $\bG_k ={\rm diag}(\bg_k)$ for $k\in[K]$, where $\bg_k = (g_{k,1}, \ldots,g_{k,d})^{\T}$. Notice that
	\begin{align*}
		\bigg[\sum_{k = 1}^K\bG_k \bB^{\T} \bB \bG_k\bigg]_{i,j} =\sum_{k = 1}^K  g_{k,i} g_{k,j}[\bB^{\T} \bB]_{i,j}  \,, \qquad i,j\in[d]\,. 
	\end{align*}
	Then we have
	\begin{align*}
		\sum_{k = 1}^K\bG_k \bB^{\T} \bB \bG_k = \bG \circ (\bB^{\T} \bB)\,,
	\end{align*}
	where $\circ$ denotes the Hadamard product. Recall $\bB=(\bb_1, \ldots, \bb_{d})$ and $\bB \odot \bA$ is a $pq \times d$ matrix.
	By Condition \ref{cd:ra}(i),  we have ${\rm rank}(\bB \odot \bA)=d$, which implies $\bb_{i}\ne \bzero$ for each $i\in[d]$. Since  $\bG$ and $\bB^{\T}\bB$  have no zero main  diagonal entries and $\max\{ \mathcal{R}(\bG) + d_2,\mathcal{R}(\bB^\T \bB) + \textup{rank}(\bG) \}>d$,   by  Theorem 4 of \citeS{horn2020rank}, we have
	\begin{align*}
		{\rm rank} \bigg( \sum_{k = 1}^K\bG_k \bB^{\T} \bB \bG_k\bigg)= {\rm rank}\{\bG \circ (\bB^{\T} \bB)\} =d\,.
	\end{align*}
	Recall $\bM_1 = \bA (\sum_{k = 1}^K\bG_k \bB^{\T} \bB \bG_k) \bA^{\T}$ with some $d\times d$ matrix $\sum_{k = 1}^K\bG_k \bB^{\T} \bB \bG_k$.  Then  ${\rm rank}(\bM_1) = {\rm rank} (\bA) =d_1$. Analogously, we can also show Proposition \ref{pro:m1-rank-con}(ii). Hence, we complete the proof of Proposition \ref{pro:m1-rank-con}.
	$\hfill\Box$

	\subsection{Proof of Proposition \ref{pro:rankomega}}\label{sec:pro:rankomega}
	Recall $\bW_\ell = \sum_{i=1}^d \theta^{i,\ell} \bC_i$ with $\bTheta^{-1}=(\theta^{i,j})_{d\times d}$ and $\bC_i=\bu_i\bv_i^{\T}$. Then
	\begin{align}\label{eq:Drs}
		\bPsi(\bW_r, \bW_s) =  \sum_{j=1}^{d}\sum_{k=1}^{d} \theta^{j,r} \theta^{k,s}  \bPsi(\bC_j, \bC_k)\,.
	\end{align} 
	Write  $\bd_{r,s} = \vec \bPsi(\bW_r, \bW_s)$ and $\bar{\bd}_{r,s} =\vec \bPsi(\bC_r, \bC_s)$ for any $r,s \in [d]$.   Let
	\begin{align}\label{eq:d-omegas-omega}
		\bD =&~ (\bar{\bd}_{1,2},\ldots,\bar{\bd}_{1,d},\bar{\bd}_{2,3},\ldots,\bar{\bd}_{2,d} ,\ldots,\bar{\bd}_{d-1,d}) \in \mathbb{R}^{d_1^2d_2^2 \times d(d-1)/2}   \,,\notag\\
		\bOmega^* =&~ (\bd_{1,1},\ldots,\bd_{1,d}, \bd_{2,1},\ldots,\bd_{2,d}, \ldots, \bd_{d,1}, \ldots,\bd_{d,d}) \in \mathbb{R}^{d_1^2d_2^2 \times d^2}\,,\\
		\bD^* =&~ (\bar{\bd}_{1,1},\ldots,\bar{\bd}_{1,d},\bar{\bd}_{2,1},\ldots,\bar{\bd}_{2,d} ,\ldots,\bar{\bd}_{d,1},\ldots,\bar{\bd}_{d,d}) \in \mathbb{R}^{d_1^2d_2^2 \times d^2} \notag\,.
	\end{align} 
	By \eqref{eq:Drs}, it holds that
	\begin{align*}
		\bd_{r,s}=\sum_{j=1}^{d}\sum_{k=1}^{d} {\theta}^{j,r} {\theta}^{k,s}  \bar{\bd}_{j,k} \,.
	\end{align*}
	Write $\tilde{\bTheta} = \bTheta^{-1} $. With direct calculation, we have $\bOmega^* =  \bD^*(\tilde{\bTheta} \otimes \tilde{\bTheta})$. Since $\tilde{\bTheta}$ is a $d \times d$ invertible matrix, we have ${\rm rank}(\tilde{\bTheta} \otimes \tilde{\bTheta})=d^2$, which implies ${\rm rank}(\bOmega^{*}) = {\rm rank}(\bD^{*})$.  By Theorem 2.1 of \citeS{de2006link}, it holds that  $\bar{\bd}_{r,r}=\bzero$ for each $r\in[d]$. By the symmetry of $\bPsi(\cdot, \cdot)$, it holds that $\bd_{r,s}=\bd_{s,r}$ and $\bar{\bd}_{r,s}=\bar{\bd}_{s,r}$ for any $r,s\in[d]$. Recall $\bOmega$ defined in \eqref{eq:omega-D}. We have $\text{rank}(\bOmega) =\text{rank}(\bOmega^*)  $  and $ \text{rank}(\bD) =\text{rank}(\bD^*) $,  which implies $  {\rm rank}(\bD) ={\rm rank}(\bOmega)$. Due to $\bD \in \mathbb{R}^{d_1^2d_2^2\times d(d-1)/2}$ with $d_1d_2\ge d$, we have Proposition \ref{pro:rankomega}(i) holds.  Furthermore, It holds that
	\begin{equation*}
		\{\bar{\bd}_{r,s}\}_{1\le r<s\le d}  \text{ are linear independent} \quad \Leftrightarrow  \quad {\rm rank}(\bOmega) =\frac{d(d-1)}{2}\,.
	\end{equation*}
	We complete the proof of Proposition \ref{pro:rankomega}(ii). 
	
	Recall $\bOmega \in \mathbb{R}^{d_1^2d_2^2\times d(d+1)/2}$ and  ${\rm ker}(\bOmega) = \{\bx \in \mathbb{R}^{d(d+1)/2} : \bOmega \bx = \mathbf{0}\}$.
	By Rank–nullity theorem, we have
	\begin{equation*}
		{\rm dim}\{{\rm ker}(\bOmega)\}
		= \frac{d(d+1)}{2} - {\rm rank}(\bOmega) \,,
	\end{equation*}
	which implies that ${\rm dim}\{{\rm ker}(\bOmega)\} = d$ if and only if ${\rm rank}(\bOmega) = d(d-1)/2$. We complete the proof of Proposition \ref{pro:rankomega}(iii). 
	$\hfill\Box$

	\subsection{Proof of Proposition \ref{pro:theta-unique}}\label{sec:pro:theta-unique}
	Our first step is to show that for a given basis of the linear space ${\rm ker}(\bOmega)$, denoted by ${\bh}_1,\ldots,{\bh}_d$, the columns of $\bTheta$ can be uniquely determined up to the reflection and permutation indeterminacy. Our second step is to show that for any two different selections of the basis of ${\rm ker}(\bOmega)$, the columns of the two associated $\bTheta$ defined based on these two selections of the basis are identical up to the reflection and permutation indeterminacy.

	\underline{{\it Proof of Step 1.}} Write $\tilde{\bS} = ( \bh_{1},\ldots,  \bh_{d})$ where $ \{\bh_1,\ldots, \bh_d$\} is a given basis of the linear space ${\rm ker}(\bOmega)$. Then ${\rm rank}(\tilde{\bS})=d$. Recall  $ {\bh}_{i}=( {h}^{i}_{1,1},\ldots,  {h}^{i}_{1,d},  {h}^{i}_{2,2},\ldots , {h}^{i}_{d,d})^{\T}$ for any $i\in[d]$. Write the symmetric matrix  $\bH_i=(h_{i,r,s})_{d\times d}$ with $h_{i,r,s} =  {h}^{i}_{r,s}/2 =h_{i,s,r}$ for $ 1\le r < s \le d$ and $h_{i,r,r} =  {h}^{i}_{r,r}$  for $r\in[d]$. Define 
	\begin{equation*}
		\begin{split}
			&\bs_{i}  = (h_{i,1,1}, \ldots, h_{i,1,d},h_{i,2,1}, \ldots, h_{i,2,d}, \ldots, h_{i,d,1}, \ldots, h_{i,d,d})^{{\T}} \in \mathbb{R}^{d^2}\,, \\
			&~~~~\acute{\bs}_{i} = (h_{i,1,1}, \ldots, h_{i,1,d},h_{i,2,2}, \ldots, h_{i,2,d}, \ldots, h_{i,d,d})^{{\T}}\in \mathbb{R}^{d(d+1)/2}.
		\end{split}
	\end{equation*}
	Let $\bphi=(\phi_1, \ldots, \phi_d)^{{\T}}$ be a vector such that $\sum_{i=1}^{d}\phi_i \bs_{i} =\mathbf{0}$ which is equivalent to satisfy $\sum_{i=1}^{d}\phi_i \acute{\bs}_{i} =\mathbf{0}$ due to $h_{i,r,s}=h_{i,s,r}$ for any $r,s\in[d]$. Let   $\acute{\bS}=(\acute{\bs}_{1}, \ldots, \acute{\bs}_{d})$ and 
	\begin{align*}
		\bJ={\rm diag}\big(1,\underbrace{0.5, \ldots, 0.5}_{d-1}, 1, \underbrace{0.5, \ldots, 0.5}_{d-2}, \ldots, 1,0.5,1\big) \in \mathbb{R}^{d(d+1)/2 \times d(d+1)/2}\,.
	\end{align*}
	Then $\acute{\bS} = \bJ \tilde{\bS}$  and ${\rm rank}(\acute{\bS})=d$, which implies $\bphi=\mathbf{0}$. Hence, $\{\bs_{i}\}_{i=1}^{d}$ are linear independent. Write $\bGamma_{i}= {\rm diag}(\bgamma_{i})$  with $\bgamma_i = (\gamma_{1,i},\ldots, \gamma_{d,i})^{\T}$ for any $i\in[d]$ and  $\bar{\bGamma} = (\bgamma_1,\ldots,\bgamma_d)$. Let $\tilde{\bphi}=(\tilde{\phi}_1, \ldots, \tilde{\phi}_d)^{{\T}}$ be a vector such that $\sum_{i = 1}^{d} \tilde{\phi}_{i} \bgamma_i =\mathbf{0}$. Due to $\sum_{i = 1}^{d} \tilde{\phi}_{i} \bgamma_i =\mathbf{0}$, we have $\sum_{i = 1}^{d} \tilde{\phi}_{i} \bGamma_i =\mathbf{0}$. Recall $\bH_{i}=\bTheta\bGamma_{i} \bTheta^{\T}$ for each $i\in[d]$. Then $\sum_{i = 1}^{d}\tilde{\phi}_{i} \bH_i = \bTheta ( \sum_{i = 1}^{d} \tilde{\phi}_{i} \bGamma_i) \bTheta^{\T}={\bf 0}$. Since ${\rm vec}(\bH_{i}) = \bs_i$ and  $\{\bs_i\}_{i=1}^{d}$ are linear independent, we have $\tilde{\bphi}=\bzero$, which implies $\text{rank}(\bar{\bGamma}) = d$.  Recall the columns of $\bTheta$ are unit vectors.  By Theorem 1 of \citeS{afsari2008sensitivity}, it holds that  by the non-orthogonal joint diagonalization, the columns of $\bTheta$  are uniquely determined up to the reflection and permutation indeterminacy.

	\underline{{\it Proof of Step 2.}} Consider two different selections of the basis of the linear space ${\rm ker}(\bOmega)$, denoted by $\{ {\bh}_1^{(1)},\ldots, {\bh}_d^{(1)}\}$ and $\{ {\bh}_1^{(2)},\ldots, {\bh}_d^{(2)}\}$. For each $j\in\{1,2\}$, we can define the symmetric matrices $\bH_1^{(j)},\ldots,\bH_d^{(j)}$  in the same manner as $\bH_1,\ldots,\bH_d$ defined below \eqref{eq:off-diagonal}  but with replacing $\{ {\bh}_i\}_{i=1}^d$  by $\{ {\bh}_i^{(j)}\}_{i=1}^d$. Let $(\bTheta^{(1)})^{-1}$ and $(\bTheta^{(2)})^{-1}$ be, respectively, the non-orthogonal joint diagonalizers of $\bH_1^{(1)},\ldots,\bH_d^{(1)}$ and $\bH_1^{(2)},\ldots,\bH_d^{(2)}$. For each $j\in\{1,2\}$, we have $\bH_{i}^{(j)} =\bTheta^{(j)} \bGamma_{i}^{(j)} \{\bTheta^{(j)}\}^{\T}$ for each $i\in[d]$, where each column  of $\bTheta^{(j)}$ is a unit vector and the  diagonal matrix $\bGamma_{i}^{(j)}={\rm diag}\{\bgamma_i^{(j)}\}$  for some $\bgamma_i^{(j)} \in \mathbb{R}^{d}$. 
	Let $\bar{\bGamma}^{(1)} = (\bgamma_1^{(1)},\ldots,\bgamma_d^{(1)})$ and $\bar{\bGamma}^{(2)} = (\bgamma_1^{(2)},\ldots,\bgamma_d^{(2)})$. Using the similar arguments for the derivation of  ${\rm rank}(\bar{\bGamma})=d$ in the proof of Step 1, we also have ${\rm rank}\{\bar{\bGamma}^{(1)}\}=d ={\rm rank}\{\bar{\bGamma}^{(2)}\}$. Due to 
	$( {\bh}_1^{(2)}, \ldots,  {\bh}_{d}^{(2)}) =( {\bh}_1^{(1)}, \ldots,  {\bh}_{d}^{(1)})\bT$ with an invertible matrix $\bT= (\tau_{k,i})_{d\times d}$, then
	\begin{align*}
		\bH_{i}^{(2)}=\sum_{k=1}^{d} \tau_{k,i} \bH_{k}^{(1)} = \bTheta^{(1)} \bigg\{ \sum_{k = 1}^{d} \tau_{k,i} \bGamma_k^{(1)} \bigg\} \{\bTheta^{(1)}\}^{\T}\,,~~~~~i\in[d]\,,
	\end{align*}
	which provides a second form for the non-orthogonal joint diagonalization of $\bH_{1}^{(2)},\ldots,\bH_d^{(2)}$. Let  $\bGamma_{i}^{*} =\sum_{k = 1}^{d} \tau_{k,i} \bGamma_k^{(1)} ={\rm diag}(\bgamma_{i}^{*})$ for some $\bgamma_{i}^{*}\in \mathbb{R}^{d}$. Write $\bar{\bGamma}^{*}=(\bgamma_{1}^{*}, \ldots, \bgamma_{d}^{*})$. Then $\bar{\bGamma}^{*}=\bar{\bGamma}^{(1)}\bT$ and ${\rm rank}(\bar{\bGamma}^{*})=d$.  
	By Theorem 1 of \citeS{afsari2008sensitivity}, we know the columns of  $\bTheta^{(1)}$ and  $\bTheta^{(2)}$ are  identical up to the reflection and permutation indeterminacy.
	$\hfill\Box$

	\subsection{Proof of Proposition \ref{pro:Psitouniqueness}}
	Recall $\bA=\bP\bU$ and $\bB=\bQ\bV$ with $\bP\in \mathbb{R}^{p\times d_1 }$ and $\bQ\in \mathbb{R}^{q \times d_2}$.  Since  ${\rm rank}(\bP) = {\rm rank}(\bA)=d_1$ and ${\rm rank}(\bQ) ={\rm rank}(\bB) =d_2$, then $\mathcal{R}(\bU) =\mathcal{R}(\bA)$ and $\mathcal{R}(\bV)=\mathcal{R}(\bB)$. Recall $\bar{\bd}_{r,s} =\vec \bPsi(\bC_r, \bC_s)$  with $\bC_i=\bu_i\bv_i^{\T}$. 
	Notice that $\bar{\bd}_{r,s} = (\bu_r \otimes \bu_s -  \bu_s \otimes \bu_r) \otimes (\bv_r \otimes \bv_s -  \bv_s \otimes \bv_r)$. Let
	\begin{align*}
		\breve{\bU} =&~ (\breve{\bu}_{1,2},\ldots,\breve{\bu}_{1,d},\breve{\bu}_{2,3},\ldots,\breve{\bu}_{2,d},\ldots,\breve{\bu}_{d-1,d})  \in \mathbb{R}^{d_1^2 \times d(d-1)/2}\,,\\
		\breve{\bV} =&~ (\breve{\bv}_{1,2},\ldots,\breve{\bv}_{1,d},\breve{\bv}_{2,3},\ldots,\breve{\bv}_{2,d},\ldots,\breve{\bv}_{d-1,d})  \in \mathbb{R}^{d_2^2 \times d(d-1)/2} \,,
	\end{align*}
	where $\breve{\bu}_{r,s}=\bu_r \otimes \bu_s -  \bu_s \otimes \bu_r$ and $\breve{\bv}_{r,s}=\bv_r \otimes \bv_s -  \bv_s \otimes \bv_r$ for $1\le r < s \le d$. 
	For $\bD$ defined in   \eqref{eq:d-omegas-omega}, we have $\bD  = \breve{\bU} \odot  \breve{\bV}$. 
	Let 
	\begin{align}\label{eq:gu-def}
		\bG_{u}=(\bg_{1,2}^{u},\ldots, \bg_{1,d}^{u},\bg_{2,3}^{u},\ldots, \bg_{2,d}^{u},\ldots,\bg_{d-1,d}^{u})  \in \mathbb{R}^{d^2 \times d(d-1)/2}\,,
	\end{align}
	where $\bg_{r,s}^{u}=(g_{r,s,1}^{u},\ldots,g_{r,s,d^2}^{u})^{{\T}}$ with $g_{r,s,s+(r-1)d}^{u} = 1 $, $g_{r,s,r+(s-1)d}^{u} = -1$ and $g_{r,s,t}^{u}=0$ for $t\in[d^2]\backslash\{s+(r-1)d, r+(s-1)d\}$. It holds that
	${\rm rank}(\bG_{u})=d(d-1)/2$ and 
	\begin{align}\label{eq:breveu-dec}
		\breve{\bU}=(\bU \otimes \bU)\bG_{u}\,,\qquad \breve{\bV}=(\bV \otimes \bV)\bG_{u}  \,.
	\end{align} 
	Due to the symmetry of $\bA$ and $\bB$ in \eqref{eq:abm}, the relationships among $d$, $d_1$ and $d_2$ can be divided into three scenarios: (i) $d_1 = d_2 = d$, (ii) $d_2 <d_1 = d$, and (iii) $\max(d_1,d_2) < d$. In the following, we will show Proposition \ref{pro:Psitouniqueness} holds in these three scenarios, respectively.  
	
	{\underline{{\it Scenario} {\rm (i)}: $d_1 = d_2 = d$.}}  
	Notice that $\mathcal{R}(\bA) = d$ and $\mathcal{R}(\bB) = d$ when $d_1=d=d_2$. We have $\mathcal{R}(\bA)+d \ge d+2$ and $\mathcal{R}(\bB)+d \ge d+2$ 
	hold automatically for  any $d\ge 2$.  Since ${\rm rank}(\bU) = {\rm rank}(\bA)  =d$ and ${\rm rank}(\bG_{u})=d(d-1)/2$, by \eqref{eq:breveu-dec}, we have ${\rm rank}(\breve{\bU})={\rm rank}(\bG_{u})=d(d-1)/2$. Analogously, we   also have ${\rm rank}(\breve{\bV})=d(d-1)/2$. Hence, we have ${\rm rank}(\bD) = {\rm rank}(\breve{\bU} \odot \breve{\bV}) =d(d-1)/2$. By Proposition \ref{pro:rankomega}(ii), it holds that ${\rm rank}(\bOmega) = d(d-1)/2$. 
	
	{\underline{{\it Scenario} {\rm (ii)}: $d_2 < d_1 = d$.}} 
	Consider a vector $\boldsymbol{\phi} =(\phi_{1,2},\ldots, \phi_{1,d},\phi_{2,3},\ldots, \phi_{2,d},\ldots, \phi_{d-1,d})^{{\T}}$ such that $\bD\boldsymbol{\phi} = \textbf{0}$. Then $\sum_{r = 1}^{d-1} \sum_{s=r+1}^{d}$ $ \phi_{r,s}\breve{\bu}_{r,s} \otimes \breve{\bv}_{r,s} = \textbf{0}$. Write $\breve{\bv}_{r,s}=(\breve{v}_{r,s,1}, \ldots, \breve{v}_{r,s,d_2^2})^{\T}$. Then
	\begin{equation*}
		\sum_{r = 1}^{d-1} \sum_{s=r+1}^{d}\phi_{r,s}\breve{v}_{r,s,k} \breve{\bu}_{r,s} = \textbf{0}\,, \qquad k\in[d_2^2]\,.
	\end{equation*}
	Since ${\rm rank}(\bU)  = {\rm rank}(\bA)   =d$, by \eqref{eq:breveu-dec}, we have  ${\rm rank}(\breve{\bU})={\rm rank}(\bG_{u}) =d(d-1)/2$. It then holds that $\phi_{r,s}\breve{v}_{r,s,k} =0 $ for any  $1\le r < s \le d$ and $k\in[d_2^2]$, which implies $\phi_{r,s}\breve{\bv}_{r,s} =\textbf{0}$ for any $1\le r< s \le d$. Recall $\mathcal{R}(\bB) + d_1 \ge d+2$.  Since $d_1=d$, we have   $\mathcal{R}(\bV) = \mathcal{R}(\bB) \ge d+2 -d_1 =  2$, which  indicates that any two columns of $\bV$ are not proportional.   Notice that $\breve{\bv}_{r,s}=\bv_r \otimes \bv_s -  \bv_s \otimes \bv_r \ne \textbf{0}$. Hence,  $\phi_{r,s}=0$ for any  $1\le r < s \le d$, which implies ${\rm rank}(\bD)=d(d-1)/2$. By Proposition \ref{pro:rankomega}(ii), it holds that ${\rm rank}(\bOmega) = d(d-1)/2$. 
	% Analogously, we can also show such result holds if $d_1 <d_2 = d$ and $ \mathcal{R}_{\bA} \ge 2$. 
	
	{\underline{{\it Scenario} 
			{\rm (iii)}: $\max{(d_1,d_2)} < d$.}} 
	Notice that $\mathcal{R}(\bA)+d_2 \ge d+2$ and $\mathcal{R}(\bB) + d_1 \ge d+2$. Due to $\max\{\mathcal{R}(\bA), \mathcal{R}(\bB)\} \le \max(d_1, d_2) < d$, we have $\min(d_1,d_2) > 2$ in this scenario. Recall ${\rm rank}(\bU)=d_1$ with $\bU=(\bu_1,\ldots, \bu_d)$. Without loss of generality,  we assume  ${\rm rank}\{(\bu_1,\ldots, \bu_{d_1})\}=d_1$. Then there exists a $d_1\times d_1$ invertible matrix $\tilde{\bG}_{u}$ such that $\tilde{\bG}_{u}\bU = (\mathbf{I}_{d_1},\bar{\bs}_{d_1+1},\ldots,\bar{\bs}_{d})$. Write $\tilde{\bU}=(\mathbf{I}_{d_1 },\bar{\bS})$ with $\bar{\bS} =(\bar{\bs}_{d_1+1},\ldots,\bar{\bs}_{d})$. Recall $\bD  = \breve{\bU} \odot  \breve{\bV}$. By \eqref{eq:breveu-dec}, we have $\breve{\bU} = (\tilde{\bG}_{u}^{-1} \otimes \tilde{\bG}_{u}^{-1}) (\tilde{\bU} \otimes \tilde{\bU}) \bG_{u}$ and 
	$\bD = \{(\tilde{\bG}_{u}^{-1} \otimes \tilde{\bG}_{u}^{-1}) \otimes \bI_{d_2^2}\} [\{(\tilde{\bU} \otimes \tilde{\bU}) \bG_{u}\}  \odot \breve{\bV}]$, which implies that 
	\begin{align}\label{eq:pro-rankD}
		{\rm rank}(\bD) = {\rm rank} [\{(\tilde{\bU} \otimes \tilde{\bU}) \bG_{u}\}  \odot \breve{\bV} ]\,.
	\end{align}
	Denote by ${\be_{i}}$ the $d_1$-dimensional unit vector with only $i$-th component being 1. For any  $i\in[d_1]$, write $\breve{\bU}_{i} =(\breve{\bu}_{i,i+1},\ldots, \breve{\bu}_{i,d}) $ with  $\breve{\bu}_{i,j} =  (\be_{ i} \otimes \be_{ j} - \be_{j} \otimes \be_{i})I(i < j\le d_1) +  (\be_{i} \otimes \bar{\bs}_{j} - \bar{\bs}_{j} \otimes \be_{i})I(d_1 < j\le d)$. Let $\breve{\bS}= (\breve{\bs}_{d_1+1,d_1+2}, \ldots, \breve{\bs}_{d_1+1,d},\breve{\bs}_{d_1+2,d_1+3}, \ldots, \breve{\bs}_{d_1+2,d},\ldots, \breve{\bs}_{d -1,d} )$ with  $\breve{\bs}_{i,j} =  \bar{\bs}_i \otimes \bar{\bs}_j -  \bar{\bs}_j \otimes \bar{\bs}_i$ for any $d_1+1\le i<j\le d$. For $\bG_{u}$ defined in \eqref{eq:gu-def}, it holds that
	\begin{align*}
		(\tilde{\bU} \otimes \tilde{\bU}) \bG_{u} =(\breve{\bU}_1, \breve{\bU}_2,\ldots, \breve{\bU}_{d_1}, \breve{\bS}) \,.
	\end{align*}
	Write  $\bar{\bs}_{i}=(\bar{s}_{1,i},\ldots, \bar{s}_{d_1,i})^{{\T}}$ and $\breve{\bs}_{i,j}=(\breve{s}_{1,i,j},\ldots, \breve{s}_{d_1^2,i,j})^{{\T}}$ for any $d_1+1\le i<j\le d$.  Notice that $[\be_{i} \otimes \be_{j} - \be_{j} \otimes \be_{i}]_{k}=I\{k=(i-1)d_1+j\} - I\{k=(j-1)d_1+i\}$, and $[\be_{i} \otimes \bar{\bs}_{j} - \bar{\bs}_{j} \otimes \be_{i}]_k = \bar{s}_{\ell_1,j}I\{k=(i-1)d_1+\ell_1\} -\bar{s}_{\ell_2,j}I\{k=(\ell_2-1)d_1+i\}$ for any $k\in[d_1^2]$. 
	% Hence, 
	% \begin{align*}
		%     [\be_{r} \otimes  \be_{s} - \be_{s} \otimes \be_{r}]_{m_1} =1\,, \quad [\be_{r} \otimes \bar{\bs}_{j} - \bar{\bs}_{j} \otimes \be_{r}]_{m_1} = \bar{s}_{s,j}\,, \quad    [\be_{s} \otimes \bar{\bs}_{j} - \bar{\bs}_{j} \otimes \be_{s}]_{m_1} = -\bar{s}_{r,j}
		% \end{align*}
	% for any $m_1 =(r-1)d_1+s$ with  $1\le r < s\le d_1$ and $d_1 +1 \le j \le d$. 
	Consider the vector $\boldsymbol{\phi} =(\phi_{1,2},\ldots, \phi_{1,d},\phi_{2,3},\ldots, \phi_{2,d},\ldots, \phi_{d-1,d})^{{\T}}$ such that $[\{(\tilde{\bU} \otimes \tilde{\bU}) \bG_{u}\}  \odot \breve{\bV} ]\boldsymbol{\phi} = \textbf{0}$. Then, we have
	\begin{align}\label{eq:bv-rs}
		\phi_{r,s}\breve{\bv}_{r,s} + \sum_{j=d_1+1}^{d} \phi_{r,j}\bar{s}_{s,j}\breve{\bv}_{r,j} -\sum_{j=d_1+1}^{d} \phi_{s,j}\bar{s}_{r,j}\breve{\bv}_{s,j} + \sum_{j=d_1+1}^{d-1}\sum_{k=j+1}^{d}\phi_{j,k} \breve{s}_{m_1,j,k} \breve{\bv}_{j,k} =\textbf{0}
	\end{align}
	for any  $m_1=(r-1)d_1 + s$ with $1\le r< s\le d_1$. Write $\bV = (\bv_1,\ldots,\bv_{d_1}, \bH_\bv)$ with $\bH_\bv \in \mathbb{R}^{d_2 \times (d-d_1)}$.  
	Since $\mathcal{R}(\bB) + d_1 \ge d+2$, we have $\mathcal{R}(\bV) = \mathcal{R}(\bB) \ge d-d_1+2 $, which implies  ${\rm rank}\{(\bv_r, \bv_{s},\bH_{\bv})\}= d-d_1+2$. Recall $\breve{\bv}_{r,s}=\bv_r \otimes \bv_s -  \bv_s \otimes \bv_r$. For any $1\le r < s \le d_1$, we have 
	$$(\breve{\bv}_{r,s},  \{\breve{\bv}_{r,j}\}_{d_1< j \le d}, \{\breve{\bv}_{s,j}\}_{d_1< j \le d}, \{\breve{\bv}_{j,k}\}_{d_1< j < k \le d}) = \{(\bv_{r}, \bv_{s}, \bH_{\bv}) \otimes (\bv_{r}, \bv_{s}, \bH_{\bv})\} \bT_{r,s}\,,$$ where $\bT_{r,s} \in \mathbb{R}^{(d-d_1+2)^2\times (d-d_1+2)(d-d_1+1)/2}$ is  defined in the similar manner as $\bG_{u}$ given in \eqref{eq:gu-def} with  ${\rm rank}(\bT_{r,s})= (d-d_1+2)(d-d_1+1)/2$.  
	Using the similar arguments for the proof of ${\rm rank}(\breve{\bU})=d(d-1)/2$ in scenario (i),
	it holds that ${\rm rank}\{(\breve{\bv}_{r,s},  \{\breve{\bv}_{r,j}\}_{d_1< j \le d}, \{\breve{\bv}_{s,j}\}_{d_1< j \le d}, \{\breve{\bv}_{j,k}\}_{d_1< j < k \le d})\} ={\rm rank}(\bT_{r,s})= (d-d_1+2)(d-d_1+1)/2$. Hence, by \eqref{eq:bv-rs}, we have $\phi_{r,s}=0$ for any $1\leq r<s\leq d_1$, 
	$\phi_{r,j}\bar{s}_{t,j}=0$ for any $r,t\in[d_1]$, $r\ne t$ and $d_1 < j\le d$, and $\phi_{j,k}\breve{s}_{m_1,j,k}=0$  with $m_1=(r-1)d_1+s$ for any $d_1<j < k \le d$.
	Recall $(\bI_{d_1}, \bar{\bS}) =\tilde{\bG}_{u}\bU$ with a $d_1\times d_1$ invertible matrix $\tilde{\bG}_{u}$ and $\bar{\bS} =(\bar{\bs}_{d_1+1},\ldots,\bar{\bs}_{d})$. Since $\mathcal{R}(\bA) + d_2 \ge d+2$ and $d_2 < d$, we have $\mathcal{R}(\bU) = \mathcal{R}(\bA) > 2$, which implies any two columns of $\tilde{\bG}_{u}\bU$ are not proportional. Hence, any two columns of $\bar{\bS}$ are not proportional  and $\bar{\bs}_j$ has at least 2 nonzero components for any $d_1< j\le d$. It then holds that $\phi_{r,j}=0$ for any $r\in[d_1]$ and $d_1 < j \le d$. Recall $[\{(\tilde{\bU} \otimes \tilde{\bU}) \bG_{u}\}  \odot \breve{\bV} ]\boldsymbol{\phi} = \textbf{0}$. Then, we have 
	\begin{align*}
		\sum_{j=d_1+1}^{d-1}\sum_{k=j+1}^{d}\phi_{j,k} \breve{\bs}_{j,k} \otimes \breve{\bv}_{j,k}=\bzero\,.
	\end{align*}
	Since $ \{\breve{\bv}_{j,k}\}_{d_1< j < k \le d}$ are linear independent, it holds that $\phi_{j,k} \breve{\bs}_{j,k}=\bzero$ for any  $d_1 < j < k \le d$. Recall $\breve{\bs}_{j,k}= \bar{\bs}_{j} \otimes \bar{\bs}_{k} - \bar{\bs}_{k} \otimes \bar{\bs}_{j}$ and any two columns of $\bar{\bS} =(\bar{\bs}_{d_1+1},\ldots,\bar{\bs}_{d})$ are not proportional. Then we have $\breve{\bs}_{j,k} \ne \textbf{0}$, which implies that $\phi_{j,k}=0$ for any  $d_1< j < k \le d$.
	Hence, it holds that $\bphi=\mathbf{0}$, which implies ${\rm rank} [\{(\tilde{\bU} \otimes \tilde{\bU}) \bG_{u}\}  \odot \breve{\bV} ] = d(d-1)/2$. By \eqref{eq:pro-rankD}, we have ${\rm rank}(\bD)  = d(d-1)/2$. By   Proposition \ref{pro:rankomega}(ii), it then holds that ${\rm rank}(\bOmega) = d(d-1)/2$. $\hfill\Box$

	\subsection{Proof of Proposition \ref{pro:Omega-impossible}}\label{sec:pro:Omega-impossible}
	Consider $(\bA,\bB)$  with  $\bA=(\ba_1, \ldots, \ba_d)$ and $\bB=(\bb_1, \ldots, \bb_d)$ satisfying  four conditions:
	(i) ${\rm rank}(\bA)=d_1$, ${\rm rank}(\bB)=d_2$, ${\rm rank}(\bB \odot \bA)=d$ and $|\ba_{\ell}|_2=1=|\bb_{\ell}|_2$ for each $\ell \in[d]$,
	(ii) $\ba_3 = |\ba_1 + \ba_2|_2^{-1}(\ba_1 + \ba_2)$ and $\bb_3 = |\bb_1 + \bb_2|_2^{-1}(\bb_1 +  \bb_2)$, (iii)  $\mathcal{R}(\bA)=2=\mathcal{R}(\bB) $, and (iv) $|\ba_1 \pm \ba_2|_2 \ge 1/2$ and $ |\bb_1 \pm \bb_2|_2 \ge 1/2$. 
	Recall $\bA=\bP\bU$ and $\bB=\bQ\bV$ with $\bP^{\T}\bP=\bI_{d_1}$ and $\bQ^{\T}\bQ=\bI_{d_2}$. Write $\bU=(\bu_1,\ldots, \bu_d)$ and  $\bV=(\bv_1, \ldots, \bv_d)$. It holds that $\bu_3 = \tau_{\bu}(\bu_1 + \bu_2)$ and  $\bv_3 = \tau_{\bv}(\bv_1 +  \bv_2)$ with $\tau_{\bu} = |\bu_1 + \bu_2|_2^{-1} $ and $\tau_{\bv} = |\bv_1 + \bv_2|_2^{-1}$. Recall $\bar{\bd}_{r,s} =\vec \bPsi(\bC_r, \bC_s)$  with $\bC_i=\bu_i\bv_i^{\T}$. Since $\bar{\bd}_{r,s} = (\bu_r \otimes \bu_s -  \bu_s \otimes \bu_r) \otimes (\bv_r \otimes \bv_s -  \bv_s \otimes \bv_r)$ for $1 \le r < s \le d$, then 
	\begin{align*}
		\bar{\bd}_{1,3} &= (\bu_1 \otimes \bu_3 -  \bu_3 \otimes \bu_1) \otimes (\bv_1 \otimes \bv_3 -  \bv_3 \otimes \bv_1) \\
		&= \tau_{\bu} \tau_{\bv} (\bu_1 \otimes \bu_2 -  \bu_2 \otimes \bu_1) \otimes (\bv_1 \otimes \bv_2 -  \bv_2 \otimes \bv_1) \\
		&= \tau_{\bu} \tau_{\bv} \bar{\bd}_{1,2}\,.
	\end{align*}
	Recall $\bD = (\bar{\bd}_{1,2},\ldots,\bar{\bd}_{1,d},\bar{\bd}_{2,3},\ldots,\bar{\bd}_{2,d} ,\ldots,\bar{\bd}_{d-1,d}) \in \mathbb{R}^{d_1^2d_2^2 \times d(d-1)/2}$. By Proposition \ref{pro:rankomega}, it holds that $ \text{rank}(\bD) < d(d-1)/2$. As shown in the proof of Proposition \ref{pro:rankomega} in Section \ref{sec:pro:rankomega},  ${\rm rank}(\bOmega) = \text{rank}(\bD)$. 
	% Thus, for such selected $(\bA,\bB) \in \mathcal{U}$, the associated  $\bOmega$ satisfies  
	% $\text{rank}(\bOmega) < d(d-1)/2$.
	Hence, $(\bA,\bB) \in \mathcal{U}$.
	
	Let
	$\tilde{\bA} = (\tilde{\ba}_1, \ldots, \tilde{\ba}_d)$ and $\tilde{\bB} = (\tilde{\bb}_1, \ldots, \tilde{\bb}_d)$ with $\tilde{\ba}_3=|\ba_1-  \ba_2|_2^{-1}(\ba_1 - \ba_2)$, $\tilde{\bb}_3=|\bb_1 -\bb_2|_2^{-1}(\bb_2- \bb_1)$, and  $\tilde{\ba}_i=\ba_i$, $\tilde{\bb}_i=\bb_i$ for $i\ne 3$.
	Analogously, we can also show $(\tilde{\bA},\tilde{\bB}) \in \mathcal{U}$. Furthermore, it holds that
	\begin{equation*}
		(\bB \odot \bA) \bXi = \tilde{\bB} \odot \tilde{\bA} 
	\end{equation*}
	with $\bXi=(\xi_{i,j})_{d\times d}$ defined as 
	\begin{align*}
		\xi_{i,j}= \left\{
		\begin{aligned}
			1\,, ~~~~~~~~~~~~~~~~ &{\rm if}~ i=j \in[d]\backslash[3] \,,\\
			\frac{|\ba_1 +\ba_2|_2  |\bb_1 + \bb_2|_2}{|\ba_1 -\ba_2|_2  |\bb_1-  \bb_2|_2} \,, ~~&{\rm if}~i=j=3 \,,\\
			-\frac{2}{|\ba_1 -\ba_2|_2  |\bb_1-  \bb_2|_2} \,, ~~&{\rm if}~i\in \{1,2\}, j=3 \,,\\
			0\,,~~~~~~~~~~~~~~~~&{\rm otherwise} \,.
		\end{aligned}
		\right.
	\end{align*}
	Recall $ \bY_t = \bA \bX_t \bB^{\T} +  \beps_t $ with $\bX_{t} ={\rm diag}(x_{t,1}, \ldots, x_{t,d})$. Let $\tilde{\bX}_t = \text{diag}(\tilde{\bx}_t)$ for $\tilde{\bx}_t = \bXi^{-1}\bx_t$ with  $\bx_t=(x_{t,1},\ldots, x_{t,d})^{\T}$. We then have $\bY_t =  \tilde{\bA} \tilde{\bX}_t \tilde{\bB}^{\T} + \beps_t$, 
	which implies  that the matrix time series $\{\bY_t\}_{t\ge 1}$ can be represented as a new matrix CP-factor model with the newly defined factor loading matrices $(\tilde{\bA}, \tilde{\bB})$.
	
	Since $|\ba_1|_2=1=|\ba_2|_2$ and $|\ba_1 \pm \ba_2|_2 \ge 1/2$, 
	then 
	\begin{align*}
		% &|\tilde{\ba}_3 - \ba_1|_2 = \bigg|\frac{\ba_1 -\ba_2}{|\ba_1 -\ba_2|_2} -\ba_1 \bigg|_2 = \bigg|\frac{(1-|\ba_1 -\ba_2|_2)\ba_1 -\ba_2}{|\ba_1 -\ba_2|_2}  \bigg|_2 \ge \frac{|1-|1-|\ba_1 -\ba_2|_2||}{2}\,,\\
		% &~~~~~~~|\tilde{\ba}_3 - \ba_2|_2 =  \bigg|\frac{\ba_1 -\ba_2}{|\ba_1 -\ba_2|_2} -\ba_2 \bigg|_2 = \bigg|\frac{\ba_1 - (1+|\ba_1 -\ba_2|_2)\ba_2}{|\ba_1 -\ba_2|_2}  \bigg|_2 \ge \frac{|\ba_1 -\ba_2|_2}{2}\,,\\ 
		|\ba_3 - \tilde{\ba}_3  |_2 =&~  \bigg| \frac{\ba_1 + \ba_2}{|\ba_1 +  \ba_2|_2} -\frac{\ba_1 -\ba_2}{|\ba_1 -\ba_2|_2}\bigg|_2 \\
		=&~ \bigg|\frac{(|\ba_1 -\ba_2|_2 -|\ba_1 +\ba_2|_2) \ba_1 +  (|\ba_1 -\ba_2|_2 +|\ba_1 +\ba_2|_2) \ba_2}{|\ba_1 +  \ba_2|_2 \cdot |\ba_1 -\ba_2|_2} \bigg|_2\\
		\ge&~  \frac{\min\{|\ba_1 +  \ba_2|_2, |\ba_1 -\ba_2|_2\}}{2} 
		\ge \frac{1}{4}\,.
	\end{align*}
	Recall $\mathscr{D}(\tilde{\bA},\bA) = \max_{\ell \in[d]} | \tilde{\ba}_{\ell} - \ba_{\ell}|_2$. We have  $\mathscr{D}(\tilde{\bA},\bA) \ge 1/4$. Using the same arguments, we can also show
	$\mathscr{D}(\tilde{\bB},\bB) \ge 1/4$.  
	% $\zeta_2 := |\bb_3-\tilde{\bb}_3|_2 \ge \min\{|\bb_1 +  \bb_2|_2, |\bb_1 -\bb_2|_2\}/2 > 0$. Recall $\mathscr{D}(\tilde{\bA},\bA) = \max_{\ell \in[d]} |\tilde{\kappa}_{1,j_{\ell}}\tilde{\ba}_{j_{\ell}} - \ba_{\ell}|_2$ and $\mathscr{D}(\tilde{\bB},\bB) = \max_{\ell \in[d]} |\tilde{\kappa}_{2,j_{\ell}}\tilde{\bb}_{j_{\ell}} - \bb_{\ell}|_2$, where $\tilde{\kappa}_{1,j_{\ell}}, \tilde{\kappa}_{2,j_{\ell}} \in\{1,-1\}$ and $\{j_1,\ldots, j_d\}$ is a permutation of $\{1,\ldots,d\}$.  
	For any $(\breve{\bA}, \breve{\bB}) \in \mathcal{G}$, we have
	\begin{align*}
		\frac{1}{4}  \le &\, \max\{\mathscr{D}(\tilde{\bA},\bA), \mathscr{D}(\tilde{\bB},\bB)\} \\ \le &\, \max\{\mathscr{D}(\bA,\breve{\bA}), \mathscr{D}(\bB,\breve{\bB})\} + \max\{\mathscr{D}(\tilde{\bA},\breve{\bA}), \mathscr{D}(\tilde{\bB},\breve{\bB})\}\,.
	\end{align*}
	By Bonferroni inequality, it holds that
	\begin{equation*}
		\P\bigg[\max\{\mathscr{D}(\bA,\breve{\bA}), \mathscr{D}(\bB,\breve{\bB})\} \ge \frac{1}{8}\bigg] + \P\bigg[\max\{\mathscr{D}(\tilde{\bA},\breve{\bA}), \mathscr{D}(\tilde{\bB},\breve{\bB})\} \ge \frac{1}{8}\bigg] \ge 1\,,
	\end{equation*}
	which implies
	\begin{equation*}
		\max\bigg\{\P\bigg[\max\{\mathscr{D}(\bA,\breve{\bA}), \mathscr{D}(\bB,\breve{\bB})\} \ge \frac{1}{8}\bigg], \P\bigg[\max\{\mathscr{D}(\tilde{\bA},\breve{\bA})\,, \mathscr{D}(\tilde{\bB},\breve{\bB})\} \ge \frac{1}{8}\bigg]\bigg\} \ge \frac{1}{2}\,.
	\end{equation*}
	Recall $(\bA,\bB) \in \mathcal{U}$ and $(\tilde{\bA},\tilde{\bB}) \in \mathcal{U}$. Hence, for any $(\breve{\bA}, \breve{\bB}) \in \mathcal{G}$, 
	\begin{equation*}
		\sup_{(\bA,\bB) \in \mathcal{U}} \P\bigg[\max\{\mathscr{D}(\bA,\breve{\bA}), \mathscr{D}(\bB,\breve{\bB})\} \ge \frac{1}{8}\bigg] \ge \frac{1}{2}\,.
	\end{equation*}
	Since the above result holds for any $(\breve{\bA},\breve{\bB}) \in \mathcal{G}$, we complete the proof of Proposition \ref{pro:Omega-impossible}. 
	$\hfill\Box$

	\subsection{Proof of Proposition \ref{pro:rotation}}\label{sec:calce-sec4}
	Recall $\bUpsilon_0$, $\bUpsilon_1$, $\bUpsilon_2$ defined in \eqref{eq:bUpsilon_k}, and $\bGamma_i={\rm diag}(\bgamma_i)$ with $\bgamma_i=(\gamma_{1,i}, \ldots, \gamma_{d,i})^{\T}$ for any $i\in[d]$. Write $\bar{\bGamma}=(\bgamma_1,\ldots, \bgamma_{d} )$ and $\bGamma^{\dagger} =\sum_{i=1}^{d}\phi_i\bGamma_i$ with any $d$-dimensional vector $(\phi_1, \ldots, \phi_d)^\T \ne \bzero$ such that $\bH$ is invertible. By direct calculation, we have 
	\begin{align}\label{eq:upsilon-123}
		\bUpsilon_0  = (\bTheta \odot \bTheta)\bar{\bGamma}\,, \quad \bUpsilon_1  = \{\bTheta \odot (\bTheta^{-1})^{\T}\}  (\bGamma^{\dagger})^{-1} \bar{\bGamma}\,, \quad \bUpsilon_2  = \{(\bTheta^{-1})^{\T} \odot \bTheta\}(\bGamma^{\dagger})^{-1} \bar{\bGamma}\,.
	\end{align}
	Recall  $({\bh}^*_1,\ldots,{\bh}^*_d) = ({\bh}_1,\ldots,{\bh}_d)\bPi$  with $\bPi = \{2(\bUpsilon_0^\T\bUpsilon_2)(\bUpsilon_1^\T\bUpsilon_2 +  \bUpsilon_2^\T\bUpsilon_1)^{-1}(\bUpsilon_2^\T\bUpsilon_0) \}^{-1/2} $.  We can define $\{\bH_i^*\}_{i = 1}^d$, $\{\bGamma_i^*\}_{i = 1}^d$ and $\bar{\bGamma}^*$ in the same manner, respectively, as $\{\bH_i\}_{i = 1}^d$, $\{\bGamma_i\}_{i = 1}^d$ and $\bar{\bGamma}$ but with replacing  $\{{\bh}_1,\ldots,{\bh}_d\}$ by $\{{\bh}^*_1,\ldots,{\bh}^*_d\}$.  Let
	\begin{align*}
		& \bUpsilon_0^{*}   = \big(\textup{vec}(\bH_1^{*}),\ldots,\textup{vec}(\bH_d^{*})\big) \, , \qquad \bUpsilon_1^{*}   = \big(\textup{vec}( \bH^{-1}\bH_1^{*}),\ldots,\textup{vec}(\bH^{-1}\bH_d^{*})\big)\,,\\
		&~~~~~~~~~~~~~~~~~~~~~~
		\bUpsilon_2^{*}   = \big(\textup{vec}(\bH_1^{*}\bH^{-1}),\ldots,\textup{vec}(\bH_d^{*}\bH^{-1})\big)\,.
	\end{align*}
	Analogously, we also have $\bUpsilon_0^{*}  = (\bTheta \odot \bTheta)\bar{\bGamma}^{*}$, $\bUpsilon_1^{*}  = \{\bTheta \odot (\bTheta^{-1})^{\T}\} (\bGamma^{\dagger})^{-1} \bar{\bGamma}^{*}$, and $\bUpsilon_2^{*}  = \{(\bTheta^{-1})^{\T} \odot \bTheta\} ( \bGamma^{\dagger})^{-1} \bar{\bGamma}^{*}$. Then
	\begin{align}\label{eq:gamma*-i}
		(\bar{\bGamma}^{*})^\T \bar{\bGamma}^{*} = 2\{(\bUpsilon_0^{*})^\T\bUpsilon_2^{*}\}\{(\bUpsilon_1^{*})^\T\bUpsilon_2^{*} +  (\bUpsilon_2^{*})^\T\bUpsilon_1^{*}\}^{-1}\{(\bUpsilon_2^{*})^\T\bUpsilon_0^{*}\}\,.
	\end{align}
	Write $\bPi = (\boldsymbol{\pi}_1, \ldots, \boldsymbol{\pi}_{d})$ with $\boldsymbol{\pi}_{i}=(\pi_{1,i}, \ldots, \pi_{d,i})^{\T}$ for any $i\in[d]$. Then ${\bh}_{i}^{*} =\sum_{j=1}^{d}\pi_{j,i}{\bh}_{j}$,  which implies that $\bH^{*}_{i} = \sum_{j=1}^{d}\pi_{j,i}\bH_{j}$ for any $j\in[d]$.
	Since $\bH^{*}_{i} =\bTheta\bGamma_{i}^{*} \bTheta^{\T}$ and  $\bH_{i} =\bTheta\bGamma_{i}\bTheta^{\T}$ for any $i\in[d]$, we have $\bGamma_{i}^{*} = \sum_{j=1}^{d}\pi_{j,i}\bGamma_{j}$ for any $i\in[d]$, which implies that $\bar{\bGamma}^{*} = \bar{\bGamma}\bPi$. By \eqref{eq:upsilon-123}, $\bUpsilon_k^* = \bUpsilon_k\bPi$  for $k\in\{0,1, 2\}$. Together with  the definition of $\bPi$, by \eqref{eq:gamma*-i}, we have 
	\begin{align*}
		(\bar{\bGamma}^{*})^\T \bar{\bGamma}^{*} =&~ 2(\bPi^{\T} \bUpsilon_{0}^{\T} \bUpsilon_{2} \bPi) (\bPi^{\T} \bUpsilon_{1}^{\T} \bUpsilon_{2} \bPi + \bPi^{\T} \bUpsilon_{2}^{\T} \bUpsilon_{1} \bPi)^{-1}(\bPi^{\T} \bUpsilon_{2}^{\T} \bUpsilon_{0} \bPi) \\
		=&~ 2\bPi^{\T} (\bUpsilon_{0}^{\T} \bUpsilon_{2}) ( \bUpsilon_{1}^{\T} \bUpsilon_{2}  + \bUpsilon_{2}^{\T} \bUpsilon_{1} )^{-1}(\bUpsilon_{2}^{\T} \bUpsilon_{0}) \bPi\\
		=&~\bI_{d} \,,
	\end{align*}
	which implies that  $\rho(\bh_1^{*}, \ldots, \bh_d^{*}) =0$ and $\eta(\bh_1^{*}, \ldots, \bh_d^{*})=2$. We complete the proof of Proposition \ref{pro:rotation}. 
	$\hfill\Box$

	\section{Proofs of Theorem \ref{thm:rank} and Proposition \ref{pro:PQW}}\label{sec:thm:rank}
	\subsection{Proof of Theorem \ref{thm:rank}}
	To prove Theorem \ref{thm:rank}, we need  Lemmas \ref{lem:M1M2}--\ref{lem:mhm}. Lemma  \ref{lem:M1M2} is same as Lemma 1 of  \citeS{chang2023modelling}.
	% The proof of Lemma \ref{lem:php} is identical  to the proof of Proposition 3 in \citeS{chang2023modelling}. Here we omit the proofs of 
	% Lemmas \ref{lem:M1M2} and \ref{lem:php}.  
	The proofs of Lemmas \ref{lem:sigmaz-h} and \ref{lem:mhm} can be found in Sections \ref{sec:sub-sigmaz-h} and \ref{sec:sub-mhm}, respectively. 
	% the following lemma which is Lemma 1 of \citeS{chang2023modelling}. 
	
	\begin{lemma}\label{lem:M1M2}
		Under Conditions \ref{cd:tail} and \ref{cd:sgm_yxi}, if the threshold level $\delta_1=\breve{C} \{n^{-1}\log(pq)\}^{1/2}$ for some sufficiently large constant $\breve{C}>0$, we have $\|\hat{\bM}_1  - \bM_1  \|_2=O_{\rm p}(\Pi_{1,n})=\|\hat{\bM}_2 - \bM_2 \|_2 $, provided that $\Pi_{1,n}=o(1)$ and  $\log(pq)=o(n^c)$ for some constant $c \in (0,1)$ depending only on $r_1$ and $r_2$ specified in Condition \ref{cd:tail}.
	\end{lemma}

	% \subsubsection{Proof of Theorem \ref{thm:rank}(ii)}
	% To prove Theorem \ref{thm:rank}(ii), we need Lemmas \ref{lem:php}--\ref{lem:mhm}.  The proof of Lemma \ref{lem:php} is identically to the proof of Proposition 3 in \citeS{chang2023modelling}. The proofs of Lemmas \ref{lem:sigmaz-h} and \ref{lem:mhm} are given in Section \ref{sec:sub-sigmaz-h} and \ref{sec:sub-mhm}, respectively. 
	% \begin{lemma}\label{lem:php}
		% 	Let Conditions \ref{cd:ra}--\ref{cd:sgm_yxi} hold and the threshold level $\delta_1=\bar{C}\{n^{-1}\log(pq)\}^{1/2}$ for some sufficiently large constant $\bar{C}>0$. If $(\hat{d}_1,\hat{d}_2)=(d_1,d_2)$, there exist some orthogonal matrices $\bE_1$ and $\bE_2$ such that $\|\hat{\bP}\bE_1  - \bP  \|_2=O_{\rm p}(\Pi_{1,n})=\|\hat{\bQ}\bE_2 - \bQ \|_2 $, provided that $\Pi_{1,n}=o(1)$ and  $\log(pq)=o(n^c)$ for some constant $c \in (0,1)$ depending only on $r_1$ and $r_2$.
		% \end{lemma}

	\begin{lemma}\label{lem:sigmaz-h}
		Under Conditions \ref{cd:tail} and \ref{cd:sgm_yxi}, it holds that $\|T_{\delta_2}\{\hat{\bSigma}_{ \vec{\bY} }(k)\} -\bSigma_{\vec{\bY}}(k)\|_2 = O_{\rm p}(\Pi_{2,n}) $,
		provided that $\log(pq)=o(n^c)$ for some constant $c \in (0,1)$ depending only on $r_1$ and $r_2$. 
	\end{lemma}

	\begin{lemma}\label{lem:mhm}
		Let Conditions \ref{cd:ra}--\ref{cd:sgm_yxi} hold and  the threshold level $\delta_2=\bar{C}\{n^{-1}\log(pq)\}^{1/2}$ for some sufficiently large constant $\bar{C} > 0$. If $(\hat{d}_1,\hat{d}_2)=(d_1,d_2)$, there exist some orthonormal matrices $\bE_1\in \mathbb{R}^{d_1\times d_1} $ and  $\bE_2\in \mathbb{R}^{d_2\times d_2}$  such that $\|(\bE_2 \otimes \bE_1)^{{\T}}\hat{\bM}(\bE_2 \otimes \bE_1) - \bM\|_2=O_{\rm p}(\Pi_{1,n}+\Pi_{2,n})$, provided that $\Pi_{1,n}+\Pi_{2,n}=o(1)$ and  $\log(pq)=o(n^c)$ for some constant $c \in (0,1)$ depending only on $r_1$ and $r_2$.
	\end{lemma}

	We first prove the consistency of $\hat{d}_1$ and $\hat{d}_2$.   Denote by $\hat{\lambda}_1 \ge \cdots \ge \hat{\lambda}_p$ and $\lambda_1 \ge \cdots \ge \lambda_p$, respectively, the eigenvalues of $\hat{\bM}_1$ and $\bM_1$. By Lemma \ref{lem:M1M2}, $\max_{j\in[p]}|\hat{\lambda}_j - \lambda_j| \le \| \hat{\bM}_1  - \bM_1  \|_2 = o_{\textup p}(c_{1,n})$ for some $c_{1,n} = o(1)$ satisfying $\Pi_{1,n} = o(c_{1,n})$. By Condition  \ref{cd:bounded-value}(ii),  we know $\lambda_{d_1} \ge C$ for some positive constant $C$, and $\lambda_{j} = 0$ for any $d_1<j\le p$.
	Hence,   $(\hat{\lambda}_{j+1}+c_{1,n})/(\hat{\lambda}_{j}+c_{1,n}) \overset{{\rm p}}{\rightarrow} \lambda_{j+1}/\lambda_{j}$ for any $1\le j< d_1$, $(\hat{\lambda}_{d_1+1}+c_{1,n})/(\hat{\lambda}_{d_1}+c_{1,n}) \overset{{\rm p}}{\rightarrow} 0$, and $(\hat{\lambda}_{j+1}+c_{1,n})/(\hat{\lambda}_{j}+c_{1,n}) \overset{{\rm p}}{\rightarrow} 1$ for any $d_1<j\le p$, which implies $(\hat{\lambda}_{d_1+1}+c_{1,n})/(\hat{\lambda}_{d_1}+c_{1,n}) = \min_{j \in [p]}(\hat{\lambda}_{j+1}+c_{1,n})/(\hat{\lambda}_{j}+c_{1,n})$ with probability approaching one. This indicates that $\P(\hat{d}_1 = d_1) \to 1$ as $n \to \infty$ provided that $\Pi_{1,n}=o(1)$ and  $\log(pq)=o(n^c)$ for some constant $c \in (0,1)$ depending only on $r_1$ and $r_2$.   Analogously, we can also show $\P(\hat{d}_2 = d_2) \to 1$ as $n \to \infty$.

	We then prove the consistency of $\hat{d}$. Notice that
	\begin{align}\label{eq:d12h-d12}
		\mathbb{P}(  \hat{d}_1=d_1,\,\hat{d}_2=d_2)  \ge  \mathbb{P}(  \hat{d}_1=d_1) +  \mathbb{P}(\hat{d}_2=d_2) - \mathbb{P}(  \{\hat{d}_1=d_1\}\cup\{\hat{d}_2=d_2\}) \to 1
	\end{align}
	as $n \to \infty$, provided that $\Pi_{1,n}=o(1)$ and  $\log(pq)=o(n^c)$ for some constant $c \in (0,1)$ depending only on $r_1$ and $r_2$.   Recall $d_1d_2\ge d$ and $\max(d_1,d_2) \le d$.  When $d=1$, then $d_1=1=d_2$. By \eqref{eq:dh} and \eqref{eq:d12h-d12}, we have
	\begin{align*}
		\mathbb{P}(\hat{d}=d)=  \mathbb{P}(\hat{d}=1)   \ge  \mathbb{P}(  \hat{d}_1\hat{d}_2  =1) = \mathbb{P}(  \hat{d}_1=1,\,\hat{d}_2=1) =\mathbb{P}(\hat{d}_1=d_1,\,\hat{d}_2=d_2) \to 1 
	\end{align*}
	as $n \to \infty$, provided that $\Pi_{1,n}=o(1)$ and  $\log(pq)=o(n^c)$ for some constant $c \in (0,1)$ depending only on $r_1$ and $r_2$. When $d\ge 2$,  it holds that
	\begin{align}\label{eq:dhd-2}
		\mathbb{P}(\hat{d}=d)=&~ \mathbb{P}(\hat{d}=d,\,  \mathcal{G}_1)  + \mathbb{P}(\hat{d}=d,\,  \mathcal{G}_1^{\rm c})  
		\ge  \mathbb{P}(\hat{d}=d \,|\,  \mathcal{G}_1)  \mathbb{P}( \mathcal{G}_1)\,,
	\end{align}
	where $\mathcal{G}_1=\{\hat{d}_1=d_1, \hat{d}_2=d_2\}$. By \eqref{eq:d12h-d12}, we have $\mathbb{P}(\mathcal{G}_1)  \to 1$ as $n\to \infty$,  provided that $\Pi_{1,n}=o(1)$ and  $\log(pq)=o(n^c)$ for some constant $c \in (0,1)$ depending only on $r_1$ and $r_2$. Recall $\hat{\bM}$ is a $(\hat{d}_1\hat{d}_2) \times (\hat{d}_1\hat{d}_2)$ matrix and we adopt the convention $\lambda_{\hat{d}_1\hat{d}_2 + 1}(\hat{\bM}) = 0$. Write  $\hat{\lambda}_{j}^{*} =\lambda_{j}(\hat{\bM})$ for any $j\in[\hat{d}_1\hat{d}_2+1]$.  Denote by  $\lambda_1^{*} \ge \cdots \ge \lambda_{d_1d_2}^{*}$ the eigenvalues of $\bM$. Due to $d_1d_2 \ge d$, we have $\hat{d}_1\hat{d}_2 \ge 2$ restricted on  $\mathcal{G}_1$. Hence, by \eqref{eq:dh}, restricted on $\mathcal{G}_1$, we have   $\hat{d}=\arg\min_{j\in [d_1d_2]}(\hat{\lambda}_{j+1}^{*}+c_{3,n})/(\hat{\lambda}_{j}^{*} +c_{3,n})$ with $c_{3,n} \to 0^{+}$ as $n\to \infty$. Since $\bE_1$ and $\bE_2$ are two orthonormal matrices, then  $\hat{\bM}$ and $(\bE_2 \otimes \bE_1)^{{\T}}\hat{\bM}(\bE_2 \otimes \bE_1)$ share the same eigenvalues.  By Lemma \ref{lem:mhm}, restricted on $\mathcal{G}_1$,  we have  $\max_{j\in[d_1d_2]}|\hat{\lambda}_j^{*} - \lambda_j^{*}| \le \| (\bE_2 \otimes \bE_1)^{{\T}}\hat{\bM}(\bE_2 \otimes \bE_1)  - \bM  \|_2 = o_{\textup p}(c_{3,n})$ for some $c_{3,n} = o(1)$ satisfying $\Pi_{1,n}+\Pi_{2,n} = o(c_{3,n})$. Under Condition \ref{cd:bounded-value}(ii), it holds that $\lambda_{d}^{*}\ge C$ for some positive constant $C$, and $\lambda_{j}^{*} =0$ for any $ d< j \le d_1d_2$. Restricted on $\mathcal{G}_1$, we have $(\hat{\lambda}^{*}_{j+1}+c_{3,n})/(\hat{\lambda}_{j}^{*}+c_{3,n}) \overset{{\rm p}}{\rightarrow} \lambda_{j+1}/\lambda_{j}$   for any $1\le j< d$,  
	$(\hat{\lambda}_{d+1}^{*}+c_{3,n})/(\hat{\lambda}_{d}^{*}+c_{3,n}) \overset{{\rm p}}{\rightarrow} 0$, and $(\hat{\lambda}_{j+1}^{*}+c_{3,n})/(\hat{\lambda}_{j}^{*}+c_{3,n}) \overset{{\rm p}}{\rightarrow} 1$ for any $ d <j \le d_1d_2$. 
	% Restricted on $\mathcal{G}_1$, if , we have $(\hat{\lambda}_{d+1}^{*}+c_{3,n})/(\hat{\lambda}_{d}^{*}+c_{3,n}) \overset{{\rm p}}{\rightarrow} 0$. And if $d< d_1d_2$, it holds that $(\hat{\lambda}_{d+1}^{*}+c_{3,n})/(\hat{\lambda}_{d}^{*}+c_{3,n}) \overset{{\rm p}}{\rightarrow} 0$ and $(\hat{\lambda}_{j+1}^{*}+c_{3,n})/(\hat{\lambda}_{j}^{*}+c_{3,n}) \overset{{\rm p}}{\rightarrow} 1$ for $ d <j \le d_1d_2$. Furthermore,   restricted on $\mathcal{G}_1$, we have   $(\hat{\lambda}^{*}_{j+1}+c_{3,n})/(\hat{\lambda}_{j}^{*}+c_{3,n}) \overset{{\rm p}}{\rightarrow} \lambda_{j+1}/\lambda_{j}$ for $1\le j< d$.
	Hence, restricted on $\mathcal{G}_1$, we have $(\hat{\lambda}_{d+1}^{*}+c_{3,n})/(\hat{\lambda}_{d}^{*}+c_{3,n}) = \min_{j \in [d_1 d_2]}(\hat{\lambda}_{j+1}^{*}+c_{3,n})/(\hat{\lambda}_{j}^{*}+c_{3,n})$ with probability approaching one, which implies $ \P(\hat{d} = d\,|\,\mathcal{G}_1)  \to 1$ as $n \to \infty$.  By \eqref{eq:dhd-2}, when $d\ge 2$, we have $\mathbb{P}(\hat{d}=d) \to 1$ as $n\to \infty$, provided that $\Pi_{1,n}=o(1)$, $\Pi_{2,n}=o(1)$ and  $\log(pq)=o(n^c)$ for some constant $c \in (0,1)$ depending only on $r_1$ and $r_2$. We complete the proof of Theorem \ref{thm:rank}.
	$\hfill\Box$

	\subsection{Proof of Proposition \ref{pro:PQW}}\label{sec:proof-PQW}
	When $(\hat{d}_1,\hat{d}_2) = (d_1,d_2)$, the convergence rates of $\|\hat{\bP}\bE_1  - \bP  \|_2 $ and $\|\hat{\bQ}\bE_2 - \bQ \|_2 $ have been established in  \eqref{eq:php-qhq} in Section \ref{sec:sub-mhm} for the proof of Lemma \ref{lem:mhm}. Now we consider the convergence rate of the estimation error for$\bW$. Recall the columns of $\hat{\bW}$ are the $\hat{d}$ orthonormal eigenvectors of $\hat{\bM}$ corresponding to its $\hat{d}$ largest eigenvalues. Since $(\hat{d}_1,\hat{d}_2,\hat{d}) = (d_1,d_2,d)$ and $\bE_2 \otimes \bE_1$ is an orthogonal matrix, we know $(\bE_2 \otimes \bE_1 )^{{\T}} \hat{\bW}$ is a $(\hat{d}_1\hat{d}_2)\times \hat{d}$ matrix of which the columns are the $\hat{d}$ orthonormal eigenvectors of $(\bE_2 \otimes \bE_1)^{{\T}} \hat{\bM}(\bE_2 \otimes \bE_1 )$ corresponding to its $\hat{d}$  largest eigenvalues. Write $ \bSigma_{\bx^{*}}(k) =(n-k)^{-1}\sum_{t=k+1}^{n} \mathbb{E} [\{ \bx^*_t - \mathbb{E}(\bar{\bx}^* ) \}\{ \bx^*_{t-k} - \mathbb{E}(\bar{\bx}^* )\}^{\T} ]$ and  $\bSigma_{\bx^{*}, \vec{\bDelta}}(k) =(n-k)^{-1}\sum_{t=k+1}^{n}\mathbb{E} [\{ \bx^*_t - \mathbb{E}(\bar{\bx}^* ) \}\vec{\bDelta}_{t-k}^{\T}]$ with $\bar{\bx}^{*}=n^{-1}\sum_{t=1}^{n}\bx_{t}^{*}$ and $\bx^{*} = \bTheta\bx_{t}$. Recall $\bM$ defined below \eqref{eq:mixinga}. Notice that $\bM=\bW\tilde{\bD}\bW^{\T}$ with $\tilde{\bD}= \sum_{k=1}^{\tilde{K}}\{\bSigma_{\bx^{*}}(k) \bW^{{\T}} +  \bSigma_{\bx^{*}, \vec{\bDelta}}(k)\}\{\bSigma_{\bx^{*}}(k) \bW^{{\T}} +  \bSigma_{\bx^{*}, \vec{\bDelta}}(k)\}^{{\T}}$ and ${\rm rank}(\bM)=d$.  Let $\bL$ be an orthogonal complement of $\bW$, i.e., $\bW^{\T}\bL = \mathbf{0}$. Notice that $\bW^{\T}\bW = \mathbf{I}_d$.  Then $\bM \bL= \mathbf{0}$ and
	\begin{align*}
		\begin{pmatrix}
			\bW^{{\T}}\\   \bL^{{\T}}
		\end{pmatrix} \bM
		\begin{pmatrix}
			\bW & \bL
		\end{pmatrix}=
		\begin{pmatrix}
			\tilde{\bD} & \bzero\\ \bzero & \bzero
		\end{pmatrix}
	\end{align*}
	with ${\rm sep}(\tilde{\bD}, \bzero):= \lambda_{\min}(\tilde{\bD}) = \lambda_{d}(\bM)$. Under Condition \ref{cd:bounded-value}(ii), by Lemma \ref{lem:mhm}, we have $\|(\bE_2^{\T} \otimes \bE_1^{\T} )\hat{\bM}(\bE_2 \otimes \bE_1 ) -\bM\|_2 \le {\rm sep}(\tilde{\bD}, \bzero)/5$ with probability approaching one, provided that $\Pi_{1,n} + \Pi_{2,n} =o(1)$ and $\log(pq)=o(n^c)$ for sufficient large $n$ and some constant $c \in (0,1)$ depending only on $r_1$ and $r_2$. 
	By Lemma 1  of \citeS{Chang2018}, there exists a $d\times d$ orthonormal matrix $\bE_3$ such that
	\begin{equation*}
		\|(\bE_2 \otimes \bE_1)^{{\T}}  \hat{\bW}  \bE_3  - \bW  \|_2 \le \frac{8}{{\rm sep}(\tilde{\bD}, \bzero)} \|(\bE_2 \otimes \bE_1 )^{{\T}} \hat{\bM}(\bE_2 \otimes \bE_1 )-\bM\|_2 
	\end{equation*}
	with probability approaching one. Hence, by Lemma \ref{lem:mhm} and Condition \ref{cd:bounded-value}(ii), we have $\|(\bE_2 \otimes \bE_1)^{{\T}}  \hat{\bW}  \bE_3 - \bW
	\|_2  = O_{\rm p}(\Pi_{1,n}+\Pi_{2,n})$ provided that $\Pi_{1,n} +\Pi_{2,n}=o(1)$ and  $\log(pq)=o(n^c)$ for some constant $c \in (0,1)$ depending only on $r_1$ and $r_2$. 	We complete the proof of Proposition \ref{pro:PQW}.  
	$\hfill\Box$

	\section{Proof of Theorem \ref{thm:hat_a_hat_b}}\label{sec:thm:hat_a_hat_b}
	
	% \begin{lemma}\label{lemma:whw}
		% 	Let Conditions \ref{cd:ra}--\ref{cd:sgm_yxi} hold. Suppose that $(\hat{d}_1, \hat{d}_2,\hat{d})=(d_1,d_2,d)$,  there exists a $(d_1d_2)\times d$ matrix $\bW^{*}$ of which columns are the $d$ orthonormal eigenvectors of $\bM$ corresponding to its $d$ largest eigenvalues such that $\|(\bE_2 \otimes \bE_1)^{{\T}} \hat{\bW}^{*} -\bW^{*} \|_2=O_{\rm p}(\Pi_{1,n}) + O_{\rm p}(\Pi_{2,n})$ with $(\bE_1,\bE_2)$ specified in Lemma \ref{lem:php},  provided that $\Pi_{1,n}=o(1)$, $\Pi_{2,n}=o(1)$ and  $\log(pq)=o(n^c)$ for some constant $c \in (0,1)$ depending only on $r_1$ and $r_2$.
		% \end{lemma}
	
	Let $\breve{\bTheta} = \bE_3\bTheta = (\breve{\btheta}_1, \ldots, \breve{\btheta}_{d})$ with $\bE_3$ specified in Proposition \ref{pro:PQW} and $\varpi(\breve{\bOmega}) = \lambda_{(d-1)d/2}(\breve{\bOmega}^{\T}\breve{\bOmega})$ with $\breve{\bOmega}$ specified in Section \ref{sec:asymptotics}. Recall   $\hat{\bTheta} $ defined in Section  \ref{sec:estimation}. Write $\hat{\bTheta} =(\hat{\btheta}_1, \ldots, \hat{\btheta}_{\hat{d}})$.   To prove Theorem \ref{thm:hat_a_hat_b}, we need the following lemma with its proof given in Section \ref{sec:sub-theta}.

	\begin{lemma}\label{lemma:theta}
		Let Conditions \ref{cd:ra}--\ref{cd:sgm_yxi} hold. If  $\textup{rank}(\bOmega) = d(d-1)/2$ and $(\hat{d}_1, \hat{d}_2, \hat{d}) = (d_1,d_2,d)$, there exists a permutation of $(1,\ldots,d)$, denoted by $(j_1,\ldots,j_d)$, 
		such that 
		\begin{align*}
			\max_{\ell \in[d]} | \kappa_{\ell} \hat{\btheta}_{j_\ell}  - \breve{\btheta}_{\ell} |_2  = \frac{O_{\rm p} (\Pi_{1,n} + \Pi_{2,n}) }{\varpi(\breve{\bOmega})}
		\end{align*}
		with some $\kappa_{\ell} \in \{-1, 1\}$, provided that $ (\Pi_{1,n}+\Pi_{2,n})\{\varpi(\breve{\bOmega})\}^{-1} =o_{\rm p}(1)$  and  $\log(pq)=o(n^c)$ for some constant $c \in (0,1)$ depending only on $r_1$ and $r_2$.
		% \begin{itemize}
			%     \item [(i)] There exists a permutation of $(1,\ldots,d)$, denoted by $(j_1,\ldots,j_d)$, such that 
			%     \begin{align*}
				%         \max_{\ell \in [d]}|\kappa_{\ell} \hat{\btheta}_{j_\ell}^{\rm (init)} - \breve{\btheta}_{\ell} |_2  = \frac{\eta(\bar{\breve{\bGamma}} ) \cdot O_{\rm p} (\Pi_{1,n} + \Pi_{2,n}) }{\varpi(\breve{\bOmega}) \{1 - \rho^2(\bar{\breve{\bGamma}}) \}}
				%     \end{align*}
			% with some $\kappa_{\ell} \in \{-1, 1\}$, provided that $ (\Pi_{1,n}+\Pi_{2,n}) \eta(\bar{\breve{\bGamma}} )[\varpi(\breve{\bOmega}) \{1 - \rho^2(\bar{\breve{\bGamma}}) \}]^{-1} =o_{\rm p}(1)$  and  $\log(pq)=o(n^c)$ for some constant $c \in (0,1)$ depending only on $r_1$ and $r_2$, where $\rho(\cdot)$ and $\eta(\cdot)$ are defined in \eqref{eq:alpha and rho}. 
			%  \item [(ii)]  There exists a permutation of $(1,\ldots,d)$, denoted by $(j^{*}_1,\ldots,j^{*}_d)$, 
			%  such that 
			%  \begin{align*}
				%      \max_{\ell \in[d]} | \tilde{\kappa}_{\ell} \hat{\btheta}_{j^{*}_\ell} - \breve{\btheta}_{\ell} |_2  = \frac{O_{\rm p} (\Pi_{1,n} + \Pi_{2,n}) }{\varpi(\breve{\bOmega})}
				%  \end{align*}
			% with some $\tilde{\kappa}_{\ell} \in \{-1, 1\}$, provided that $ (\Pi_{1,n}+\Pi_{2,n})\{\varpi(\breve{\bOmega})\}^{-1} =o_{\rm p}(1)$  and  $\log(pq)=o(n^c)$ for some constant $c \in (0,1)$ depending only on $r_1$ and $r_2$.
			% \end{itemize}
	\end{lemma}

	% \subsection{Proof of Theorem \ref{thm:hat_a_hat_b}(i)}\label{sec:thm:hat_a_hat_b-1}
	Recall ${\rm vec}(\hat{\bC}_{\ell} ) =\sum_{i=1}^{\hat{d}}\hat{\theta}_{i,\ell} {\rm vec}(\hat{\bW}_{i})$ given in Section \ref{sec:estimation}, where  $\hat{\bTheta}=(\hat{\theta}_{i,j})_{\hat{d} \times \hat{d}}$ and $\hat{\bW} =({\rm vec}(\hat{\bW}_1), \ldots, {\rm vec}(\hat{\bW}_{\hat{d}}) )$.  Recall $(\hat{d}_1, \hat{d}_2, \hat{d}) =(d_1,d_2,d)$. Let $\tilde{\bW} = \hat{\bW}\hat{\bTheta}  = (\tilde{\bw}_1,\ldots,\tilde{\bw}_{d})$ with  $\tilde{\bw}_{\ell}=(\tilde{w}_{1,\ell}, \ldots, \tilde{w}_{d_1d_2,\ell})^{{\T}}$ for any $\ell\in[d]$.  Then  $ 
	\textup{vec}(\hat{\bC}_{\ell}) =\tilde{\bw}_{\ell} $ and  $ \text{vec}(\bE_1^{\T} \hat{\bC}_{\ell}\bE_2)=(\bE_2^{\T} \otimes \bE_1^{\T})\tilde{\bw}_\ell = (\bE_2^{\T} \otimes \bE_1^{\T})\hat{\bW}\hat{\btheta}_{\ell}  $ for any  $\ell \in [d]$, where $\bE_1$ and $\bE_2$ are specified in Proposition \ref{pro:PQW}. Recall  $\bC =\bV \odot \bU =(\vec\bC_1, \ldots, \vec\bC_d)$  with $\bU=(\bu_1, \ldots, \bu_d)$ and $\bV=(\bv_1, \ldots, \bv_d)$,  $\bC=\bW\bTheta$  with $\bTheta=(\btheta_{1}, \ldots, \btheta_{d})$, and $\bC_\ell = \bu_\ell \bv_\ell^{\T}$ for any $\ell \in[d]$. It holds that $\vec\bC_\ell =\bW \btheta_{\ell}  $ for any $\ell \in[d]$. Notice that $\|\bE_2  \otimes \bE_1\|_2=1=  \|\hat{\bW}\|_2 $ and $|\btheta_{\ell}|_2=1$. As shown in  Section \ref{sec:sub-theta} for the proof of Lemma \ref{lemma:theta}, we know $\varpi(\breve{\bOmega}) \le 8$.  For the permutation $(j_1,\ldots,j_d)$ of $(1,\ldots,d)$ specified in Lemma \ref{lemma:theta},  by Proposition \ref{pro:PQW} and Lemma  \ref{lemma:theta},  we have
	\begin{align}\label{eq:C-Chat-Fnorm}
		\| \kappa_{\ell} \bE_1^{\T}\hat{\bC}_{j_\ell} \bE_2 - \bC_\ell \|_{\text{F}} =&~ | \kappa_{\ell} \text{vec}( \bE_1^{\T}\hat{\bC}_{j_\ell}  \bE_2) - \text{vec}(\bC_\ell) |_{2} \notag \\ 
		=&~	| \kappa_{\ell} (\bE_2 \otimes \bE_1)^{{\T}} \hat{\bW}  \bE_{3}\bE_{3}^{\T}  \hat{\btheta}_{j_\ell}  -  \bW  \btheta_{\ell}  |_\text{2}  \notag \\
		\le&~ | (\bE_2 \otimes \bE_1)^{{\T}} \hat{\bW} \bE_{3}(\kappa_{\ell}\bE_{3}^{\T} \hat{\btheta}_{j_\ell}  - \btheta_{\ell}) |_\text{2} \notag\\
		&+ | \{(\bE_2 \otimes \bE_1)^{{\T}}\hat{\bW} \bE_{3} - \bW  \}\btheta_{\ell} |_\text{2} \notag\\
		\le&~ | \kappa_{\ell} \hat{\btheta}_{j_\ell}   - \breve{\btheta}_{\ell} |_2 +  \| (\bE_2  \otimes \bE_1 )^{{\T}} \hat{\bW} \bE_{3} - \bW  \|_2 \notag\\
		=&~  \frac{ O_{\rm p} (\Pi_{1,n} + \Pi_{2,n}) }{\varpi(\breve{\bOmega}) }   
	\end{align}
	for any $\ell \in [d]$ with some $\kappa_{\ell} \in \{-1, 1\}$, provided that $ (\Pi_{1,n}+\Pi_{2,n}) \{\varpi(\breve{\bOmega}) \}^{-1} =o_{\rm p}(1)$  and  $\log(pq)=o(n^c)$ for some constant $c \in (0,1)$ depending only on $r_1$ and $r_2$. For any $\ell \in [d]$,
	let $\hat{\bbeta}_{\bu,j_\ell}$ and $ \bbeta_{\bu,\ell}$ be, respectively, the eigenvectors of $\hat{\bC}_{j_\ell}\hat{\bC}_{j_\ell}^{\T}$ and $\bC_\ell \bC_\ell^{\T}$ with unit $L_2$-norm corresponding to their largest eigenvalues. Recall $\bC_\ell = \bu_\ell \bv_\ell^\T$ with $|\bu_{\ell}|_2=1=|\bv_{\ell}|_2$ for any $\ell\in[d]$. We have $  \bbeta_{\bu,\ell} =\bu_\ell$ for any $\ell \in[d]$. As mentioned in Section \ref{sec:estimation}, we also have $    \hat{\bbeta}_{\bu,{j_\ell}} = \hat{\bu}_{j_\ell}$ for any $\ell \in[d]$. 
	Let $\hat{\bbeta}^*_{\bu,j_\ell} = \bE_1^{\T}\hat{\bbeta}_{\bu,j_\ell}$ with $\bE_1$ specified in Proposition \ref{pro:PQW}. Hence, $\hat{\bbeta}^*_{\bu,j_\ell}$ is the eigenvector of  $\bE_1^{\T}\hat{\bC}_{j_{\ell}}\hat{\bC}_{j_{\ell}}^{\T}\bE_1$   corresponding to its largest eigenvalue.  Notice that ${\rm rank}(\bC_{\ell}\bC_{\ell}^{{\T}})=1$ and its  nonzero eigenvalue is 1. 
	Let $\br_{{\rm c},\ell,j}$ for $j\in[d_1-1]$ be the 
	$d_1-1$ orthogonal eigenvectors of $\bC_{\ell}\bC_{\ell}^{{\T}}$ corresponding to the zero eigenvalues such that $\br_{{\rm c},\ell,j}^{\T}\bu_{\ell}=0$.
	Write $\bR_{{\rm c},\ell}=(\br_{{\rm c},\ell,1}, \ldots, \br_{{\rm c},\ell,d_1-1}) \in \mathbb{R}^{d_1\times (d_1-1)}$. We have ${\rm rank}(\bR_{{\rm c},\ell})=d_1-1$ and $\bR_{{\rm c},\ell}^{{\T}}\bu_{\ell}=\mathbf{0}$.  Since   $\bC_{\ell}\bC_{\ell}^{{\T}}=(\bu_{\ell}, \bR_{{\rm c},\ell}) \bLambda_{{\rm c},\ell} (\bu_{\ell}, \bR_{{\rm c},\ell})^{{\T}}$ with  $\bLambda_{{\rm c},\ell}={\rm diag}(1,0, \ldots, 0)$, we have $\bR_{{\rm c},\ell}^{{\T}} \bC_{\ell}\bC_{\ell}^{{\T}} \bR_{{\rm c},\ell} -\bI_{d_1-1} = -\bI_{d_1-1}$, which implies the smallest singular value of $\bR_{{\rm c},\ell}^{{\T}} \bC_{\ell}\bC_{\ell}^{{\T}} \bR_{{\rm c},\ell} -\bI_{d_1-1}$ is 1. Applying Corollary 7.2.6 of \citeS{golub2013matrix}, it holds that
	\begin{equation*}
		| \kappa_{1,\ell}  \hat{\bbeta}^*_{\bu,{j_\ell}} - \bbeta_{\bu,\ell} |_2 \lesssim \| \bE_1^{\T}\hat{\bC}_{j_\ell}\hat{\bC}_{j_\ell}^{\T}  \bE_1 - \bC_\ell  \bC^{\T}_\ell \|_2
	\end{equation*}
	for any $\ell \in [d]$ with some $\kappa_{1,\ell} \in \{-1,1\}$, provided that $5\| \bE_1^{\T} \hat{\bC}_{j_\ell}\hat{\bC}_{j_\ell}^{\T} \bE_1 - \bC_\ell\bC^{\T} _\ell \|_{\text{F}} \le 1$.
	By Triangle inequality and \eqref{eq:C-Chat-Fnorm}, we have 
	\begin{align*}
		&\| \bE_1^{\T}\hat{\bC}_{j_\ell}\hat{\bC}_{j_\ell}^{\T}\bE_1 - \bC _\ell \bC^{\T} _\ell \|_{\text{F}} \\
		&~~~~~~~\le \| (\kappa_{\ell} \bE_1^{\T}\hat{\bC}_{j_\ell}\bE_2 - \bC_\ell)(\kappa_{\ell}  \bE_1^{\T}\hat{\bC}_{j_\ell}\bE_2 - \bC_\ell)^{\T} \|_{\text{F}} + 2 \| \bC_\ell (\kappa_{\ell}  \bE_1^{\T}\hat{\bC}_{j_\ell}\bE_2 - \bC_\ell)^{\T} \|_{\text{F}}\\
		&~~~~~~~\le \| \kappa_{\ell}  \bE_1^{\T}\hat{\bC}_{j_\ell}\bE_2 - \bC_\ell \|^2_{\text{F}} + 2 \|\bC_\ell \|_2 \| \kappa_{\ell} \bE_1^{\T}\hat{\bC}_{j_\ell}\bE_2 - \bC_\ell \|_{\text{F}}\\
		&~~~~~~~ = \frac{O_{\rm p} (\Pi_{1,n} + \Pi_{2,n}) }{\varpi(\breve{\bOmega}) } \,, 
	\end{align*} 
	provided that $ (\Pi_{1,n} + \Pi_{2,n})\{\varpi(\breve{\bOmega})\}^{-1} = o_{\rm p}(1)$   and  $\log(pq)=o(n^c)$ for some constant $c \in (0,1)$ depending only on $r_1$ and $r_2$. Recall $\hat{\bbeta}^*_{\bu,{j_\ell}} =\bE_{1}^{{\T}} \hat{\bbeta}_{\bu,{j_\ell}} = \bE_1^{\T} \hat{\bu}_{j_\ell}  $ and $\bbeta_{\bu,\ell}=\bu_\ell$. Hence, it holds that
	\begin{equation}\label{eq:keuh-u}
		\max_{\ell\in[d]}|\kappa_{1,\ell}\bE_1^{\T} \hat{\bu}_{j_\ell}  -\bu_\ell |_2 = \max_{\ell\in[d]}| \kappa_{1,\ell}  \hat{\bbeta}^*_{\bu,{j_\ell}} - \bbeta_{\bu,\ell} |_2 =  \frac{ O_{\rm p} (\Pi_{1,n} + \Pi_{2,n}) }{\varpi(\breve{\bOmega}) }  \,,
	\end{equation}
	provided that $ (\Pi_{1,n} + \Pi_{2,n}) \{\varpi(\breve{\bOmega})\}^{-1}= o_{\rm p}(1)$  and  $\log(pq)=o(n^c)$ for some constant $c \in (0,1)$ depending only on $r_1$ and $r_2$. Recall $(\hat{d}_1, \hat{d}_2,\hat{d})=(d_1,d_2,d)$. Write $\hat{\bA} = (\hat{\ba}_1,\ldots,\hat{\ba}_{d})$ and $\hat{\bU} =(\hat{\bu}_{1}, \ldots, \hat{\bu}_{d})$.  Recall $\bA=(\ba_1, \ldots, \ba_{d})$ and $\bU=(\bu_1, \ldots, \bu_{d})$ with $|\ba_{\ell}|_2=1 = |\bu_{\ell}|_2$ for any $\ell \in [d]$.  Since $\bA=\bP\bU$,  $\hat{\bA}=\hat{\bP}\hat{\bU}$  and $\|\bP\|_2=1$,  by Proposition \ref{pro:PQW}, we have  
	\begin{align*}
		\max_{\ell\in[d]}|\kappa_{1,\ell}\hat{\ba}_{j_\ell} -\ba_\ell |_2=&~ \max_{\ell\in[d]}|\kappa_{1,\ell}\hat{\bP}\hat{\bu}_{j_\ell}  -\bP\bu_\ell |_2  \\
		\le&~ \|\hat{\bP}\bE_1 - \bP\|_2 \times \max_{\ell\in[d]}|\kappa_{1,\ell}\bE_1^{\T} \hat{\bu}_{j_\ell}  -\bu_{\ell}|_2 \\
		&+\|\bP\|_2 \times \max_{\ell\in[d]}|\kappa_{1,\ell}\bE_1^{\T}\hat{\bu}_{j_\ell}  -\bu_\ell |_2 + \|\hat{\bP}\bE_1 - \bP\|_2 \times \max_{\ell\in[d]}|\bu_\ell |_2\\
		= &~ \frac{ O_{\rm p} (\Pi_{1,n} + \Pi_{2,n}) }{\varpi(\breve{\bOmega})  } \,,
	\end{align*}
	provided that $ (\Pi_{1,n} + \Pi_{2,n})\{ \varpi(\breve{\bOmega})\}^{-1}= o_{\rm p}(1)$  and  $\log(pq)=o(n^c)$ for some constant $c \in (0,1)$ depending only on $r_1$ and $r_2$. 
	% Recall $|{\ba}_{j}^{\rm (init)}|_{2} = 1=|\ba_{j}|_{2}$ for any $j\in[d]$. Since  $| \kappa_{1,\ell}\hat{\ba}_{j_\ell}^{\rm (init)}-\ba_\ell |^2_2 \ge 2 - 2 |(\hat{\ba}_{j_\ell}^{\rm (init)})^{\T}\ba_\ell  |$, then $1 - |(\hat{\ba}_{j_\ell}^{\rm (init)})^{\T}\ba_\ell  | = O_{\rm p} \{ d^6(\Pi_{1,n} + \Pi_{2,n})^2 \}\alpha^2(\bar{\breve{\bGamma}} )[\varpi(\breve{\bOmega}) \{1 - \rho^2(\bar{\breve{\bGamma}}) \}]^{-2}$, which implies that
	% \begin{equation*}
		% 	1 - |(\hat{\ba}_{j_\ell}^{\rm init})^{\T}\ba_\ell  |^2 \le 2\{ 1 - |(\hat{\ba}_{j_\ell}^{\rm init})^{\T}\ba_\ell  |\} = \frac{\alpha^2(\bar{\breve{\bGamma}} ) \cdot O_{\rm p} \{ d^6(\Pi_{1,n} + \Pi_{2,n})^2 \}  }{\varpi^2(\breve{\bOmega}) \{1 - \rho^2(\bar{\breve{\bGamma}})\}^2 } \,,
		% \end{equation*}   
	% provided that $ d^{3}(\Pi_{1,n} + \Pi_{2,n})\eta(\bar{\breve{\bGamma}} )[\varpi(\breve{\bOmega}) \{1 - \rho^2(\bar{\breve{\bGamma}}) \}]^{-1} = o_{\rm p}(1)$ and  $\log(pq)=o(n^c)$ for some constant $c \in (0,1)$ depending only on $r_1$ and $r_2$.  
	Recall $\bB=(\bb_1,\ldots, \bb_d)$ and $(\hat{d}_1, \hat{d}_2,\hat{d})=(d_1,d_2,d)$. Write $\hat{\bB}  = (\hat{\bb}_1,\ldots,\hat{\bb}_{d} )$.
	Analogously, we can also show 
	\begin{align*}
		\max_{\ell\in[d]}|\kappa_{2,\ell}\hat{\bb}_{j_\ell} -\bb_\ell |_2 =\frac{ O_{\rm p} (\Pi_{1,n} + \Pi_{2,n}) }{\varpi(\breve{\bOmega}) } 
	\end{align*} 
	with $\kappa_{2,\ell}\in\{1,-1\}$, 
	provided that  $ (\Pi_{1,n} + \Pi_{2,n})\{\varpi(\breve{\bOmega})\}^{-1} = o_{\rm p}(1)$  and  $\log(pq)=o(n^c)$ for some constant $c \in (0,1)$ depending only on $r_1$ and $r_2$. Recall $\varpi(\breve{\bOmega}) = \lambda_{(d-1)d/2}(\breve{\bOmega}^{\T}\breve{\bOmega})$. Hence,  by Condition \ref{cd:eigen-bomega},  we complete the proof of Theorem \ref{thm:hat_a_hat_b}. 
	$\hfill\Box$

	\section{Proofs of Auxiliary Lemmas}
	%\subsection{Proof of Lemma \ref{lem:php}}
	%Denote by $\lambda_1^* \ge \ldots \ge \lambda_p^*$ and $\lambda_1^\dag \ge \ldots \ge \lambda_q^\dag$, respectively, the eigenvalues of $\bM_1$ and $\bM_2$. It follows from Condition \ref{cd:rm12} that $\lambda_1^* \ge \ldots \ge \lambda_{d_1}^* > 0 =  \lambda_{d_1 + 1}^*  \ge \ldots \ge \lambda_p^*$ and $\lambda_1^\dag \ge \ldots \ge \lambda_{d_2}^\dag > 0 = \lambda_{d_2+1}^\dag \ldots \ge \lambda_q^\dag$. Notice that $\lambda_{d_1}^*$ and $\lambda_{d_2}^\dag$ are uniformly bounded away from zero. Lemma 1 of \citeS{Chang2018} implies that $\| \hat{\bP}\bE_1 - \bP \|_2 \le C \|\hat{\bM}_1 - \bM_1 \|_2$ and $\| \hat{\bQ}\bE_2 - \bQ \|_2 \le C \|\hat{\bM}_2 - \bM_2 \|_2$, where $\bE_1$ and $\bE_2$ are two orthogonal matrices. Together with Lemma \ref{lem:M1M2}, we complete the proof of Lemma \ref{lem:php}.

	\subsection{Proof of Lemma \ref{lem:sigmaz-h}}\label{sec:sub-sigmaz-h}
	Recall  
	\begin{align*}
		&\bSigma_{\vec{\bY}}(k) = \frac{1}{n-k}\sum_{t=k+1}^{n} \mathbb{E}[\{\vec{\bY}_{t}-\mathbb{E}(\bar{\vec{\bY}})\}\{\vec{\bY}_{t-k}-\mathbb{E}(\bar{\vec{\bY}})\} ^{\T}]\,,\\
		&~~~~~~~~\hat{\bSigma}_{\vec{\bY}}(k) = \frac{1}{n-k}\sum_{t=k+1}^{n} (\vec{\bY}_{t}-\bar{\vec{\bY}})(\vec{\bY}_{t-k}-\bar{\vec{\bY}})^{\T}
	\end{align*}
	with  $\bar{\vec{\bY}}= n^{-1}\sum_{t=1}^{n}\vec{\bY}_{t}$. Write $\bSigma_{\vec{\bY}}(k)=(\sigma_{i,j}^{(k)})_{pq\times pq}$ and $\hat{\bSigma}_{\vec{\bY}}(k)=(\hat{\sigma}_{i,j}^{(k)})_{pq\times pq}$.  Under Condition \ref{cd:tail}, using the similar arguments for the proof of Lemma 1 in \citeS{chang2023modelling}, it holds that
	\begin{align}\label{eq:sigmah-sigma-tail}
		\max_{i,j \in [pq]}\mathbb{P}\big\{|\hat{\sigma}_{i,j}^{(k)} - \sigma_{i,j}^{(k)}| \ge x\big\} \lesssim \exp{(-C_1nx^2)} +\exp{(-C_1n^{\tilde{r}}x^{\tilde{r}})} + \exp{(-C_1n^{\breve{r}}x^{\breve{r}})}
	\end{align}
	for any $x \in (0,1)$, where $\tilde{r}^{-1}=1+2r_{1}^{-1}+r_2^{-1}$ and $\breve{r}^{-1}=2+|r_1^{-1}-1|_{+}+r_2^{-1}$. Hence,
	\begin{align}\label{eq:sigmah-sigma-cov}
		\max_{i, j\in[pq]} |\hat{\sigma}_{i,j}^{(k)}-\sigma_{i,j}^{(k)}|= O_{\rm p}[\{n^{-1}\log(pq)\}^{1/2}]
	\end{align}
	provided that $\log(pq)=o(n^c)$ for some constant $c \in (0,1)$ depending only on $r_1$ and $r_2$. 
	By Triangle inequality, we have
	\begin{align}\label{eq:td2-sigy}
		\|T_{\delta_2}\{\hat{\bSigma}_{\vec{\bY}}(k)\} -\bSigma_{\vec{\bY}}(k)\|_2 \le \|T_{\delta_2}\{\hat{\bSigma}_{\vec{\bY}}(k)\} - T_{\delta_2}\{\bSigma_{\vec{\bY}}(k)\}\|_2  + \|T_{\delta_2}\{\bSigma_{\vec{\bY}}(k)\} -\bSigma_{\vec{\bY}}(k)\|_2\,.
	\end{align}
	On the one hand, by Condition \ref{cd:sgm_yxi}{\rm (ii)}, it holds that
	\begin{align}\label{eq:td2-sigy-1}
		\|T_{\delta_2}\{\bSigma_{\vec{\bY}}(k)\} -\bSigma_{\vec{\bY}}(k)\|_2^2 \le&~ \bigg[\max_{i\in[pq]} \sum_{j=1}^{pq}|\sigma_{i,j}^{(k)}|I\{|\sigma_{i,j}^{(k)}| < \delta_2\}\bigg]\bigg[\max_{j\in[pq]} \sum_{i=1}^{pq}|\sigma_{i,j}^{(k)}|I\{|\sigma_{i,j}^{(k)}| < \delta_2\}\bigg] \notag \\
		\le&~ \big(\delta_2^{1-\iota}\big)^2s_3s_4 \,.
	\end{align}
	On the other hand, 	  
	\begin{align}\label{eq:r1-dep}
		\|T_{\delta_2}\{\hat{\bSigma}_{ \vec{\bY}}(k)\} -T_{\delta_2}\{\bSigma_{ \vec{\bY}}(k)\}\|_2^2 
		\le&~\underbrace{
			\bigg[\max_{i\in[pq]} \sum_{j=1}^{pq}|\hat{\sigma}_{i,j}^{(k)}I\{|\hat{\sigma}_{i,j}^{(k)}| \ge \delta_2\} - \sigma_{i,j}^{(k)}I\{|\sigma_{i,j}^{(k)}| \ge  \delta_2\}|\bigg] }_{\text{R}_{1}} \notag \\
		&~\times \underbrace{\bigg[\max_{j\in[pq]} \sum_{i=1}^{pq}|\hat{\sigma}_{i,j}^{(k)}I\{|\hat{\sigma}_{i,j}^{(k)}| \ge \delta_2\} - \sigma_{i,j}^{(k)}I\{|\sigma_{i,j}^{(k)}| \ge  \delta_2\}|\bigg] }_{\text{R}_{2}} \,.
	\end{align}
	By Triangle inequality, we have
	\begin{align*}
		{\rm R}_{1}\le &~ \underbrace{\max_{i\in[pq]} \sum_{j=1}^{pq}|\hat{\sigma}_{i,j}^{(k)} - \sigma_{i,j}^{(k)}|I\{|\hat{\sigma}_{i,j}^{(k)}| \ge \delta_2, |\sigma_{i,j}^{(k)}| \ge  \delta_2\} }_{\text{R}_{11}} + 
		\underbrace{\max_{i\in[pq]} \sum_{j=1}^{pq}|\hat{\sigma}_{i,j}^{(k)} |I\{|\hat{\sigma}_{i,j}^{(k)}| \ge \delta_2, |\sigma_{i,j}^{(k)}| <\delta_2\} }_{\text{R}_{12}} \\
		&+ \underbrace{\max_{i\in[pq]} \sum_{j=1}^{pq}|\sigma_{i,j}^{(k)} |I\{| \sigma_{i,j}^{(k)}| \ge \delta_2, |\hat{\sigma}_{i,j}^{(k)}| <\delta_2\} }_{\text{R}_{13}}\,.
	\end{align*}
	By Condition \ref{cd:sgm_yxi}{\rm(ii)} and \eqref{eq:sigmah-sigma-cov}, it holds that
	\begin{align*}
		{\rm R}_{11} \le \max_{i, j\in[pq]} |\hat{\sigma}_{i,j}^{(k)}-\sigma_{i,j}^{(k)}|\delta_2^{-\iota}s_3 = O_{\rm p}[\delta_2^{-\iota}s_3 \{n^{-1}\log(pq)\}^{1/2} ]\,,
	\end{align*}
	provided that $\log(pq)=o(n^c)$ for some constant $c \in (0,1)$ depending only on $r_1$ and $r_2$.  Taking $\theta \in (0,1)$, by Triangle inequality and Condition \ref{cd:sgm_yxi}{\rm(ii)},
	\begin{align*}
		{\rm R}_{12} \le&~ \max_{i\in[pq]} \sum_{j=1}^{pq}|\sigma_{i,j}^{(k)} |I\{ |\sigma_{i,j}^{(k)}| <\delta_2\} + \max_{i\in[pq]} \sum_{j=1}^{pq}|\hat{\sigma}_{i,j}^{(k)}-\sigma_{i,j}^{(k)} |I\{|\hat{\sigma}_{i,j}^{(k)}| \ge \delta_2, |\sigma_{i,j}^{(k)}| <\delta_2\} \\
		\le &~ \delta_2^{1-\iota}s_3 + \max_{i\in[pq]} \sum_{j=1}^{pq}|\hat{\sigma}_{i,j}^{(k)}-\sigma_{i,j}^{(k)} |I\{|\hat{\sigma}_{i,j}^{(k)}| \ge \delta_2, |\sigma_{i,j}^{(k)}| \le \theta \delta_2\}\\
		&+ \max_{i\in[pq]} \sum_{j=1}^{pq}|\hat{\sigma}_{i,j}^{(k)}-\sigma_{i,j}^{(k)} |I\{|\hat{\sigma}_{i,j}^{(k)}| \ge \delta_2, \theta \delta_2<|\sigma_{i,j}^{(k)}| <\delta_2\}\\
		\le &~ \delta_2^{1- \iota}s_3   
		+ \max_{i, j\in[pq]} |\hat{\sigma}_{i,j}^{(k)}-\sigma_{i,j}^{(k)}| \max_{i\in[pq]}\sum_{j=1}^{pq}I\{|\hat{\sigma}_{i,j}^{(k)} - \sigma_{i,j}^{(k)}|  \ge (1-\theta)\delta_2\}\\
		&+\max_{i, j\in[pq]} |\hat{\sigma}_{i,j}^{(k)}- \sigma_{i,j}^{(k)}|\theta^{-\iota}\delta_2^{-\iota}s_3\,.
	\end{align*}
	Selecting $\delta_2 = \bar{C} \{n^{-1}\log(pq)\}^{1/2}$ for some sufficiently large constant $\bar{C} >0$, by \eqref{eq:sigmah-sigma-tail}, Bonferroni's inequality and  Markov inequality, it holds that
	\begin{align*}
		\mathbb{P}\bigg[ \max_{i\in[pq]}\sum_{j=1}^{pq}I\{|\hat{\sigma}_{i,j}^{(k)} - \sigma_{i,j}^{(k)}|  \ge (1-\theta)\delta_2\} > x \bigg]\le 
		\frac{1}{x} \sum_{i=1}^{pq}\sum_{j=1}^{pq} \mathbb{P}\big\{ |\hat{\sigma}_{i,j}^{(k)} - \sigma_{i,j}^{(k)}|  \ge (1-\theta)\delta_2\big\} \le \frac{C_2}{x}
	\end{align*}
	for any $x>1$, provided that $\log(pq)=o(n^c)$ for some constant $c \in (0,1)$ depending only on $r_1$ and $r_2$. Hence, $\max_{i\in[pq]}\sum_{j=1}^{pq}I\{|\hat{\sigma}_{i,j}^{(k)} - \sigma_{i,j}^{(k)}|  \ge (1-\theta)\delta_2\}=O_{\rm p}(1)$. By \eqref{eq:sigmah-sigma-cov}, we have
	\begin{align*}
		{\rm R}_{12} =  O_{\rm p}[s_3 \{n^{-1}\log(pq)\}^{(1-\iota)/2} ] \,,
	\end{align*}
	provided that $\log(pq)=o(n^c)$ for some constant $c \in (0,1)$ depending only on $r_1$ and $r_2$. Furthermore, by Triangle inequality, \eqref{eq:sigmah-sigma-cov} and Condition \ref{cd:sgm_yxi}{\rm(ii)} again, 
	\begin{align*}
		{\rm R}_{13} \le &~  \max_{i\in[pq]} \sum_{j=1}^{pq}|\hat{\sigma}_{i,j}^{(k)}-\sigma_{i,j}^{(k)} |I\{|\hat{\sigma}_{i,j}^{(k)}| <\delta_2, |\sigma_{i,j}^{(k)}| \ge\delta_2\} + \max_{i\in[pq]} \sum_{j=1}^{pq}|\hat{\sigma}_{i,j}^{(k)} |I\{ |\hat{\sigma}_{i,j}^{(k)}| <\delta_2,|\sigma_{i,j}^{(k)}| \ge\delta_2\} \\
		\le&~ \max_{i, j\in[pq]} |\hat{\sigma}_{i,j}^{(k)}-\sigma_{i,j}^{(k)}|\delta_2^{-\iota}s_3+ \delta_2^{1-\iota}s_3= O_{\rm p}\big[s_3 \{n^{-1}\log(pq)\}^{(1-\iota)/2} \big] \,,
	\end{align*}
	provided that $\log(pq)=o(n^c)$ for some constant $c \in (0,1)$ depending only on $r_1$ and $r_2$. Hence,
	\begin{align*}
		{\rm R}_{1}\le {\rm R}_{11} + {\rm R}_{12}+ {\rm R}_{13} = O_{\rm p}[s_3 \{n^{-1}\log(pq)\}^{(1-\iota)/2} ] \,,
	\end{align*}
	provided that $\log(pq)=o(n^c)$ for some constant $c \in (0,1)$ depending only on $r_1$ and $r_2$. Analogously, we can also show  ${\rm R}_{2} = O_{\rm p}[s_4 \{n^{-1}\log(pq)\}^{(1-\iota)/2} ] $. By \eqref{eq:r1-dep}, 
	\begin{align*} 
		\|T_{\delta_2}\{\hat{\bSigma}_{ \vec{\bY}}(k)\} -T_{\delta_2}\{\bSigma_{ \vec{\bY}}(k)\}\|_2^2 \le{\rm R}_{1} \times {\rm R}_{2} =O_{\rm p}\big[s_3s_4 \{n^{-1}\log(pq)\}^{1-\iota} \big] \,,
	\end{align*}
	provided that $\log(pq)=o(n^c)$ for some constant $c \in (0,1)$ depending only on $r_1$ and $r_2$. Recall $\Pi_{2,n} = (s_3s_4)^{1/2} \{n^{-1}\log(pq)\}^{(1-\iota)/2}$. Hence, together with \eqref{eq:td2-sigy-1}, by \eqref{eq:td2-sigy}, we have
	\begin{align*}
		\|T_{\delta_2}\{\hat{\bSigma}_{ \vec{\bY}}(k)\} -\bSigma_{ \vec{\bY}}(k)\|_2 =O_{\rm p}(\Pi_{2,n}) \,,
	\end{align*}
	provided that $\log(pq)=o(n^c)$ for some constant $c \in (0,1)$ depending only on $r_1$ and $r_2$. We complete the proof of Lemma \ref{lem:sigmaz-h}.
	$\hfill\Box$

	\subsection{Proof of Lemma \ref{lem:mhm}}\label{sec:sub-mhm}
	Recall $\bZ_{t} = \bP^{\T}\bY_{t}\bQ$,   $\hat{\bSigma}_{ \vec{\bZ}}(k)= (\hat{\bQ} \otimes \hat{\bP})^{{\T}} T_{\delta_2}\{\hat{\bSigma}_{ \vec{\bY}}(k)\}  (\hat{\bQ} \otimes \hat{\bP})$ and  $\bSigma_{\vec{\bY}}(k) = (n-k)^{-1}\sum_{t=k+1}^{n} \mathbb{E}[\{\vec{\bY}_{t}-\mathbb{E}(\vec{\bY})\}\{\vec{\bY}_{t-k}-\mathbb{E}(\bar{\vec{\bY}})\} ^{\T}]$ with $\bar{\vec{\bY}} =n^{-1}\sum_{t=1}^{n}\vec{\bY}_{t}$. Notice that  $\bSigma_{\vec{\bZ}}(k) =(n-k)^{-1}\sum_{t=k+1}^{n} \mathbb{E}  [\{ \vec{\bZ}_t - \mathbb{E}(\bar{\vec{ 
			\bZ}} )  \} \{ \vec{\bZ}_{t-k} - \mathbb{E}(\bar{\vec{\bZ}} )\}^{\T} ] =(\bQ \otimes \bP )^{{\T}} \bSigma_{ \vec{\bY}}(k) (\bQ \otimes \bP)$ for $k\ge 1$, where $\bar{\vec{\bZ}} = n^{-1}\sum_{t=1}^{n}\vec{\bZ}_t$.  Then
	\begin{align*}
		&(\bE_2 \otimes \bE_1)^{{\T}} \hat{\bSigma}_{ \vec{\bZ}}(k)(\bE_2 \otimes \bE_1) - \bSigma_{ \vec{\bZ}}(k) \\ 
		&~~~~~=  \{(\hat{\bQ}\bE_2) \otimes (\hat{\bP}\bE_1) \}^{\T} T_{\delta_2}\{\hat{\bSigma}_{ \vec{\bY}}(k)\} \{(\hat{\bQ}\bE_2) \otimes (\hat{\bP}\bE_1)\} - (\bQ \otimes \bP )^{{\T}} \bSigma_{ \vec{\bY}}(k) (\bQ \otimes \bP) \\
		&~~~~~ = \big\{(\hat{\bQ}\bE_2)^{\T} \otimes (\hat{\bP}\bE_1)^{{\T}}- \bQ^{\T} \otimes \bP^{\T} \big\}\big[T_{\delta_2}\{\hat{\bSigma}_{ \vec{\bY}}(k)\} - \bSigma_{ \vec{\bY}}(k) \big] \big\{(\hat{\bQ}\bE_2) \otimes (\hat{\bP}\bE_1)- \bQ \otimes \bP\big\} \\
		&~~~~~~~~+  (\bQ^{\T} \otimes \bP^{\T} ) \big[T_{\delta_2}\{\hat{\bSigma}_{ \vec{\bY}}(k)\} - \bSigma_{ \vec{\bY}}(k) \big] \big\{(\hat{\bQ}\bE_2) \otimes (\hat{\bP}\bE_1)- \bQ \otimes \bP\big\} \\
		&~~~~~~~~+\big\{(\hat{\bQ}\bE_2)^{\T} \otimes (\hat{\bP}\bE_1)^{{\T}}- \bQ^{\T} \otimes \bP^{\T} \big\} \bSigma_{ \vec{\bY}}(k)   \big\{(\hat{\bQ}\bE_2) \otimes (\hat{\bP}\bE_1)- \bQ \otimes \bP\big\} \\
		&~~~~~~~~+\big\{(\hat{\bQ}\bE_2)^{\T} \otimes (\hat{\bP}\bE_1)^{{\T}}- \bQ^{\T} \otimes \bP^{\T}  \big\}\big[T_{\delta_2}\{\hat{\bSigma}_{ \vec{\bY}}(k)\} - \bSigma_{ \vec{\bY}}(k) \big] (\bQ \otimes \bP) \\
		&~~~~~~~~+  (\bQ^{\T} \otimes \bP^{\T} ) \big[T_{\delta_2}\{\hat{\bSigma}_{ \vec{\bY}}(k)\} - \bSigma_{ \vec{\bY}}(k) \big] (\bQ \otimes \bP) \\
		&~~~~~~~~+  (\bQ^{\T} \otimes \bP^{\T} )  \bSigma_{ \vec{\bY}}(k) \big\{(\hat{\bQ}\bE_2 )\otimes (\hat{\bP}\bE_1)- \bQ \otimes \bP\big\} \\
		&~~~~~~~~+ \big\{(\hat{\bQ}\bE_2)^{\T} \otimes (\hat{\bP}\bE_1)^{{\T}}- \bQ^{\T} \otimes \bP^{\T} \big\}  \bSigma_{ \vec{\bY}}(k)   (\bQ \otimes \bP)   
	\end{align*}
	with $(\bE_1,\bE_2)$ specified in Proposition \ref{pro:PQW}. Due to $\|\bQ\otimes \bP\|_2=1$, we have 
	\begin{align}\label{eq:vysig}
		&\|(\bE_2 \otimes \bE_1)^{{\T}} \hat{\bSigma}_{ \vec{\bZ}}(k)(\bE_2 \otimes \bE_1) - \bSigma_{ \vec{\bZ}}(k)\|_2 \notag\\ 
		&~~~~~\le  \| (\hat{\bQ}\bE_2)\otimes (\hat{\bP}\bE_1)- \bQ \otimes \bP  \|_2^2 \|T_{\delta_2}\{\hat{\bSigma}_{ \vec{\bY}}(k)\} - \bSigma_{ \vec{\bY}}(k) \|_2  \\
		&~~~~~~~~+ 2\|T_{\delta_2}\{\hat{\bSigma}_{ \vec{\bY}}(k)\} - \bSigma_{ \vec{\bY}}(k) \|_2 \|(\hat{\bQ}\bE_2) \otimes (\hat{\bP}\bE_1)- \bQ \otimes \bP\|_2 
		+ \|T_{\delta_2}\{\hat{\bSigma}_{ \vec{\bY}}(k)\} - \bSigma_{ \vec{\bY}}(k) \|_2 \notag\\
		&~~~~~~~~+\|(\hat{\bQ}\bE_2)\otimes (\hat{\bP}\bE_1)- \bQ \otimes \bP \|_2^2 \|\bSigma_{ \vec{\bY}}(k)\|_2  +2\|(\hat{\bQ}\bE_2)\otimes (\hat{\bP}\bE_1)- \bQ \otimes \bP \|_2 \|\bSigma_{ \vec{\bY}}(k)\|_2\notag \,.
	\end{align}   
	Under Conditions \ref{cd:ra}--\ref{cd:sgm_yxi}, repeating the proof of Proposition 3 in \citeS{chang2023modelling},  if $(\hat{d}_1,\hat{d}_2)=(d_1,d_2)$, there exist some orthogonal matrices $\bE_1$ and $\bE_2$ such that 
	\begin{align}\label{eq:php-qhq}
		\|\hat{\bP}\bE_1  - \bP  \|_2=O_{\rm p}(\Pi_{1,n})=\|\hat{\bQ}\bE_2 - \bQ \|_2\,, 
	\end{align}
	provided that $\Pi_{1,n}=o(1)$ and  $\log(pq)=o(n^c)$ for some constant $c \in (0,1)$ depending only on $r_1$ and $r_2$. Since $\|\bP\|_2=1=\|\bQ\|_2$,  it holds that
	\begin{align}\label{eq:qe-q}
		&\|(\hat{\bQ}\bE_2) \otimes (\hat{\bP}\bE_1) - \bQ \otimes \bP \|_2 \notag\\
		% &~~~~~~~~\le  \|(\hat{\bQ}\bE_2 - \bQ) \otimes \hat{\bP}\bE_1\|_2 + \|\bQ \otimes(\hat{\bP}\bE_1 - \bP)  \|_2 \notag\\
		&~~~~~~~~\le   \|(\hat{\bQ}\bE_2 - \bQ) \otimes (\hat{\bP}\bE_1 - \bP)\|_2 + \|(\hat{\bQ}\bE_2 - \bQ) \otimes  \bP\|_2 + \|\bQ \otimes(\hat{\bP}\bE_1 - \bP)  \|_2 \notag\\
		&~~~~~~~~\le \|\hat{\bQ} \bE_2 - \bQ\|_2 \|\hat{\bP} \bE_1 - \bP\|_2 +  \|\hat{\bQ} \bE_2 - \bQ\|_2 + \|\hat{\bP} \bE_1 - \bP\|_2 \notag\\ 
		&~~~~~~~~=O_{\rm p}(\Pi_{1,n}) \,,
	\end{align}
	provided that $\Pi_{1,n}=o(1)$ and  $\log(pq)=o(n^c)$ for some constant $c \in (0,1)$ depending only on $r_1$ and $r_2$. Under Condition \ref{cd:sgm_yxi}(i), by Lemma \ref{lem:sigmaz-h} and \eqref{eq:vysig}, we have 
	\begin{align*}
		\|(\bE_2 \otimes \bE_1)^{{\T}} \hat{\bSigma}_{ \vec{\bZ}}(k)(\bE_2 \otimes \bE_1)-\bSigma_{ \vec{\bZ}}(k)\|_2=O_{\rm p}(\Pi_{1,n} +\Pi_{2,n})\,,
	\end{align*}
	provided that $\Pi_{1,n} + \Pi_{2,n}=o(1)$ and $\log(pq)=o(n^c)$ for some constant $c \in (0,1)$ depending only on $r_1$ and $r_2$. 
	
	Recall $\bM=\sum_{k=1}^{\tilde{K}} \bSigma_{ \vec{\bZ}}(k)\bSigma_{ \vec{\bZ}}(k)^{{\T}}$ and  $\hat{\bM}=\sum_{k=1}^{\tilde{K}}\hat{\bSigma}_{ \vec{\bZ}}(k)\hat{\bSigma}_{ \vec{\bZ}}(k)^{{\T}}$. Due to $\|\bSigma_{ \vec{\bZ}}(k)\|_2 = \|(\bQ \otimes \bP )^{{\T}} \bSigma_{ \vec{\bY}}(k) (\bQ \otimes \bP)\|_2$, under Condition \ref{cd:sgm_yxi}(i), by Triangle inequality, it holds that
	\begin{align*}
		\|(\bE_2 \otimes \bE_1)^{{\T}}\hat{\bM}(\bE_2 \otimes \bE_1) - \bM\|_2 \le&~ \sum_{k=1}^{\tilde{K} }\|(\bE_2 \otimes \bE_1)^{{\T}} \hat{\bSigma}_{ \vec{\bZ}}(k)(\bE_2 \otimes \bE_1) - \bSigma_{ \vec{\bZ}}(k)\|_2^2 \\
		&~+ 2 \sum_{k=1}^{\tilde{K}} \|\bSigma_{ \vec{\bZ}}(k)\|_2 \| (\bE_2 \otimes \bE_1)^{{\T}} \hat{\bSigma}_{ \vec{\bZ}}(k)(\bE_2 \otimes \bE_1) - \bSigma_{ \vec{\bZ}}(k)\|_2\\
		=&~O_{\rm p}(\Pi_{1,n} + \Pi_{2,n}) \,,
	\end{align*}
	provided that $\Pi_{1,n}+\Pi_{2,n}=o(1)$ and $\log(pq)=o(n^c)$ for some constant $c \in (0,1)$ depending only on $r_1$ and $r_2$. We complete the proof of Lemma \ref{lem:mhm}.
	$\hfill\Box$

	\subsection{Proof of Lemma \ref{lemma:theta}}\label{sec:sub-theta}
	% \subsubsection{Proof of Lemma \ref{lemma:theta}(i)}\label{sec:lemma:theta-i} 
	Recall the 4-way tensor $\bPsi(\cdot,\cdot)$ defined in Section \ref{sec:theta-iden}, $\breve{\bOmega}$ given in Section \ref{sec:asymptotics}, and $ \breve{\bW}  =(\bE_2 \otimes \bE_1) \bW \bE_3^{\T}$ with some orthogonal matrices $(\bE_1,\bE_2,\bE_3)$ specified in Proposition \ref{pro:PQW}. Write $\breve{\bW}   = (\vec{\breve{\bW}}_1, \ldots, \vec{\breve{\bW}}_{d})$,  $\breve{\bW}_i=(\breve{w}_{i,j,k})_{d_1\times d_2}$,  $\bI_{d_1}=(\tilde{\be}_{1}, \ldots, \tilde{\be}_{d_1})$ and $\bI_{d_2}=(\breve{\be}_{1}, \ldots, \breve{\be}_{d_2})^{\T}$.  Notice that
	\begin{align*}
		\bPsi(\breve{\bW}_{r},\breve{\bW}_{s} )  =\Psi(\bI_{d_1}  \breve{\bW}_{r}  \bI_{d_2},\bI_{d_1} \breve{\bW}_{s}  \bI_{d_2}) =\sum_{i=1}^{d_1}\sum_{k=1}^{d_2}\sum_{j=1}^{d_1}\sum_{l=1}^{d_2} \breve{w}_{r, i,k} \breve{w}_{s, j,l}   \Psi(\tilde{\be}_{i}\breve{\be}_{k}^{{\T}} , \tilde{\be}_{j}\breve{\be}_{l}^{{\T}}) \,.
	\end{align*}
	For any $i,j\in[d_1]$ and $k,l\in[d_2]$, define a $(d_1^2d_2^2)$-dimensional vector  $\breve{\bg}_{i,j,k,l}$ as $$[\breve{\bg}_{i,j,k,l}]_{(i'-1)d_1d_2^2+(j'-1)d_2^2+(k'-1)d_2+l'} = [\Psi(\tilde{\be}_{i}\breve{\be}_{k}^{{\T}} , \tilde{\be}_{j}\breve{\be}_{l}^{{\T}})]_{i',j',k',l'}\,, \qquad i',j' \in[d_1], \, k',l'\in[d_2]\,.$$ 
	It holds that $\breve{\bg}_{i,j,k,l} = (\tilde{\be}_i \otimes \tilde{\be}_j -  \tilde{\be}_j \otimes \tilde{\be}_i) \otimes (\breve{\be}_k \otimes \breve{\be}_l -  \breve{\be}_l \otimes \breve{\be}_k )$ for any  $i,j\in[d_1]$ and $k,l\in[d_2]$.  
	Let $\breve{\bG}$ be a $(d_1^2d_2^2) \times (d_1^2d_2^2)$ matrix of which the $\{(k-1)d_1^2d_2  + (i-1)d_1d_2 + (l-1)d_1 + j\}$-th column is $\breve{\bg}_{i,j,k,l}$ for any  $i,j\in[d_1]$ and $k,l\in[d_2]$. Write $[\breve{\bd}_{r,s}]_{ (i'-1) d_1 d_2^2 + (j'-1)d_2^2 + (k'-1)d_2 + l'} = [\bPsi(\breve{\bW}_{r},\breve{\bW}_{s} ) ]_{i',j',k',l'}$ for any $i',j' \in[d_1]$ and $ k',l'\in[d_2]$. We have
	\begin{align*}
		\breve{\bd}_{r,s} =\sum_{i=1}^{d_1}\sum_{k=1}^{d_2}\sum_{j=1}^{d_1}\sum_{l=1}^{d_2}  \breve{w}_{r, i,k} \breve{w}_{s, j,l} \breve{\bg}_{i,j,k,l} = \breve{\bG} (\vec{\breve{\bW}}_{r} \otimes \vec{\breve{\bW}}_{s} )  \,.
	\end{align*}
	Write $\breve{\bOmega}=(\breve{\bd}_{1,1}, \ldots, \breve{\bd}_{1,d}, \breve{\bd}_{2,2}, \ldots, \breve{\bd}_{2,d},\ldots, \breve{\bd}_{d,d})$. It holds that 
	\begin{align}\label{eq:omegab-dec}
		\breve{\bOmega} = \breve{\bG}  (\breve{\bW} \otimes \breve{\bW} )  \bT\,,
	\end{align} 
	where $\bT$ is a $d^2\times d(d+1)/2$ matrix defined as $\bT={\rm diag} (\bF_{0}, \bF_{1}, \ldots, \bF_{d-1})$  with $\bF_{0} =\bI_{d}$ and $\bF_{k} = (\bzero_{(d-k) \times k}, \bI_{d-k} )^{\T} $ for any $ k \in [d-1]$. Recall   
	$\hat{\bW}$  is the $(\hat{d}_1\hat{d}_2) \times \hat{d}$ matrix    of which the columns are the $\hat{d}$  orthonormal eigenvectors of $\hat{\bM}$  corresponding to its $\hat{d}$ 
	largest eigenvalues, where $\hat{\bM}$ is defined in \eqref{eq:mh}. 
	Due to  $(\hat{d}_1, \hat{d}_2,\hat{d})=(d_1,d_2,d)$,  for $\hat{\bOmega}$ given in \eqref{eq:omega-h},  it also  holds that
	\begin{align}\label{eq:omegah-dec}
		\hat{\bOmega} = \breve{\bG} (\hat{\bW} \otimes \hat{\bW} ) \bT\,.
	\end{align}
	Notice that $\breve{\bg}_{i,j,k,l}=\mathbf{0}$ for any  $i=j\in[d_1]$ or $k=l\in[d_2]$, and 
	\begin{align*}
		\breve{\bg}_{i_1,j_1,k_1,l_1}^{{\T}} \breve{\bg}_{i_2,j_2,k_2,l_2} 
		=&~ \big\{(\tilde{\be}_{i_1} \otimes \tilde{\be}_{j_1} -  \tilde{\be}_{j_1} \otimes \tilde{\be}_{i_1})^{{\T}} \otimes (\breve{\be}_{k_1} \otimes \breve{\be}_{l_1} -  \breve{\be}_{l_1} \otimes \breve{\be}_{k_1} )^{{\T}} \big\}\\
		&\times \big\{(\tilde{\be}_{i_2} \otimes \tilde{\be}_{j_2} -  \tilde{\be}_{j_2} \otimes \tilde{\be}_{i_2}) \otimes (\breve{\be}_{k_2} \otimes \breve{\be}_{l_2} -  \breve{\be}_{l_2} \otimes \breve{\be}_{k_2} )\big\}\\
		=&~  \big\{(\tilde{\be}_{i_1} \otimes \tilde{\be}_{j_1} -  \tilde{\be}_{j_1} \otimes \tilde{\be}_{i_1})^{{\T}} (\tilde{\be}_{i_2} \otimes \tilde{\be}_{j_2} -  \tilde{\be}_{j_2} \otimes \tilde{\be}_{i_2}) \big\}\\
		&\otimes \big\{ (\breve{\be}_{k_1} \otimes \breve{\be}_{l_1} -  \breve{\be}_{l_1} \otimes \breve{\be}_{k_1} )^{{\T}} (\breve{\be}_{k_2} \otimes \breve{\be}_{l_2} -  \breve{\be}_{l_2} \otimes \breve{\be}_{k_2} ) \big\}\\
		=&~ \big( \tilde{\be}_{i_1}^{{\T}} \tilde{\be}_{i_2} \times \tilde{\be}_{j_1}^{{\T}}\tilde{\be}_{j_2} -  \tilde{\be}_{i_1}^{{\T}} \tilde{\be}_{j_2} \times \tilde{\be}_{j_1}^{{\T}}\tilde{\be}_{i_2}-  \tilde{\be}_{j_1}^{{\T}} \tilde{\be}_{i_2} \times \tilde{\be}_{i_1}^{{\T}}\tilde{\be}_{j_2} +  \tilde{\be}_{j_1}^{{\T}} \tilde{\be}_{j_2} \times \tilde{\be}_{i_1}^{{\T}}\tilde{\be}_{i_2} \big)\\
		&\times  \big( \breve{\be}_{k_1}^{{\T}} \breve{\be}_{k_2} \times \breve{\be}_{l_1}^{{\T}}\breve{\be}_{l_2} -  \breve{\be}_{k_1}^{{\T}} \breve{\be}_{l_2} \times \breve{\be}_{l_1}^{{\T}}\breve{\be}_{k_2}-  \breve{\be}_{l_1}^{{\T}} \breve{\be}_{k_2} \times \breve{\be}_{k_1}^{{\T}}\breve{\be}_{l_2} +  \breve{\be}_{l_1}^{{\T}} \breve{\be}_{l_2} \times\breve{\be}_{k_1}^{{\T}}\breve{\be}_{k_2} \big)\\
		=&~ 4\big(\tilde{\be}_{i_1}^{{\T}} \tilde{\be}_{i_2} \times \tilde{\be}_{j_1}^{{\T}}\tilde{\be}_{j_2} -  \tilde{\be}_{i_1}^{{\T}} \tilde{\be}_{j_2} \times \tilde{\be}_{j_1}^{{\T}}\tilde{\be}_{i_2}\big) \big( \breve{\be}_{k_1}^{{\T}} \breve{\be}_{k_2} \times \breve{\be}_{l_1}^{{\T}}\breve{\be}_{l_2} -  \breve{\be}_{k_1}^{{\T}} \breve{\be}_{l_2} \times \breve{\be}_{l_1}^{{\T}}\breve{\be}_{k_2}\big)\,.
	\end{align*}
	For any $i_1 \ne j_1$ and $k_1 \ne l_1$,  it holds that
	\begin{align*}
		\breve{\bg}_{i_1,j_1,k_1,l_1}^{{\T}} \breve{\bg}_{i_2,j_2,k_2,l_2} = \left\{
		\begin{aligned}
			4\,, ~~& \text{if}~ (i_2, j_2, k_2, l_2) \in \{(i_1,j_1,k_1,l_1), (j_1,i_1,l_1,k_1)\} \,,\\
			-4\,, ~~& \text{if}~ (i_2, j_2, k_2, l_2) \in \{(j_1,i_1,k_1,l_1), (i_1,j_1,l_1,k_1)\}\,, \\
			0\,, ~~& \text{otherwise}\,.
		\end{aligned}
		\right.
	\end{align*}
	% \begin{align*}
		% 		\breve{\bg}_{i_1,j_1,k_1,l_1}^{{\T}} \breve{\bg}_{i_2,j_2,k_2,l_2} = \left\{
		% 	\begin{aligned}
			% 		4\,, ~~& \text{if}~ i_1=i_2,j_1=j_2 , k_1=k_2,l_1=l_2~\text{or}~ i_1=j_2,i_2=j_1, k_1=l_2,k_2=l_1\,,\\
			% 	   -4\,, ~~& \text{if}~ i_1=j_2,i_2=j_1 , k_1=k_2,l_1=l_2~\text{or}~ i_1=i_2,j_1=j_2, k_1=l_2,k_2=l_1\,, \\
			%     0\,, ~~& \text{otherwise}\,,
			% 	\end{aligned}
		% 	\right.
		% \end{align*}
	% and 
	% \begin{align*}
		% 	\breve{\bg}_{i_1,j_1,k_1,l_1}^{{\T}} \breve{\bg}_{j_2,i_2,k_2,l_2}=&~  - \breve{\bg}_{i_1,j_1,k_1,l_1}^{{\T}} \breve{\bg}_{i_2,j_2,k_2,l_2}\,,\\
		% 	 \breve{\bg}_{i_1,j_1,k_1,l_1}^{{\T}} \breve{\bg}_{i_2,j_2,l_2,k_2}=&~ -\breve{\bg}_{i_1,j_1,k_1,l_1}^{{\T}} \breve{\bg}_{i_2,j_2,k_2,l_2}\,,\\
		%   \breve{\bg}_{i_1,j_1,k_1,l_1}^{{\T}} \breve{\bg}_{j_2,i_2,l_2,k_2} =&~ \breve{\bg}_{i_1,j_1,k_1,l_1}^{{\T}} \breve{\bg}_{i_2,j_2,k_2,l_2}\,.
		% \end{align*} 
	Let $\bx$ be a $(d_1^2d_2^2)\times 1$ vector of which the $\{(k-1)d_1^2d_2  + (i-1)d_1d_2 + (l-1)d_1 + j\}$-th component is $x_{i,j,k,l}$. Then
	\begin{align*}
		\|\breve{\bG}^{{\T}} \breve{\bG}\|_2^2 = &~\sup_{|\bx|_2=1} |\breve{\bG}^{{\T}} \breve{\bG} \bx|_2^2 = \sup_{|\bx|_2=1} \sum_{i_1=1}^{d_1}\sum_{j_1=1}^{d_1}\sum_{k_1=1}^{d_2} \sum_{l_1=1}^{d_2} (\breve{\bg}_{i_1,j_1,k_1,l_1}^{\T}\breve{\bG} \bx) ^2\\
		=&~  \sup_{|\bx|_2=1} \sum_{i_1=1}^{d_1}\sum_{j_1=1}^{d_1}\sum_{k_1=1}^{d_2} \sum_{l_1=1}^{d_2} \bigg( \sum_{i_2=1}^{d_1}\sum_{j_2=1}^{d_1}\sum_{k_2=1}^{d_2} \sum_{l_2=1}^{d_2}\breve{\bg}_{i_1,j_1,k_1,l_1}^{\T}\breve{\bg}_{i_2,j_2,k_2,l_2} x_{i_2,j_2,k_2,l_2}\bigg) ^2\\
		\le&~  16 \sup_{|\bx|_2=1} \sum_{i_1=1}^{d_1}\sum_{j_1=1}^{d_1}\sum_{k_1=1}^{d_2} \sum_{l_1=1}^{d_2} (x_{i_1,j_1,k_1,l_1} + x_{j_1,i_1,l_1,k_1} -x_{j_1,i_1,k_1,l_1} - x_{i_1,j_1,l_1,k_1})^2  \\
		\le &~ 16 \times \sup_{|\bx|_2=1} 4 |\bx|_2^2 = 64 \,,
	\end{align*}
	which implies $\|\breve{\bG}\|_2 \le 2\sqrt{2}$.
	% Write $\breve{\bG}^{{\T}} \breve{\bG} =(\tilde{\bg}_{1,1,1,1},\ldots, \tilde{\bg}_{1,1,1,d_2}, \ldots, \tilde{\bg}_{d_1,d_1,d_2,d_2})^{{\T}}$. We have $\tilde{\bg}_{i,j,k,l}, \tilde{\bg}_{j,i,k,l}, \tilde{\bg}_{i,j,l,k}, \tilde{\bg}_{j,i,l,k}$ are identical
	% if we ignore the sign of columns for any $i,j\in[d_1]$ and $k,l\in[d_2]$. Let $\tilde{\bG}$ be a $(d_1^2d_2^2)\times \{d_1d_2(d_1-1)(d_2-1)/4\}$ matrix that consists of $\tilde{\bg}_{i,j,k,l}$ for $1\le i < j\le d_1$ and $1\le k < l\le d_2$. Then
	% \begin{align*}
		% 	\|\breve{\bG}^{{\T}} \breve{\bG}\|_2^2 = &~\sup_{|\bx|_2=1} |\breve{\bG}^{{\T}} \breve{\bG} \bx|_2^2 =\sup_{|\bx|_2=1} \sum_{i=1}^{d_1}\sum_{j=1}^{d_1}\sum_{k=1}^{d_2} \sum_{l=1}^{d_2} (\tilde{\bg}_{i,j,k,l}^{{\T}}\bx)^2 \\
		% 	=&~ 4\sup_{|\bx|_2=1} \sum_{1\le i < j\le d_1} \sum_{1\le k < l\le d_2} (\tilde{\bg}_{i,j,k,l}^{{\T}}\bx)^2 = 4 \sup_{|\bx|_2=1} |\tilde{\bG} \bx|_2^2 = 4\|\tilde{\bG}\|_2^2\,.
		% \end{align*} 
	% Since $\tilde{\bG}^{\T} \tilde{\bG} $ is a diagonal matrix and  the diagonal elements are all $64$, it holds that $\|\breve{\bG}^{{\T}} \breve{\bG}\|_2=16$.
	Recall $\breve{\bW}  =(\bE_2 \otimes \bE_1)\bW \bE_3^{\T}$ with some orthogonal matrices $(\bE_1, \bE_2,\bE_3)$ specified in Proposition \ref{pro:PQW}. Notice that  $\|\hat{\bW}\|_2 =1$ and $\|\breve{\bW}\|_2 =\|(\bE_2\otimes \bE_1) \bW \bE_3^{\T}\|_2 \le \|\bW\|_2 =1$.  For $\breve{\bOmega}$ defined in \eqref{eq:omegab-dec}, it holds that $\|\breve{\bOmega}\|_2 \le \|\breve{\bG}\|_2 \| \breve{\bW} \otimes \breve{\bW}  \|_2\|\bT\|_2 \le 2\sqrt{2}$. Together with \eqref{eq:omegah-dec}, by Proposition \ref{pro:PQW}, we have
	\begin{align*}
		\|\hat{\bOmega} - \breve{\bOmega}\|_2 \le &~\|\breve{\bG}\|_2\|\bT\|_2  \|\hat{\bW}\otimes \hat{\bW} - \breve{\bW} \otimes \breve{\bW} \|_2 \\
		\lesssim &~  \| \hat{\bW} \otimes (\hat{\bW}  - \breve{\bW}) \|_2 + \|  (\hat{\bW}  - \breve{\bW} ) \otimes \breve{\bW} \|_2 \\
		\lesssim  &~  \|  \hat{\bW} - (\bE_2 \otimes \bE_1) \bW \bE_3^{\T} \|_2   =O_{\rm p}(\Pi_{1,n} +\Pi_{2,n})\,, 
	\end{align*}
	which implies
	\begin{align}\label{eq:omegah-omegab}
		\|\hat{\bOmega}^{{\T}} \hat{\bOmega} - \breve{\bOmega}^{{\T}} \breve{\bOmega}\|_2 \le   \|\hat{\bOmega} - \breve{\bOmega}\|_2^2 + 2 \|\hat{\bOmega} - \breve{\bOmega}\|_2 \| \breve{\bOmega}\|_2 =O_{\rm p}(\Pi_{1,n} +\Pi_{2,n})\,, 
	\end{align} 
	provided that $\Pi_{1,n}+\Pi_{2,n}=o(1)$ and  $\log(pq)=o(n^c)$ for some constant $c \in (0,1)$ depending only on $r_1$ and $r_2$.

	Recall  $\bV\odot \bU = \bC =\bW\bTheta$,  $\breve{\bW} = (\bE_2 \otimes \bE_1) \bW \bE_3^{\T}$,  $\breve{\bTheta} = \bE_3\bTheta$,  and $\breve{\bOmega}$ defined in Section \ref{sec:asymptotics}, where $(\bE_1,\bE_2,\bE_3)$ is specified in Proposition \ref{pro:PQW}.  Define  $\breve{\bC}  = (\bE_2\bV) \odot (\bE_1\bU)$. Then  $\breve{\bC} = \breve{\bW}\breve{\bTheta}$.  We derive $(\breve{\bD},\breve{\bOmega}^{*},\breve{\bD}^{*})$ in the same manner as $(\bD,\bOmega^{*},\bD^{*})$ given in \eqref{eq:d-omegas-omega} but with replacing $(\bC, \bW, \bTheta)$ by 
	$(\breve{\bC}, \breve{\bW},  \breve{\bTheta} )$.  Using the similar arguments in Section \ref{sec:pro:rankomega} for the proof of Proposition \ref{pro:rankomega}, it holds that $\breve{\bOmega}^{*} = \breve{\bD}^{*} (\breve{\bTheta}^{-1} \otimes \breve{\bTheta}^{-1})$. Since $\breve{\bD}^{*} = \{(\bE_1\otimes \bE_1) \otimes (\bE_2\otimes \bE_2)\} \bD^{*}$, we have $ {\rm rank}(\breve{\bOmega}^{*}) = {\rm rank}(\bD^{*})$. Parallel to the proof of ${\rm rank}(\bOmega) = {\rm rank}(\bOmega^{*})$ in Section  \ref{sec:pro:rankomega} for the proof of Proposition \ref{pro:rankomega}, we can also show  ${\rm rank}(\breve{\bOmega}) = {\rm rank}(\breve{\bOmega}^{*})$. Notice that ${\rm rank}(\bD^{*})= {\rm rank}(\bOmega)$ which has been shown in Section \ref{sec:pro:rankomega} for the proof of Proposition \ref{pro:rankomega}. Then  ${\rm rank}(\breve{\bOmega})={\rm rank}(\bOmega) =d(d-1)/2$.  By the singular value decomposition, we have
	\begin{align}\label{eq:omgga-dec}
		\breve{\bOmega} =  \bL \left( 
		\begin{array}{cc}
			\bLambda_{\breve{\Omega}} & \mathbf{0} \\
			\mathbf{0} &  \mathbf{0}\\
		\end{array}
		\right)  \bR^{{\T}}= \bL 
		\left( 
		\begin{array}{cc}
			\bLambda_{\breve{\Omega}} & \mathbf{0} \\
			\mathbf{0} &  \mathbf{0}\\
		\end{array}
		\right)  \left(
		\begin{array}{c}
			\bR_{\text{e}}^{\T} \\
			\bR_{\text{o}}^{\T} \\
		\end{array}
		\right)  \,,
	\end{align}
	where  $\bL^{{\T}}\bL=\bI_{d_1^2d_2^2 }$, $\bR^{{\T}}\bR=\bI_{d(d+1)/2}$, $\bLambda_{\breve{\Omega}}$ is a $\{d(d-1)/2\} \times \{d(d-1)/2\}$ diagonal matrix with the elements in the main diagonal being the nonzero singular values of $\breve{\bOmega}$, and $\bR_{\rm o} \in \mathbb{R}^{d(d+1)/2 \times d}$ with $\bR_{\rm o}^{\T}\bR_{\rm o}=\bI_{d}$, where  the  columns of $\bR_{\rm o}$ are the $d$  right-singular vectors of $\breve{\bOmega} $ corresponding to its $d$ zero singular values. 
	It then holds that
	\begin{align*}
		\left(
		\begin{array}{c}
			\bR_{\text{o}}^{\T} \\
			\bR_{\text{e}}^{\T} \\
		\end{array}
		\right) \breve{\bOmega}^{\T}\breve{\bOmega} (\bR_{\rm o}, \bR_{\rm e})= 
		\left(
		\begin{array}{cc}
			\mathbf{0} & \mathbf{0} \\
			\mathbf{0} & \bLambda_{\breve{\Omega}}^{2} \\
		\end{array}
		\right)  \,.
	\end{align*}
	Recall $(\hat{d}_1, \hat{d}_2,\hat{d})=(d_1,d_2,d)$ and $\{\tilde{\bh}_1, \ldots, \tilde{\bh}_{\hat{d}}\}$ defined in \eqref{eq:init-basis}. Write $ \hat{\bR}_{\rm o} = (\tilde{\bh}_{1}, \ldots, \tilde{\bh}_{d})$.  Then  the columns of $\hat{\bR}_{\rm o}$ are the $d$ orthonormal eigenvectors of $\hat{\bOmega}^{{\T}}\hat{\bOmega}$ corresponding to its $d$ smallest eigenvalues. Recall $\varpi(\breve{\bOmega}) = \lambda_{d(d-1)/2}(\breve{\bOmega}^{\T}\breve{\bOmega})$. 
	By  Lemma 1 of \citeS{Chang2018}, there exists a $d\times d$  orthogonal  matrix $\bE_4$  such that 
	\begin{align*}
		\|\hat{\bR}_\text{o}\bE_4   - \bR_\text{o}\|_2 \lesssim \frac{\| \hat{\bOmega}^{\T}\hat{\bOmega} - \breve{\bOmega}^{\T}\breve{\bOmega} \|_2}{\varpi(\breve{\bOmega})}  
	\end{align*}
	provided that $\Pi_{1,n} +\Pi_{2,n} \le \varpi(\breve{\bOmega})/5$. By  \eqref{eq:omegah-omegab}, we have
	\begin{align}\label{eq:roh-roe}
		\|\hat{\bR}_\text{o}  - \bR_\text{o}\bE_4^{\T} \|_2   = \frac{ O_{\rm p}(\Pi_{1,n}+\Pi_{2,n}) }{\varpi(\breve{\bOmega})}\,,
	\end{align}   
	provided that $(\Pi_{1,n} +\Pi_{2,n})\{\varpi(\breve{\bOmega})\}^{-1}=o_{\rm p}(1)$ and  $\log(pq)=o(n^c)$ for some constant $c \in (0,1)$ depending only on $r_1$ and $r_2$.  Due to $\breve{\bOmega}\bR_{\rm o}\bE_4^{\T} =\bzero$ and ${\rm rank}(\bR_{\rm o}\bE_4^{\T}) =d$, we know the columns of $\bR_{\rm o}\bE_4^{\T}$ provide a basis of ${\rm ker}(\breve{\bOmega})$.  Let $\breve{\bh}_i$ be the $i$-th column of $\bR_{\rm o}\bE_4^{\T}$. 
	By \eqref{eq:roh-roe}, when $\hat{d}=d$, $\breve{\bh}_i$  can be consistently estimated by $ \tilde{\bh}_i$  for any $i\in[d]$.  We  define $d\times d$ symmetric matrices $\{\breve{\bH}_i\}_{i=1}^{d}$ in the same manner as $\{\bH_i\}_{i=1}^{d}$ given below \eqref{eq:off-diagonal} but with replacing $\{\bh_i\}_{i=1}^{d}$ by $\{\breve{\bh}_i\}_{i=1}^{d}$. Recall $\breve{\bC} =  \breve{\bW} \breve{\bTheta}$.  Analogous to the discussion below \eqref{eq:off-diagonal},  by Proposition \ref{pro:theta-unique}, it holds that   $\breve{\bH}_{i} = \breve{\bTheta} \breve{\bGamma}_{i} \breve{\bTheta}^{\T} $ with    diagonal matrix  $\breve{\bGamma}_i$ for any $i\in[d]$. Recall $(\hat{d}_1, \hat{d}_2,\hat{d})=(d_1,d_2,d)$ and $\tilde{\bH}_i$ defined in below \eqref{eq:init-basis}.  By \eqref{eq:roh-roe}, we have
	\begin{align}\label{eq:H-int-Hb}
		\max_{i\in[d]} \|\tilde{\bH}_i  - \breve{\bH}_{i}\|_{\rm F}  \le  \max_{i\in[d]} |\tilde{\bh}_{i} - \breve{\bh}_i|_2 \le \|\hat{\bR}_\text{o}  - \bR_\text{o}\bE_4 \|_2 = \frac{O_{\rm p}(\Pi_{1,n}+\Pi_{2,n}) }{\varpi(\breve{\bOmega})}\,,
	\end{align}
	provided that $(\Pi_{1,n} +\Pi_{2,n})\{\varpi(\breve{\bOmega})\}^{-1}=o_{\rm p}(1)$ and  $\log(pq)=o(n^c)$ for some constant $c \in (0,1)$ depending only on $r_1$ and $r_2$. 

	Recall $ (\breve{\bh}_1, \ldots, \breve{\bh}_{d}) ={\rm \bR}_{\rm o}\bE_{4}^{\T} $ with $({\rm \bR}_{\rm o}\bE_{4}^{\T})^{\T}{\rm \bR}_{\rm o}\bE_{4}^{\T} =\bI_{d}$. Write  $\breve{\bh}_{i} =(\breve{h}_{i,1,1}, \ldots, \breve{h}_{i,1,d},\breve{h}_{i,2,2}, \ldots,  \breve{h}_{i,d,d})^{\T}$ for any $i\in[d]$.  For any $\bx=(x_1, \ldots, x_d)^{\T}$ with $|\bx|_2=1$, we have
	\begin{align*}
		1 = \bx^{\T} (\breve{\bh}_1,\ldots, \breve{\bh}_d)^{\T}(\breve{\bh}_1,\ldots, \breve{\bh}_d)\bx =  \bigg|\sum_{i=1}^{d} \breve{\bh}_ix_{i}\bigg|_2^2 = \sum_{j=1}^{d}\bigg(\sum_{i=1}^{d}\breve{h}_{i,j,j}x_i\bigg)^2 +   \sum_{1 \le j< k\le d} \bigg(\sum_{i=1}^{d}\breve{h}_{i,j,k}x_i\bigg)^2  \,.
	\end{align*}
	Notice that $\breve{\bH}_{i}=(\acute{h}_{i,r,s})_{d\times d}$  with $\acute{h}_{i,r,s} = \breve{h}_{i,r,s}/2 =\acute{h}_{i,s,r}$ for $1\le  r <s \le d$  and $\acute{h}_{i,r,r} =\breve{h}_{i,r,r}$ for $r\in[d]$. Let $\breve{\bUpsilon}_0   = (\textup{vec}(\breve{\bH}_1),\ldots,\textup{vec}(\breve{\bH}_d))$. It holds that 
	\begin{align*}
		\sigma_{d}^2 (\breve{\bUpsilon}_0) =&~  \inf_{|\bx|_2=1} |\breve{\bUpsilon}_0\bx|_2^2 = \inf_{|\bx|_2=1} \bigg|\sum_{i=1}^{d} {\rm vec}(\breve{\bH}_i)x_{i}\bigg|_2^2 \\
		=&~  \inf_{|\bx|_2=1}  \sum_{j=1}^{d}\sum_{k=1}^{d}\bigg(\sum_{i=1}^{d}\acute{h}_{i,j,k}x_i\bigg)^2  \\
		=&~  \inf_{|\bx|_2=1} \bigg\{\sum_{j=1}^{d}\bigg(\sum_{i=1}^{d}\breve{h}_{i,j,j}x_i\bigg)^2 +  \frac{1}{2} \sum_{1 \le j< k\le d} \bigg(\sum_{i=1}^{d}\breve{h}_{i,j,k}x_i\bigg)^2  \bigg\}\\
		\ge &~ 1- \frac{1}{2}\sup_{|\bx|_2=1}  \sum_{1 \le j< k\le d} \bigg(\sum_{i=1}^{d}\breve{h}_{i,j,k}x_i\bigg)^2  \ge \frac{1}{2}\,.
	\end{align*}
	On the other hand, 
	\begin{align*}
		\sigma_{1}^2 (\breve{\bUpsilon}_0) =&~  \sup_{|\bx|_2=1} |\breve{\bUpsilon}_0\bx|_2^2 = \sup_{|\bx|_2=1} \bigg|\sum_{i=1}^{d} {\rm vec}(\breve{\bH}_i)x_{i}\bigg|_2^2 \\
		=&~  \sup_{|\bx|_2=1} \bigg\{\sum_{j=1}^{d}\bigg(\sum_{i=1}^{d}\breve{h}_{i,j,j}x_i\bigg)^2 +  \frac{1}{2} \sum_{1 \le j< k\le d}\bigg(\sum_{i=1}^{d}\breve{h}_{i,j,k}x_i\bigg)^2  \bigg\} \le 1\,.
		% \le &~ \sup_{|\bx|_2=1} \bigg\{\sum_{j=1}^{d}\bigg(\sum_{i=1}^{d}\tilde{h}_{i,j,j}x_i\bigg)^2 +  \sum_{1 \le j< k\le d}^{d}\bigg(\sum_{i=1}^{d}\tilde{h}_{i,j,k}x_i\bigg)^2  \bigg\}\\
	\end{align*}  
	Hence, all the nonzero singular values of  $ \breve{\bUpsilon}_0$ are finite and uniformly  bounded away from zero and infinity.  Let  $\bar{\breve{\bGamma}}  =(\breve{\bgamma}_{1} , \ldots, \breve{\bgamma}_{d} )$ with $\breve{\bGamma}_{i}  ={\rm diag}(\breve{\bgamma}_{i}) $ for any $i\in[d]$.  Parallel to the calculation of $\bUpsilon_0$ in Section \ref{sec:calce-sec4} for the proof of Proposition \ref{pro:rotation}, it holds that $\breve{\bUpsilon}_0  = (\breve{\bTheta} \odot \breve{\bTheta})\bar{\breve{\bGamma}}$. Then  
	\begin{align*}
		\sigma_i^2(\breve{\bUpsilon}_0) = \lambda_i(\breve{\bUpsilon}_0^{\T}\breve{\bUpsilon}_0)=\lambda_i \{\bar{\breve{\bGamma}}^{\T} (\breve{\bTheta} \odot \breve{\bTheta})^{\T}(\breve{\bTheta} \odot \breve{\bTheta})\bar{\breve{\bGamma}}\} = \lambda_i \{ (\breve{\bTheta} \odot \breve{\bTheta})^{\T}(\breve{\bTheta} \odot \breve{\bTheta}) \bar{\breve{\bGamma}}\bar{\breve{\bGamma}}^{\T}\}\,, \quad i\in[d] \,.
	\end{align*}
	Analogous to the proof of ${\rm rank}(\bar{\bGamma})$ in Section \ref{sec:pro:theta-unique} for the proof of Proposition \ref{pro:theta-unique}, we have ${\rm rank}(\bar{\breve{\bGamma}} )=d$.  Due to  ${\rm rank}(\breve{\bTheta} \odot \breve{\bTheta})=d$, by Theorem 2.2 of \citeS{anderson1963some}, it holds that 
	\begin{align}\label{eq:gamma_0-dec}
		\lambda_d\{(\breve{\bTheta} \odot \breve{\bTheta})^{\T}(\breve{\bTheta} \odot \breve{\bTheta})\}\lambda_i(\bar{\breve{\bGamma}}\bar{\breve{\bGamma}}^{\T}) \le \sigma_i^2(\breve{\bUpsilon}_0)  \le \lambda_1\{(\breve{\bTheta} \odot \breve{\bTheta})^{\T}(\breve{\bTheta} \odot \breve{\bTheta})\} \lambda_i(\bar{\breve{\bGamma}}\bar{\breve{\bGamma}}^{\T}) \,, \quad i\in[d] \,.
	\end{align}
	Recall $\sigma_{i}(\breve{\bTheta}) $  is  finite and uniformly bounded away
	from zero and infinity for any $i\in[d]$. Since $(\breve{\bTheta} \odot \breve{\bTheta})^{\T}(\breve{\bTheta} \odot \breve{\bTheta}) = (\breve{\bTheta}^{\T} \breve{\bTheta})\circ (\breve{\bTheta}^{\T} \breve{\bTheta})$, $\breve{\bTheta} = \bE_{3}\bTheta$ with $\bE_3^{\T}\bE_3=\bI_{d}$, and the $L_2$-norm of each column in $\bTheta$ is equal to 1, by Proposition 6.3.4 of \citeS{rao1998matrix}, we have 
	\begin{align*}
		\lambda_{d}(\breve{\bTheta}^{\T} \breve{\bTheta}) \le \lambda_{i} \{(\breve{\bTheta}^{\T}  \breve{\bTheta})\circ (\breve{\bTheta}^{\T} \breve{\bTheta})\}  \le \lambda_{1}(\breve{\bTheta}^{\T} \breve{\bTheta}) \,, \qquad i\in[d]\,,
	\end{align*}
	which implies  $\lambda_{i}\{(\breve{\bTheta} \odot \breve{\bTheta})^{\T}(\breve{\bTheta} \odot \breve{\bTheta}) \}$ is  finite and uniformly bounded away
	from zero and infinity for any $i\in[d]$. By \eqref{eq:gamma_0-dec},  we have $ \lambda_i(\bar{\breve{\bGamma}}\bar{\breve{\bGamma}}^{\T}) $ is  finite and uniformly bounded away
	from zero and infinity for any $i\in[d]$, which implies \begin{align}\label{eq:ass-eigen-b}
		C_{3} \le  \sigma_i(\bar{\breve{\bGamma}} )  \le C_{4}\,, \qquad i\in[d] \,.
	\end{align} 
	Recall $\breve{\bH}_{i} =\breve{\bTheta}\breve{\bGamma}_{i}\breve{\bTheta}^{\T}$,  $\bar{\breve{\bGamma}}  =(\breve{\bgamma}_{1} , \ldots, \breve{\bgamma}_{d} )$ with $\breve{\bGamma}_{i}  ={\rm diag}(\breve{\bgamma}_{i}) $ for any $i\in[d]$.
	Define $\breve{\bGamma}^{\dagger}$ in the same manner as  $\bGamma^{\dagger}$ specified in Section \ref{sec:calce-sec4} but with replacing $(\bh_{1},\ldots,  \bh_{d}, \bphi)$   by $(\breve{\bh}_{1},\ldots, \breve{\bh}_{d}, \bphi)$,  where $\{\bh_{i}\}_{i=1}^{d}$ and  $\bphi$ are given in Sections \ref{sec:theta-iden} and \ref{sec:theta-hat-est}, respectively. Then $\breve{\bGamma}^{\dagger} =\sum_{i=1}^{d}\breve{\bGamma}_i {\phi}_i $ is a diagonal matrix with  constant vector $ {\bphi} =( {\phi}_1, \ldots,  {\phi}_{d})^{\T}$. 
	Write $\breve{\bGamma}^{\dagger}={\rm diag}(\breve{\gamma}_{1}^{\dagger}, \ldots, \breve{\gamma}_{d}^{\dagger})$. Notice that $(\breve{\gamma}_{1}^{\dagger}, \ldots, \breve{\gamma}_{d}^{\dagger})^{\T} =\bar{\breve{\bGamma}}  \bphi$ with fixed $d$. By \eqref{eq:ass-eigen-b}, there exists some $\bphi$ such that the elements in the main diagonal of $\breve{\bGamma}^{\dagger}$ are finite and  uniformly bounded away from zero and infinity.  With the  selected $\bphi$, define $\breve{\bH}$  in the same manner as $\bH$ but with replacing $(\bH_{1},\ldots, \bH_{d}, \bphi)$ by $(\breve{\bH}_{1},\ldots, \breve{\bH}_{d}, \bphi)$. We have $\breve{\bH}  = \sum_{i=1}^{d} {\phi}_{i}  \breve{\bH}_{i}  =\breve{\bTheta}\breve{\bGamma}^{\dagger} \breve{\bTheta}^{\T}$ and ${\rm rank}(\breve{\bH})=d$.
	Recall $\bB\odot \bA = (\bQ \otimes \bP)(\bV\odot \bU)$ with $(\bQ \otimes \bP)^{\T} (\bQ \otimes \bP) = \bI_{d_1d_2}$. We have $\sigma_{i} (\bV \odot \bU) =\sigma_{i} (\bB \odot \bA)$ for any $i\in[d]$.  Since $\breve{\bC} = \breve{\bW}\breve{\bTheta}$ with $\breve{\bC} = (\bE_2\bV) \odot (\bE_1\bU)$ and $\breve{\bW}^{\T}\breve{\bW}=\bI_{d}$, it also holds that $\sigma_{i}(\breve{\bTheta}) = \sigma_{i} (\bV \odot \bU) $ for any $i\in[d]$.  Recall  $\bU=(\bu_1, \ldots, \bu_{d})$, $\bV=(\bv_1, \ldots, \bv_{d})$. Since $|\bu_{\ell}|_2=1=|\bv_{\ell}|$ for any $\ell\in[d]$, we have $\|\bV \odot \bU\|_2 \le \sqrt{d}$. By Conditions \ref{cd:ra}(i) and  \ref{cd:bounded-value}(i),  it holds that $\sigma_{i}(\breve{\bTheta})$ is finite and uniformly bounded away from zero and infinity for any $i\in[d]$.   Since $\sigma_{d}^2(\breve{\bTheta}) \sigma_{i}( \breve{\bGamma}^{\dagger} ) \le \sigma_{i}(\breve{\bH}) \le \sigma_{1}^2(\breve{\bTheta}) \sigma_{i}(\breve{\bGamma}^{\dagger} )$, then $ \sigma_{i}(\breve{\bH} )  $ is finite and uniformly bounded away
	from zero and infinity for any $i\in[d]$.  
	Recall  $(\hat{d}_1, \hat{d_2}, \hat{d}) =(d_1, d_2, d)$ and  $\tilde{\bH}= \sum_{i = 1}^{d} {\phi}_i \tilde{\bH}_{i} $. By \eqref{eq:H-int-Hb},  
	\begin{align*}
		\|\tilde{\bH}  - \breve{\bH} \|_{\rm F} \le \max_{i\in[d]}\|\tilde{\bH}_{i}  - \breve{\bH}_{i}\|_{\rm F} \times \sum_{i=1}^{d}| {\phi}_i| = \frac{O_{\rm p} (\Pi_{1,n}+\Pi_{2,n}) }{\varpi(\breve{\bOmega})}\,,
	\end{align*}
	provided that  $ (\Pi_{1,n}+\Pi_{2,n})\{\varpi(\breve{\bOmega})\}^{-1}=o_{\rm p}(1)$ and  $\log(pq)=o(n^c)$ for some constant $c \in (0,1)$ depending only on $r_1$ and $r_2$. Since  $\max_{i\in[d]}|\sigma_i(\tilde{\bH} ) - \sigma_i(\breve{\bH}) | \le \|\tilde{\bH} - \breve{\bH} \|_{\rm F}$,  then  $\| \tilde{\bH}  ^{-1} \|_2$  is bounded with probability approaching one provided that   $ (\Pi_{1,n}+\Pi_{2,n})\{\varpi(\breve{\bOmega})\}^{-1}=o_{\rm p}(1)$ and  $\log(pq)=o(n^c)$ for some constant $c \in (0,1)$ depending only on $r_1$ and $r_2$. Hence,    
	\begin{align*}
		\|\tilde{\bH}^{-1}  - \breve{\bH}^{-1} \|_{\rm F} =&~  \|\tilde{\bH}^{-1} (\tilde{\bH} - \breve{\bH})\breve{\bH}^{-1}  \|_{\rm F}  
		\le  \|\tilde{\bH}^{-1} \|_2 \|(\tilde{\bH}- \breve{\bH}) \breve{\bH}^{-1}  \|_{\rm F} \\
		\le&~  \|\tilde{\bH}^{-1} \|_2 \|\breve{\bH}^{-1} \|_2  \| \tilde{\bH} - \breve{\bH} \|_{\rm F} = \frac{O_{\rm p} (\Pi_{1,n}+\Pi_{2,n}) }{\varpi(\breve{\bOmega})}\,,
	\end{align*}     
	provided that  $ (\Pi_{1,n}+\Pi_{2,n}) \{\varpi(\breve{\bOmega}))\}^{-1} =o_{\rm p}(1)$ and  $\log(pq)=o(n^c)$ for some constant $c \in (0,1)$ depending only on $r_1$ and $r_2$. Due to $\|\breve{\bH}_{i}\|_2 \le |\breve{\bh}_{i}|_2 =1$ for any $i\in[d]$, by Triangle inequality and \eqref{eq:H-int-Hb}, we have 
	\begin{align}\label{eq:hdhh-hbhb}
		&\max_{i\in[d]}\| \tilde{\bH}^{-1} \tilde{\bH}_i - \breve{\bH}^{-1} \breve{\bH}_i \|_{\rm F} \notag \\
		&~~~~~~~~\le  \|\tilde{\bH}^{-1}  - \breve{\bH}^{-1}\|_{2} \times \max_{i\in[d]}\| \tilde{\bH}_i - \breve{\bH}_i \|_{\rm F} \notag\\
		&~~~~~~~~~~~+ \| \breve{\bH}^{-1} \|_{2}\times \max_{i\in[d]}\| \tilde{\bH}_i - \breve{\bH}_i \|_{\rm F}  + \max_{i\in[d]}\| \breve{\bH}_i\|_{2} \times \| \tilde{\bH}^{-1} - \breve{\bH}^{-1} \|_{\rm F} \notag\\
		&~~~~~~~~=  \frac{ O_{\rm p} (\Pi_{1,n}+\Pi_{2,n}) }{\varpi(\breve{\bOmega})}\,,
	\end{align}  
	provided that  $ (\Pi_{1,n}+\Pi_{2,n}) \{\varpi(\breve{\bOmega}))\}^{-1} =o_{\rm p}(1)$ and  $\log(pq)=o(n^c)$ for some constant $c \in (0,1)$ depending only on $r_1$ and $r_2$.  Recall $\breve{\bUpsilon}_0   = (\textup{vec}(\breve{\bH}_1),\ldots,\textup{vec}(\breve{\bH}_d))$ and   $\hat{\bUpsilon}_0   = (\textup{vec}(\tilde{\bH}_1),\ldots,\textup{vec}(\tilde{\bH}_d))$.  By \eqref{eq:H-int-Hb}, it holds that
	\begin{align}\label{eq:ups0-ups0h}
		\|\hat{\bUpsilon}_0 - \breve{\bUpsilon}_0 \|_{\rm F} =&~ \sqrt{ \|(\textup{vec}(\tilde{\bH}_1 ),\ldots,\textup{vec}(\tilde{\bH}_d))   - (\textup{vec}(\breve{\bH}_1),\ldots,\textup{vec}(\breve{\bH}_d))\|_{\rm F}^2}  \\
		=&~ \sqrt{\sum_{i=1}^{d} \|\tilde{\bH}_{i}   - \breve{\bH}_{i} \|_{\rm F}^2 } \le d^{1/2} \max_{i\in[d]} \|\tilde{\bH}_{i}  -\breve{\bH}_{i} \|_{\rm F} = \frac{O_{\rm p} (\Pi_{1,n}+\Pi_{2,n}) }{\varpi(\breve{\bOmega})} \notag \,,
	\end{align}
	provided that $(\Pi_{1,n}+\Pi_{2,n})\{\varpi(\breve{\bOmega})\}^{-1}=o_{\rm p}(1)$ and  $\log(pq)=o(n^c)$ for some constant $c \in (0,1)$ depending only on $r_1$ and $r_2$. For $k\in\{1,2\}$, we define $\breve{\bUpsilon}_{k}$ in the same manner as $\bUpsilon_{k}$ but with replacing $(\bH_1, \ldots, \bH_d, \bH )$ by $(\breve{\bH}_1, \ldots, \breve{\bH}_d, \breve{\bH})$.  Notice that $\breve{\bUpsilon}_1   = (\textup{vec}(\breve{\bH}^{-1}\breve{\bH}_1),\ldots,\textup{vec}(\breve{\bH}^{-1}\breve{\bH}_d))$ and 
	$\hat{\bUpsilon}_1   = (\textup{vec}(\tilde{\bH}^{-1}\tilde{\bH}_1),\ldots, \textup{vec}(\tilde{\bH}^{-1}\tilde{\bH}_d) )$.
	Using the similar arguments for the derivation of \eqref{eq:ups0-ups0h}, by \eqref{eq:hdhh-hbhb}, we have
	\begin{align*}%\label{eq:ups1-ups1h}
		\| \hat{\bUpsilon}_1 - \breve{\bUpsilon}_1 \|_{\rm F} \le d^{1/2} \max_{i\in[d]}  \| \tilde{\bH}^{-1} \tilde{\bH}_i - \breve{\bH}^{-1} \breve{\bH}_i \|_{\rm F}   = \frac{O_{\rm p} (\Pi_{1,n}+\Pi_{2,n}) }{\varpi(\breve{\bOmega})} \,,
	\end{align*}
	provided that  $  (\Pi_{1,n}+\Pi_{2,n})\{\varpi(\breve{\bOmega})\}^{-1}=o_{\rm p}(1)$ and  $\log(pq)=o(n^c)$ for some constant $c \in (0,1)$ depending only on $r_1$ and $r_2$. Analogously, we also have 
	\begin{align*}
		\| \hat{\bUpsilon}_2 - \breve{\bUpsilon}_2 \|_{\rm F} = \frac{ O_{\rm p} (\Pi_{1,n}+\Pi_{2,n}) }{\varpi(\breve{\bOmega})} \,.
	\end{align*}  
	Parallel to the calculation of $\bUpsilon_{1}$ in Appendix \ref{sec:calce-sec4}, we have  $\breve{\bUpsilon}_1  = \{\breve{\bTheta} \odot (\breve{\bTheta}^{-1})^{\T}\} (\breve{\bGamma}^{\dagger})^{-1}  \bar{\breve{\bGamma}}$.  Using the similar arguments for the derivation of \eqref{eq:gamma_0-dec}, it holds that
	\begin{align}\label{eq:sigma-ups-1}
		&\lambda_d [\{\breve{\bTheta} \odot (\breve{\bTheta}^{-1})^{\T}\}^{\T}\{\breve{\bTheta} \odot (\breve{\bTheta}^{-1})^{\T}\} ]  \lambda_{i}\{ (\breve{\bGamma}^{\dagger})^{-1} \bar{\breve{\bGamma}} \bar{\breve{\bGamma}}^{\T} (\breve{\bGamma}^{\dagger})^{-1}\}  \le \sigma_i^2(\breve{\bUpsilon}_1)  \notag\\
		&~~~~~~~~~~~~~~~~~~~~~~~\le \lambda_1 [\{\breve{\bTheta} \odot (\breve{\bTheta}^{-1})^{\T}\}^{\T}\{\breve{\bTheta} \odot (\breve{\bTheta}^{-1})^{\T}\} ] \lambda_{i}\{ (\breve{\bGamma}^{\dagger})^{-1} \bar{\breve{\bGamma}} \bar{\breve{\bGamma}}^{\T} (\breve{\bGamma}^{\dagger})^{-1}\} \,.
	\end{align}
	Recall $\breve{\bTheta} = \bE_{3}\bTheta$ with $\bE_3^{\T}\bE_3=\bI_{d}$, and the $L_2$-norm of each column in $\bTheta$ is equal to 1. Since $\{\breve{\bTheta} \odot (\breve{\bTheta}^{-1})^{\T}\}^{\T}\{\breve{\bTheta} \odot (\breve{\bTheta}^{-1})^{\T}\}  =  (\breve{\bTheta}^{\T} \breve{\bTheta})  \circ \{\breve{\bTheta}^{-1}  (\breve{\bTheta}^{-1})^{\T}\} = \{ \breve{\bTheta}^{-1}  (\breve{\bTheta}^{-1})^{\T}\}  \circ (\breve{\bTheta}^{\T} \breve{\bTheta}) $, by Proposition 6.3.4 of \citeS{rao1998matrix}, we have $\lambda_{d}\{\breve{\bTheta}^{-1} (\breve{\bTheta}^{-1} )^{\T}\} \le \lambda_{i} [\{ \breve{\bTheta}^{-1}  (\breve{\bTheta}^{-1})^{\T}\}  \circ (\breve{\bTheta}^{\T} \breve{\bTheta})]  \le \lambda_{1}\{\breve{\bTheta}^{-1} (\breve{\bTheta}^{-1})^{\T} \} $ for any $i\in[d]$. Due to $  \sigma_{i}(\breve{\bTheta}) $ is  finite and uniformly bounded away
	from zero and infinity for any $i\in[d]$, then  $ \lambda_{i}[\{\breve{\bTheta} \odot (\breve{\bTheta}^{-1})^{\T}\}^{\T}\{\breve{\bTheta} \odot (\breve{\bTheta}^{-1})^{\T}\} ] $ is  finite and uniformly bounded away
	from zero and infinity for any $i\in[d]$.  Since $\lambda_{i}\{ (\breve{\bGamma}^{\dagger})^{-1} \bar{\breve{\bGamma}} \bar{\breve{\bGamma}}^{\T} (\breve{\bGamma}^{\dagger})^{-1}\} = \lambda_{i}\{ \bar{\breve{\bGamma}} \bar{\breve{\bGamma}}^{\T} (\breve{\bGamma}^{\dagger})^{-2}\}$ and $\breve{\bGamma}^{\dagger}$ is a diagonal matrix with the elements in the main diagonal being finite and uniformly bounded away from zero and infinity, by 
	$\lambda_{d}(\bar{\breve{\bGamma}} \bar{\breve{\bGamma}}^{\T}) \lambda_{i}\{ (\breve{\bGamma}^{\dagger})^{-2}\} \le  \lambda_{i}\{ (\bar{\breve{\bGamma}} \bar{\breve{\bGamma}}^{\T} (\breve{\bGamma}^{\dagger})^{-2}\} \le \lambda_{1}(\bar{\breve{\bGamma}} \bar{\breve{\bGamma}}^{\T}) \lambda_{i}\{ (\breve{\bGamma}^{\dagger})^{-2}\}$ and \eqref{eq:ass-eigen-b}, we have  
	$ \lambda_{i}\{ (\breve{\bGamma}^{\dagger})^{-1} \bar{\breve{\bGamma}} \bar{\breve{\bGamma}}^{\T} (\breve{\bGamma}^{\dagger})^{-1}\} $ is  finite and uniformly bounded away
	from zero and infinity for any $i\in[d]$. By \eqref{eq:sigma-ups-1}, $   \sigma_{i}^2(\breve{\bUpsilon}_{1}) $ is  finite and uniformly bounded away
	from zero and infinity for any $i\in[d]$.  Analogously, we can also show $ \sigma_i^2(\breve{\bUpsilon}_2) $ is  finite and uniformly bounded away from zero and infinity for any $i\in[d]$.  Recall  all the nonzero singular values of  $ \breve{\bUpsilon}_0$ are finite and uniformly  bounded away from zero and infinity. Hence, by Triangle inequality, it holds that 
	\begin{align}\label{eq:ups02-ups02h}
		\| \hat{\bUpsilon}_0^\T\hat{\bUpsilon}_2 -\breve{\bUpsilon}_0^\T \breve{\bUpsilon}_2  \|_{\rm F} &\le \| \breve{\bUpsilon}_0\|_2 \| \hat{\bUpsilon}_2 - \breve{\bUpsilon}_2 \|_{\rm F} + \| \breve{\bUpsilon}_2\|_2 \| \hat{\bUpsilon}_0 - \breve{\bUpsilon}_0 \|_{\rm F} + \| \hat{\bUpsilon}_0 - \breve{\bUpsilon}_0\|_{\rm F} \| \hat{\bUpsilon}_2 - \breve{\bUpsilon}_2 \|_{\rm F} \notag\\
		& = \frac{O_{\rm p}  (\Pi_{1,n}+\Pi_{2,n}) }{\varpi(\breve{\bOmega})} \,,
	\end{align}
	provided that  $ (\Pi_{1,n}+\Pi_{2,n}) \{\varpi(\breve{\bOmega})\}^{-1}=o_{\rm p}(1)$ and  $\log(pq)=o(n^c)$ for some constant $c \in (0,1)$ depending only on $r_1$ and $r_2$.  Analogously, we have 
	\begin{align*}
		\| \hat{\bUpsilon}_1^\T\hat{\bUpsilon}_2 - \breve{\bUpsilon}_1^\T \breve{\bUpsilon}_2  \|_{\rm F} &\le \| \breve{\bUpsilon}_1\|_2 \| \hat{\bUpsilon}_2 - \breve{\bUpsilon}_2 \|_{\rm F} + \| \breve{\bUpsilon}_2\|_2 \| \hat{\bUpsilon}_1 - \breve{\bUpsilon}_1 \|_{\rm F} + \| \hat{\bUpsilon}_1 - \breve{\bUpsilon}_1\|_{\rm F} \| \hat{\bUpsilon}_2 - \breve{\bUpsilon}_2 \|_{\rm F} \\
		&= \frac{O_{\rm p} (\Pi_{1,n}+\Pi_{2,n}) }{\varpi(\breve{\bOmega})}\,.
	\end{align*} 
	Notice that  $\breve{\bUpsilon}_1^\T\breve{\bUpsilon}_2 = \bar{\breve{\bGamma}}^{\T} (\breve{\bGamma}^{\dagger})^{-2} \bar{\breve{\bGamma} }=  \breve{\bUpsilon}_2^\T \breve{\bUpsilon}_1$. Since  $\breve{\bGamma}^{\dagger}$ is a diagonal matrix with the elements in the main diagonal are finite and uniformly bounded away from zero and infinity, by  $ \sigma_{d}^2(\bar{\breve{\bGamma}})\sigma_{i}\{(\breve{\bGamma}^{\dagger})^{-2}\} \le \sigma_{i}(\bar{\breve{\bGamma}}^{\T} (\breve{\bGamma}^{\dagger})^{-2} \bar{\breve{\bGamma} }) \le \sigma_{1}^2(\bar{\breve{\bGamma}})\sigma_{i}\{(\breve{\bGamma}^{\dagger})^{-2}\} $ and \eqref{eq:ass-eigen-b}, it holds that  $    \sigma_i(\breve{\bUpsilon}_1^\T\breve{\bUpsilon}_2 +  \breve{\bUpsilon}_2^\T \breve{\bUpsilon}_1) $ is finite and uniformly bounded away from zero and infinity for any $i\in [d]$,   which implies  $ \sigma_i(\hat{\bUpsilon}_1^\T\hat{\bUpsilon}_2 +  \hat{\bUpsilon}_2^\T \hat{\bUpsilon}_1) $ is finite and uniformly bounded away from zero and infinity for any $i\in[d]$ with probability approaching one, provided that  $ (\Pi_{1,n}+\Pi_{2,n})\{\varpi(\breve{\bOmega})\}^{-1}=o_{\rm p}(1)$ and  $\log(pq)=o(n^c)$ for some constant $c \in (0,1)$ depending only on $r_1$ and $r_2$. Hence, 
	\begin{align}\label{eq:ups12-ups12h}
		&\|(\hat{\bUpsilon}_1^\T\hat{\bUpsilon}_2 +  \hat{\bUpsilon}_2^\T \hat{\bUpsilon}_1)^{-1} - (\breve{\bUpsilon}_1^\T\breve{\bUpsilon}_2 +  \breve{\bUpsilon}_2^\T \breve{\bUpsilon}_1)^{-1} \|_{\rm F} \notag\\   &~~~~~ \le \|(\hat{\bUpsilon}_1^\T\hat{\bUpsilon}_2 +  \hat{\bUpsilon}_2^\T \hat{\bUpsilon}_1)^{-1} \|_2    \| (\hat{\bUpsilon}_1^\T\hat{\bUpsilon}_2 +  \hat{\bUpsilon}_2^\T \hat{\bUpsilon}_1) - (\breve{\bUpsilon}_1^\T\breve{\bUpsilon}_2 +  \breve{\bUpsilon}_2^\T \breve{\bUpsilon}_1) \|_{\rm F} \| (\breve{\bUpsilon}_1^\T\breve{\bUpsilon}_2 +  \breve{\bUpsilon}_2^\T \breve{\bUpsilon}_1)^{-1} \|_{2} \notag\\
		&~~~~~=\frac{ O_{\rm p} (\Pi_{1,n}+\Pi_{2,n}) }{\varpi(\breve{\bOmega})} \,,
	\end{align}  
	provided that  $ (\Pi_{1,n}+\Pi_{2,n})\{\varpi(\breve{\bOmega})\}^{-1}=o_{\rm p}(1)$ and  $\log(pq)=o(n^c)$ for some constant $c \in (0,1)$ depending only on $r_1$ and $r_2$.  Write  $\bK = (\breve{\bUpsilon}_0^\T \breve{\bUpsilon}_2)(\breve{\bUpsilon}_1^\T\breve{\bUpsilon}_2 +  \breve{\bUpsilon}_2^\T \breve{\bUpsilon}_1)^{-1}(\breve{\bUpsilon}_2^\T\breve{\bUpsilon}_0)$ and $\hat{\bK} = (\hat{\bUpsilon}_0^\T\hat{\bUpsilon}_2)(\hat{\bUpsilon}_1^\T\hat{\bUpsilon}_2 +  \hat{\bUpsilon}_2^\T\hat{\bUpsilon}_1)^{-1}(\hat{\bUpsilon}_2^\T\hat{\bUpsilon}_0)$. Notice that $\breve{\bUpsilon}_0^\T\breve{\bUpsilon}_1 = \bar{\breve{\bGamma}}^{\T}(\breve{\bGamma}^{\dagger})^{-1} \bar{\breve{\bGamma}} =  \breve{\bUpsilon}_0^\T \breve{\bUpsilon}_2$. Analogously, we have $ \sigma_{i}(\breve{\bUpsilon}_0^\T\breve{\bUpsilon}_1)  $ and $ \sigma_{i}(\breve{\bUpsilon}_0^\T\breve{\bUpsilon}_2)  $ are finite and uniformly bounded away from zero and infinity for any $i\in[d]$. Since 
	\begin{align*}
		\hat{\bK} - \bK =&~ \big(\hat{\bUpsilon}_0^\T\hat{\bUpsilon}_2 -\breve{\bUpsilon}_0^\T \breve{\bUpsilon}_2 \big) \big\{(\hat{\bUpsilon}_1^\T\hat{\bUpsilon}_2 +  \hat{\bUpsilon}_2^\T\hat{\bUpsilon}_1)^{-1} - (\breve{\bUpsilon}_1^\T\breve{\bUpsilon}_2 +  \breve{\bUpsilon}_2^\T \breve{\bUpsilon}_1)^{-1}\big\}\big(\hat{\bUpsilon}_2^\T\hat{\bUpsilon}_0 - \breve{\bUpsilon}_2^\T\breve{\bUpsilon}_0\big)\\
		&+\big(\breve{\bUpsilon}_0^\T \breve{\bUpsilon}_2 \big) \big\{(\hat{\bUpsilon}_1^\T\hat{\bUpsilon}_2 +  \hat{\bUpsilon}_2^\T\hat{\bUpsilon}_1)^{-1} - (\breve{\bUpsilon}_1^\T\breve{\bUpsilon}_2 +  \breve{\bUpsilon}_2^\T \breve{\bUpsilon}_1)^{-1}\big\}\big(\hat{\bUpsilon}_2^\T\hat{\bUpsilon}_0 - \breve{\bUpsilon}_2^\T\breve{\bUpsilon}_0\big)\\
		&+\big(\hat{\bUpsilon}_0^\T\hat{\bUpsilon}_2 -\breve{\bUpsilon}_0^\T \breve{\bUpsilon}_2 \big)  (\breve{\bUpsilon}_1^\T\breve{\bUpsilon}_2 +  \breve{\bUpsilon}_2^\T \breve{\bUpsilon}_1)^{-1} \big(\hat{\bUpsilon}_2^\T\hat{\bUpsilon}_0 - \breve{\bUpsilon}_2^\T\breve{\bUpsilon}_0\big)\\
		&+\big(\hat{\bUpsilon}_0^\T\hat{\bUpsilon}_2 -\breve{\bUpsilon}_0^\T \breve{\bUpsilon}_2 \big) \big\{(\hat{\bUpsilon}_1^\T\hat{\bUpsilon}_2 +  \hat{\bUpsilon}_2^\T\hat{\bUpsilon}_1)^{-1} - (\breve{\bUpsilon}_1^\T\breve{\bUpsilon}_2 +  \breve{\bUpsilon}_2^\T \breve{\bUpsilon}_1)^{-1}\big\}\big( \breve{\bUpsilon}_2^\T\breve{\bUpsilon}_0\big)\\
		&+ \big( \breve{\bUpsilon}_0^\T \breve{\bUpsilon}_2 \big)  (\breve{\bUpsilon}_1^\T\breve{\bUpsilon}_2 +  \breve{\bUpsilon}_2^\T \breve{\bUpsilon}_1)^{-1} \big(\hat{\bUpsilon}_2^\T\hat{\bUpsilon}_0 - \breve{\bUpsilon}_2^\T\breve{\bUpsilon}_0\big)
		\\&+ \big( \breve{\bUpsilon}_0^\T \breve{\bUpsilon}_2 \big) \big\{(\hat{\bUpsilon}_1^\T\hat{\bUpsilon}_2 +  \hat{\bUpsilon}_2^\T\hat{\bUpsilon}_1)^{-1} - (\breve{\bUpsilon}_1^\T\breve{\bUpsilon}_2 +  \breve{\bUpsilon}_2^\T \breve{\bUpsilon}_1)^{-1}\big\}\big( \breve{\bUpsilon}_2^\T\breve{\bUpsilon}_0\big)\\
		&+ \big(\hat{\bUpsilon}_0^\T\hat{\bUpsilon}_2 -\breve{\bUpsilon}_0^\T \breve{\bUpsilon}_2 \big)  (\breve{\bUpsilon}_1^\T\breve{\bUpsilon}_2 +  \breve{\bUpsilon}_2^\T \breve{\bUpsilon}_1)^{-1} \big( \breve{\bUpsilon}_2^\T\breve{\bUpsilon}_0\big)\,,
	\end{align*}
	by \eqref{eq:ups02-ups02h} and \eqref{eq:ups12-ups12h}, it holds that
	\begin{align*}
		\|\hat{\bK} - \bK \|_{\rm F} \le &~\big\|\hat{\bUpsilon}_0^\T\hat{\bUpsilon}_2 -\breve{\bUpsilon}_0^\T \breve{\bUpsilon}_2 \big\|^2_{\rm F} \big\|(\hat{\bUpsilon}_1^\T\hat{\bUpsilon}_2 +  \hat{\bUpsilon}_2^\T\hat{\bUpsilon}_1)^{-1} - (\breve{\bUpsilon}_1^\T\breve{\bUpsilon}_2 +  \breve{\bUpsilon}_2^\T \breve{\bUpsilon}_1)^{-1}\big\|_{\rm F}\\
		&+ 2  \big\|(\hat{\bUpsilon}_1^\T\hat{\bUpsilon}_2 +  \hat{\bUpsilon}_2^\T\hat{\bUpsilon}_1)^{-1} - (\breve{\bUpsilon}_1^\T\breve{\bUpsilon}_2 +  \breve{\bUpsilon}_2^\T \breve{\bUpsilon}_1)^{-1}\big\|_{\rm F}\big\|\hat{\bUpsilon}_2^\T\hat{\bUpsilon}_0 - \breve{\bUpsilon}_2^\T\breve{\bUpsilon}_0\big\|_{\rm F}  \|\breve{\bUpsilon}_2^\T\breve{\bUpsilon}_0\big\|_{2}\\
		&+ \big\|\hat{\bUpsilon}_2^\T\hat{\bUpsilon}_0 - \breve{\bUpsilon}_2^\T\breve{\bUpsilon}_0\big\|_{\rm F}^2 \|(\breve{\bUpsilon}_1^\T\breve{\bUpsilon}_2 +  \breve{\bUpsilon}_2^\T \breve{\bUpsilon}_1)^{-1}\|_2\\
		&+2 \|\breve{\bUpsilon}_2^\T\breve{\bUpsilon}_0\big\|_{2} \|(\breve{\bUpsilon}_1^\T\breve{\bUpsilon}_2 +  \breve{\bUpsilon}_2^\T \breve{\bUpsilon}_1)^{-1}\|_2  \big\|\hat{\bUpsilon}_2^\T\hat{\bUpsilon}_0 - \breve{\bUpsilon}_2^\T\breve{\bUpsilon}_0\big\|_{\rm F}\\
		&+ \|\breve{\bUpsilon}_2^\T\breve{\bUpsilon}_0\big\|_{2}^2 \big\|(\hat{\bUpsilon}_1^\T\hat{\bUpsilon}_2 +  \hat{\bUpsilon}_2^\T\hat{\bUpsilon}_1)^{-1} - (\breve{\bUpsilon}_1^\T\breve{\bUpsilon}_2 +  \breve{\bUpsilon}_2^\T \breve{\bUpsilon}_1)^{-1}\big\|_{\rm F} \\
		=&~\frac{O_{\rm p} (\Pi_{1,n}+\Pi_{2,n}) }{\varpi(\breve{\bOmega})}\,,
	\end{align*}
	provided that  $ (\Pi_{1,n}+\Pi_{2,n})\{\varpi(\breve{\bOmega})\}^{-1}=o_{\rm p}(1)$ and  $\log(pq)=o(n^c)$ for some constant $c \in (0,1)$ depending only on $r_1$ and $r_2$. Since $\bK =2^{-1} \bar{\breve{\bGamma}}^{\T}\bar{\breve{\bGamma}}$, by \eqref{eq:ass-eigen-b}, it holds that $ \lambda_i(\bK)  $ is finite and uniformly bounded away from zero and infinity for any $i \in [d]$, which implies $ \lambda_i(\hat \bK) $ is finite and uniformly bounded away from zero and infinity for any $i \in [d]$ with probability approaching one, provided that $ (\Pi_{1,n} + \Pi_{2,n})\{\varpi(\breve{\bOmega})\}^{-1} = o_{\rm p}(1)$  and  $\log(pq)=o(n^c)$ for some constant $c \in (0,1)$ depending only on $r_1$ and $r_2$. By Theorem 6.2 of \citeS{higham2008functions}, it holds that 
	\begin{equation*}
		\| \hat{\bK}^{1/2} - \bK^{1/2} \|_{\rm F} \le \frac{ \| \hat{\bK}  - \bK  \|_{\rm F} }{\lambda^{1/2}_{d}(\hat{\bK}) + \lambda^{1/2}_{d}(\bK)} \lesssim \frac{O_{\rm p} (\Pi_{1,n}+\Pi_{2,n})  }{\varpi(\breve{\bOmega})}\,.
	\end{equation*} 
	Recall $\hat{\bPi} =(2\hat{\bK})^{-1/2}$. Let $\breve{\bPi} =(2\bK)^{-1/2}$.
	Hence,  
	\begin{align*}
		\| \hat{\bPi} - \breve{\bPi} \|_{\rm F}  
		\le \frac{\sqrt{2}}{2} \| \hat{\bK}^{-1/2}\|_{2} \| \hat{\bK}^{1/2} - \bK^{1/2} \|_{\rm F} \| \bK^{-1/2}\|_{2}  = \frac{O_{\rm p} (\Pi_{1,n}+\Pi_{2,n}) }{\varpi(\breve{\bOmega})} \,,
	\end{align*}
	provided that $ (\Pi_{1,n} + \Pi_{2,n})\{\varpi(\breve{\bOmega})\}^{-1} = o_{\rm p}(1)$  and  $\log(pq)=o(n^c)$ for some constant $c \in (0,1)$ depending only on $r_1$ and $r_2$.  Recall $(\hat{\bh}_1, \ldots, \hat{\bh}_{d}) = (\tilde{\bh}_1 , \ldots, \tilde{\bh}_{d} )\hat{\bPi}$ with $\hat{\bR}_{\rm o} = (\tilde{\bh}_1 , \ldots, \tilde{\bh}_{d} )$.  Let $(\breve{\bh}^*_1, \ldots, \breve{\bh}^*_{d}) = (\breve{\bh}_1, \ldots, \breve{\bh}_{d})\breve{\bPi}$, where  $ (\breve{\bh}_1, \ldots, \breve{\bh}_{d}) ={\rm \bR}_{\rm o}\bE_{4}^{\T} $ for $\bR_{\rm o}$ and  $\bE_4$ specified in \eqref{eq:roh-roe}.  Notice 
	that $ \sigma_{i}(\breve{\bPi}) $ is finite and uniformly bounded away from zero and infinity for any $i\in[d]$. Together with \eqref{eq:roh-roe}, we have \begin{align*}
		\max_{i\in[d]}|\hat{\bh}_i- \breve{\bh}^*_i|_2 \le&~ \|\hat{\bR}_{\rm o}\hat{\bPi} -   \bR_{\rm o}\bE_4^{\T}\breve{\bPi} \|_2\\
		\le &~ \|\hat{\bR}_{\rm o} -   \bR_{\rm o}\bE_4^{\T}  \|_2 \| \hat{\bPi} - \breve{\bPi} \|_2 + \|\bR_{\rm o}\bE_4^{\T}\|_2\|\hat{\bPi} - \breve{\bPi} \|_2 + \|\hat{\bR}_{\rm o} -   \bR_{\rm o}\bE_4^{\T}\|_2 \|\breve{\bPi}\|_2 \\
		\le&~ \frac{O_{\rm p} (\Pi_{1,n}+\Pi_{2,n}) }{\varpi(\breve{\bOmega})} \,,
	\end{align*}  
	provided that $ (\Pi_{1,n} + \Pi_{2,n})\{\varpi(\breve{\bOmega})\}^{-1} = o_{\rm p}(1)$  and  $\log(pq)=o(n^c)$ for some constant $c \in (0,1)$ depending only on $r_1$ and $r_2$.  For any $i\in[d]$, we derive $\breve{\bH}_i^{*}$ and $\hat{\bH}_{i}$ in the same manner as $\breve{\bH}_{i}$ but with replacing $\breve{\bh}_i$ by $\breve{\bh}_{i}^{*}$ and $\hat{\bh}_{i}$, respectively.  It holds that
	\begin{align}\label{eq:convergence Hstar}
		\max_{i\in[d]}\|\hat{\bH}_{i} - \breve{\bH}^*_{i}\|_{\rm F} \le  \max_{i\in[d]}|\hat{\bh}_i- \breve{\bh}^*_i|_2 = \frac{O_{\rm p} (\Pi_{1,n}+\Pi_{2,n}) }{\varpi(\breve{\bOmega})} \,,
	\end{align}
	provided that $ (\Pi_{1,n} + \Pi_{2,n})\{\varpi(\breve{\bOmega})\}^{-1} = o_{\rm p}(1)$  and  $\log(pq)=o(n^c)$ for some constant $c \in (0,1)$ depending only on $r_1$ and $r_2$.  Notice that $\breve{\bH}^*_i = \breve{\bTheta} \breve{\bGamma}^*_i \breve{\bTheta}^{\T}$, where  $ \breve{\bGamma}^*_i =\textup{diag}(\breve{\bgamma}^*_i)$ with  $\breve{\bgamma}^*_i = (\breve{\gamma}^*_{i,1,1},\ldots,\breve{\gamma}^*_{i,d,d})^\top$. By \eqref{eq:convergence Hstar}, $\hat{\bH}_i$ can be written as
	\begin{equation*}
		\hat{\bH}_i = \breve{\bTheta} \breve{\bGamma}^*_i \breve{\bTheta}^{\T} + \tilde{\delta}_n \tilde{\bN}_i\,,\qquad i\in[d]\,,
	\end{equation*}
	where $\tilde{\delta}_n = (\Pi_{1,n}+\Pi_{2,n})\{\varpi(\breve{\bOmega})\}^{-1}$, and $\tilde{\bN}_i$ is a $d \times d$ random matrix with  $\max_{i\in[d]}\|\tilde{\bN}_i\|_{2} =  O_{\rm p}(1)$.  Write $\bar{\breve{\bGamma}}^* = (\breve{\bgamma}^*_1,\ldots,\breve{\bgamma}^*_d)$ and $\bPhi = \hat{\bTheta}^{-1}$.  
	Following Theorem 3 of \citeS{afsari2008sensitivity}, for small enough $\tilde{\delta}_n$, by the non-orthogonal joint diagonalization,
	\begin{equation*}%\label{eq:theta-breve}
		\bPhi  = (\mathbf{I}_d + \tilde{\delta}_n \tilde{\boldsymbol{\aleph}})(\breve{\bTheta}\breve{\bD}_1\breve{\bD}_2 )^{-1} + o_{\rm p}(\tilde{\delta}_n)
	\end{equation*}
	with a diagonal matrix $\breve{\bD}_1$, a permutation matrix $\breve{\bD}_{2}$ and
	\begin{equation*}%\label{eq:aleph-dec}
		\|\tilde{\boldsymbol{\aleph}}\|_{\textup{F}} \le  \frac{d \cdot \eta(\breve{\bh}^*_1, \ldots, \breve{\bh}^*_{d})}{1 - \rho^2(\breve{\bh}^*_1, \ldots, \breve{\bh}^*_{d})} \|(\breve{\bTheta} \breve{\bD}_1\breve{\bD}_2  )^{-1}\|_2^2\sum_{i = 1}^d\| \tilde{\bN}_i \|_2 \| \breve{\bGamma}^*_i \|_2\,,
	\end{equation*}
	where  the elements in the main diagonal of $\breve{\bD}_1$  are  $\pm 1$, $\eta(\breve{\bh}^*_1, \ldots, \breve{\bh}^*_{d})$ and $\rho(\breve{\bh}^*_1, \ldots, \breve{\bh}^*_{d})$ are defined, respectively, in the same manner as $\eta(\bh_1, \ldots, \bh_d)$ and  $\rho(\bh_1, \ldots, \bh_d)$  given in \eqref{eq:alpha and rho} but with replacing $\bh_1, \ldots, \bh_d$ by $\breve{\bh}^*_1, \ldots, \breve{\bh}^*_{d}$.   Since $\|(\breve{\bh}_1, \ldots, \breve{\bh}_{d})\|_2=1$,   $\|\breve{\bPi}\|_2 \le C_2 $ and $C_3 \le \sigma_{i}(\breve{\bTheta})\le C_{4}$ for any $i\in[d]$, then  $\| \breve{\bGamma}^*_i \|_2 \le \|\breve{\bTheta}^{-1}\|_2^{2} \| \breve{\bH}^*_i\|_2 \lesssim  |\breve{\bh}^*_i|_2 \le   \|(\breve{\bh}_1, \ldots, \breve{\bh}_{d})\breve{\bPi}\|_2 \le   \|(\breve{\bh}_1, \ldots, \breve{\bh}_{d})\|_2\|\breve{\bPi}\|_2 \le C_{5}$.  It holds that
	\begin{align*}
		\| \bPhi  - (\breve{\bTheta} \breve{\bD}_1\breve{\bD}_2 )^{-1} \|_{\rm F} = &~ \|\tilde{\delta}_n \tilde{\boldsymbol{\aleph}}(\breve{\bTheta}\breve{\bD}_1\breve{\bD}_2)^{-1} + o_{\rm p}(\tilde{\delta}_n)\|_{\rm F}  
		\le   \tilde{\delta}_{n} \| \breve{\bTheta}^{-1}\|_2 \|\tilde{\boldsymbol{\aleph}}\|_{\rm F}    + o_{\rm p}(\tilde{\delta}_{n})   \notag \\
		\le &~ \frac{d \tilde{\delta}_{n}\cdot  \eta(\breve{\bh}^*_1, \ldots, \breve{\bh}^*_{d})}{1 - \rho^2(\breve{\bh}^*_1, \ldots, \breve{\bh}^*_{d})}  \|\breve{\bTheta}^{-1}\|_2^3  \max_{i\in[d]} \| \tilde{\bN}_i \|_2 \times 
		\sum_{i = 1}^d\| \breve{\bGamma}_i^{*} \|_2 + o_{\rm p}( \tilde{\delta}_{n}) \\
		= &~ \frac{\eta(\breve{\bh}^*_1, \ldots, \breve{\bh}^*_{d}) \cdot    O_{\rm p}( \tilde{\delta}_{n})}{1 - \rho^2(\breve{\bh}^*_1, \ldots, \breve{\bh}^*_{d}) }   + o_{\rm p} ( \tilde{\delta}_{n} ) \,.
	\end{align*}
	Recall $\tilde{\delta}_{n} = (\Pi_{1,n} +\Pi_{2,n})\{\varpi(\breve{\bOmega})\}^{-1}$  and $(\hat{d}_1, \hat{d}_2,\hat{d})=(d_1,d_2,d)$. We have $ \sigma_{i}(\bPhi) $ is finite and uniformly bounded away from zero and infinity for any $i\in[d]$ with probability approaching one, provided that $ (\Pi_{1,n}+\Pi_{2,n})\{\varpi(\breve{\bOmega})\}^{-1}[1+ \eta(\breve{\bh}^*_1, \ldots, \breve{\bh}^*_{d} )  / \{1 - \rho^2(\breve{\bh}^*_1, \ldots, \breve{\bh}^*_{d}) \}]  =o_{\rm p}(1)$  and  $\log(pq)=o(n^c)$ for some constant $c \in (0,1)$ depending only on $r_1$ and $r_2$. Hence,
	\begin{align*}
		\| \bPhi^{-1}  - \breve{\bTheta}\breve{\bD}_1\breve{\bD}_2  \|_{\rm F}  \le&~ \|\bPhi^{-1} \|_2 \|\breve{\bTheta}\breve{\bD}_1\breve{\bD}_2  \|_2  \| \bPhi  - (\breve{\bTheta}\breve{\bD}_1\breve{\bD}_2 )^{-1} \|_{\rm F}\\
		=&~ \frac{\eta(\breve{\bh}^*_1, \ldots, \breve{\bh}^*_{d}) \cdot O_{\rm p} (\Pi_{1,n} +\Pi_{2,n}) }{\varpi(\breve{\bOmega}) \{1 - \rho^2(\breve{\bh}^*_1, \ldots, \breve{\bh}^*_{d})\}}  + \frac{o_{\rm p} (\Pi_{1,n} +\Pi_{2,n}) }{ \varpi(\breve{\bOmega})} \,,
	\end{align*}
	provided that $ (\Pi_{1,n}+\Pi_{2,n})\{\varpi(\breve{\bOmega})\}^{-1}[1+ \eta(\breve{\bh}^*_1, \ldots, \breve{\bh}^*_{d} )  / \{1 - \rho^2(\breve{\bh}^*_1, \ldots, \breve{\bh}^*_{d}) \}] =o_{\rm p}(1)$  and  $\log(pq)=o(n^c)$ for some constant $c \in (0,1)$ depending only on $r_1$ and $r_2$.  Parallel to the calculation of $\rho(\bh_1^{*}, \ldots, \bh_d^{*})$ and $\eta(\bh_1^{*}, \ldots,\bh_d^{*})$ given in Section \ref{sec:calce-sec4} for the proof of Proposition \ref{pro:rotation}, it also holds that $\rho(\breve{\bh}^*_1, \ldots, \breve{\bh}^*_{d}) =0$ and $\eta(\breve{\bh}^*_1, \ldots, \breve{\bh}^*_{d})=2$, 
	% Due to $\varpi(\breve{\bOmega})  \le 8$, then
	% \begin{align*}
		%     \frac{\eta(\bar{\breve{\bGamma}}^{*}) }{\varpi(\breve{\bOmega}) \{1 - \rho^2(\bar{\breve{\bGamma}}^{*})\}}  \ge \frac{1}{4}\,,
		% \end{align*}
	which implies 
	\begin{equation*}
		\| \bPhi^{-1} - \breve{\bTheta}\breve{\bD}_1\breve{\bD}_2  \|_{\textup F}=  \frac{   O_{\rm p} (\Pi_{1,n} + \Pi_{2,n})  }{\varpi(\breve{\bOmega}) } \,, 
	\end{equation*}
	provided that $ (\Pi_{1,n}+\Pi_{2,n})   \{\varpi(\breve{\bOmega})\}^{-1} =o_{\rm p}(1)$  and  $\log(pq)=o(n^c)$ for some constant $c \in (0,1)$ depending only on $r_1$ and $r_2$. Recall  
	$\breve{\bTheta}=(\breve{\btheta}_1,\ldots, \breve{\btheta}_{d})$ and $\bPhi = \hat{\bTheta}^{-1} $. Write $ \hat{\bTheta}=(\hat{\btheta}_{1},\ldots, \hat{\btheta}_{d})$. Then, there exists a permutation of $(1,\ldots,d)$, denoted by $(j_1,\ldots,j_d)$,  such that  
	\begin{equation*}
		\max_{\ell\in[d]}| {\kappa}_{\ell} \hat{\btheta}_{j_\ell}  - \breve{\btheta}_{\ell} |_2 \le \| \hat{\bTheta} - \breve{\bTheta}\breve{\bD}_1\breve{\bD}_2 \|_{\textup F}=  \frac{ O_{\rm p} (\Pi_{1,n} + \Pi_{2,n}) }{\varpi(\breve{\bOmega}) }  
	\end{equation*}
	with some $  {\kappa}_{\ell}  \in \{-1, 1\}$, provided that $ (\Pi_{1,n}+\Pi_{2,n})\{\varpi(\breve{\bOmega})\}^{-1} =o_{\rm p}(1)$  and  $\log(pq)=o(n^c)$ for some constant $c \in (0,1)$ depending only on $r_1$ and $r_2$. We complete the proof of Lemma \ref{lemma:theta}.
	$\hfill\Box$

	\setcounter{table}{0}
	\renewcommand{\thetable}{S\arabic{table}}
	\setcounter{figure}{0}
	\renewcommand{\thefigure}{S\arabic{figure}}

	\section{Additional simulations}
\begin{figure}[p] 
		\centerline{\includegraphics[width= 16.6cm]{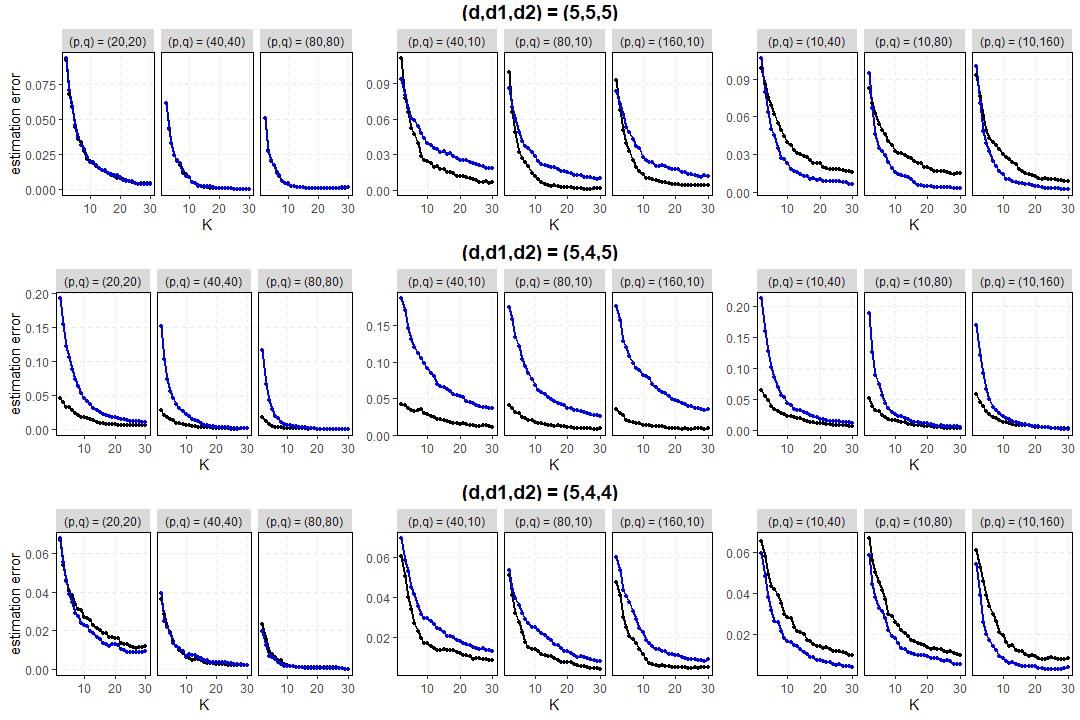}}
		\caption{\footnotesize{The lineplots for the averages of $\varpi^2(\bA,\hat{\bA})$ and $\varpi^2(\bB,\hat{\bB})$ with respect to $\tilde{K} = 10 $ and $K \in \{2,3,\ldots,30\}$ in Scenarios R1--R3 with $d=5$ based on 2000 repetitions. The legend is defined as follows:  $\varpi^2(\bA,\hat{\bA})$ ($\color{black}{-\bullet-}$) and   $\varpi^2(\bB,\hat{\bB})$ ($\color{blue}{-\bullet-}$).}}
		\label{fig:K-estimation}
	\end{figure}

\begin{figure}[htbp]
		\centerline{\includegraphics[width= 16.6cm]{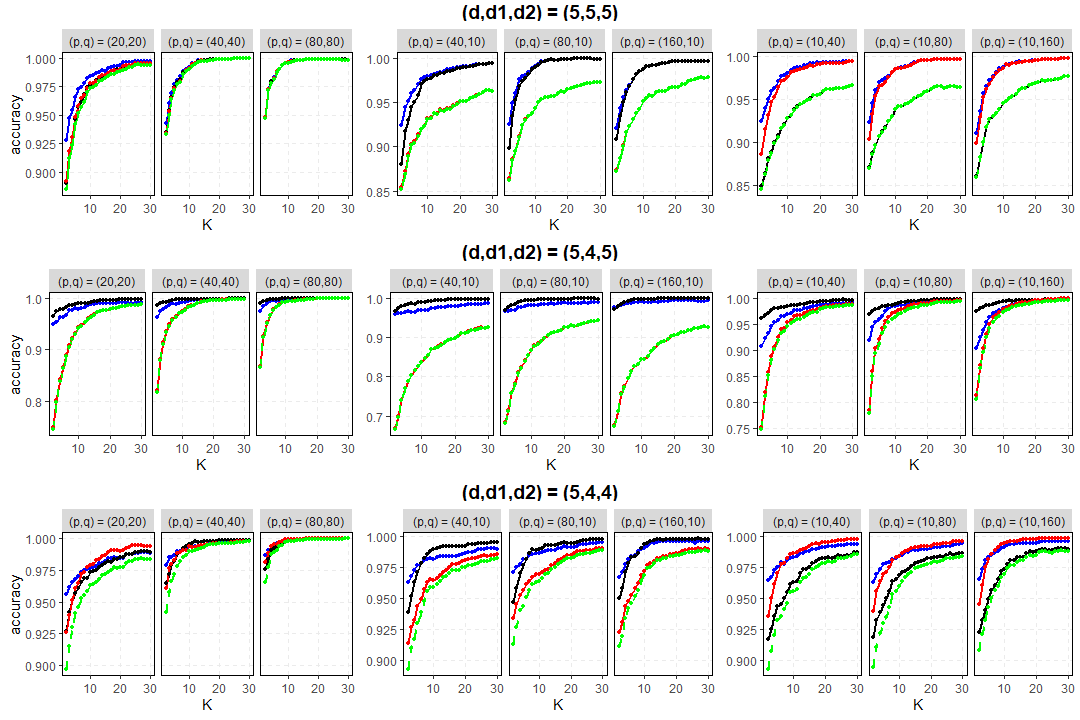}}
		\caption{\footnotesize{The lineplots for the relative frequency estimates of $\P( \hat{d} = d )$, $\P( \hat{d}_1 = d_1 )$, $ \P( \hat{d}_2 = d_2 )$ and $\P(\hat{d} = d, \hat{d}_1 = d_1, \hat{d}_2 = d_2)$ with respect to $\tilde{K} = 10$ and $K \in \{2,3,\ldots,30\}$  in Scenarios R1--R3 with $d=5$  based on 2000 repetitions. The legend is defined as follows:  \textup{(i)} the relative frequency estimate of $\P(\hat{d}_1 = d_1)$ ($\color{black}{-\bullet-}$); \textup{(ii)} the relative frequency estimate of $\P(\hat{d}_2 = d_2)$ ($\color{red}{-\bullet-}$); \textup{(iii)} the relative frequency estimate of $\P(\hat{d} = d)$ ($\color{blue}{-\bullet-}$); \textup{(iv)} the relative frequency estimate of $\P\{(\hat{d}_1,\hat{d}_2,\hat{d}) =(d_1, d_2,d) \}$ ($\color{green}{-\bullet-}$).}}
		\label{fig:K-rank}
	\end{figure}

\begin{figure}[htbp] 
		\centerline{\includegraphics[width= 16.6cm]{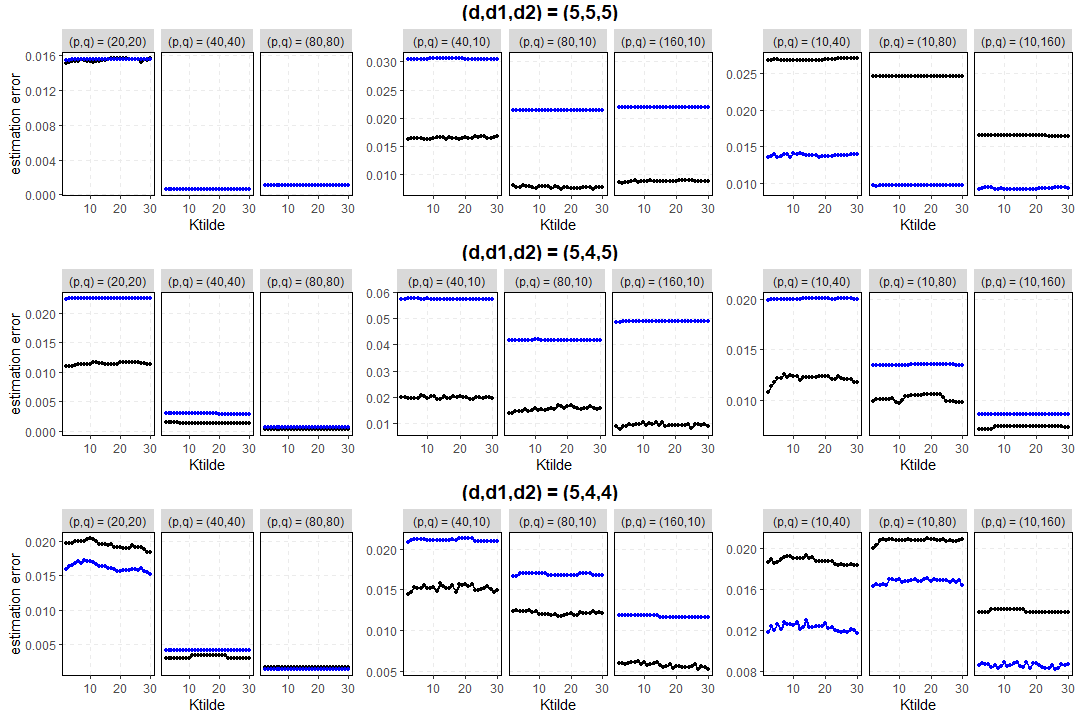}}
		\caption{\footnotesize{The lineplots for the averages of $\varpi^2(\bA,\hat{\bA})$ and $\varpi^2(\bB,\hat{\bB})$ with respect to $\tilde{K} \in \{2,3,\ldots,30\}$ and $K = 20$ in Scenarios R1--R3 with $d=5$ based on 2000 repetitions. The legend is defined as follows:  $\varpi^2(\bA,\hat{\bA})$ ($\color{black}{-\bullet-}$) and   $\varpi^2(\bB,\hat{\bB})$ ($\color{blue}{-\bullet-}$).}}
		\label{fig:Ktilde-estimation}
	\end{figure}

\begin{figure}[htbp]
		\centerline{\includegraphics[width= 16.6cm]{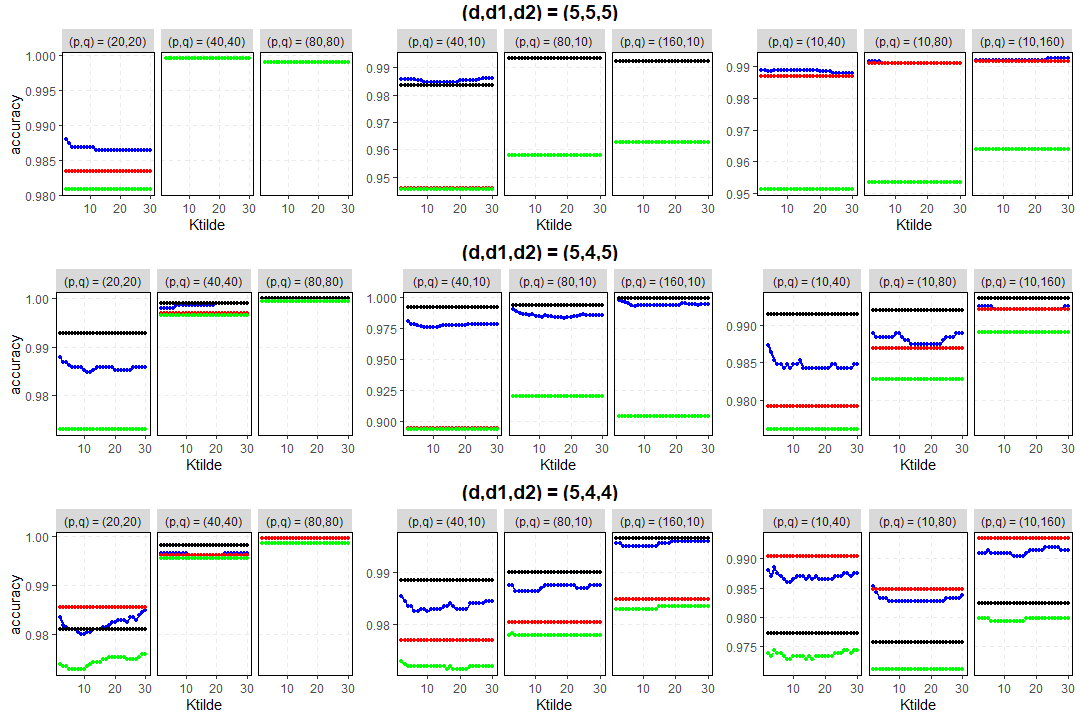}}
		\caption{\footnotesize{The lineplots for the relative frequency estimates of $\P( \hat{d} = d )$, $\P( \hat{d}_1 = d_1 )$, $ \P( \hat{d}_2 = d_2 )$ and $\P(\hat{d} = d, \hat{d}_1 = d_1, \hat{d}_2 = d_2)$ with respect to $\tilde{K} \in \{2,3,\ldots,30\}$ and $K = 20$ in Scenarios R1--R3  with $d=5$ based on 2000 repetitions. The legend is defined as follows:  \textup{(i)} the relative frequency estimate of $\P(\hat{d}_1 = d_1)$ ($\color{black}{-\bullet-}$); \textup{(ii)} the relative frequency estimate of $\P(\hat{d}_2 = d_2)$ ($\color{red}{-\bullet-}$); \textup{(iii)} the relative frequency estimate of $\P(\hat{d} = d)$ ($\color{blue}{-\bullet-}$); \textup{(iv)} the relative frequency estimate of $\P\{(\hat{d}_1,\hat{d}_2,\hat{d}) =(d_1, d_2,d) \}$ ($\color{green}{-\bullet-}$).}}
		\label{fig:Ktilde-rank}
	\end{figure}

	This section presents some additional results for the numerical studies considered in Sections \ref{sec:estimation} and \ref{section:simulation}. Figure \ref{fig: compare-W} reports the comparison between the two-stage approach introduced in Remark \ref{rek:two-stage} and our proposed method, in terms of both estimation error and computational cost.  The averages and standard deviations of the estimation errors $\varpi^2(\bA,\hat{\bA})$  and $\varpi^2(\bB,\hat{\bB})$ in Scenarios R1--R3 are, respectively,  reported in Tables \ref{table:varpi-case1}--\ref{table:varpi-case3}.  The averages of two-step ahead forecast RMSE are shown in Figure \ref{fig:fore2-case1}. Figures \ref{fig:K-estimation}--\ref{fig:Ktilde-rank} present sensitivity analysis with respect to the tuning parameters $K$ and $\tilde{K}$, including their impact on estimation errors and the accuracy of rank estimation.

	\begin{landscape}
		$ $\\
		$ $\\
		\begin{figure}[htbp]
			\centering
			\subfigure[The averages of estimation errors $\varpi^2(\bA,\hat{\bA})$ and $\varpi^2(\bB,\hat{\bB})$]{\includegraphics[width=23cm]{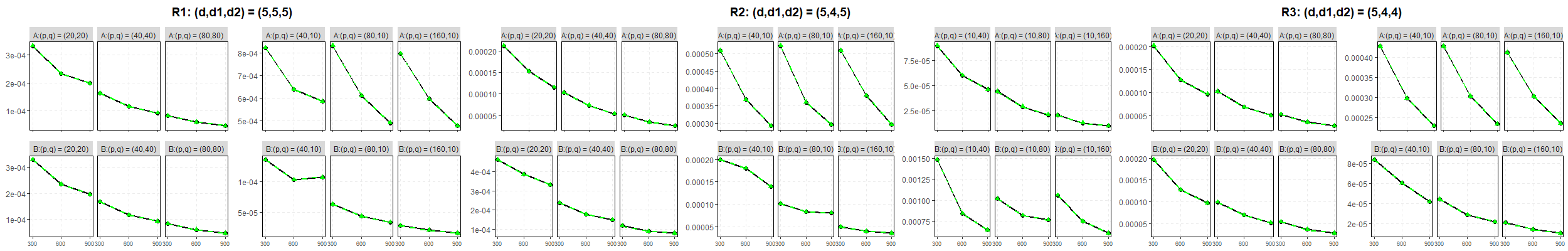}} \\
			\subfigure[The averages of CPU times (seconds)]{\includegraphics[width=23cm]{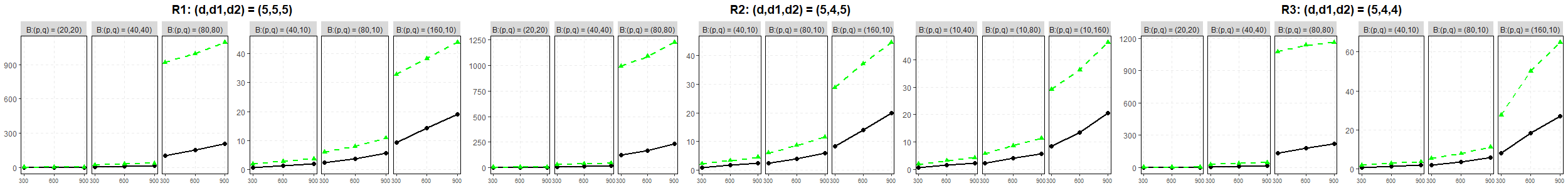}}
			\caption{The lineplots for the averages of estimation errors $\varpi^2(\bA,\hat{\bA})$ and  $\varpi^2(\bB,\hat{\bB})$,  and CPU times (seconds) based on 2000 repetitions in Scenarios R1--R3 with $d = 5$.  The black solid line represents solving $\bW$ with our proposed one-step approach, while the green dashed line represents solving $\bW$ with the alternative two-stage approach introduced in Remark \ref{rek:two-stage}. }
			\label{fig: compare-W}
		\end{figure}
	\end{landscape}

	% \begin{figure}[htbp]
		%   \centering
		%   \subfigure[The averages of estimation errors $\varpi^2(\bA,\hat{\bA})$ and $\varpi^2(\bB,\hat{\bB})$]{\includegraphics[width=0.8\textwidth]{plot/est-compare-W.png}} \\
		%   \subfigure[The averages of CPU times (seconds)]{\includegraphics[width=0.8\textwidth]{plot/time-compare-W.png}}
		%   \caption{The lineplots for the averages of estimation errors $\varpi^2(\bA,\hat{\bA})$ and $\varpi^2(\bB,\hat{\bB})$  and CPU times (seconds) based on 2000 repetitions with $(d,d_1,d_2) = (5,5,5)$.  The black solid line represents solving $\bW$ with our proposed method, while the green dashed line represents solving $\bW$ with the two-stage approach described in Remark \ref{rek:two-stage}. The data-generating process follows that described in Section \ref{sec:simulation setting up} of the main text with $(d,d_1,d_2) = (5,5,5)$ }
		%     \label{fig: compare-W}
		% \end{figure} 

	\begin{landscape}
		$ $\\
		$ $\\
		\begin{table}[htbp]
			
			\caption{
				The averages and standard deviations (in parentheses) of estimation errors   $\varpi^2(\bA,\hat{\bA})$ and $\varpi^2(\bB,\hat{\bB})$ based on 2000 repetitions in Scenario R1. All numbers reported below are multiplied by 100.
			}
			\renewcommand\tabcolsep{1.5pt}
			\label{table:varpi-case1}
			\resizebox{22.9cm}{!}{
				\begin{tabular}{c|c|cccccccccc|cccccccccc}
					\hline\hline
					\multirow{3}{*}{$d$} & \multirow{3}{*}{$n$} & \multicolumn{10}{c|}{$\varpi^2(\bA,\hat{\bA})$}                                                                                                                              & \multicolumn{10}{c}{$\varpi^2(\bB,\hat{\bB})$}                                                                                                                               \\ \cline{3-22} 
					&                      & \multicolumn{5}{c|}{$p = q$}                                                                    & \multicolumn{5}{c|}{$p > q$}                                                & \multicolumn{5}{c|}{$p = q$}                                                                    & \multicolumn{5}{c}{$p > q$}                                                 \\
					&                      & $(p,q)$                  & Proposed  & CP-refined & cPCA      & \multicolumn{1}{c|}{HOPE}       & $(p,q)$                   & Proposed  & CP-refined & cPCA      & HOPE       & $(p,q)$                  & Proposed  & CP-refined & cPCA      & \multicolumn{1}{c|}{HOPE}       & $(p,q)$                   & Proposed  & CP-refined & cPCA      & HOPE       \\ \hline
					\multirow{9}{*}{3}   & 300                  & \multirow{3}{*}{(20,20)} & 3.0(16.5) & 8.0(21.7)  & 5.1(21.2) & \multicolumn{1}{c|}{6.7(22.2)}  & \multirow{3}{*}{(40,10)}  & 3.3(17.5) & 8.3(21.3)  & 4.4(20.2) & 6.3(21.7)  & \multirow{3}{*}{(20,20)} & 3.0(16.6) & 8.1(21.9)  & 5.2(21.4) & \multicolumn{1}{c|}{6.9(22.6)}  & \multirow{3}{*}{(40,10)}  & 4.0(18.2) & 6.6(19.7)  & 4.1(19.0) & 6.0(20.6)  \\
					& 600                  &                          & 2.3(14.4) & 5.8(18.4)  & 3.5(17.7) & \multicolumn{1}{c|}{4.9(19.2)}  &                           & 2.9(16.4) & 6.6(19.5)  & 3.9(19.0) & 4.9(19.6)  &                          & 2.3(14.8) & 5.9(18.7)  & 3.6(18.1) & \multicolumn{1}{c|}{4.9(19.5)}  &                           & 3.7(17.2) & 5.0(17.5)  & 3.5(17.4) & 4.4(17.9)  \\
					& 900                  &                          & 2.0(13.5) & 4.5(15.6)  & 2.5(15.0) & \multicolumn{1}{c|}{3.6(16.1)}  &                           & 2.1(14.1) & 5.5(17.4)  & 2.9(16.4) & 4.1(17.8)  &                          & 2.0(13.2) & 4.5(15.6)  & 2.5(14.9) & \multicolumn{1}{c|}{3.5(15.9)}  &                           & 2.7(14.6) & 4.0(15.4)  & 2.6(15.0) & 3.4(15.6)  \\ \cline{2-22} 
					& 300                  & \multirow{3}{*}{(40,40)} & 1.2(10.8) & 5.1(16.1)  & 2.1(14.2) & \multicolumn{1}{c|}{2.8(14.9)}  & \multirow{3}{*}{(80,10)}  & 3.0(16.8) & 7.7(20.5)  & 3.9(19.2) & 5.1(20.1)  & \multirow{3}{*}{(40,40)} & 1.2(10.7) & 5.1(16.1)  & 2.2(14.3) & \multicolumn{1}{c|}{2.9(15.1)}  & \multirow{3}{*}{(80,10)}  & 4.0(18.2) & 5.7(18.4)  & 3.6(17.8) & 4.9(19.2)  \\
					& 600                  &                          & 0.5(6.7)  & 2.9(11.0)  & 1.0(9.5)  & \multicolumn{1}{c|}{1.4(10.2)}  &                           & 2.3(14.9) & 5.6(17.8)  & 2.9(16.8) & 3.8(17.7)  &                          & 0.4(6.4)  & 3.0(11.2)  & 1.0(9.8)  & \multicolumn{1}{c|}{1.4(9.8)}   &                           & 2.6(14.4) & 4.0(15.2)  & 2.6(14.8) & 3.2(15.2)  \\
					& 900                  &                          & 0.2(4.2)  & 2.1(9.0)   & 0.6(7.7)  & \multicolumn{1}{c|}{1.0(8.2)}   &                           & 1.3(11.3) & 4.3(15.2)  & 2.0(13.9) & 3.3(15.8)  &                          & 0.2(4.3)  & 2.1(9.2)   & 0.6(7.7)  & \multicolumn{1}{c|}{1.0(8.1)}   &                           & 2.2(13.4) & 2.9(13.1)  & 1.8(12.6) & 2.9(14.5)  \\ \cline{2-22} 
					& 300                  & \multirow{3}{*}{(80,80)} & 0.2(4.9)  & 2.9(11.2)  & 0.9(9.0)  & \multicolumn{1}{c|}{0.9(8.5)}   & \multirow{3}{*}{(160,10)} & 1.9(13.5) & 6.9(18.7)  & 3.1(17.1) & 4.0(18.1)  & \multirow{3}{*}{(80,80)} & 0.2(4.4)  & 2.9(11.3)  & 0.9(9.0)  & \multicolumn{1}{c|}{1.0(8.8)}   & \multirow{3}{*}{(160,10)} & 2.7(15.0) & 4.7(16.0)  & 2.7(15.4) & 3.5(16.3)  \\
					& 600                  &                          & 0.0(0.1)  & 2.1(9.1)   & 0.5(6.9)  & \multicolumn{1}{c|}{0.8(8.1)}   &                           & 1.0(9.7)  & 4.5(14.6)  & 1.6(12.5) & 2.1(13.2)  &                          & 0.0(0.1)  & 2.1(9.0)   & 0.5(6.9)  & \multicolumn{1}{c|}{0.9(8.4)}   &                           & 1.7(11.7) & 3.0(12.5)  & 1.5(11.6) & 2.0(12.2)  \\
					& 900                  &                          & 0.0(0.1)  & 1.9(9.1)   & 0.4(6.6)  & \multicolumn{1}{c|}{0.7(7.1)}   &                           & 0.9(9.2)  & 4.1(14.0)  & 1.5(12.0) & 2.0(12.9)  &                          & 0.0(0.1)  & 1.9(9.1)   & 0.4(6.5)  & \multicolumn{1}{c|}{0.6(7.0)}   &                           & 1.6(11.4) & 2.7(11.6)  & 1.3(10.6) & 1.8(11.5)  \\ \hline
					\multirow{9}{*}{5}   & 300                  & \multirow{3}{*}{(20,20)} & 2.1(14.1) & 15.3(23.9) & 5.4(21.7) & \multicolumn{1}{c|}{12.1(25.3)} & \multirow{3}{*}{(40,10)}  & 1.8(12.9) & 17.4(24.2) & 5.1(21.3) & 13.9(26.8) & \multirow{3}{*}{(20,20)} & 2.2(14.1) & 15.1(23.6) & 5.3(21.6) & \multicolumn{1}{c|}{12.4(25.7)} & \multirow{3}{*}{(40,10)}  & 3.7(16.9) & 12.7(21.8) & 4.5(19.4) & 12.6(24.3) \\
					& 600                  &                          & 1.6(12.1) & 11.5(21.1) & 3.7(18.0) & \multicolumn{1}{c|}{10.6(23.7)} &                           & 1.5(11.7) & 12.6(20.3) & 3.0(16.5) & 11.6(24.1) &                          & 1.6(12.1) & 11.5(20.9) & 3.6(17.8) & \multicolumn{1}{c|}{10.4(23.2)} &                           & 2.9(14.6) & 8.6(17.3)  & 2.6(14.6) & 10.5(22.0) \\
					& 900                  &                          & 1.6(12.1) & 10.3(20.3) & 3.7(18.2) & \multicolumn{1}{c|}{10.0(23.3)} &                           & 1.3(10.9) & 10.9(18.8) & 2.7(15.7) & 9.3(21.4)  &                          & 1.6(12.1) & 10.1(20.1) & 3.7(18.0) & \multicolumn{1}{c|}{10.0(23.1)} &                           & 2.3(12.7) & 7.3(15.6)  & 2.2(13.3) & 8.1(19.1)  \\ \cline{2-22} 
					& 300                  & \multirow{3}{*}{(40,40)} & 0.4(6.2)  & 10.8(18.8) & 1.8(12.8) & \multicolumn{1}{c|}{4.9(16.3)}  & \multirow{3}{*}{(80,10)}  & 1.1(9.8)  & 15.8(22.4) & 3.5(17.9) & 9.2(22.5)  & \multirow{3}{*}{(40,40)} & 0.4(6.2)  & 10.8(18.7) & 1.8(12.8) & \multicolumn{1}{c|}{4.8(16.1)}  & \multirow{3}{*}{(80,10)}  & 2.9(14.9) & 10.4(19.3) & 3.1(16.1) & 8.1(20.0)  \\
					& 600                  &                          & 0.5(7.1)  & 8.3(16.8)  & 1.2(10.6) & \multicolumn{1}{c|}{3.7(13.9)}  &                           & 0.8(8.6)  & 12.0(19.5) & 2.4(14.9) & 7.6(19.8)  &                          & 0.5(6.7)  & 8.2(16.8)  & 1.2(10.3) & \multicolumn{1}{c|}{3.7(14.2)}  &                           & 1.9(11.4) & 7.6(16.0)  & 1.9(12.6) & 6.7(18.1)  \\
					& 900                  &                          & 0.3(5.3)  & 6.0(13.9)  & 0.8(8.8)  & \multicolumn{1}{c|}{3.5(13.4)}  &                           & 0.6(7.4)  & 9.4(17.1)  & 1.5(12.0) & 6.1(17.6)  &                          & 0.3(5.4)  & 6.0(13.8)  & 0.8(8.9)  & \multicolumn{1}{c|}{3.3(13.2)}  &                           & 1.6(10.8) & 5.8(13.7)  & 1.2(10.1) & 5.4(15.6)  \\ \cline{2-22} 
					& 300                  & \multirow{3}{*}{(80,80)} & 0.1(3.1)  & 9.0(18.0)  & 1.7(12.2) & \multicolumn{1}{c|}{2.1(10.2)}  & \multirow{3}{*}{(160,10)} & 0.9(9.0)  & 14.6(21.0) & 2.6(15.6) & 7.2(19.8)  & \multirow{3}{*}{(80,80)} & 0.1(3.1)  & 9.1(18.2)  & 1.7(12.8) & \multicolumn{1}{c|}{2.4(11.2)}  & \multirow{3}{*}{(160,10)} & 2.0(12.0) & 9.0(17.0)  & 2.1(13.3) & 6.2(17.7)  \\
					& 600                  &                          & 0.1(2.2)  & 5.9(14.4)  & 0.8(8.7)  & \multicolumn{1}{c|}{1.8(10.1)}  &                           & 0.9(8.8)  & 10.8(18.5) & 1.9(13.5) & 5.7(18.1)  &                          & 0.1(2.2)  & 6.0(14.4)  & 0.9(9.0)  & \multicolumn{1}{c|}{1.9(10.6)}  &                           & 1.6(10.7) & 6.4(14.5)  & 1.5(11.1) & 4.7(15.5)  \\
					& 900                  &                          & 0.1(2.2)  & 5.1(12.7)  & 0.6(7.5)  & \multicolumn{1}{c|}{1.5(9.1)}   &                           & 0.4(5.5)  & 9.1(16.2)  & 1.3(11.3) & 4.2(14.5)  &                          & 0.1(2.2)  & 5.1(12.8)  & 0.6(7.8)  & \multicolumn{1}{c|}{1.5(9.1)}   &                           & 1.3(9.3)  & 5.0(11.9)  & 0.9(8.4)  & 3.5(12.4)  \\ \hline
					\multirow{9}{*}{7}   & 300                  & \multirow{3}{*}{(20,20)} & 1.6(12.1) & 24.1(24.8) & 5.4(21.3) & \multicolumn{1}{c|}{24.5(29.6)} & \multirow{3}{*}{(40,10)}  & 1.7(12.1) & 26.7(24.6) & 7.0(23.7) & 29.5(31.2) & \multirow{3}{*}{(20,20)} & 1.7(12.5) & 24.4(24.7) & 5.3(21.1) & \multicolumn{1}{c|}{23.7(29.8)} & \multirow{3}{*}{(40,10)}  & 4.1(16.6) & 18.6(21.9) & 5.0(19.2) & 25.4(27.5) \\
					& 600                  &                          & 0.9(8.6)  & 18.2(21.8) & 3.2(16.5) & \multicolumn{1}{c|}{18.5(26.0)} &                           & 1.1(9.9)  & 21.1(22.2) & 5.4(21.2) & 25.6(29.9) &                          & 1.0(9.6)  & 18.2(21.9) & 3.1(16.4) & \multicolumn{1}{c|}{18.9(26.6)} &                           & 3.8(16.3) & 14.2(18.9) & 3.6(15.9) & 22.4(26.5) \\
					& 900                  &                          & 0.8(8.4)  & 16.1(20.7) & 3.1(16.2) & \multicolumn{1}{c|}{18.6(26.9)} &                           & 1.3(10.8) & 17.9(20.9) & 5.0(20.4) & 23.7(29.0) &                          & 0.8(8.4)  & 15.8(20.4) & 3.1(16.2) & \multicolumn{1}{c|}{18.0(26.0)} &                           & 3.4(15.9) & 12.2(17.9) & 3.3(15.3) & 20.2(25.2) \\ \cline{2-22} 
					& 300                  & \multirow{3}{*}{(40,40)} & 0.2(3.8)  & 20.7(24.7) & 3.9(18.2) & \multicolumn{1}{c|}{10.6(21.4)} & \multirow{3}{*}{(80,10)}  & 1.4(11.0) & 26.0(24.6) & 7.2(24.5) & 23.6(30.4) & \multirow{3}{*}{(40,40)} & 0.2(3.8)  & 20.8(24.7) & 4.0(18.5) & \multicolumn{1}{c|}{10.9(21.9)} & \multirow{3}{*}{(80,10)}  & 3.8(15.8) & 17.3(21.0) & 4.6(18.0) & 20.4(26.3) \\
					& 600                  &                          & 0.1(3.2)  & 16.1(21.4) & 2.3(13.9) & \multicolumn{1}{c|}{9.0(20.0)}  &                           & 0.9(8.5)  & 19.6(21.9) & 4.6(20.1) & 18.5(26.9) &                          & 0.1(3.1)  & 16.1(21.6) & 2.4(14.4) & \multicolumn{1}{c|}{8.9(19.9)}  &                           & 2.9(14.0) & 12.7(18.4) & 2.9(14.3) & 15.4(23.8) \\
					& 900                  &                          & 0.1(3.5)  & 13.1(19.7) & 1.9(12.9) & \multicolumn{1}{c|}{7.6(18.8)}  &                           & 0.7(7.7)  & 17.5(20.6) & 4.2(19.3) & 17.1(26.2) &                          & 0.1(2.8)  & 13.1(19.6) & 1.8(12.9) & \multicolumn{1}{c|}{7.8(19.1)}  &                           & 2.8(13.6) & 10.7(16.1) & 2.5(12.7) & 14.5(22.2) \\ \cline{2-22} 
					& 300                  & \multirow{3}{*}{(80,80)} & 0.0(0.1)  & 19.2(26.1) & 4.7(20.1) & \multicolumn{1}{c|}{5.5(16.0)}  & \multirow{3}{*}{(160,10)} & 0.6(6.9)  & 25.1(23.8) & 6.5(23.6) & 14.9(24.2) & \multirow{3}{*}{(80,80)} & 0.0(0.1)  & 19.3(26.1) & 4.7(20.4) & \multicolumn{1}{c|}{5.2(15.5)}  & \multirow{3}{*}{(160,10)} & 2.7(13.0) & 15.9(20.0) & 3.7(15.3) & 13.3(21.3) \\
					& 600                  &                          & 0.1(2.2)  & 14.2(22.5) & 2.1(13.8) & \multicolumn{1}{c|}{4.3(14.7)}  &                           & 0.4(5.1)  & 18.4(21.4) & 4.3(19.5) & 13.2(23.4) &                          & 0.1(2.2)  & 14.1(22.4) & 2.2(14.1) & \multicolumn{1}{c|}{4.0(13.9)}  &                           & 2.7(13.2) & 10.8(16.7) & 2.5(12.7) & 11.3(19.8) \\
					& 900                  &                          & 0.0(0.1)  & 12.8(21.6) & 1.6(12.3) & \multicolumn{1}{c|}{4.0(13.8)}  &                           & 0.1(2.2)  & 15.8(19.5) & 3.3(17.3) & 12.1(23.4) &                          & 0.0(0.1)  & 12.7(21.5) & 1.7(12.4) & \multicolumn{1}{c|}{4.0(13.8)}  &                           & 1.9(10.8) & 9.1(15.2)  & 1.9(11.0) & 10.3(19.2) \\ \hline\hline
				\end{tabular}
				
			} 
			
		\end{table}    
	\end{landscape}

	\begin{landscape}
		$ $\\
		$ $\\
		\begin{table}[htbp]
			\caption{
				The averages and standard deviations (in parentheses) of estimation error   $\varpi^2(\bA,\hat{\bA})$ based on 2000 repetitions in Scenario R2. All numbers reported below are multiplied by 100.
			}
			\renewcommand\tabcolsep{1.5pt}
			\label{table:varpi-case2-a}
			\resizebox{22.9cm}{!}{
				\begin{tabular}{c|c|ccccccccccccccc}
					\hline\hline
					\multirow{3}{*}{$d$} & \multirow{3}{*}{$n$} & \multicolumn{15}{c}{$\varpi^2(\bA,\hat{\bA})$}                                                                                                                                                                                                                                      \\ \cline{3-17} 
					&                      & \multicolumn{5}{c|}{$p = q$}                                                                     & \multicolumn{5}{c|}{$p > q$}                                                                      & \multicolumn{5}{c}{$p < q$}                                                  \\
					&                      & $(p,q)$                  & Proposed  & CP-refined & cPCA       & \multicolumn{1}{c|}{HOPE}       & $(p,q)$                   & Proposed  & CP-refined & cPCA       & \multicolumn{1}{c|}{HOPE}       & $(p,q)$                   & Proposed  & CP-refined & cPCA       & HOPE       \\ \hline
					\multirow{9}{*}{3}   & 300                  & \multirow{3}{*}{(20,20)} & 4.6(16.5) & 10.9(23.4) & 20.4(25.6) & \multicolumn{1}{c|}{35.3(31.0)} & \multirow{3}{*}{(40,10)}  & 6.1(17.3) & 32.8(27.1) & 29.5(24.5) & \multicolumn{1}{c|}{35.0(28.3)} & \multirow{3}{*}{(40,10)}  & 3.9(14.7) & 8.4(20.3)  & 18.4(23.3) & 33.3(28.8) \\
					& 600                  &                          & 3.2(12.6) & 7.7(18.3)  & 18.3(22.5) & \multicolumn{1}{c|}{33.6(30.0)} &                           & 5.7(16.2) & 32.3(26.6) & 28.5(23.8) & \multicolumn{1}{c|}{33.4(27.8)} &                           & 3.2(13.1) & 6.7(17.8)  & 17.0(22.3) & 32.4(27.9) \\
					& 900                  &                          & 3.9(13.0) & 7.7(17.8)  & 18.7(22.6) & \multicolumn{1}{c|}{33.2(28.9)} &                           & 5.2(15.3) & 33.0(26.9) & 29.6(24.3) & \multicolumn{1}{c|}{34.4(27.2)} &                           & 2.5(10.7) & 5.9(16.7)  & 17.8(21.9) & 33.3(27.2) \\ \cline{2-17} 
					& 300                  & \multirow{3}{*}{(40,40)} & 2.1(9.9)  & 6.9(17.8)  & 18.9(22.9) & \multicolumn{1}{c|}{33.5(28.5)} & \multirow{3}{*}{(80,10)}  & 5.7(16.6) & 34.3(27.1) & 29.8(24.1) & \multicolumn{1}{c|}{35.6(28.1)} & \multirow{3}{*}{(80,10)}  & 2.7(12.2) & 7.3(19.4)  & 18.1(22.7) & 31.6(55.8) \\
					& 600                  &                          & 2.1(9.4)  & 5.1(14.3)  & 17.9(21.6) & \multicolumn{1}{c|}{31.4(26.4)} &                           & 5.6(16.5) & 32.9(26.8) & 29.1(23.5) & \multicolumn{1}{c|}{35.3(27.9)} &                           & 2.4(11.2) & 5.9(16.8)  & 17.3(21.7) & 31.4(27.0) \\
					& 900                  &                          & 2.0(9.2)  & 4.9(13.8)  & 17.3(20.6) & \multicolumn{1}{c|}{31.6(26.5)} &                           & 5.4(15.0) & 32.6(26.6) & 29.1(23.7) & \multicolumn{1}{c|}{34.4(27.0)} &                           & 2.3(10.3) & 5.0(14.7)  & 17.4(21.1) & 31.0(27.6) \\ \cline{2-17} 
					& 300                  & \multirow{3}{*}{(80,80)} & 1.1(6.5)  & 4.4(13.2)  & 17.4(21.4) & \multicolumn{1}{c|}{30.5(27.5)} & \multirow{3}{*}{(160,10)} & 5.3(16.3) & 34.2(27.0) & 30.6(24.2) & \multicolumn{1}{c|}{36.0(27.4)} & \multirow{3}{*}{(160,10)} & 2.4(11.6) & 6.7(19.0)  & 17.7(23.0) & 32.3(28.7) \\
					& 600                  &                          & 1.1(6.2)  & 3.8(12.4)  & 17.7(21.2) & \multicolumn{1}{c|}{27.9(24.6)} &                           & 5.1(15.0) & 33.4(26.9) & 29.8(24.0) & \multicolumn{1}{c|}{35.5(27.6)} &                           & 1.7(8.8)  & 4.7(15.2)  & 17.4(21.6) & 31.4(30.3) \\
					& 900                  &                          & 0.9(5.1)  & 3.0(10.5)  & 16.9(20.9) & \multicolumn{1}{c|}{28.7(27.3)} &                           & 5.0(14.6) & 32.1(26.4) & 29.3(23.5) & \multicolumn{1}{c|}{34.3(30.9)} &                           & 1.8(9.1)  & 4.7(15.0)  & 17.1(21.3) & 29.1(26.7) \\ \hline
					\multirow{9}{*}{5}   & 300                  & \multirow{3}{*}{(20,20)} & 1.3(10.1) & 15.3(23.9) & 22.4(30.1) & \multicolumn{1}{c|}{63.8(24.3)} & \multirow{3}{*}{(40,10)}  & 2.3(12.9) & 62.5(22.8) & 57.1(22.2) & \multicolumn{1}{c|}{67.3(22.6)} & \multirow{3}{*}{(40,10)}  & 1.9(12.5) & 12.8(22.5) & 20.6(27.6) & 60.1(23.9) \\
					& 600                  &                          & 0.9(8.2)  & 11.6(20.2) & 20.4(27.8) & \multicolumn{1}{c|}{62.9(23.5)} &                           & 2.0(11.7) & 61.8(23.5) & 57.4(22.9) & \multicolumn{1}{c|}{67.4(22.2)} &                           & 1.8(12.1) & 9.7(19.4)  & 20.0(27.0) & 59.9(23.9) \\
					& 900                  &                          & 1.6(11.2) & 12.4(23.0) & 21.7(29.3) & \multicolumn{1}{c|}{63.2(23.9)} &                           & 1.3(8.8)  & 62.5(23.2) & 57.3(22.6) & \multicolumn{1}{c|}{67.3(22.3)} &                           & 1.4(9.4)  & 7.8(16.6)  & 19.4(25.7) & 58.6(40.5) \\ \cline{2-17} 
					& 300                  & \multirow{3}{*}{(40,40)} & 0.5(5.6)  & 10.6(18.9) & 19.1(27.5) & \multicolumn{1}{c|}{62.8(24.1)} & \multirow{3}{*}{(80,10)}  & 1.6(10.6) & 65.0(22.7) & 59.8(22.6) & \multicolumn{1}{c|}{68.8(21.8)} & \multirow{3}{*}{(80,10)}  & 1.2(10.0) & 11.4(20.8) & 20.3(27.0) & 59.5(24.6) \\
					& 600                  &                          & 0.3(4.3)  & 8.5(16.9)  & 18.8(27.2) & \multicolumn{1}{c|}{62.3(24.7)} &                           & 1.5(10.2) & 63.0(22.6) & 59.1(21.9) & \multicolumn{1}{c|}{68.4(21.3)} &                           & 1.0(8.9)  & 7.6(16.6)  & 19.4(25.8) & 58.2(24.3) \\
					& 900                  &                          & 0.4(5.6)  & 6.9(15.1)  & 17.8(26.4) & \multicolumn{1}{c|}{60.7(24.5)} &                           & 1.2(8.9)  & 63.9(23.0) & 59.1(22.3) & \multicolumn{1}{c|}{69.3(21.4)} &                           & 0.9(8.3)  & 6.6(15.5)  & 19.0(25.5) & 57.5(24.1) \\ \cline{2-17} 
					& 300                  & \multirow{3}{*}{(80,80)} & 0.1(3.1)  & 9.3(17.7)  & 17.7(27.0) & \multicolumn{1}{c|}{60.2(24.8)} & \multirow{3}{*}{(160,10)} & 1.1(9.2)  & 65.1(22.7) & 60.3(22.0) & \multicolumn{1}{c|}{69.2(21.8)} & \multirow{3}{*}{(160,10)} & 1.2(9.9)  & 9.3(18.4)  & 20.0(26.3) & 58.1(24.4) \\
					& 600                  &                          & 0.1(2.0)  & 6.6(14.6)  & 17.5(26.5) & \multicolumn{1}{c|}{58.4(24.7)} &                           & 1.2(9.2)  & 64.2(23.3) & 59.7(22.6) & \multicolumn{1}{c|}{68.4(21.7)} &                           & 0.9(8.4)  & 6.8(15.8)  & 19.4(26.0) & 56.0(24.4) \\
					& 900                  &                          & 0.0(1.5)  & 5.0(12.3)  & 17.6(26.7) & \multicolumn{1}{c|}{56.9(24.7)} &                           & 0.9(7.7)  & 63.3(23.0) & 59.9(22.0) & \multicolumn{1}{c|}{67.5(21.4)} &                           & 0.5(6.1)  & 5.1(12.9)  & 18.2(25.4) & 55.8(23.2) \\ \hline
					\multirow{9}{*}{7}   & 300                  & \multirow{3}{*}{(20,20)} & 0.9(8.9)  & 23.1(24.1) & 17.8(30.3) & \multicolumn{1}{c|}{73.5(18.3)} & \multirow{3}{*}{(40,10)}  & 1.9(12.2) & 73.9(18.0) & 70.3(19.1) & \multicolumn{1}{c|}{79.1(16.1)} & \multirow{3}{*}{(40,10)}  & 1.4(10.9) & 19.1(22.4) & 18.0(28.0) & 68.5(22.2) \\
					& 600                  &                          & 0.4(5.9)  & 18.1(21.7) & 16.4(28.5) & \multicolumn{1}{c|}{73.7(18.5)} &                           & 1.6(10.7) & 73.4(18.1) & 69.7(19.5) & \multicolumn{1}{c|}{78.9(16.8)} &                           & 1.2(9.9)  & 15.2(20.7) & 16.1(26.7) & 68.2(19.6) \\
					& 900                  &                          & 0.6(6.9)  & 16.6(20.7) & 15.7(28.0) & \multicolumn{1}{c|}{73.6(18.4)} &                           & 1.7(11.4) & 74.8(17.9) & 70.9(19.2) & \multicolumn{1}{c|}{78.8(15.7)} &                           & 1.1(9.5)  & 12.9(18.4) & 14.0(25.0) & 67.8(19.1) \\ \cline{2-17} 
					& 300                  & \multirow{3}{*}{(40,40)} & 0.0(0.1)  & 20.3(23.4) & 15.9(28.9) & \multicolumn{1}{c|}{75.0(18.0)} & \multirow{3}{*}{(80,10)}  & 1.1(9.1)  & 75.7(17.7) & 72.1(19.2) & \multicolumn{1}{c|}{79.9(16.0)} & \multirow{3}{*}{(80,10)}  & 1.2(10.1) & 16.6(20.9) & 16.9(26.9) & 68.2(20.6) \\
					& 600                  &                          & 0.1(1.3)  & 14.8(19.5) & 13.6(26.5) & \multicolumn{1}{c|}{73.5(19.3)} &                           & 1.5(10.8) & 76.0(17.6) & 71.8(18.8) & \multicolumn{1}{c|}{80.7(15.1)} &                           & 1.1(9.6)  & 13.1(18.8) & 14.3(25.4) & 66.9(19.1) \\
					& 900                  &                          & 0.2(3.8)  & 13.4(19.3) & 12.5(26.2) & \multicolumn{1}{c|}{72.4(21.1)} &                           & 1.0(7.9)  & 76.7(17.5) & 72.4(19.0) & \multicolumn{1}{c|}{79.6(16.5)} &                           & 0.5(6.4)  & 10.5(16.4) & 13.8(24.4) & 65.8(19.3) \\ \cline{2-17} 
					& 300                  & \multirow{3}{*}{(80,80)} & 0.1(2.2)  & 18.0(23.3) & 15.2(28.7) & \multicolumn{1}{c|}{71.9(20.8)} & \multirow{3}{*}{(160,10)} & 1.0(8.1)  & 77.1(17.2) & 73.0(18.9) & \multicolumn{1}{c|}{80.7(15.5)} & \multirow{3}{*}{(160,10)} & 0.6(6.7)  & 15.7(20.0) & 17.0(26.9) & 66.1(20.2) \\
					& 600                  &                          & 0.1(2.2)  & 13.2(19.8) & 12.6(26.4) & \multicolumn{1}{c|}{71.6(20.3)} &                           & 1.2(9.1)  & 77.5(17.1) & 73.5(18.1) & \multicolumn{1}{c|}{80.6(15.6)} &                           & 0.5(5.9)  & 11.5(16.9) & 13.7(24.2) & 65.1(20.1) \\
					& 900                  &                          & 0.0(0.1)  & 11.1(18.5) & 12.7(26.4) & \multicolumn{1}{c|}{69.3(21.4)} &                           & 1.0(7.9)  & 77.1(17.2) & 73.6(17.7) & \multicolumn{1}{c|}{80.3(15.4)} &                           & 0.5(6.1)  & 9.4(15.4)  & 14.2(24.5) & 64.2(20.1) \\ \hline\hline
				\end{tabular}
			}
		\end{table}    
	\end{landscape}

	\begin{landscape}
		$ $\\
		$ $\\
		\begin{table}[htbp]
			\caption{
				The averages and standard deviations (in parentheses) of estimation error   $\varpi^2(\bB,\hat{\bB})$ based on 2000 repetitions in Scenario R2. All numbers reported below are multiplied by 100.
			}
			\renewcommand\tabcolsep{1.5pt}
			\label{table:varpi-case2-b}
			\resizebox{22.9cm}{!}{
				\begin{tabular}{c|c|ccccccccccccccc}
					\hline\hline
					\multirow{3}{*}{$d$} & \multirow{3}{*}{$n$} & \multicolumn{15}{c}{$\varpi^2(\bB,\hat{\bB})$}                                                                                                                                                                                                                                         \\ \cline{3-17} 
					&                      & \multicolumn{5}{c|}{$p = q$}                                                                      & \multicolumn{5}{c|}{$p > q$}                                                                       & \multicolumn{5}{c}{$p < q$}                                                   \\
					&                      & $(p,q)$                  & Proposed   & CP-refined & cPCA       & \multicolumn{1}{c|}{HOPE}       & $(p,q)$                   & Proposed   & CP-refined & cPCA       & \multicolumn{1}{c|}{HOPE}       & $(p,q)$                   & Proposed   & CP-refined & cPCA       & HOPE       \\ \hline
					\multirow{9}{*}{3}   & 300                  & \multirow{3}{*}{(20,20)} & 13.1(31.7) & 27.2(36.0) & 72.7(34.3) & \multicolumn{1}{c|}{70.0(24.9)} & \multirow{3}{*}{(40,10)}  & 18.0(34.4) & 80.0(18.7) & 79.8(19.3) & \multicolumn{1}{c|}{77.0(19.0)} & \multirow{3}{*}{(40,10)}  & 13.0(32.5) & 25.6(36.9) & 75.8(35.1) & 72.3(25.0) \\
					& 600                  &                          & 12.2(30.7) & 24.0(34.5) & 71.1(35.9) & \multicolumn{1}{c|}{68.3(24.8)} &                           & 17.9(33.7) & 79.9(18.7) & 80.1(19.1) & \multicolumn{1}{c|}{76.7(19.1)} &                           & 11.6(30.9) & 23.0(35.6) & 74.5(36.4) & 70.8(26.0) \\
					& 900                  &                          & 14.2(32.7) & 24.9(35.3) & 72.3(34.9) & \multicolumn{1}{c|}{70.0(24.8)} &                           & 16.4(32.6) & 80.2(18.4) & 80.1(19.3) & \multicolumn{1}{c|}{76.9(18.6)} &                           & 10.1(29.0) & 20.3(34.2) & 74.7(36.7) & 70.8(25.1) \\ \cline{2-17} 
					& 300                  & \multirow{3}{*}{(40,40)} & 8.7(27.4)  & 21.5(34.1) & 74.6(36.2) & \multicolumn{1}{c|}{70.1(28.9)} & \multirow{3}{*}{(80,10)}  & 17.2(33.4) & 78.9(19.4) & 79.3(19.7) & \multicolumn{1}{c|}{76.6(19.9)} & \multirow{3}{*}{(80,10)}  & 9.0(27.9)  & 21.6(35.6) & 79.2(34.9) & 71.5(25.7) \\
					& 600                  &                          & 8.8(27.4)  & 19.0(32.3) & 73.3(37.0) & \multicolumn{1}{c|}{70.5(35.7)} &                           & 16.7(32.7) & 78.9(18.9) & 78.7(19.6) & \multicolumn{1}{c|}{75.9(19.6)} &                           & 9.0(28.0)  & 19.7(34.7) & 76.8(37.4) & 71.7(25.1) \\
					& 900                  &                          & 8.4(26.9)  & 18.5(32.4) & 73.9(37.2) & \multicolumn{1}{c|}{70.9(23.8)} &                           & 17.7(33.2) & 79.6(18.7) & 79.9(19.2) & \multicolumn{1}{c|}{77.4(18.0)} &                           & 9.5(28.7)  & 18.5(34.2) & 77.6(37.1) & 71.2(49.2) \\ \cline{2-17} 
					& 300                  & \multirow{3}{*}{(80,80)} & 5.5(22.4)  & 15.8(30.3) & 74.5(38.1) & \multicolumn{1}{c|}{71.4(23.9)} & \multirow{3}{*}{(160,10)} & 15.1(31.7) & 79.4(19.1) & 79.4(19.6) & \multicolumn{1}{c|}{76.3(19.7)} & \multirow{3}{*}{(160,10)} & 8.0(26.6)  & 19.7(34.5) & 78.1(36.7) & 72.5(27.6) \\
					& 600                  &                          & 5.6(22.5)  & 14.3(29.1) & 75.6(38.0) & \multicolumn{1}{c|}{71.0(24.0)} &                           & 15.0(31.2) & 79.9(18.3) & 80.0(19.0) & \multicolumn{1}{c|}{76.1(18.9)} &                           & 8.1(26.9)  & 17.0(33.1) & 77.5(37.8) & 71.8(25.5) \\
					& 900                  &                          & 5.0(21.3)  & 13.1(28.0) & 73.7(39.4) & \multicolumn{1}{c|}{69.6(24.3)} &                           & 17.6(33.8) & 80.2(18.4) & 80.1(18.9) & \multicolumn{1}{c|}{77.7(18.5)} &                           & 8.7(27.8)  & 18.2(34.6) & 78.8(37.3) & 72.1(23.9) \\ \hline
					\multirow{9}{*}{5}   & 300                  & \multirow{3}{*}{(20,20)} & 2.3(13.9)  & 20.3(26.7) & 45.3(40.7) & \multicolumn{1}{c|}{76.5(18.7)} & \multirow{3}{*}{(40,10)}  & 5.9(20.3)  & 71.1(18.3) & 70.5(19.6) & \multicolumn{1}{c|}{71.4(17.4)} & \multirow{3}{*}{(40,10)}  & 2.5(15.1)  & 20.1(26.3) & 48.7(43.2) & 79.2(16.8) \\
					& 600                  &                          & 2.3(13.7)  & 16.3(23.6) & 42.1(40.5) & \multicolumn{1}{c|}{76.7(16.6)} &                           & 5.6(19.4)  & 72.5(18.0) & 71.3(19.1) & \multicolumn{1}{c|}{71.8(17.4)} &                           & 2.3(14.5)  & 16.0(23.9) & 46.2(43.7) & 79.2(17.1) \\
					& 900                  &                          & 2.8(15.4)  & 16.3(25.3) & 42.5(41.3) & \multicolumn{1}{c|}{76.8(17.2)} &                           & 5.4(18.6)  & 71.9(18.1) & 70.5(19.7) & \multicolumn{1}{c|}{72.4(16.6)} &                           & 1.7(12.5)  & 13.9(22.1) & 46.4(43.9) & 78.7(16.9) \\ \cline{2-17} 
					& 300                  & \multirow{3}{*}{(40,40)} & 1.0(9.4)   & 15.3(22.2) & 42.5(42.5) & \multicolumn{1}{c|}{78.9(17.4)} & \multirow{3}{*}{(80,10)}  & 5.2(19.0)  & 72.0(17.8) & 71.5(19.1) & \multicolumn{1}{c|}{71.0(17.7)} & \multirow{3}{*}{(80,10)}  & 1.7(12.5)  & 19.0(25.7) & 49.5(44.7) & 80.7(16.9) \\
					& 600                  &                          & 0.6(6.9)   & 12.2(20.0) & 40.9(43.1) & \multicolumn{1}{c|}{78.5(16.7)} &                           & 4.7(17.7)  & 72.4(17.9) & 71.2(19.3) & \multicolumn{1}{c|}{72.6(16.4)} &                           & 1.7(12.6)  & 14.1(22.4) & 47.7(45.5) & 79.5(16.6) \\
					& 900                  &                          & 0.8(8.7)   & 10.4(18.6) & 39.1(43.0) & \multicolumn{1}{c|}{78.1(16.4)} &                           & 4.6(17.4)  & 72.5(17.9) & 71.0(19.5) & \multicolumn{1}{c|}{72.5(17.3)} &                           & 1.0(9.3)   & 12.6(21.2) & 47.2(45.6) & 79.3(16.7) \\ \cline{2-17} 
					& 300                  & \multirow{3}{*}{(80,80)} & 0.3(4.9)   & 13.8(21.1) & 39.1(43.1) & \multicolumn{1}{c|}{78.8(16.5)} & \multirow{3}{*}{(160,10)} & 4.1(16.4)  & 71.9(18.2) & 71.2(19.5) & \multicolumn{1}{c|}{71.6(17.5)} & \multirow{3}{*}{(160,10)} & 1.4(11.2)  & 16.5(23.6) & 50.9(45.3) & 80.1(16.8) \\
					& 600                  &                          & 0.3(5.2)   & 10.0(17.7) & 38.1(44.2) & \multicolumn{1}{c|}{77.8(16.3)} &                           & 4.5(17.1)  & 72.3(18.2) & 71.2(19.4) & \multicolumn{1}{c|}{71.7(17.2)} &                           & 1.3(10.8)  & 13.2(21.7) & 47.5(46.3) & 79.8(16.3) \\
					& 900                  &                          & 0.2(3.7)   & 8.1(15.8)  & 37.5(44.3) & \multicolumn{1}{c|}{75.9(17.2)} &                           & 4.6(16.6)  & 72.1(17.8) & 70.8(19.3) & \multicolumn{1}{c|}{71.5(17.5)} &                           & 0.7(8.0)   & 10.2(18.5) & 45.4(46.5) & 78.8(16.1) \\ \hline
					\multirow{9}{*}{7}   & 300                  & \multirow{3}{*}{(20,20)} & 1.1(9.7)   & 25.5(24.5) & 29.2(36.7) & \multicolumn{1}{c|}{80.0(13.9)} & \multirow{3}{*}{(40,10)}  & 4.4(17.3)  & 67.3(17.5) & 66.3(18.9) & \multicolumn{1}{c|}{72.0(15.7)} & \multirow{3}{*}{(40,10)}  & 1.5(11.6)  & 27.3(25.1) & 36.1(41.4) & 82.7(13.6) \\
					& 600                  &                          & 0.7(7.6)   & 20.5(22.8) & 27.3(36.7) & \multicolumn{1}{c|}{80.0(13.3)} &                           & 4.5(16.9)  & 67.9(16.8) & 66.9(18.2) & \multicolumn{1}{c|}{72.4(16.2)} &                           & 1.0(9.4)   & 22.4(23.6) & 32.5(41.1) & 82.5(13.5) \\
					& 900                  &                          & 1.0(8.8)   & 18.3(21.5) & 26.8(36.7) & \multicolumn{1}{c|}{79.5(13.5)} &                           & 4.1(16.3)  & 68.3(16.9) & 66.2(18.7) & \multicolumn{1}{c|}{72.4(15.6)} &                           & 1.2(10.4)  & 20.6(22.6) & 28.9(39.7) & 82.3(13.1) \\ \cline{2-17} 
					& 300                  & \multirow{3}{*}{(40,40)} & 0.1(2.8)   & 23.1(24.2) & 28.1(38.5) & \multicolumn{1}{c|}{82.3(13.3)} & \multirow{3}{*}{(80,10)}  & 3.7(14.8)  & 67.6(17.5) & 65.8(19.1) & \multicolumn{1}{c|}{72.9(14.9)} & \multirow{3}{*}{(80,10)}  & 1.2(10.4)  & 26.0(24.5) & 36.2(42.7) & 83.7(13.8) \\
					& 600                  &                          & 0.2(3.8)   & 17.2(21.0) & 24.2(37.0) & \multicolumn{1}{c|}{81.7(13.7)} &                           & 4.2(16.3)  & 67.9(16.9) & 65.5(18.5) & \multicolumn{1}{c|}{72.1(16.7)} &                           & 1.0(9.4)   & 21.1(22.7) & 30.3(41.6) & 83.0(14.2) \\
					& 900                  &                          & 0.2(4.5)   & 15.8(20.4) & 21.8(36.3) & \multicolumn{1}{c|}{81.0(13.6)} &                           & 4.1(15.8)  & 68.7(16.7) & 66.4(18.5) & \multicolumn{1}{c|}{72.4(15.4)} &                           & 0.4(5.4)   & 18.0(21.3) & 29.7(41.6) & 82.7(14.1) \\ \cline{2-17} 
					& 300                  & \multirow{3}{*}{(80,80)} & 0.1(2.2)   & 21.4(24.5) & 26.6(38.7) & \multicolumn{1}{c|}{82.4(13.3)} & \multirow{3}{*}{(160,10)} & 3.7(14.5)  & 67.8(17.0) & 65.4(18.5) & \multicolumn{1}{c|}{71.7(15.4)} & \multirow{3}{*}{(160,10)} & 0.4(5.5)   & 25.8(24.3) & 36.0(43.7) & 83.7(14.2) \\
					& 600                  &                          & 0.1(2.2)   & 16.0(21.3) & 22.3(37.2) & \multicolumn{1}{c|}{81.3(13.8)} &                           & 4.2(16.6)  & 68.7(17.4) & 66.7(18.7) & \multicolumn{1}{c|}{72.9(15.7)} &                           & 0.4(5.9)   & 19.9(21.8) & 31.3(42.5) & 83.9(13.3) \\
					& 900                  &                          & 0.0(0.1)   & 13.6(19.9) & 22.1(37.6) & \multicolumn{1}{c|}{80.1(13.9)} &                           & 3.8(15.5)  & 69.0(17.0) & 66.5(18.9) & \multicolumn{1}{c|}{72.0(15.4)} &                           & 0.3(5.0)   & 16.8(20.7) & 31.1(43.3) & 82.7(13.9) \\ \hline\hline
				\end{tabular}
			}
		\end{table}    
	\end{landscape}

	\begin{landscape}
		$ $\\
		$ $\\
		\begin{table}[htbp]
			
			\caption{
				The averages and standard deviations (in parentheses) of estimation errors   $\varpi^2(\bA,\hat{\bA})$ and $\varpi^2(\bB,\hat{\bB})$ based on 2000 repetitions in Scenario R3. All numbers reported below are multiplied by 100.
			}
			\renewcommand\tabcolsep{1.5pt}
			\label{table:varpi-case3}
			\resizebox{22.9cm}{!}{
				\begin{tabular}{c|c|cccccccccc|cccccccccc}
					\hline \hline
					\multirow{3}{*}{$d$} & \multirow{3}{*}{$n$} & \multicolumn{10}{c|}{$\varpi^2(\bA,\hat{\bA})$}                                                                                                                                   & \multicolumn{10}{c}{$\varpi^2(\bB,\hat{\bB})$}                                                                                                                                    \\ \cline{3-22} 
					&                      & \multicolumn{5}{c|}{$p = q$}                                                                      & \multicolumn{5}{c|}{$p > q$}                                                  & \multicolumn{5}{c|}{$p = q$}                                                                      & \multicolumn{5}{c}{$p > q$}                                                   \\
					&                      & $(p,q)$                  & Proposed   & CP-refined & cPCA       & \multicolumn{1}{c|}{HOPE}       & $(p,q)$                   & Proposed   & CP-refined & cPCA       & HOPE       & $(p,q)$                  & Proposed   & CP-refined & cPCA       & \multicolumn{1}{c|}{HOPE}       & $(p,q)$                   & Proposed   & CP-refined & cPCA       & HOPE       \\ \hline
					\multirow{9}{*}{3}   & 300                  & \multirow{3}{*}{(20,20)} & 26.1(25.8) & 36.6(29.4) & 42.0(31.9) & \multicolumn{1}{c|}{40.7(29.6)} & \multirow{3}{*}{(40,10)}  & 26.2(26.1) & 36.2(29.5) & 41.6(32.0) & 39.5(31.0) & \multirow{3}{*}{(20,20)} & 24.3(24.9) & 36.6(28.8) & 42.3(31.7) & \multicolumn{1}{c|}{40.4(29.7)} & \multirow{3}{*}{(40,10)}  & 26.1(25.9) & 35.8(28.9) & 39.8(31.2) & 39.3(29.7) \\
					& 600                  &                          & 24.7(24.8) & 35.6(28.8) & 40.0(31.2) & \multicolumn{1}{c|}{39.0(29.4)} &                           & 24.6(25.2) & 35.2(28.7) & 41.3(31.5) & 39.5(29.4) &                          & 24.0(24.9) & 34.3(28.4) & 39.9(31.5) & \multicolumn{1}{c|}{37.8(32.1)} &                           & 23.6(24.4) & 35.0(28.7) & 39.3(31.0) & 37.8(32.4) \\
					& 900                  &                          & 23.2(23.7) & 34.7(28.5) & 40.5(31.5) & \multicolumn{1}{c|}{38.3(29.2)} &                           & 23.8(25.0) & 36.4(29.3) & 42.9(32.1) & 38.9(34.3) &                          & 22.5(23.8) & 34.3(28.2) & 39.7(31.1) & \multicolumn{1}{c|}{37.7(29.4)} &                           & 23.8(24.3) & 35.2(29.2) & 39.2(31.5) & 38.0(29.5) \\ \cline{2-22} 
					& 300                  & \multirow{3}{*}{(40,40)} & 24.9(24.5) & 37.7(29.6) & 42.7(31.7) & \multicolumn{1}{c|}{41.0(29.6)} & \multirow{3}{*}{(80,10)}  & 25.5(25.7) & 36.6(29.2) & 43.0(32.1) & 39.2(54.6) & \multirow{3}{*}{(40,40)} & 24.6(25.0) & 35.5(28.8) & 40.9(31.7) & \multicolumn{1}{c|}{38.6(29.5)} & \multirow{3}{*}{(80,10)}  & 24.8(25.9) & 35.4(29.7) & 39.4(31.6) & 38.1(30.4) \\
					& 600                  &                          & 23.4(23.4) & 34.6(28.3) & 40.0(31.1) & \multicolumn{1}{c|}{38.2(29.1)} &                           & 24.1(24.7) & 35.3(28.6) & 41.6(31.7) & 39.2(29.1) &                          & 23.4(24.1) & 34.0(27.8) & 39.1(30.5) & \multicolumn{1}{c|}{37.9(29.1)} &                           & 23.6(24.4) & 34.8(28.7) & 39.0(31.4) & 37.2(30.1) \\
					& 900                  &                          & 22.8(24.0) & 34.1(28.0) & 40.1(31.0) & \multicolumn{1}{c|}{38.3(30.7)} &                           & 24.1(24.7) & 34.8(28.4) & 41.0(31.2) & 38.8(31.0) &                          & 22.5(23.6) & 33.2(28.0) & 38.7(31.1) & \multicolumn{1}{c|}{37.6(29.6)} &                           & 23.2(24.3) & 33.7(28.7) & 37.8(31.2) & 36.7(29.3) \\ \cline{2-22} 
					& 300                  & \multirow{3}{*}{(80,80)} & 24.0(24.4) & 35.8(28.6) & 41.5(31.4) & \multicolumn{1}{c|}{40.1(29.6)} & \multirow{3}{*}{(160,10)} & 26.5(26.1) & 36.7(29.0) & 42.4(31.6) & 39.9(29.4) & \multirow{3}{*}{(80,80)} & 24.1(24.3) & 36.4(28.9) & 41.4(31.4) & \multicolumn{1}{c|}{39.6(29.9)} & \multirow{3}{*}{(160,10)} & 25.1(25.3) & 35.5(28.9) & 39.5(31.0) & 38.3(30.0) \\
					& 600                  &                          & 21.7(22.8) & 33.7(28.3) & 38.5(30.8) & \multicolumn{1}{c|}{38.4(29.7)} &                           & 24.9(25.1) & 35.4(28.5) & 41.4(31.5) & 39.2(29.7) &                          & 22.9(23.5) & 34.0(27.7) & 39.3(30.8) & \multicolumn{1}{c|}{37.9(28.6)} &                           & 24.5(24.7) & 34.4(28.7) & 39.3(31.3) & 37.5(30.2) \\
					& 900                  &                          & 22.1(23.2) & 35.8(29.2) & 41.6(31.9) & \multicolumn{1}{c|}{40.0(29.3)} &                           & 22.9(23.9) & 34.7(28.4) & 41.4(31.8) & 39.4(30.1) &                          & 22.0(23.5) & 34.6(27.9) & 40.2(31.2) & \multicolumn{1}{c|}{38.9(28.9)} &                           & 23.2(23.4) & 34.8(28.8) & 39.4(31.3) & 37.6(29.1) \\ \hline
					\multirow{9}{*}{5}   & 300                  & \multirow{3}{*}{(20,20)} & 1.6(12.1)  & 62.8(23.5) & 65.8(25.4) & \multicolumn{1}{c|}{69.4(21.5)} & \multirow{3}{*}{(40,10)}  & 1.9(12.6)  & 62.7(23.2) & 68.0(24.4) & 69.2(22.1) & \multirow{3}{*}{(20,20)} & 1.7(12.0)  & 63.1(23.1) & 67.1(24.0) & \multicolumn{1}{c|}{69.6(21.4)} & \multirow{3}{*}{(40,10)}  & 2.5(14.3)  & 58.2(23.1) & 60.7(24.8) & 66.5(21.1) \\
					& 600                  &                          & 0.9(8.2)   & 61.1(23.4) & 65.2(24.6) & \multicolumn{1}{c|}{68.8(21.8)} &                           & 1.4(10.6)  & 61.7(23.5) & 67.5(24.6) & 69.5(21.5) &                          & 1.0(9.5)   & 61.1(23.5) & 64.9(25.1) & \multicolumn{1}{c|}{68.5(21.7)} &                           & 1.9(12.6)  & 59.4(23.3) & 61.5(24.1) & 66.0(21.8) \\
					& 900                  &                          & 1.2(10.2)  & 62.6(22.8) & 65.7(24.4) & \multicolumn{1}{c|}{69.8(21.3)} &                           & 1.2(9.7)   & 62.4(23.1) & 68.0(24.8) & 69.6(21.2) &                          & 1.1(10.1)  & 62.7(22.8) & 66.7(24.1) & \multicolumn{1}{c|}{69.8(21.2)} &                           & 1.7(11.5)  & 57.9(23.3) & 60.7(24.9) & 65.2(22.1) \\ \cline{2-22} 
					& 300                  & \multirow{3}{*}{(40,40)} & 0.5(7.1)   & 63.7(23.0) & 67.2(25.0) & \multicolumn{1}{c|}{70.4(21.0)} & \multirow{3}{*}{(80,10)}  & 1.2(9.9)   & 64.8(22.9) & 69.1(24.1) & 70.6(21.5) & \multirow{3}{*}{(40,40)} & 0.4(6.1)   & 63.8(23.1) & 67.9(24.6) & \multicolumn{1}{c|}{69.9(21.4)} & \multirow{3}{*}{(80,10)}  & 1.7(11.9)  & 59.0(23.1) & 61.6(24.6) & 66.1(21.7) \\
					& 600                  &                          & 0.3(5.6)   & 63.8(23.6) & 67.3(24.6) & \multicolumn{1}{c|}{70.0(21.3)} &                           & 0.8(8.0)   & 62.5(22.8) & 67.2(24.6) & 70.3(21.0) &                          & 0.2(4.3)   & 63.4(23.4) & 67.9(24.9) & \multicolumn{1}{c|}{70.4(21.8)} &                           & 1.5(10.9)  & 59.0(23.0) & 61.3(24.3) & 65.2(21.8) \\
					& 900                  &                          & 0.3(5.6)   & 63.9(23.1) & 68.3(24.2) & \multicolumn{1}{c|}{69.3(21.3)} &                           & 0.9(8.4)   & 63.7(23.2) & 69.9(24.5) & 71.3(20.6) &                          & 0.4(5.6)   & 63.3(23.3) & 67.5(24.4) & \multicolumn{1}{c|}{69.7(21.2)} &                           & 1.3(10.2)  & 58.9(23.3) & 61.7(24.5) & 66.4(21.5) \\ \cline{2-22} 
					& 300                  & \multirow{3}{*}{(80,80)} & 0.2(4.4)   & 65.2(23.0) & 68.2(24.9) & \multicolumn{1}{c|}{70.9(20.3)} & \multirow{3}{*}{(160,10)} & 0.9(8.7)   & 65.0(22.8) & 69.7(23.7) & 71.2(20.8) & \multirow{3}{*}{(80,80)} & 0.2(3.6)   & 65.0(23.2) & 68.2(24.4) & \multicolumn{1}{c|}{70.3(21.1)} & \multirow{3}{*}{(160,10)} & 1.3(10.5)  & 58.9(23.4) & 61.5(24.5) & 65.6(21.6) \\
					& 600                  &                          & 0.0(0.1)   & 65.6(23.0) & 69.1(24.5) & \multicolumn{1}{c|}{70.9(21.3)} &                           & 0.6(6.9)   & 64.0(23.5) & 69.2(24.6) & 70.7(21.6) &                          & 0.0(0.1)   & 65.7(22.8) & 68.6(24.3) & \multicolumn{1}{c|}{70.8(21.2)} &                           & 1.4(10.4)  & 58.8(23.7) & 61.4(24.8) & 66.0(21.5) \\
					& 900                  &                          & 0.0(0.1)   & 65.3(22.6) & 68.2(24.1) & \multicolumn{1}{c|}{70.4(21.2)} &                           & 0.8(8.4)   & 62.9(23.3) & 69.0(24.6) & 70.3(20.8) &                          & 0.0(2.0)   & 65.5(22.9) & 68.8(24.2) & \multicolumn{1}{c|}{70.8(21.7)} &                           & 1.7(11.7)  & 58.7(22.8) & 61.4(24.1) & 65.3(21.9) \\ \hline
					\multirow{9}{*}{7}   & 300                  & \multirow{3}{*}{(20,20)} & 0.9(9.0)   & 71.9(17.6) & 72.1(19.4) & \multicolumn{1}{c|}{77.6(15.9)} & \multirow{3}{*}{(40,10)}  & 0.8(8.2)   & 73.8(17.9) & 75.1(19.4) & 79.4(15.6) & \multirow{3}{*}{(20,20)} & 0.7(8.2)   & 71.1(17.5) & 71.2(19.7) & \multicolumn{1}{c|}{77.3(15.8)} & \multirow{3}{*}{(40,10)}  & 1.2(10.0)  & 64.0(18.8) & 63.7(20.3) & 71.9(17.3) \\
					& 600                  &                          & 0.2(4.4)   & 70.5(18.5) & 70.8(20.1) & \multicolumn{1}{c|}{77.9(15.4)} &                           & 0.7(7.4)   & 73.3(18.1) & 74.4(19.6) & 79.5(15.8) &                          & 0.3(4.7)   & 71.4(18.1) & 71.3(19.9) & \multicolumn{1}{c|}{78.1(15.3)} &                           & 1.0(8.9)   & 63.9(18.5) & 63.8(20.1) & 71.7(16.4) \\
					& 900                  &                          & 0.5(6.8)   & 71.1(18.1) & 71.1(19.8) & \multicolumn{1}{c|}{78.1(15.0)} &                           & 1.0(9.7)   & 74.5(18.0) & 74.4(19.5) & 80.1(14.9) &                          & 0.4(6.3)   & 71.3(17.9) & 71.6(19.5) & \multicolumn{1}{c|}{77.7(15.7)} &                           & 1.3(10.3)  & 64.4(18.6) & 62.9(20.6) & 72.1(16.6) \\ \cline{2-22} 
					& 300                  & \multirow{3}{*}{(40,40)} & 0.0(0.1)   & 76.0(17.1) & 75.7(19.0) & \multicolumn{1}{c|}{79.4(15.3)} & \multirow{3}{*}{(80,10)}  & 0.6(6.9)   & 75.5(18.0) & 76.6(19.4) & 81.1(15.4) & \multirow{3}{*}{(40,40)} & 0.0(0.1)   & 75.1(17.5) & 75.9(19.0) & \multicolumn{1}{c|}{79.5(15.4)} & \multirow{3}{*}{(80,10)}  & 1.0(9.0)   & 64.1(18.9) & 63.8(20.7) & 72.9(16.3) \\
					& 600                  &                          & 0.1(3.1)   & 75.3(17.5) & 76.0(18.8) & \multicolumn{1}{c|}{80.1(15.2)} &                           & 0.7(7.8)   & 75.9(17.8) & 76.2(19.4) & 81.2(14.9) &                          & 0.1(2.0)   & 75.6(17.4) & 75.5(19.2) & \multicolumn{1}{c|}{80.2(15.2)} &                           & 1.0(8.9)   & 63.9(18.6) & 63.1(20.0) & 73.0(16.1) \\
					& 900                  &                          & 0.0(1.5)   & 75.6(17.4) & 75.6(19.1) & \multicolumn{1}{c|}{80.1(15.2)} &                           & 0.4(5.2)   & 76.5(17.5) & 76.6(19.5) & 80.9(15.3) &                          & 0.0(0.1)   & 75.3(17.3) & 75.3(19.3) & \multicolumn{1}{c|}{80.5(15.1)} &                           & 0.7(7.2)   & 65.0(18.1) & 63.7(19.9) & 73.3(15.7) \\ \cline{2-22} 
					& 300                  & \multirow{3}{*}{(80,80)} & 0.0(0.1)   & 77.9(17.1) & 77.4(19.6) & \multicolumn{1}{c|}{80.9(15.3)} & \multirow{3}{*}{(160,10)} & 0.3(4.3)   & 76.8(17.1) & 76.3(19.9) & 80.5(15.1) & \multirow{3}{*}{(80,80)} & 0.0(0.1)   & 77.5(17.4) & 77.2(19.3) & \multicolumn{1}{c|}{81.3(15.2)} & \multirow{3}{*}{(160,10)} & 0.5(5.8)   & 64.2(18.6) & 63.6(19.9) & 71.5(16.8) \\
					& 600                  &                          & 0.0(0.1)   & 78.1(17.1) & 77.9(19.5) & \multicolumn{1}{c|}{80.5(15.8)} &                           & 0.3(4.4)   & 77.3(17.4) & 77.5(19.1) & 81.1(15.7) &                          & 0.0(0.1)   & 77.9(16.8) & 77.7(19.2) & \multicolumn{1}{c|}{80.7(15.5)} &                           & 1.0(8.7)   & 65.1(18.9) & 64.1(20.4) & 73.2(15.9) \\
					& 900                  &                          & 0.0(0.1)   & 77.8(16.9) & 78.2(19.1) & \multicolumn{1}{c|}{80.3(15.4)} &                           & 0.2(3.0)   & 76.8(17.4) & 77.5(18.6) & 81.3(14.5) &                          & 0.0(0.1)   & 77.6(17.3) & 78.6(19.0) & \multicolumn{1}{c|}{80.9(15.8)} &                           & 0.6(5.8)   & 65.0(18.5) & 63.4(20.1) & 72.5(16.4) \\ \hline \hline
				\end{tabular}
				
			} 
			
		\end{table}    
	\end{landscape}

	\begin{landscape}
		$ $\\
		$ $\\
		\begin{figure}[htbp]
			\centering
			\centerline{\includegraphics[width= 23cm]{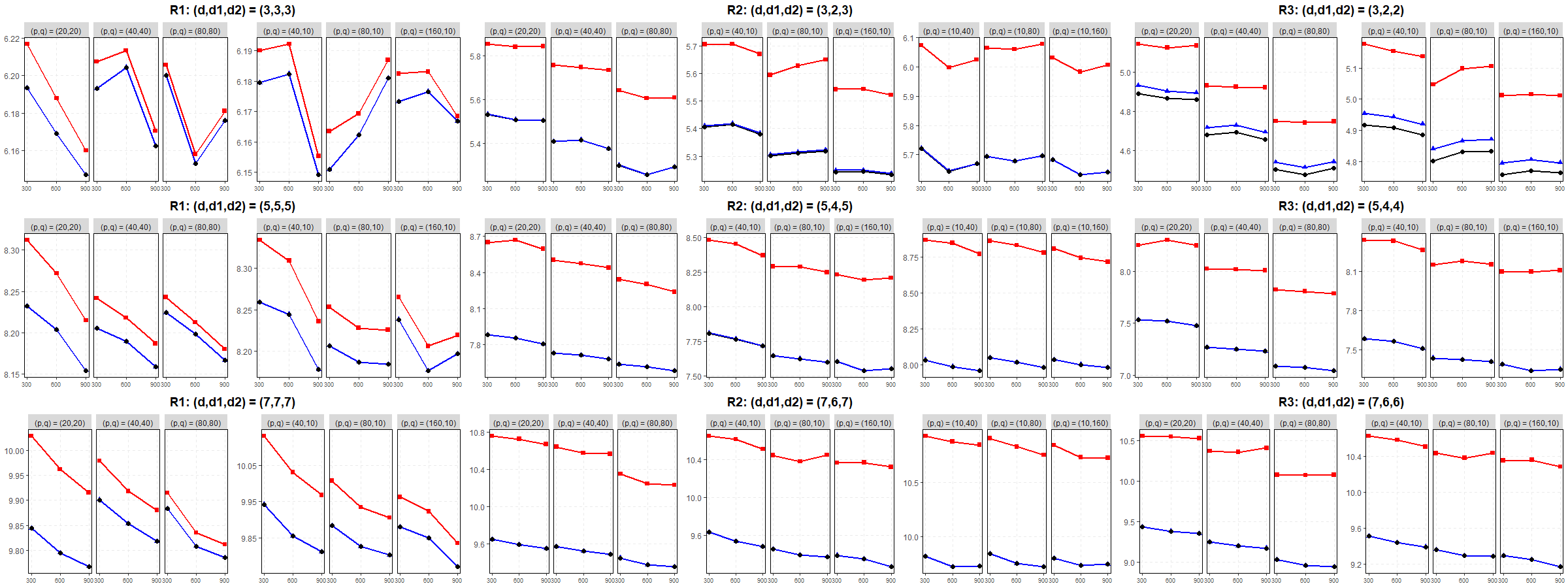}}
			\caption{The lineplots for the averages of two-step ahead forecast RMSE based on 2000 repetitions in Scenarios R1--R3. The legend is defined as follows: \textup{(i)} our proposed prediction method ($\color{black}{-\bullet-}$), \textup{(ii)} the prediction method introduced in \cite{chang2023modelling} with $(\hat{\bA},\hat{\bB})$ selected as our proposed  estimate ($\color{blue}{-\blacktriangle-}$), \textup{(iii)} the prediction method introduced in \cite{chang2023modelling} with $(\hat{\bA},\hat{\bB})$ selected as the CP-refined estimate ($\color{red}{-\blacksquare-}$).}
			\label{fig:fore2-case1}
		\end{figure}
	\end{landscape}

	\bibliographystyleS{jasa}
	\spacingset{0.95}\selectfont
	\bibliographyS{mybibfile-app}

\end{document}